\newif\iffigs\figstrue
\documentclass[10pt,a4paper]{article}
\usepackage{latexsym,amssymb,lscape,graphics,setspace}
\usepackage{amsmath}
\usepackage{graphicx}        
\usepackage{longtable}
\usepackage{multirow}
\usepackage{booktabs}
\usepackage{multirow}
\usepackage{color}
\usepackage{slashed,epsfig}
\usepackage{amsfonts}
\usepackage{cite}
\usepackage{verbatim}
\usepackage{colortbl}
\usepackage[table]{xcolor}
\usepackage{hyperref}       
\usepackage{array}
\usepackage{fancyhdr}
\usepackage{tikz}
\usepackage{multicol}
\usepackage{lscape}
\graphicspath{{images/}}

\newcommand{\mathsym}[1]{{}}

\newtheorem{definizione}{Definition}[section]
\newtheorem{teorema}{Theorem}[section]

\newtheorem{lemma}{Lemma}[section]
\newtheorem{proof}{Proof}[section]

\newcommand{\bd}{\begin{definizione}}
\newcommand{\ed}{\end{definizione}}

\def\IC{\relax\,\hbox{$\inbar\kern-.3em{\rm C}$}}
\def\IG{\relax\,\hbox{$\inbar\kern-.3em{\rm G}$}}
\def\IB{\relax{\rm I\kern-.18em B}}
\def\ID{\relax{\rm I\kern-.18em D}}
\def\IL{\relax{\rm I\kern-.18em L}}
\def\IF{\relax{\rm I\kern-.18em F}}
\def\IH{\relax{\rm I\kern-.18em H}}
\def\II{\relax{\rm I\kern-.17em I}}
\def\IN{\relax{\rm I\kern-.18em N}}
\def\IP{\relax{\rm I\kern-.18em P}}
\def\IQ{\relax\,\hbox{$\inbar\kern-.3em{\rm Q}$}}
\def\bfzero{\relax\,\hbox{$\inbar\kern-.3em{\rm 0}$}}
\def\IK{\relax{\rm I\kern-.18em K}}
\def\IG{\relax\,\hbox{$\inbar\kern-.3em{\rm G}$}}
 \font\cmss=cmss10 \font\cmsss=cmss10 at 7pt
\def\IR{\relax{\rm I\kern-.18em R}}
\def\ZZ{\relax\ifmmode\mathchoice
{\hbox{\cmss Z\kern-.4em Z}}{\hbox{\cmss Z\kern-.4em Z}}
{\lower.9pt\hbox{\cmsss Z\kern-.4em Z}} {\lower1.2pt\hbox{\cmsss
Z\kern-.4em Z}}\else{\cmss Z\kern-.4em Z}\fi}
\def\bfone{\relax{\rm 1\kern-.35em 1}}

\def\Solv{\mathop{\rm Solv}\nolimits}
\def\Riem{\mathop{\rm Riem}\nolimits}
\def\inbar{\vrule height1.5ex width.4pt depth0pt}
\def\bfzero{\relax{\rm I\kern-.18em 0}}
\def\bfone{\relax{\rm 1\kern-.35em 1}}

\DeclareFontFamily{U}{rsf}{} \DeclareFontShape{U}{rsf}{m}{n}{
  <5> <6> rsfs5 <7> <8> <9> rsfs7 <10-> rsfs10}{}
\DeclareMathAlphabet\Scr{U}{rsf}{m}{n}

\newcommand{\Sp}{\mathop{\rm {}Sp}}

\newcommand{\ft}[2]{{\textstyle\frac{#1}{#2}}}
\def\tilde{\widetilde}

\def\1bar{1\hskip -.275cm -}
\def\2bar{2\hskip -.275cm -}
\def\3bar{3\hskip -.275cm -}

\newsavebox{\uuunit}
\sbox{\uuunit}
{\setlength{\unitlength}{0.825em}
\begin{picture}(0.6,0.7)
\thinlines
\put(0,0){\line(1,0){0.5}}
\put(0.15,0){\line(0,1){0.7}}
\put(0.35,0){\line(0,1){0.8}}
\multiput(0.3,0.8)(-0.04,-0.02){10}{\rule{0.5pt}{0.5pt}}
\end {picture}}

\makeatletter \@addtoreset{equation}{section} \makeatother

\def\bfone{\relax{\rm 1\kern-.35em 1}}

\def\bfone{\relax{\rm 1\kern-.35em 1}}
\font\cmss=cmss10 \font\cmsss=cmss10 at 7pt

\newcommand{\so}{\mathfrak{so}}
\newcommand{\su}{\mathfrak{su}}

\newcommand{\uu}{\mathfrak{u}}
\newcommand{\sym}{\mathfrak{sp}}
\newcommand{\slal}{\mathfrak{sl}}

\begin{document}
\begin{titlepage}
\begin{center}
\vskip 0.2cm
{{\large {\sc   Thermodynamics {\`a} la Souriau on K\"ahler Non Compact Symmetric Spaces for  Cartan Neural Networks
  ${}^\dagger$}} }\\
 \vskip 1cm {\sc Pietro Fr\'e\,$^{a,b}$,  \\ Alexander S. Sorin\,$^{c}$
 and Mario Trigiante\,$^{d, e}$} \vskip 0.5cm
\smallskip
{\sl \small \frenchspacing ${}^a\,$ {\tt Emeritus Professor of}  Dipartimento di Fisica, 
Universit\`a di Torino, Via P. Giuria 1, I-10125 Torino, Italy \\[2pt]
${}^{b}\,${\tt Senior Consultant of } Additati\&Partners Consulting s.r.l, 
Via Filippo Pacini 36, I-51100 Pistoia, Italy \\[2pt]
${}^c\,$ Sasha affiliation, Israel \\[2pt]
${}^d\,$Dipartimento DISAT, Politecnico di Torino,
C.so Duca degli Abruzzi 24, I-10129 Torino, Italy\\[2pt]
${}^e\,$INFN, Sezione di Torino\\[2pt]

E-mail:  {\tt pietro.fre@unito.it, asorin@tauex.tau.ac.il,\\
 mario.trigiante@polito.it, } }

\begin{abstract}
In this paper, we clarify several issues concerning the abstract geometrical formulation of thermodynamics on non compact symmetric spaces $\mathrm{U/H}$ that are the mathematical model of hidden layers in the new paradigm of Cartan Neural Networks. We introduce a clearcut distinction between the generalized thermodynamics associated with Integrable Dynamical Systems and the challenging proposal of Gibbs probability distributions on $\mathrm{U/H}$ provided by generalized thermodynamics {\`a} la Souriau. Our main result is the proof that $\mathrm{U/H}$.s supporting such Gibbs distributions are only the K\"ahler ones. Furthermore, for the latter, we solve the problem of determining the space of temperatures, namely of Lie algebra elements for which the partition function converges. The space of generalized temperatures is the orbit under the adjoint action of $\mathrm{U}$ of a positivity domain in the Cartan subalgebra $C_c\subset\mathbb{H}$ of the maximal compact subalgebra $\mathbb{H}\subset\mathbb{U}$. We illustrate how our explicit constructions for the Poincar\'e and Siegel planes might be extended to the whole class of Calabi-Vesentini manifolds utilizing Paint Group symmetry.
Furthermore we claim that  Rao's, Chentsov's, Amari's Information Geometry and the thermodynamical geometry of Ruppeiner and Lychagin are the very same thing. In particular we provide an explicit study of thermodynamical geometry for the case of the Poincar\'e plane. The most important property of the Gibbs probability distributions provided by  the here introduced setup is their covariance with respect to the action of the full group of symmetries $\mathrm{U}$. The partition function is invariant against $\mathrm{U}$ transformations and the set of its  arguments, namely the generalized temperatures,  can be always reduced to a minimal set  whose cardinality is equal to the rank of the compact denominator group $\mathrm{H}\subset \mathrm{U}$.    
\end{abstract}
\vfill
\end{center}
\noindent \parbox{175mm}{\hrulefill}
\par
${}^\dagger$ P.G. Fr\'e acknowledges support by the Company \textit{Additati\&Partners Consulting s.r.l} during the development of the present research.
\par 
The work of A. Sorin is supported in part by the Center for Integration in Science of the Israel Ministry of Aliyah and Integration
\\[5pt]
\end{titlepage}
\section{Introduction} \label{introibo}
The purpose of the present paper is to clarify all the proper relations, identifications and, when necessary, clear-cut distinctions among several mathematical constructions that have been recently  introduced by different researchers  into the mathematical formulation of Machine Learning and that admit, as their own conceptual pivot, the notions of Lie group, hamiltonian dynamical system and the geometrical rephrasing of thermodynamical equilibrium states. This is particularly relevant in view of the new paradigm of Cartan Neural Networks\cite{pgtstheory,TSnaviga,naviga,tassellandum}, intrinsically characterized by the identification of the network hidden layers with as many non-compact symmetric spaces $\mathrm{U_i/H_i}$, each metrically equivalent to an appropriate solvable Lie group. The generalized notion of Gibbs state, provides the proper way of introducing gaussian like probability distributions on the non-compact symmetric spaces that in the Cartan Neural Network architectures constitute the hidden layers.  Hence we start with a short summary of the new  CaNN paradigm, highlighting its geometrical and group-theoretical strategical aspects. After that we introduce the other mathematical ingredients of our discussion and clarification plan.
\par Let us remark since the very beginning that the generalized Gibbs probability distributions, whose structure, properties and appropriate construction principles are the main target of the present investigation, have been introduced into Machine Learning about a decade ago as a preferred method of Machine Learning analysis of electromagnetic signals such as those involved in radar technologies \cite{radarone}(see also the review \cite{ondoso}). More generally these probability distributions, that are covariant with respect to the full group of isometries of the manifold on which they are defined and fit into geometrical thermodynamics (alias information geometry), are promising tools for Machine Learning architectures concerned with all kinds of electromagnetic signals and also all kinds of temporal sequences.
\subsection{Cartan Neural Networks: A New Paradigm}
In \cite{pgtstheory}, whose authors' list includes two of us, a  new mathematical paradigm was introduced for the engineering of neural network architectures under the name of \textbf{PGTS\footnote{PGTS is an acronym for Paint Group Tits Satake} theory of non-compact symmetric spaces}. The essential points of this paradigm are:
\begin{enumerate}
\item The systematic substitution of the Euclidean $\mathbb{R}^n$ space with a non-compact symmetric coset manifold $\mathrm{U/H}$ where $\mathrm{U}$ is a
    simple non-compact Lie group  and $\mathrm{H}$ the  maximal compact subgroup of  $\mathrm{U}$. All these manifolds are Cartan-Hadamard manifolds and are  metrically equivalent to a specific \textbf{solvable Lie group manifold}
    $\mathcal{S}_{U/H}$.
\item The grouping of these manifolds into Tits-Satake universality classes, which provides an ideal mathematical definition of neural layers.
\item The systematic \textit{suppression of point-wise activation functions} like the sigmoid and its close relatives, the
necessary non-linearity being universally provided by  \textit{generalized exponential maps} from Lie Algebras to the corresponding Lie Groups and
the \textit{generalized logarithm maps} that are the inverse of the former.
\end{enumerate}
In a twin pair of papers \cite{TSnaviga,naviga}, it was shown how a generic multi-layer neural network can be cast into a form that implements points 1), 2), and 3) of the above paradigm. This class of neural network architectures based on the above principles was named \textbf{Cartan neural networks} (\textbf{CaNN}) in honour of the monumental achievement of Èlie Cartan, who obtained the complete classification of all symmetric spaces and their one-to-one correspondence with the classification of real forms of simple complex Lie algebras \cite{cartan_1926,
Helgasonobook,fre2023book,magnea_introduction_2002}.
In \cite{TSnaviga,naviga}  the general scheme of \textit{supervised learning} for a  \textit{classification task}
within \textbf{CaNN} was addressed. In \cite{tassellandum} many additional mathematical general conceptions and constructions were included in the tool-box for the engineering of \textbf{CaNN}.s encompassing, in particular, the general theory of codimension one separators, harmonic analysis on non-compact $\mathrm{U/H}$, Tits Satake fibre bundles and tessellation groups. We refer the reader to \cite{tassellandum} for further details on such topics.
\paragraph{Covariance of CaNNs}. In \cite{TSnaviga} an algorithm
was described where each \textit{datum} is linearly mapped (with a matrix $W_0$ target of learning) to the solvable coordinate vector $\boldsymbol{\Upsilon}$ labeling a point in a non-compact symmetric space $\mathrm{U_1/H_1}$. The latter is the first layer in a sequence of similar layers $\mathrm{U_i/H_i}$,
each being some non-compact symmetric space with some (a priori different) dimension $d_i$.
As discussed at length in \cite{TSnaviga}, the general scheme also allows the non-compact rank and the type of the various $\mathrm{U_i/H_i}$ to be different yet, thanks to the fundamental property of metric equivalence with a suitable solvable group $\mathcal{S}_i$, we can constrain the map from one layer to the next one to be a group homomorphism derived from a linear homomorphism of the corresponding solvable Lie algebras. More specifically, denoting by $K_i$ the map from the  $i^{th}$-layer, described by the space $\mathrm{U_i/H_i}$, to the subsequent one $\mathrm{U_{i+1}/H_{i+1}}$, this map is  described by a group homomorphism between $\mathcal{S}_i$ and $\mathcal{S}_{i+1}$ while the corresponding \emph{push forward} map $K_{i*}$ is a linear homomorphism between the solvable Lie algebras $Solv_i,\,Solv_{i+1}$ generating $\mathcal{S}_i$ and  $\mathcal{S}_{i+1}$, respectively:
$$\forall\,X,\,Y\,\in\, Solv_i\,:\,\,\,\,K_{i*}[X,Y]=[K_{i*}X,\,K_{i*}Y]\,\in \,Solv_{i+1}\,.$$
If $d_i\le d_{i+1}$, $K_i$ can be characterized as an \emph{isometric inclusion} \cite{KN1} having the defining property  that, if $g_i$ and $g_{i+1}$ denote the Riemannian metrics on  $\mathrm{U_i/H_i}$ and  $\mathrm{U_{i+1}/H_{i+1}}$, respectively,
$$\forall\,X,\,Y\,\in\, Solv_i\,:\,\,\,\,g_i(X,Y)=g_{i+1}(K_{i*}X,\,K_{i*}Y)\,.$$
As shown in \cite{KN1}, the mapping $K_{i*}$ between the tangent spaces at corresponding points on the two manifolds, separately isomorphic to $Solv_i$ and $Solv_{i+1}$, is injective. If, on the other hand, $d_i>d_{i+1}$, as a mapping between a linear space $Solv_i$ and a lower-dimensional one  $Solv_{i+1}$, $K_{i*}$ has a non-trivial kernel. If we define the metric $g_i^{(0)}$ on $\mathcal{S}_i$ by the property:
$$\forall\,X,\,Y\,\in\, Solv_i\,:\,\,\,\,g_i^{(0)}(X,Y)=g_{i+1}(K_{i*}X,\,K_{i*}Y)\,,$$
 $g_i^{(0)}$ is singular and thus does not coincide with $g_i$. Nevertheless, being $K_{i*}$  a homomorphism between Lie algebras, ${\rm Ker}(K_{i*})$, of dimension $d_{i}-d_{i+1}$, is an ideal of $Solv_i$ consisting of the \emph{zero-norm} vectors with respect to $g_i^{(0)}$, orthogonal to all the other vectors with respect to the same singular metric.
 Restricted to the solvable Lie algebra $Solv'_i\,\equiv\, Solv_i\,\ominus\, {\rm Ker}(K_{i*})$, $g_i^{(0)}$ coincides with $g_i$ and, therefore, when $d_i>d_{i+1}$, $K_i$ can be characterized as an \emph{isometry} between $\mathcal{S}'_i\equiv \exp(Solv'_i)$, with metric $g_i$ restricted to $Solv'_i\times Solv'_i$, and $\mathcal{S}_{i+1}$.\par
This general characterization of $K_i \, : \, \mathrm{U_i/H_i} \to \mathrm{U_{i+1}/H_{i+1}}$ as an isometric mapping implies its general covariance with respect to the transformations of both the $\mathrm{U_{i}}$ and the $\mathrm{U_{i+1}}$ group.
The action of the two groups on $K_i$ can be formally described as follows:
\begin{equation}\label{cuvarione}
  K_i\,\,\rightarrow\,\,\mathrm{U_{i+1}}\circ K_i\circ \mathrm{U_{i}}\,.
\end{equation}

\paragraph{The relevance of covariance}.
Covariance, as expressed in eq.(\ref{cuvarione}), is the fundamental conceptual and architectural advance provided by CaNN.s that, as explained in \cite{TSnaviga}, are not just one proposal among others, rather they constitute the unique available scheme, allowed by basic theorems of differential geometry, able  \textbf{to dispose off} the \textit{point-wise activation functions} (contradictory with any sort of covariance), to maintain the existence of a unique distance function on each network layer and to preserve indispensable non-linearity.
\subsection{The  Mathematical Basis of CaNN}
Having introduced the new paradigm we summarize the mathematical
key items that constitute its fundament.
\paragraph{The strategic metric equivalence with solvable groups}. As discussed at length in the foundational paper \cite{pgtstheory} and already recalled above, the strategic element that allows the construction of CaNN.s, with all the properties mentioned above, is the \textit{metric equivalence} of all non-compact symmetric spaces
$\mathrm{U/H}$ with a suitable solvable Lie subgroup $\mathcal{S}_{\mathrm{U/H}} \subset \mathrm{U}$ which is a generalization, in each different $\mathrm{U}$-case, of the Borel subgroup, applying to the case $\mathrm{U}=\mathrm{SL(N,\mathbb{R})}$. This metric equivalence amounts to the statement that all
non-compact symmetric spaces are \textit{Alekseveskian Normal Riemannian Manifolds}.
\paragraph{Alekseevsky Normal Manifolds and Solvable Lie Groups}
Following the original viewpoint of Alekseevsky \cite{Alekseevsky1975,Cortes} we say that a Riemannian manifold $\left ( \mathcal{M},g \right)$ is
\textit{normal} if it admits a completely solvable Lie group $\mathcal{S}_{\mathcal{M}}\simeq\exp[\Solv_{\mathcal{M}}]$ of isometries that acts on the manifold in a simply transitive manner (i.e. for every 2 points in the manifold there is one and only one group element connecting them).
The group $\mathcal{S}_{\mathcal{M}}$ is then generated by a so-called \textit{normal metric Lie algebra}, that is a completely
solvable Lie algebra $\Solv_{\mathcal{M}}$ endowed with an Euclidian, positive definite, $Solv$-invariant, symmetric form $< \, , \, >$ . The
main tool to classify and study the normal homogeneous spaces is provided by the theorem \cite{BorelTits}, \cite{Helgasonobook} that
states that if a Riemannian manifold $\left ( \mathcal{M},g \right)$ admits a transitive normal solvable group of isometries
$\exp[\Solv_{\mathcal{M}}]$, then it is metrically equivalent to this solvable group manifold
\begin{eqnarray}\label{identifico}
\mathcal{M} & \simeq & \exp \left[ \Solv_{\mathcal{M}}\,
\right]~,\nonumber  \\ g\mid_{e \in \mathcal{M}} & = & <,> \,
\end{eqnarray}
where $<,>$ is the Euclidian metric defined on the normal solvable Lie algebra $\Solv_{\mathcal{M}}$.
\par
The original conjecture of Alekseevsky was just restricted  to quaternionic K\"ahler manifolds and  stated that any such manifold
$\mathcal{M}$ that was also homogeneous and  of negative Ricci curvature  should   be normal, in the  over mentioned sense, namely a transitive solvable group of isometries $\exp \left[\Solv_{\mathcal{M}}\right ]$ should exist, that could  be
identified with the manifold itself. Note that the actual group of isometries $\mathrm{U}$ of $\mathcal{M}$ could be much larger than the solvable group,
\begin{equation}\label{fattispecie}
    \mathrm{U} \, \supset \, \exp \left
    [\Solv_{\mathcal{M}}\right]~,
\end{equation}
as it is for instance the case for all symmetric spaces
$\mathcal{M} \, = \, \frac{\mathrm{U}}{\mathrm{H}}$,
yet the solvable normed Lie algebra $\left (\Solv_{\mathcal{M}}\,
, \, <\, ,\, >\right)$ had to exist.  The problem of classifying
the considered manifolds was turned in this way into the problem
of classifying the \textit{normal metric solvable Lie algebras}
$\left(Solv, <\, ,\, >\right)$. Note that in Alekseevsky's
case the symmetric form $<\, ,\, >$ was not only required to be
positive definite but also quaternionic K\"ahler. Alekseevsky's
conjecture  actually applies to much more general homogeneous
Riemannian manifolds than the quaternionic ones: for instance it
applies to all those endowed with a special K\"ahler geometry or
with a real special one as the classification of de Wit et. al.
\cite{specHomgeoA1,specHomgeoA2,deWit:1995tf} demonstrated. It also applies to the symmetric spaces appearing in the scalar sector of extended supergravities with more than eight supercharges. For all these manifolds there exists the corresponding normal metric  algebra
$\left(Solv, <\, ,\, >\right)$, in other words they are
\textit{normal}. Actually by explicit construction, as discussed in \cite{pgtstheory}, all non-compact symmetric spaces $\mathrm{U/H}$ with $\mathrm{U}$ a simple  non-compact Lie group and $\mathrm{H\subset U}$ its maximal compact subgroup are normal Aleksveskian manifolds.
\subsection{The Link with Symplectic Geometry and
Generalized Thermodynamics}
Analysing the mathemathical foundations of CaNNs we are naturally lead to observe a natural link with symplectic geometry and generalized thermodynamics. In this subsection we unveil, also historically, such a conceptual path.
\paragraph{Integrability of geodesic equations, Poisson and symplectic manifolds.}
The mathematical theory that links \textit{non compact symmetric manifolds} to \textit{Normed Solvable Lie Algebras} was pioneered by mathematicians but then it was extensively developed within the context of supergravity, since all scalar manifolds of extended supergravity lagrangians happen
to be non compact symmetric spaces and their solvable representations played a decisive role in the systematic resolution of several problems, in particular the construction of cosmic billiards and extremal black-hole solutions. It was in this framework that the integrability of geodesic
equations for such manifolds was analysed by Fr\'e and Sorin in terms of a Lax pair equation \cite{sashaebog} in 2006 and also a Poisson manifold viewpoint underlying such integrability was introduced by the same authors in 2009 \cite{dualborel}. The complete integrability and the explicit integration of geodesic equations in $\mathrm{U/H}$ symmetric spaces is an essential brick in the construction of Cartan Neural Network architectures
and it is discussed at length in the foundational paper \cite{pgtstheory}, where it is shown that the explicit result for the solvable coordinates, as functions $\boldsymbol{\Upsilon}(t)$ of the affine parameter $t$, can be obtained directly in terms of the initial data, namely the starting point and the so named matrix of conserved Noether charges $Q$, bypassing the solution of Lax equation. It might seem from this that the Poisson manifold viewpoint \cite{dualborel} is interesting yet unnecessary in the context of Machine Learning, but such a conclusion is too hasty and incorrect for the following reason. The virtue of the \textit{Poisson/symplectic approach} to the geodesic problem is that it puts it into the perspective of dynamical systems and at the same time of geometric thermodynamics, creating a triple link among the geodesics on $\mathrm{U/H}$, the symplectic/contact geometry of thermodynamics and the contact structures of fluid-dynamics\cite{Fre:2015mla},\cite{mariopietroinpress}. As we are going to see, a recent research line in Machine Learning introduces in a different  setup and under the name of \textit{Gibbs States for Lie Groups}, basic structures of geometric thermodynamics, so that a conceptual clarification of
all the relations is quite appropriate and useful in order to combine Cartan Neural Networks with  statistical conceptions as those advocated in the mentioned research-line.
\paragraph{Generalized geometric thermodynamics}
A process of fundamental importance in Chemistry  is the separation of different substances that happen to be in gas mixtures and multi-component liquids. From a conceptual point of view any separation method is based on the thermodynamics of mixtures of different components coexisting in different phases. Since the 19th century this situation was carefully conceptualized by great chemists, physicists and mathematicians and its study became one of the focal points of statistical mechanics and classical thermodynamics. Gibbs' rule of phases and the use at the level of statistical mechanics of the grand canonical ensemble (see Appendix \ref{tuttinsieme}) with the introduction of the chemical potential are two fundamental junctures in this affair. However, the discouraging and critical aspect in this area of knowledge is that the exact calculation of canonical and grand canonical partition functions is of extreme difficulty when the particles forming the chemicals of our interest interact with each other, i.e., always, in the case of real and non-ideal substances. The cases of exact computation of the partition function are isolated and rare, reducing essentially to those of the classical ideal gas, the quantum free gases of bosons or fermions, and the two-dimenisonal Ising model for ferromagnetism. In all other cases there is a plethora of
approximation methods and sophisticated perturbative or approximate computational techniques. The object of primary interest for thermodynamic calculations is the equation of state, i.e., the relation between both extensional (such
as volume $V$, entropy $S$ and internal energy $U$) and intensive (such as temperature $T$, and chemical potentials $\mu_i$) thermodynamic quantities that is valid in equilibrium states (see appendix \ref{richiamatermo} for a summary of classical thermodynamics). Equations of state can be derived exactly from the partition function if one knows the latter.
Thus alternating with attempts at direct calculations of certain partition functions there has been, over the last century, a great deal of modeling activity, both theoretical and experimental aimed at constructing mathematical formulations of equations of state in want of the missing partition function.
However, such equations of state  are just phenomenological models, and a deeper
understanding of their reason for being is  necessary.
Thanks to the work, hitherto little known outside a small community of specialists, of an even smaller number of low-temperature physicists and mathematical physicists, there exists a surprisingly innovative geometric view of classical thermodynamics that provides a more intrinsic view of thermodynamic states and seems able, by classical geometric means to provide mesoscopic information about
real gases and liquids, while also defining a conceptual frame of reference in which phenomenological equations of state can be evaluated and possibly modified in a more systematic and profound way, particularly taking into account the possible isometries of the Riemannian metric surprisingly associated with the space of thermodynamic variables.
\par
The problem is quite general: also in systems of different nature, for example Big Data sets, if one arrives at a thermodynamical limit description one can wonder about the advantages of a geometrical formulation of thermodynamics.
\par In view of  applications  to the research compound  of Geometric Deep Learning,  which implies the use of predetermined metrics it is a stimulating perspective to compare the metric setup of Geometrical Thermodynamics and that of Information Geometry, in particular focusing on the symplectic structure that can migrate from Thermodynamics to Data Science.
\par
The small group of low-temperature physicists and mathematical physicists to whom we owe the entire body of developments related to the geometric view of thermodynamics we have referred to consists of three \textit{senior founders} the American \textit{George Ruppeiner} affiliated with the New College of Florida in Sarasota and the two Russians \textit{Valentin Lychagin} and \textit{Mikhail Roop}, plus a small cohort of adherents consisting of their occasional collaborators, Ph.D. students, postdocs, and so on.
It is very interesting to read Ruppeiner's autobiographical article \cite{Ruppeiner_2020} written in April 2016 for the commemorative volume in honor of Horst Meyer's 90th birthday that unfortunately  passed away a few months later. In this article, the author recounts how in the years 1975-1980 when he was a Ph.D. student at Duke University conducting low-temperature gas experiments in Meyer's laboratory, devoting most of his efforts to perfecting himself in Low Temperature experimental physics, he had nonetheless  a drive to take General Relativity Courses and deepen his knowledge of differential geometry. A spark was ignited in his mind, he recounts, when he read in Physics Today an article by Frank Weinhold\cite{vinoldo}  in which a metric form was introduced in the context of thermodynamic variables, something hitherto considered peregrine and absurd. Ruppeiner, on the other hand,  regarded that as a serious suggestion and deemed that a Riemannian view of thermodynamics should be able to be constructed and could also be useful in the analysis of critical phenomena. Gradually, continuing on the path  he had taken, he arrived at constructing two-dimensional metrics in the temperature-density plane that were consistent with the principles of thermodynamics and went so far as to calculate the curvature scalar R of such metrics. A salient moment in the development of his thought was when he arrived at the physical interpretation of $R$:
\begin{equation}\label{interpretoR}
    |R| \, \propto \, \xi^3
\end{equation}
where $\xi$ is the statistical correlation length, which, as everyone knows, tends to infinity in the vicinity of critical points and phase transitions. Of course, ideal gases correspond to flat metrics with zero curvature $R=0$  and no critical points. Thus thermodynamic curvature became a classical indicator of molecular interactions at the mesoscopic level, and in the first two decades of the 21st century Ruppeiner contributed a series of very interesting papers on the use of Riemannian geometry in the study of thermodynamics and its critical phenomena: \cite{Ruppeiner_2010},\cite{Ruppeiner_2012},\cite{Ruppeiner_2012b}
\cite{Ruppeiner_2013},\cite{ruppoRdiag}, \cite{Ruppeiner_2020}.
More recently, but following reflexions developed over the years and expounded in his 2018 lectures at Wisla
\cite{lychaginlecture}, Valentin Lychagin, a Russian mathematician for  a longtime professor at Tromso University in Norway, identified and systematically expounded within the framework of information theory, an interesting connection between contact geometry and thermodynamics, characterizing possible thermodynamic equilibrium states as Legendrian subvarieties of contact varieties. Because of the complex and general relationships between contact varieties and symplectic varieties (see Appendix \ref{geoprobo} for a summary of contact and symplectic geometry), thermodynamic states can also be interpreted as Lagrangian subvarieties of symplectic varieties, and the canonical symplectic form on them naturally connects to a Riemannian metric, which is the one hypothesized and studied by Ruppeiner.
\subsection{Gibbs States and Lie Group Generalized Thermodynamics}
\label{topolinothermo}
In a series of papers of which we quote only a small selection,
\cite{marlentropia,caldobarbaresco,barbaresco2,barbaresco3,marlegibbs}, that is most informative about the main idea, a group of  French authors, including  Charles-Michel Marle, Fr{\'e}d{\'e}ric Barbaresco, Yann Cabanes and Pierre-Yves Lagrave, relying on old ideas of late Jean-Marie Souriau, have introduced the notion of \textit{Gibbs States of Mechanical Systems with Symmetries} and of \textit{Lie Group Thermodynamics} which bears a close similarity with the geometrical formulation of thermodynamics as expounded in Lychagin's lectures \cite{lychaginlecture}, yet is
quite distinct from it. As we explain in the main body of the present article, the original distinctive idea of \textit{Lie Group Thermodynamics} is the definition of a subspace $\Omega \subset \mathbb{G}$ of the Lie algebra of symplectic Killing vector fields $\boldsymbol{X} \in \mathbb{G}$ that leave invariant a symplectic manifold $\left( \mathcal{M},\omega\right)$ in the sense that the Lie derivative along them of the symplectic form vanishes
\begin{equation}
\label{killa}
\mathcal{L}_{\mathbf{X}} \, \omega \, = \,0
\end{equation}(compare eq.(\ref{killa}) with the definition \ref{liuvillocampillo} in appendix \ref{pescione} of Liouville vector fields) such that the following integral (the partition function) is convergent:
\begin{equation}\label{candelabro}
 \forall \boldsymbol{\beta} \in \Omega \quad : \quad \quad Z(\boldsymbol{\beta}) \, \equiv \, \int_\mathcal{M} \, \exp\left[-\boldsymbol{\beta}\cdot \boldsymbol{\mathfrak{P}}(\Phi) \right] \, d\lambda(\Phi) \, < \, \infty
\end{equation}
where $\Phi$ are the coordinates on the the 2n-dimensional differentiable manifold $\mathcal{M}$,
\begin{equation}\label{grafomane}
  d\lambda(\Phi) \, \equiv \, \underbrace{\omega \wedge \omega \wedge \dots \wedge \omega}_{n-\text{times}}
\end{equation}
is its Liouville integration measure and $\boldsymbol{\mathfrak{P}}(\Phi)$ is the \textbf{moment map} (see below for the discussion of such a concept):
\begin{equation}\label{lagradisca}
\forall \boldsymbol{k} \in \mathbb{G} \quad\quad  \boldsymbol{\mathfrak{P}}(\Phi) \quad : \quad \boldsymbol{k} \, \longrightarrow \, \boldsymbol{\mathfrak{P}}_{\mathbf{k}}(\Phi) \, \in \, \mathbb{C}^\infty (\mathcal{M})
\end{equation}
The subspace $\Omega \subset \mathbb{G}$ of the symmetry Lie algebra is named the space of \textbf{generalized temperatures}.
\subsubsection{Symplectic Moment Map} {On a d-dimensional Riemannian space $\mathcal{M}$, admitting an isometry Lie algebra $\mathbb{U}$, moment maps can be defined as linear mappings between $\mathbb{U}$ and  smooth functions on $\mathcal{M}$, valued in the holonomy algebra $\mathrm{Hol}(\mathcal{M})$:
\begin{align}
&X\in \mathbb{U}\,\,\longrightarrow \,\,\,\mathfrak{P}_X\,\in \,\,\mathrm{Hol}(\mathcal{M})\times \mathbb{C}^\infty(\mathcal{M})\,,\nonumber\\
&\forall \boldsymbol{\mathfrak{k}}^i_X\,\,{\rm Killing\,\,vector}\,\longrightarrow\,\,\,\,(\mathfrak{P}_X)_i{}^j\equiv \nabla_i\boldsymbol{\mathfrak{k}}^j_X\in \mathrm{Hol}(\mathcal{M})\,.\end{align}
satisfying certain equivariance conditions. \footnote{{ For a general characterization of the moment maps see, for instance,} \url{https://arxiv.org/abs/1605.05559}.} For several applications, one considers moment maps with values in a specific subalgebra $\mathbb{H}_0$ of $\mathrm{Hol}(\mathcal{M})$. For instance, if $\mathcal{M}$ is  K\"ahler, Special K\"ahler or Hyper-K\"ahler, $\mathbb{H}_0=\mathfrak{u}(1)$, Lie algebra of the ${\rm U}(1)$ group of K\"ahler transformations. In this case, the compact $\mathfrak{u}(1)$ generator has an invertible action on the tangent space to $\mathcal{M}$. The curvature $\boldsymbol{\mathcal{K}}$ of the corresponding ${\rm U}(1)$-connection is a non-singular closed 2-form which provides a symplectic structure on the manifold, namely a maximal rank, closed 2-form $\omega=\boldsymbol{\mathcal{K}}$. In general, the existence of a symplectic 2-form is not related to the metric properties of the manifold, or to the very existence of a metric. On a Riemannian manifold, endowed with a symplectic 2-form, we require the latter to be consistent with the Riemannian structure and, in particular, with its isometries. This is only possible if the Lie derivative of $\omega$, with respect to all Killing vectors, vanishes. The only possibility for this is that the manifold be of K\"ahler type and $\omega$ is proportional to the K\"ahler 2-form $\boldsymbol{\mathcal{K}}$, since, in this case, the $\mathfrak{u}(1)$ subalgebra, defining $\boldsymbol{\mathcal{K}}$, is in the center of Holonomy algebra. K\"ahler, Special K\"ahler and  Hyper-K\"ahler manifolds, are an important ingredient in  supergravity/supersymmetric gauge theories.}  { Indeed, they }play a very important role in the whole landscape of supersymmetric field theories: in particular they are the basic building blocks in the construction of scalar potentials (see for instance the Physics Report \cite{Trigiante_2017}, the book \cite{advancio}
and the general paper \cite{pupallo2}). Moment maps play also a fundamental role in the resolution of singularities via
K\"ahler and HyperK\"ahler quotients {\`a} la Kronheimer (for a review see for instance the lecture notes \cite{freletturepotsdam} and all the vast literature there quoted). In the series of papers \cite{caldobarbaresco,barbaresco2,barbaresco3,marlegibbs,marlentropia}, the authors rely on  moment maps as a fundamental ingredient in the construction of partition functions for symplectic manifolds and the explicit examples they present, namely the cases
of the hyperbolic plane and of the Siegel plane (see also \cite{pgtstheory} for the role that the latter might play in Machine Learning), the symplectic structure utilized to define the moment maps is the one provided by the K\"ahler 2-form; hence, it applies to the very manifold $\mathcal{M}$, which in the considered examples is indeed K\"ahlerian,  rather then to its tangent bundle $\mathcal{TM}$ which,  instead, is the geometrical substratum of the  geodesic dynamical system that can be defined on every Riemannian
manifold $\mathcal{M}$.
\subsubsection{Coadjoint Orbits}
In the discussion of the thermodynamics that might be associated with symmetric spaces  $\mathrm{U/H}$, another source of possible conceptual confusions  is given by the symplectic structure, named after Kirillov-Kostant-Souriau, that can be defined on coadjoint orbits of any Lie group $\mathrm{G}$.  This matter is presented in a crystal clear form in chapter 5 of the book\cite{coaggiungo}.
\par
As we just anticipated and we show systematically in sections \ref{brillantone} and \ref{geostructa} we can define generalized temperatures and partition functions on
a smooth manifold $\mathcal{M}$, whenever the latter is endowed with a bona fide symplectic structure, namely a closed antisymmetric $2$-form $\omega$ of maximal rank, and we have a Lie group $\mathrm{G}$ acting on $\mathcal{M}$ by means of diffeomorphisms:
\begin{equation}\label{actionG}
  \forall g \in \mathrm{G} \quad  \mathcal{D}(g) \quad: \quad \mathcal{M} \longrightarrow \mathcal{M} \quad ; \quad
  \forall g_1,g_2 \in \mathrm{G} \quad \mathcal{D}(g_1\cdot g_2)\, = \, \mathcal{D}(g_1) \circ \mathcal{D}(g_2)
\end{equation}
 that are generated by hamiltonian vector fields, namely Killing vector fields $\boldsymbol{\mathfrak{k}}_A$ (see the definition  \ref{killa} of symplectic Killing vector fields) satisfying its Lie Algebra $\mathbb{G}$:
  \begin{equation}\label{cartilagine}
    \left[{\boldsymbol{\mathfrak{k}}}_B \, , \, {\boldsymbol{\mathfrak{k}}}_C\right] \, = \, f_{BC}^{\phantom{BC}A} \, {{\boldsymbol{\mathfrak{k}}}_A} \quad; \quad A,B,C = 1,2,\dots,\text{dim}\mathbb{G}
\end{equation}
Indead each symplectic Killing  vector field $\boldsymbol{\mathfrak{k}}$ on the symplectic manifold $\mathcal{M}$ is associated with a moment map $\mathfrak{P}_{\boldsymbol{\mathfrak{k}}}$ as already anticipated in eq.(\ref{lagradisca}) that is a function on the manifold $\mathcal{M}$:
 \begin{equation}\label{momentus}
   \mathfrak{P}_{\boldsymbol{\mathfrak{k}}} \quad : \quad \mathcal{M} \, \longrightarrow \, \mathbb{R}
 \end{equation}
  locally satisfying  the condition:
 \begin{equation}\label{hamiltoncondo}
   \mathrm{i}_{\boldsymbol{\mathfrak{k}}}\cdot \omega \, = \, \mathrm{d} \, \mathfrak{P}_{\boldsymbol{\mathfrak{k}}}
 \end{equation}
 where $\mathrm{i}_{\boldsymbol{X}}\cdot $ is the contraction operation along the vector field $\boldsymbol{X}$ acting on any $p$-form and $\mathrm{d}$ is the exterior derivative also acting on any $p$-form.
\par
 The moment maps are hamiltonians and can be used to define partitions functions as in eq. (\ref{candelabro}) where a candidate generalized temperature  is any element of the  Lie algebra  $\mathbb{G}$.
 \par
 Keeping this essential point in mind we turn to coadjoint orbits of a Lie group $\mathrm{G}$.
 \par
 As explained in \cite{coaggiungo}, for any Lie group $\mathrm{G}$ one can define the dual $\mathbb{G}^\star$ of its Lie algebra $\mathbb{G}$, which is also a vector space of the same dimension, and from the adjoint action of the group on $\mathbb{G}$:

 \begin{equation}\label{aggiunta}
   \forall g\in \mathrm{G}, \, \forall \mathbf{X} \in \mathbb{G} \quad : \quad \text{Adj}_g (\mathbf{X}) \, \equiv \, g^{-1} \, \mathbf{X} \, g \, \in \, \mathbb{G}
 \end{equation}
 we obtain the coadjoint action of $\mathrm{G}$ on $\mathbb{G}^\star$ as follows:
 \begin{equation}\label{crapitto}
   \forall g\in \mathrm{G}, \,\, \forall \mathbf{X} \in \mathbb{G}, \,\, \forall \boldsymbol{\lambda} \in \mathbb{G}^* \quad ; \quad
   \text{CoAdj}_g (\boldsymbol{\lambda})\left( \mathbf{X}\right) \, \equiv \, \boldsymbol{\lambda}\left(\text{Adj}_g \left(\mathbf{X}\right) \right)\, = \,
   \boldsymbol{\lambda}\left( g^{-1} \, \mathbf{X} \, g \right)
 \end{equation}
 Fixing a particular element $\boldsymbol{\lambda} \in \mathbb{G}^\star$ the coadjoint orbit $\mathcal{O}_{\boldsymbol{\lambda}}$
 is defined as the subset of elements of $\mathbb{G}^\star$ that are images of $\boldsymbol{\lambda} $ under the coadjoint action of some group element of $\mathrm{G}$:
 \begin{equation}\label{cranberrysauce}
  \forall \boldsymbol{\lambda}\in \mathbb{G}^\star \quad ; \quad  \mathcal{O}_{\boldsymbol{\lambda}} \, = \, \left\{
  \boldsymbol{\mu} \in \mathbb{G}^\star \, \mid \, \exists g \in \mathrm{G} \, / \, \text{CoAdj}_g (\boldsymbol{\lambda}) \, = \, \boldsymbol{\mu} \right\}
 \end{equation}
 Eq. (\ref{cranberrysauce}) can be decoded in a more friendly and usable way if on the Lie algebra $\mathbb{G}$, which is a vector space, we introduce a non degenerate positive definite symmetric scalar product $\langle\, ,\, \rangle$ so that any element of $\mathbb{G}^\star$, by definition  a linear functional on $\mathbb{G}$, can be described as follows:
 \begin{equation}\label{frido}
   \forall \boldsymbol{\lambda} \in \mathbb{G}^\star\, , \, \forall \mathbf{X} \in \mathbb{G} \quad : \quad \boldsymbol{\lambda}(\mathbf{X} ) \, = \, \langle \boldsymbol{\lambda}^\dagger\, ,\, \mathbf{X} \rangle \quad \text{where $\boldsymbol{\lambda}^\dagger
   \in \mathbb{G}$}
 \end{equation}
 Given a set of generators $T_A$ that form a basis for the vector space $\mathbb{G}$, we have the symmetric invertible, positive definite matrix $\kappa_{AB}$ defined below together with its inverse:
 \begin{alignat}{4}\label{carnevaledioslo}
   \kappa_{AB} & \equiv & \langle T_A\, ,\, T_B \rangle &\quad ; \quad &\kappa^{AB} \equiv \left (\kappa^{-1}\right)^{AB}
 \end{alignat}
and equation (\ref{frido}) becomes:
\begin{equation}\label{corteorzo}
  \boldsymbol{\lambda}^\dagger \, = \, \lambda^A \, T_A \quad ; \quad \mathbf{X} \, =\, X^B\, T_B  \quad ; \quad
  \boldsymbol{\lambda}(\mathbf{X}) \, = \lambda^A \, X^B \, \kappa_{AB}
\end{equation}
The adjoint representation of the Lie group is explicitly given by the adjoint matrix defined below:
\begin{equation}\label{coccode}
  g^{-1}\,  T_A \, g \, = \,\mathcal{ A}(g)_{A}^{\phantom{A}B} \, T_B
\end{equation}
and the coadjoint representation is defined by the position:
\begin{equation}\label{fischiacurvo}
  \langle \mathcal{CA}(g)_{A}^{\phantom{A}P} \,T_P \, , \, T_R \rangle \, = \, \langle
  \,T_A \, , \, \mathcal{A}(g)_{R}^{\phantom{R}Q}\, T_Q \rangle
\end{equation}
which matrix-wise implies:
\begin{equation}\label{frescobaldo}
  \mathcal{CA}(g) \, = \, \kappa\cdot \mathcal{A}(g) \cdot \kappa^{-1}
\end{equation}
If $\lambda^A$ are the coefficients of the element $\boldsymbol{\lambda}^\dagger \in \mathbb{G}$ that defines the orbit, the latter is formed by all those elements of $\mathbb{G}$ that have the following form:
\begin{equation}\label{crugone}
  \mu^\dagger(g) \, = \, \mu^A(g) \, T_A \, \equiv \,\underbrace{\lambda^B \,\mathcal{CA}(g)_B^{\phantom{B} A}}_{\mu^A(g)} \, T_A \in \mathbb{G} \quad ; \quad g\in \mathrm{G}
\end{equation}
One might think that as $g$ varies in the group $\mathrm{G}$ for a generic choice of $\lambda^\dagger$  the element $\mu^\dagger(g)$ will span the entire Lie algebra $\mathbb{G}$. If this were true the coadjoint orbit would be diffeomorphic to the  Lie group $\mathrm{G}$. However this is not true, since there is always a non trivial Lie subgroup $\mathfrak{S}(\boldsymbol{\lambda}^\dagger) \subset \mathrm{G}$ for which the adjoint action on $\boldsymbol{\lambda}^\dagger$ is trivial:
\begin{equation}\label{cratino}
 g^{-1} \, \boldsymbol{\lambda}^\dagger \, g \, = \, \boldsymbol{\lambda}^\dagger \quad \, \text{iff}  \, g\in\mathfrak{S}(\boldsymbol{\lambda}^\dagger)
 \end{equation}
That in eq.(\ref{cratino}) is the very definition of the stabilizer subgroup of the Lie algebra element $\boldsymbol{\lambda}^\dagger$ and such subgroup is never trivial since  it includes, at least, the one-dimensional subgroup generated by $\boldsymbol{\lambda}^\dagger$ itself;  for special choices of $\boldsymbol{\lambda}^\dagger$, the stabilizer can be much larger.
This means that the coadjoint orbit $\mathcal{O}_{\boldsymbol{\lambda}}$ is always diffeomorphic to a coset manifold, namely
$\mathrm{G}/\mathfrak{S}(\boldsymbol{\lambda}^\dagger)$. The non-degenerate symplectic form $\omega$ of Kirillov-Kostant-Souriau is not defined on $\mathrm{G}$, rather it is defined on each coadjoint orbit labeled by a Lie algebra element $\boldsymbol{\lambda}$ namely on  a coset manifold $\mathrm{G}/\mathfrak{S}(\boldsymbol{\lambda}^\dagger)$. {Hence the symplectic manifolds that admit a hamiltonian
action of the group $\mathrm{G}$  are already all captured by the scan of all coset manifolds $\mathrm{G/H}$ where the subgroup $\mathrm{H}\subset \mathrm{G}$ stabilizes some non trivial element $\boldsymbol{\lambda}^\dagger$ of the Lie algebra $\mathbb{G}$.}
\par
Just for completeness let us mention that the Kirillov-Kostant-Souriau symplectic 2-form $\omega^{KKS}$ defined on a coadjoint orbit is very simply given. Let $\mathbf{t}^\sharp_A$ be the realization on the orbit $\mathcal{O}_{\boldsymbol{\lambda}}$, namely on the coset manifold $\mathrm{G}/\mathfrak{S}(\boldsymbol{\lambda}^\dagger)$ of the invariant vector fields $\mathbf{t}_A$ spanning the Lie algebra $\mathbb{G}$. They form a basis of sections of the tangent bundle $\mathcal{T}\mathcal{O}_{\boldsymbol{\lambda}}$. Hence a $2$-form $\omega$ is completely defined if we give its value on any pair of such vector fields. The Kirillov-Kostant-Souriau form is defined by setting:
\begin{equation}\label{KKSdefi}
  \omega^{KKS}_{\boldsymbol{\lambda}}\left(\mathbf{t}^\sharp_A\, , \,\mathbf{t}^\sharp_B \right) \, = \, f_{AB}^{\phantom{AB}C} \, \kappa_{CE} \lambda^E
\end{equation}
{We will restrict ourselves to the case in which ${\rm G=U}$ is a semisimple, isometry group of a symmetric space ${\rm U/H}$. In this case $\kappa_{AB}$ is non-singular and $H$ is the maximal compact subgroup of ${\rm U}$. The KKS symplectic form is invariant under ${\rm U}$ only if the element $\lambda^\dagger$ is central in $\mathbb{H}$, Lie algebra of ${\rm H}$, namely if ${\rm U/H}$ is K\"ahler and $\lambda^\dagger$ corresponds to the K\"ahler $\mathfrak{u}(1)$ generator.}
\subsection{Clearcut Distinctions}
\label{cromatillo}
Notwithstanding whether the manifold $\mathcal{M}$ has a symplectic structure or not, its tangent bundle $\mathcal{TM}$ has always the symplectic structure associated with the hamiltonian description of geodesic equations on $\mathcal{M}$.
Here comes a first essential distinction. Whenever we have a canonical dynamical system on a symplectic manifold $\left(\mathcal{SM}_{2n},\omega\right)$, like the geodesic one where $\mathcal{SM}_{2n}\, = \, \mathcal{TM}_{n}$, we can construct standard thermodynamics in geometrical formulation, starting from the minimization of the Shannon entropy functional and arriving at Gibbs states of the form:
\begin{eqnarray}\label{predone}
  \mathrm{G}\left(\boldsymbol{\lambda},\Phi\right)&=&
  \frac{\exp\left[-\boldsymbol{\lambda}\cdot \boldsymbol{\mathfrak{H}}(\Phi)\right]}{Z(\boldsymbol{\lambda})}\nonumber\\
  Z(\boldsymbol{\lambda}) & \equiv & \int_{\mathcal{SM}}\,  \exp\left[\,-\,\boldsymbol{\lambda}\cdot \boldsymbol{\mathfrak{H}}(\Phi)\right]\,  d\lambda(\Phi)
\end{eqnarray}
where $\boldsymbol{\mathfrak{H}}(\Phi)$ is the multiplet of $k$
hamiltonians in involution admitted by the dynamical system\footnote{If the dynamical system is Liouville integrable, one has $k=n$, the dimension of the symplectic manifold  being $2n$. In general $1 \leq k<n$ and it is just $1$ for a generic dynamical system without conserved charges.}
(including the standard one defined by Legendre transform of the Lagrangian):
\begin{equation}\label{larvatarlo}
 \boldsymbol{\mathfrak{H}}(\Phi) \, = \, \left\{\mathfrak{H}_1(\Phi),\dots \, , \mathfrak{H}_k (\Phi)\right\} \quad ; \quad \underbrace{\left\{\mathfrak{H}_i \, , \, \mathfrak{H}_j\right\}}_{\text{Poisson bracket}}\, = \, 0 \quad \forall\, i,j
\end{equation}
$\boldsymbol{\lambda} \in \mathbb{R}^k$ is a vector of generalized temperatures and $Z(\boldsymbol{\lambda})$ is the partition function.
\par
Following the conception reviewed in section \ref{parcondicio} and making reference to eq.s(\ref{muccacarolina}) and following ones, we  should also note that the stochastic vector variable $\mathbf{X}(q)$, of which we fix the average value in order to define a probability distribution that extremizes the Shannon functional with constraints, does not need to be a set of hamiltonians in involution and there is no need of integrability of any dynamical system in order to introduce a \textbf{generalized thermodynamics}. In the case the space of events (see Appendix \ref{fondiprobi} for the basic definitions of probability theory) is a symplectic manifold endowed with hamiltonian vector fields we can use the moment-maps of the latter as a convenient set of stochastic variables in order to introduce a generalized thermodynamics by fixing their average values. Yet this is only a subclass of examples in a general class.
\par
The Lie group thermodynamics advocated in \cite{caldobarbaresco,barbaresco2,barbaresco3,marlegibbs,marlentropia}
leads to Gibbs states of the form:
\begin{eqnarray}\label{ciromerlotti}
  \mathrm{G}\left(\boldsymbol{\beta},\Upsilon\right)&=&
  \frac{\exp\left[-\boldsymbol{\beta}\cdot \boldsymbol{\mathfrak{P}}(\Upsilon )\right]}{Z(\boldsymbol{\beta})}\nonumber\\
  Z(\boldsymbol{\beta}) & \equiv & \int_{\mathcal{M}}\,  \exp\left[-\boldsymbol{\beta}\cdot \boldsymbol{\mathfrak{P}}(\Upsilon)\right]\,  \mathrm{dg}\left[\Upsilon\right]
\end{eqnarray}
where  $\Upsilon$ denotes the coordinates of the Riemannian manifold $\left(\mathcal{M},g\right)$, the generalized temperature
$\boldsymbol{\beta}\in \Omega$ is an element of the Lie algebra $\mathbb{G}$ of the isometry group $\mathrm{G}$ such that the integral defining the partition function is convergent. Let us observe that in eq.(\ref{ciromerlotti}) $\mathrm{dg}\left[\Upsilon\right]$ is the Riemannian integration measure which coincides with the Liouville measure (\ref{grafomane}) if $\left(\mathcal{M},g\right)$ is a K\"ahler manifold and the K\"ahler $2$-form $\boldsymbol{\mathcal{K}}$ is utilized to define the symplectic structure on $\mathcal{M}$.
\par
A third possibility, which is the conceptual framework  underlining \cite{caldobarbaresco,barbaresco2,barbaresco3,marlegibbs,marlentropia} is to use the setup of eq. (\ref{ciromerlotti}) using, however, as
substratum manifold $\mathcal{M}$ some coadjoint orbit of $\mathcal{O}_{\boldsymbol{\lambda}^\dagger}$ under the action of a group $\mathrm{G}$ of  some special Lie algebra element $\boldsymbol{\lambda}^\dagger\in \mathbb{G}$. As extensively discussed in the previous subsection coadjoint orbits are anyhow coset manifolds and it is conceptually much more economic to start from the coset manifold structure asking one-self the question: \textit{Given $\mathrm{G/H}$, what is the Lie algebra element $\boldsymbol{\lambda}^\dagger
\in \mathbb{G}$ that is stabilized by the chosen subgroup $\mathrm{H}$?} The answer is fairly simple. In view of the discussion of the previous section, it is clear that $\boldsymbol{\lambda}^\dagger \in \mathbb{H} \subset \mathbb{G}$ since the one-parameter group generated by
$\boldsymbol{\lambda}^\dagger$ must be contained in $\mathrm{H}$. Therefore the condition is just:
\begin{equation}\label{carpacarpionata}
  \left[ \mathbb{H} \, , \, \boldsymbol{\lambda}^\dagger\right] \, = \, 0
\end{equation}
The solution of the constraint (\ref{carpacarpionata}) is immediate. The Lie algebra $\mathbb{H}$ must have the following structure:
\begin{equation}\label{crunacammello}
  \mathbb{H} \, = \, \mathbb{H}^\prime \oplus \mathbb{H}_0 \quad ; \quad \mathbb{H}_0 = \text{span}[\boldsymbol{\lambda}^\dagger]
\end{equation}
and the unidimensional Lie algebra $\mathbb{H}_0$ is either $\mathbb{R}$ or $\uu(1)$ depending on whether the Lie algebra element
$\boldsymbol{\lambda}^\dagger$ is non compact or compact.
\subsubsection{K\"ahler Non-Compact Symmetric Spaces}
In view of Cartan Neural Networks where the relevant manifolds are non-compact symmetric spaces $\mathrm{U/H}$, with $\mathrm{H}\subset \mathrm{U}$, the maximal compact subgroup of a non-compact simple Lie group, it follows that $\mathbb{H}$ is compact and $\mathbb{H}_0 \, = \, \uu(1)$. This has a universal and simple interpretation: the presence in the isotropy subgroup $\mathrm{H}$ of a factor $\mathrm{U(1)}$ simply means that $\mathrm{U/H}$ is a K\"ahler manifold and that the symplectic structure is provided by the K\"ahler $2$-form.
\subsubsection{Hence Two Cases}
Summarizing the previous discussion we conclude that there are just two distinct cases of geometrical thermodynamics related with non-compact symmetric spaces $\mathrm{U/H}$
\begin{description}
  \item[A)] The thermodynamics associated with the Geodesic Dynamical System (GDS) on $\mathrm{U/H}$ where the symplectic structure is that provided by phase-space of the GDS, existing for all manifolds and in particular for all symmetric spaces  $\mathrm{U/H}$.
  \item[B)] K\"ahler  thermodynamics on symmetric spaces $\mathrm{U/H}$ defined by
  \begin{eqnarray}
  \mathrm{G}_{K}\left(\boldsymbol{\beta},\Upsilon\right)&=&
  \frac{\exp\left[-\boldsymbol{\beta}\cdot \boldsymbol{\mathfrak{P}}(\Upsilon )\right]}{Z_K(\boldsymbol{\beta})}\label{gibboneto}\\
  Z_K(\boldsymbol{\beta}) & \equiv & \int_{\mathrm{U/H}}\,  \exp\left[-\boldsymbol{\beta}\cdot \boldsymbol{\mathfrak{P}}(\Upsilon)\right]\,  \underbrace{\mathcal{K}\wedge\mathcal{K}\wedge\dots\wedge\mathcal{K}}_{\text{$n$-times}} \label{partofungo}\\
  \text{dim} \frac{\mathrm{U}}{\mathrm{H}} & = & 2 \, n \quad\quad ; \quad  n\in \mathbb{N}\label{parisono}\\
  \mathcal{K} & = & \text{K\"ahler $2$-form} \label{kallerone}\\
 \mathrm{ i}_{\boldsymbol{\mathfrak{k}}_A} \mathcal{K} &=& \mathrm{d}\mathfrak{P}_A \label{momentico}
\end{eqnarray}
where $\boldsymbol{\mathfrak{P}}(\Upsilon)$ denotes the vector of moment maps $\mathfrak{P}_A(\Upsilon)$ associated with
a basis of Killing vectors $\boldsymbol{\mathfrak{k}}_A$ that correspond to a basis $T_A$ of generators of the $\mathbb{U}$ Lie algebra and $\boldsymbol{\beta} \, = \, \beta^A \, T_A \in \Omega \subset \mathbb{U}$ is a generalized temperature vector such that the partition function integral (\ref{partofungo}) converges.
\end{description}
\subsection{Relevance for Cartan Neural Networks}
\label{rilevancio}
In the new paradigm of Cartan Neural Networks all the manifolds that model the hidden layers and to which data are injected are diffeomorphic to as many solvable Lie groups $\mathcal{S}$. Furthermore they have a simple group $\mathrm{U}$ of isometries leading to a Lie algebra $\mathbb{U}$. For all these manifolds the geodesic dynamical system is completely integrable and one can construct a nice algebraic resetting of the corresponding hamiltonian setup liable to be used in the study of Gibbs states of the conventional type defined by eq.(\ref{predone}). The use of such Gibbs states in Machine Learning algorithms based on the CaNNs paradigm is a perspective to be investigated with care. However it must be noted that, as we show in the sequel, the Gibbs probability distribution depends only the momenta (velocities) and not on the positions in the manifold $\mathrm{U/H}$. Hence if one is interested in probability distributions on the very manifold to which data are mapped, the Geodesic Dynamical System thermodynamics seems to be of little use.
\par
On the other hand in the organization of non compact symmetric spaces into Tits Satake universality classes, there are entire
classes consisting of K\"ahler manifolds, for instance the $r=2$ class (see \cite{pgtstheory} for details on the classification):
\begin{equation}\label{r2class}
  \mathcal{M}^{[2,q]} \, \equiv \, \frac{\mathrm{SO(2,2+q)}}{\mathrm{SO(2)}\times \mathrm{SO(2+q)}}
\end{equation}
Hence for such cases the possibility of defining K\"ahler thermodynamics and corresponding Gibbs states
 {\`a} la Souriau as in eq.(\ref{gibboneto}) arises. Once again the use of such generalized Gibbs states in Machine Learning algorithms has to be studied, yet its viability  is  guaranteed by the already existent applications to radar signal analysis and to other time series discussed in the thesis \cite{capanna} and in all references quoted therein (in particular \cite{radarone}). Indeed such Gibbs states provide a gaussian like probability distribution on the very manifold $\mathrm{U/H}$, rather than on the fibres of its tangent bundle.
\par
The perspective of Gibbs states {\`a} la Souriau for non compact symmetric spaces $\mathrm{U/H}$ that are K\"ahlerian requires a systematic theoretical study which seems to be so far missing, namely that of an intrinsic characterization of the subspace $\Omega$ of generalized temperatures inside the relevant $\mathbb{U}$ algebras. We consider such a study an interesting priority that we address in the present paper in its general form and in the case study of two examples. We found a general answer that was so far missing: the space of generalized temperature is the adjoint orbit of positivity domain in the space of the Cartan subalgebra of the compact subalgebra $\mathbb{H}$ (see section \ref{ciurlacco})
\par
Furthermore let us also recall that in \cite{caldobarbaresco} Barbaresco has claimed that the geometry of Lie Group thermodynamics is to be identified with the \textbf{Riemannian Geometry of Information} introduced several years ago by  Rao \cite{raone} and Chentsov \cite{cenzone} (see the review paper \cite{Nielsen_2020} and the book \cite{amarone} whose author is frequently credited for the introduction of Information Geometry in the Data Science community). Once the conceptual framework is clarified as we hope to have done with the present paper, the relation between K\"ahler thermodynamics and the Riemannian metric naturally associated with equilibrium states, via the general setup of \textbf{generalized geometrical thermodynamics}, becomes clear and universal as we are going to show.
\subsection{Outline of  This Paper}
The present paper is organized as follows.    First we briefly recall the basic principles of generalized thermodynamics and of its link with  Shannon's information functional. Next we analyse the general structure of the Geodesic Dynamical System and its specialization to the case of symmetric spaces $\mathrm{U/H}$. This is instrumental to complete the Poissonian structure on dual solvable Lie algebras to a full symplectic structure on the tangent bundle to non-compact symmetric spaces. In this perspective we can study examples of generalized thermodynamics associated with integrable dynamical systems and show that it is too simple and of little interest in Machine Learning applications. We come next to discuss generalized thermodynamics {\`a} la Souriau and it is in this context that we obtain our most relevant results that are summarized in the conclusive section \ref{concludo}. We do not anticipate them here. We just say that, according to our opinion, thanks to a strategic use of the metric equivalence of non compact symmetric spaces $\mathrm{U/H}$ with appropriate solvable Lie groups we have established generalized thermodynamics
{\`a} la Souriau on clear general principles for all non compact symmetric spaces that are K\"ahler manifolds, introducing, in this way, a new powerful weapon for Machine Learning algorithms.
\par
We have equipped our paper with several mathematical and physical appendices in order to make it self-contained and readable to a larger audience.
\section{Shannon Information Entropy and the Partition Function}
\label{brillantone}
In his celebrated 1948 paper \cite{shannone}, Claude Elwood Shannon introduced what is called the entropy of information relative to a probability density $\rho$ defined on some measurable space $\Omega$ (see appendix \ref{fondiprobi} for
a summary of the fundamental principle and concepts of probability theory as exposed in standard textbooks like \cite{sinaikarolo})
\par
Let $\mathbf{q}\in \Omega$  be a point in the stochastic space we
consider, let $\mathrm{d}\mu(\mathbf{q})$ be the  integration measure on
$\Omega$ and let $\rho(\mathbf{q}) \in [0,1]$  be the value  in
$\mathbf{q}$ of the probability density  that is obviously normalized
as follows:
\begin{equation}\label{normaliz}
 \mathrm{N}[\rho] \,\equiv \,   \int_\Omega \, \rho(\mathbf{q}) \,d\mu(\mathbf{q}) \, = \, 1
\end{equation}
The measure of  information contained in the probability distribution
$\rho$ was  defined by Shannon by means of the following functional
\begin{equation}\label{shaninfo}
    \mathcal{I}\left[\rho\right] \, \equiv \, - \, \int_{\Omega}\,
    \rho(\mathbf{q}) \, \log \, \left[ \rho(\mathbf{q})\right] \,d\mu(\mathbf{q})
\end{equation}
\subsection{Conditional Minimalization of Information and the Partition Function}\label{parcondicio}
The precise conceptual connection between Information Theory and Statistical Mechanics and thus with Thermodynamics can be made through the notion of conditional minimization introduced by Jaynes in 1957 who, in the papers\cite{gianno1,gianno2}, clarified transparently and definitively the logical relationship between Shannon's theory and Statistical Thermodynamics. Thanks to recent works \cite{lychaginlecture,ludaed,ludaed2,Lychagin_2020,Kushner_2020}
this relationship is further clarified in geometric terms and completes the design of the conceptual framework
in which the association of a Riemannian metric with Thermodynamics obtains a solid foundation.
\par
We pose the following problem: determine the probability distribution that extremizes the functional
$\mathcal{I}\left[\rho\right]$ under the following two conditions:
\begin{description}
  \item[A)] The correct normalization (\ref{normaliz}) should hold true.
  \item[B)] The average value of a certain stochastic vector $\mathbf{X}$
  should be fixed to a certain precise vector $\mathbf{x}\in
  \mathbb{V}$:
  \begin{equation}\label{muccacarolina}
    \langle \mathbf{X}\rangle \, \equiv \, \int_{\Omega}
    \,\mathbf{X}(\mathbf{q})\, \rho({\mathbf{q}})\,d\mu(\mathbf{q}) \, =
    \,\mathbf{x}\in\mathbb{V}
  \end{equation}
\end{description}

The classical way to solve this problem is to use variational calculus in the presence of Lagrange multipliers. One
introduces $r+1$ multipliers: $\lambda_0$ associated with the
normalization constraint (\ref{normaliz}) and $r=\text{dim} \mathbb{V}$
multipliers $\lambda^i$ that we can regard as the components
of a vector in $\pmb{\lambda}\in\mathbb{V}^\star$ that are associated with
the constraints (\ref{muccacarolina}). Thus the new functional to be extremized is
\begin{equation}\label{functlag}
    \mathcal{F}[\rho]\, = \, -\,\mathcal{I}\left[\rho\right] \, - \,
    \lambda_0 \, \left(\mathrm{N}[\rho] \, -\, 1 \right) +
    \pmb{\lambda}\cdot\left(\langle \mathbf{X}\rangle -\mathbf{x}
    \right)
\end{equation}
The variation of the functional   in $\delta\rho$  yields:
\begin{equation}\label{ralloppo}
    \frac{\delta\mathcal{F}[\rho]}{\delta\rho} \, = \,\log[\rho] +1
    -\lambda_0 +\pmb{\lambda}\cdot\mathbf{X} \, = \,0
\end{equation}
which implies:
\begin{equation}\label{calindro}
    \rho(\mathbf{q}) \, = \,\exp\left[\lambda_0 - 1-
    \pmb{\lambda}\cdot \mathbf{X}(\mathbf{q})\right]
\end{equation}
Imposing the normalization constraint(\ref{normaliz}) fixes the value of
$\lambda_0$  so that the  final expression of the extremal probability distribution is the following :
\begin{equation}\label{distribuzionepesci}
    \rho_{ex}(\mathbf{q}) \, = \,
    \frac{\exp\left[-\pmb{\lambda}\cdot\mathbf{X}\left(\mathbf{q}
    \right)\right]}{Z\left(\pmb{\lambda}\right)}
\end{equation}
where:
\begin{equation}\label{partifungo}
    Z\left(\pmb{\lambda}\right)\, \equiv \, \int_{\Omega}
    \,\exp\left[-\pmb{\lambda}\cdot\mathbf{X}\left(\mathbf{q}\right)
    \right]
    \,d\mu(\mathbf{q})
\end{equation}
is the \textbf{Partition Function} and, for reasons that will become immediately clear,
the following object:
\begin{equation}\label{tremoamillo}
    \mathcal{H}^{stoch}\left(\pmb{\lambda}\right) \, = \, - \log
    \left[Z\left(\pmb{\lambda}\right)\right]
\end{equation}
is named the \textbf{stochastic Hamiltonian}.
As a consequence of the
definition (\ref{tremoamillo}) the value  $\mathbf{x}$ imposed to
the stochastic vector $\mathbf{X}$  is obtained from the  hamiltonian
by means of a derivative:
\begin{equation}\label{variabiliestensive}
    \mathbf{x}\,= \, \mathrm{d}_{\pmb{\lambda}}\mathcal{H}^{stoch}\left(\pmb{\lambda}
    \right) \,
    \Rightarrow\,
    \text{short hand for }\null \, x_i \, = \, \frac{\partial}{\partial
    \lambda^i} \,\mathcal{H}^{stoch}\left(\lambda_1,\,\dots \, , \, \lambda_r\right)
\end{equation}
Calculating Shannon Entropy Functional (\ref{shaninfo}) on the extremal
probability distribution
(\ref{distribuzionepesci}) with elementary algebra we obtain:
\begin{equation}
   - \mathcal{I}\left[\rho_{ex}\right]\, = \,
    \mathcal{H}^{stoch}\left(\pmb{\lambda}\right) - \pmb{\lambda}\cdot
    \mathbf{x} \, = \, \mathcal{H}^{stoch}\left(\pmb{\lambda}\right)\, -\,
    \lambda^i \, \frac{\partial}{\partial
    \lambda^i} \,\mathcal{H}^{stoch}\left(\pmb{\lambda}\right)
\label{leggendoleggo}
\end{equation}
that  has the  form of a  Legendre transform. Hence the Shannon functional plays the same role as that of a Lagrangian,  the stochastic hamiltonian is indeed a hamiltonian, the intensive variables of Thermodynamics (\textit{i.e.} the Lagrange multipliers $\pmb{\lambda}$ ) are
the momenta and the average values $x^i$ are the coordinates.
\par
Next we turn to classical thermodynamics. We refer to appendix
\ref{richiamatermo} for a recollection of its basic concepts and constructions which are presented in order to fix notation and also for the benefit of those readers who are not physicists by education. In the next section \ref{geostructa}
we illustrate the geometric reformulation of classical thermodynamics in the contest of contact and symplectic geometry, which leads to the introduction of the new notion of thermodynamical curvature.
\par
Note that what is named metric of Information Geometry in the Machine Learning literature is the following Hessian obtained from a parameterized by a vector $\pmb{\lambda} \, =\,\{\lambda_1,\dots, \lambda_r\}$  probability distribution $\rho_{\pmb{\lambda}}(\mathbf{X}(\mathbf{q}))$ of the stochastic variable $\mathbf{X}(\mathbf{q})$ over the manifold of events:
\begin{equation}
\label{parlatore}
ds^{2}_{info} \equiv \frac{\partial^2}{\partial \lambda^i \, \partial \lambda^j} \log \left[\rho_{\pmb{\lambda}}(\mathbf{X}(\mathbf{q}))\right] \, \mathrm{d}\lambda^i \times \mathrm{d}\lambda^j
\end{equation}
When the probability distribution is the generalized Gibbs one of eq.(\ref{distribuzionepesci}) we find:
\begin{equation}
\label{zittatore}
ds^{2}_{info} \equiv \frac{\partial^2}{\partial \lambda^i \, \partial \lambda^j}\, \boldsymbol{\mathcal{H}}^{stoch}(\pmb{\lambda} ) \, \mathrm{d}\lambda^i \times \mathrm{d}\lambda^j
\end{equation}
As we are going to show in next section the metric (\ref{zittatore}) coincides with the thermodynamics metric introduced in geometrical thermodynamics much before the advent of Machine Learning contributions.
\section{Geometrical Structure of Thermodynamics} \label{geostructa}
In this section we present the reformulation of classical thermodynamical laws in geometrical terms on the basis of what  we have
explained in section \ref{brillantone}, where we enlightened the relation between information theory and
statistical mechanics.  It is now time to  show how classical thermodynamical laws are linked
with the notion of a contact manifold  $\left(\mathcal{M}^{2n+1},\xi_\alpha\right)$, defined by a suitable
contact $1$-form $\alpha$ and certain legendrian submanifolds $\mathcal{L}_n\subset
\mathcal{M}^{2n+1}$ of the latter, specifically defined and identifiable with lagrangian submanifolds
$\mathfrak{L}_n \subset \mathcal{S}_{2n}$  of a symplectic manifold $\left(
\mathcal{S}_{2n},\omega \right)$ that is canonically associated with the contact one
according with the scheme \ref{diagrammus}. The lagrangian vision leads to the definition of a canonical
Riemannian metric induced on $\mathfrak{L}_n$, which is the most relevant novelty of the introduced
conceptual framework.
\par
Indeed calculating the \textit{thermodynamical curvature}
is a new powerful investigation tool in all applications.
\par
The intuition of the relevance of \textit{thermodynamical curvature} as a probe
of molecular interactions at the mesoscopic level is indeed, as we stressed in the introductory
section \ref{introibo}, particularly due to Ruppeiner.
\par
There is however something even more perspective that should be stressed.
Recalling the fundamental relation between Information Theory and Statistical Mechanics outlined in section \ref{brillantone}
and contextually illustrated in appendix \ref{richiamatermo} it appears that the geometric reformulation of classical thermodynamics has a much wider scope than physical or chemical systems. Indeed any conditioned probability distribution describing whatever phenomena and fitting to whatever Big Data system defines a thermodynamical setup and would lead to equations of state if we knew the probability distribution and were able to calculate the partition function. The geometrical formulation of the equations of state
as embedding functions of lagrangian submanifolds is a scheme that can be utilized in an inverse engineering procedure to work out the probability distribution and possibly learn it from Data Behavior. This is a challenging possibility for Deep Learning completely  unexplored at the present moment.
\subsection{The Geometric Reformulation}
\label{geoformullo}
To develop the program announced above, we need only to collect the ideas already introduced, focusing on the standard
Darboux expression of a contact form given in eq.(\ref{formalpt}) and
on eq.(\ref{leggendoleggo}) which shows that the Functional measuring Information $\mathcal{I}\left[\rho_{ex}\right]$ is
related to the stochastic Hamiltonian $\mathcal{H}\left(\pmb{\lambda}\right)$
by a Legendre transform. Summarizing we can say that in thermodynamics we have $n+1\geq 3$ extensive variables
collectively denoted $x_0,x_i$ ($i=1,\dots,n$) which explicitly might be identified as:
\begin{enumerate}
  \item Internal Energy $U$
  \item Entropy $S$
  \item Volume $V$
  \item molar fractions $N_\ell$  ($\ell=1,\dots,n-3$)
\end{enumerate}
and $n$-intensive variables  collectively denoted $\lambda^i$
and explicitly identified as:
\begin{enumerate}
  \item Temperature $T$
  \item Pressure $P$
  \item Chemical Potentials  $\mu^\ell$  ($\ell=1,\dots,n-3$)
\end{enumerate}
The first principle of thermodynamics, combined with the second can be formulated by stating
that the following  differential form vanishes:
\begin{equation}\label{coronello}
0 \, \approx \, \tilde{\alpha} \, \equiv \, dU  -T dS \, + P \, dV
\, - \, \sum_{\ell=1}^{n-3}
 \, \mu^\ell \, dN_\ell \, = \, dx_0 \, + \, \sum_{i=1}^n
 \,\lambda^i \, dx_i
\end{equation}
The last  form  \`{a} la Darboux of $\tilde{\alpha}$ follows from the identification
of $x_0$ with the internal energy $U$ and the remaining coordinates
as follows $\pmb{\lambda}\, =
\,\left\{-T,P,-\mu^\ell\right\}$ e $x^i \, =\,
\left\{S,V,N_\ell\right\}$.
Obviously because of its form
\`{a} la Darboux $\tilde{\alpha}$ satisfies the defining condition in order to be
a contact $1$-form, namely:
\begin{equation}\label{crucco}
\underbrace{\mathrm{d}\tilde{\alpha}\wedge
\mathrm{d}\tilde{\alpha}\wedge \dots \wedge
\mathrm{d}\tilde{\alpha}}_{ \text{$n$ times}}\wedge \tilde{\alpha}\,
\neq \, 0
\end{equation}
To better conjugate the emerging contact geometry underlying thermodynamics with the equation (\ref{leggendoleggo}) that identifies, minus a multiplicative factor, the information measure with thermodynamic entropy (see
eq.\ref{ramificato}), it is convenient to multiply the form $\tilde{\alpha}$  introduced in equation
(\ref{coronello}) times a factor
 $ 1/(k_B T)$, obtaining in this way:
\begin{eqnarray}\label{caporale}
   \alpha & = & (k_B T)^{-1}\, \tilde{\alpha} \, = \, -k_B \, \mathrm{d}S +
   (k_B T)^{-1}\, dU +  (k_B T)^{-1}\, P \, \mathrm{d}V \, - \,
   \sum_{\ell=1}^{n-3} \, (k_B T)^{-1}\mu^\ell \,
   \mathrm{d}N_\ell\nonumber\\
   &=&  d\mathcal{I} \, - \, \sum_{i=1}^n \lambda^i dx_i
\end{eqnarray}
where the new definition of the $2n+1$ coordinates is  the following one:
\begin{equation}\label{kirgiso}
    \pmb{\lambda} \, = \,  \left\{-\frac{1}{k_B T},-\frac{P}{k_B T},\frac{\mu^\ell}{k_B
    T}\right\} \quad ; \quad \mathbf{x} \, = \, \left\{ U,
    V,N_\ell\right\} \quad;\quad x_0 \, = \, \mathcal{I}
\end{equation}
Obviously the  $1$-form $\alpha$ satisfies the same condition
(\ref{crucco}) as $\tilde{\alpha}$  and therefore it is also a contact $1$-form.
 Thus we have defined a contact manifold
 $\left(\mathcal{M}_{2n+1},\xi\right)$ where
$\xi=\text{ker}\,\alpha$  is the contact structure
$\mathcal{M}_{2n+1}=\mathbb{R}^{2n+1}$ has $2n+1$ coordinates
$\left\{ I,\pmb{\lambda},\mathbf{x}\right\}$, the variable $I$  being, at this stage, a free
coordinate, just as
$\pmb{\lambda}$ and $\mathbf{x}$.
\subsubsection{Legendrian Submanifolds}
The definition \ref{leggiadro} of legendrian submanifolds given in appendix \ref{geoprobo} states that they are
isotropic submanifolds of maximal dimension $n$ of a contact manifold $\mathcal{M}_{2n+1}$.  On the other hand we recall that
an isotropic submanifold is a submanifold such that its tangent bundle is in the kernel
of the contact form, namely the $1$-form $\alpha$ vanishes on each legendrian submanifold.  The great intuition of the authors
of \cite{ludaed,ludaed2} has been that of identifying, independently from the utilized coordinates and hence in an intrinsic way,
 the \textbf{thermodynamic equilibrium states} with the points
of  specific \textbf{legendrian submanifolds} of the ambient space.
 In terms of the theory  of conditional minimization
discussed in section \ref{parcondicio} it is very simple
to define  the legendrian submanifolds that represents the termodynamic equilibrius states.
\begin{definizione}
\label{defingolegend} Any admissible thermodynamic state can be  identified with a point
in the following  $n$-dimensional submanifold  of the contact manifold:
\begin{equation}
\label{crocolo}
    \mathcal{L}_n \, = \, \left\{ I =
    \mathcal{I}\left(\pmb{\lambda},\mathbf{x}\right)\, , \, x_i \, = \,
    \frac{\partial}{\partial \lambda^i}
    \mathcal{H}\left(\pmb{\lambda}\right)\right\} \, \subset \,\mathcal{M}_{2n+1}
\end{equation}
\end{definizione}
\par
\begin{teorema}
\label{gargantua} The submanifold $\mathcal{L}_n$ defined by means of equation
(\ref{crocolo}) is isotropic and hence legendrian.
\end{teorema}
\begin{proof}
The proof is extremely simple. It suffices to recall equation
(\ref{leggendoleggo}). Using that relation we can evaluate the
total differential $dI$, as it follows:
\begin{equation}
dI \, = \,
\mathrm{d}\mathcal{I}\left(\pmb{\lambda},\mathbf{x}\right)\, =
\,\mathrm{d}\left(H(\pmb{\lambda})-\pmb{\lambda}\cdot\mathbf{x}\right)\,
=\, \left(\frac{\partial}{\partial
\lambda^i}H(\pmb{\lambda})-x^i\right)d\lambda^i \, - \, \lambda^i
\,dx_i \, = \, - \, \lambda^i \,dx_i
\end{equation}
Hence on the submanifold (\ref{crocolo}) we have
$dI+\lambda^i \,dx_i=0$
\end{proof}
\subsubsection{The Lagrangian Submanifold and Its Metric}
\label{logranofino}
Given the original contact variety $\mathcal{M}^{2n+1}$
with the contact $1$-form given by the presentation
(\ref{caporale}), we see at once that the Reeb vector field is
\begin{equation}\label{Reebtermo}
    \mathbf{R}\, = \, \frac{\partial}{\partial \mathcal{I}}
\end{equation}
In fact, it satisfies the two conditions:
\begin{equation}\label{cominternus}
    \alpha\left(\mathbf{R}\right) \, = \, 1 \quad ; \quad
    d\alpha\left(\mathbf{R}, \mathbf{X}\right)\, = \, 0 \quad
    \forall \mathbf{X} \in \Gamma[\mathcal{TM}^{2n+1},\mathcal{M}^{2n+1}]
\end{equation}
On the other hand, from the general discussion in appendix
\ref{liasone}, we know that every $2n$-dimensional submanifold
of a contact manifold $\mathcal{M}^{2n+1}$  that is
transverse to the Reeb vector field of the latter is a symplectic
manifold $\mathcal{S}^{2n}$ whose symplectic  $2$-form is the restriction
to $\mathcal{S}^{2n}$ of the exterior differential of the contact 1-form:
\begin{equation}\label{rovereto}
    \omega \, = \,
   \mathrm{d}\alpha \mid_{\mathcal{S}^{2n}}
\end{equation}
Hence applying these general notions to the case at hand, we see that the  symplectic variety transverse to Reeb's vector
(\ref{Reebtermo}) is given by the following projection map:
\begin{equation}\label{trasversus}
    \pi \, : \, \mathcal{M}^{2n+1} \, \rightarrow \,
    \mathcal{S}^{2n} \quad ; \quad
    \pi(\mathcal{I},\pmb{\lambda},\mathbf{x})\, = \, (\pmb{\lambda},\mathbf{x})
\end{equation}
and the  symplectic $2$-form  is as follows
\begin{equation}\label{adige}
    \omega \, = \, - \, \sum_{i=1}^{n} d\lambda^i \wedge dx_i \, \quad
    ;\quad d\alpha \, = \, \pi^\star(\omega)
\end{equation}
Let us now consider the legendrian submanifold $\mathcal{L}_n
\subset \mathcal{M}^{2n+1}$ which contains the thermodynamic equilibrium states introduced in Definition \ref{defingolegend}.
It is obvious that we can consider its image through the projection map (\ref{trasversus}):
\begin{equation}\label{corneliatutta}
    \mathcal{S}^{2n} \supset \mathfrak{L}_n \, \equiv \, \pi\left(\mathcal{L}_n\right)
\end{equation}
The important result is that the submanifold $\mathfrak{L}_n$
thus defined is a lagrangian submanifold, namely one on which the symplectic form completely vanishes. The demonstration of this fact follows immediately from the definition. In fact, we can translate equation
(\ref{corneliatutta}) into the following constructive definition:
\begin{equation}\label{tarapizo}
    \mathfrak{L}_n \, = \,\left\{ x^i \, = \, \frac{\partial}{\partial \lambda^i}
    \mathcal{H}\left(\pmb{\lambda}\right)\right\}
\end{equation}
Using the latter we find:
\begin{equation}\label{adige}
    \left.\omega\right|_{\mathfrak{L}_n} \, = \,
    - \,\sum_{i,j=1}^{n} d\lambda^i \wedge d\lambda^j \, \partial_i\partial_j
    \mathcal{H}\left(\pmb{\lambda}\right) \, = \,0
\end{equation}
which follows because of the commutativity of the partial derivatives. We can therefore conclude that the thermodynamic  equilibrium states are immersed in a lagrangian submanifold of the thermodynamic symplectic space  $\left(\mathcal{S}^{2n},\Omega\right)$ the
coordinates of this latter being $\pmb{\lambda}$ and $\mathbf{x}$,
in terms of traditional thermodynamic variables, those specified by relations
(\ref{kirgiso}).
\subsubsection{The Canonical  Riemannian Metric on the Lagrangian Submanifold}
The lagrangian submanifold $\mathfrak{L}_n$ is naturally
equipped with a riemannian canonical metric which is the image
\begin{equation}\label{barlengo}
    ds^2_{\mathfrak{L}_n}\, = \, \iota^\star\left(ds^2_{can}\right)
\end{equation}
through the pull-back $\iota^\star$ of the immersion map:
\begin{equation}\label{apnea}
    \iota \, : \, \mathfrak{L}_n \,
    \stackrel{\iota}{\hookrightarrow} \mathcal{S}^{2n}
\end{equation}
of the canonical flat metric on the symplectic ambient manifold:
\begin{equation}\label{perdinotte}
    ds^2_{can} \,=\, \ft 12 \, \sum_{i=1}^n \left(
    d\lambda^i\otimes dx^i\, +\, dx^i \otimes d\lambda^i\right)
\end{equation}
The canonical riemannian metric (\ref{perdinotte}) on the ambient symplectic space corresponds to assuming the standard complex structure and standard symplectic  2-form matrices
\begin{equation}\label{standcompstruc}
  \mathfrak{I}\, = \,\left(
                       \begin{array}{c|c}
                         -\mathit{i} \,\mathbf{1}_{n\times n} &  0\\
                         \hline
                       0 &   \, \mathit{i} \,\mathbf{1}_{n\times n}\\
                       \end{array}
                     \right) \quad ; \quad \mathfrak{K} \, = \, \ft 12 \,\left(
                       \begin{array}{c|c}
                       0 &  \mathit{i} \,\mathbf{1}_{n\times n} \\
                       \hline
                       - \, \mathit{i} \,\mathbf{1}_{n\times n} & 0 \\
                       \end{array}
                     \right)
\end{equation}
so that, according to general theory,  the canonical hermitian metric  is indeed given   by the matrix:
\begin{equation}\label{standhermitmet}
  \mathfrak{G} \, = \, \mathfrak{K}\cdot\mathfrak{I}
\end{equation}
The riemannian metric (\ref{barlengo}) is the one that was promoted to an investigation tool
of mesoscopic physical chemistry of critical phenomena
in particular by Ruppeiner and collaborators.
This  metric would be perfectly defined and calculable if we explicitly knew
the stochastic  Hamiltonian
$\mathcal{H}(\pmb{\lambda})$, since by means of the immersion equations
we get:
\begin{equation}\label{cessiano}
     ds^2_{\mathfrak{L}_n} \, = -\, \mathcal{H}_{ij}
     (\pmb{\lambda}) d\lambda^i
     \times d\lambda^j \quad ; \quad \mathcal{H}_{ij}
     (\pmb{\lambda})\, \equiv \,\partial_i\partial_j\mathcal{H}
\end{equation}
were $\mathcal{H}_{ij}$  is named the hessian. As we already anticipated above, the metric (\ref{cessiano}) exactly coincides with the metric (\ref{zittatore}) and hence with metric (\ref{parlatore}) named the Information Geometry metric in Machine Learning literature, when the probability distribution is the generalized Gibbs distribution (\ref{distribuzionepesci}). This is also reminiscent of the AMSY symplectic formulation of Toric K\"ahler Geometry in the action angle coordinates, for which we refer the reader to \cite{Bianchi_2021,bruzzo2023d3brane} and, for a review, to \cite{fre2023lectures} and to the original references there cited. The problem is that the stochastic Hamiltonian is by definition the negative of the logarithm of the partition function $Z(\pmb{\lambda})$ and generally beyond the reach of explicit analytical computation, as we have repeatedly pointed out. In the absence of this generally inaccessible tool there was an extensive research activity in  devising and proposing phenomenological equations of state, each of which provides an explicit recipe for calculating the metric (\ref{barlengo}). In the next subsection we will examine this methodology in general for the case where
$n=2$, corresponds to the canonical ensemble, and is the most frequently utilized in the field of phenomenological equations of state.
\subsubsection{The Lagrangian Submanifold in the Two-Dimensional Case
and Its Riemannian Structure}
\label{biroccio}
In order to illustrate the general concepts and as a term of comparison with the subsequent examples of
geometrical thermodynamics on Riemannian manifolds and in particular on symmetric spaces, we provide here a brief
sketch of the geometrical treatment for physical thermodynamics. In the simplest situation we deal with a  $4$-dimensional symplectic space
with coordinates $T,P,U,V$, where the former two are intensive quantities (temperature and pression) and the latter two are extensive ones: internal energy and volume (we recall that the fifth coordinate to complete the odd-dimensional contact manifold is the entropy $S$). In the even dimensional space the symplectic $2$-form
$\omega$  is the following:
\begin{equation}\label{simpletformgen}
    \omega \, \equiv \,\mathrm{d}\left[T^{-1}\right ]\wedge \mathrm{d}U\, + \, \mathrm{d}\left[T^{-1}\,P\right]\wedge \mathrm{d}V
\end{equation}
 The following two are assumed to be the embedding equations of the two-dimensional Lagrangian variety:
\begin{description}
  \item[A)] The thermic equation
  \begin{equation}\label{termiceos}
    f(P,T,V,U)\, \equiv \, P\, - \,\mathcal{A}(T,V)
  \end{equation}\label{caloreos}
  \item[B)]The caloric equation
\begin{equation}\label{caloreos}
  g(P,T,V,U) \, \equiv \, U \, -\, \mathcal{B}(T,V)
\end{equation}
\end{description}
The first condition is the  true equation of state. The second must be found in agreement with the
constraint that the hypersurface cut out by the two costraints in the symplectic manifold should be lagrangian (namely should make the $2$-form $\omega$ vanish).
Introducing the short hand
$\mathfrak{w}\equiv\{P,T,V,U\}$ we can write
\begin{equation}\label{sottocoperta}
    \mathcal{SM}_4 \, \supset \, \mathfrak{L}_2 \, = \, \left\{\mathfrak{w}
    \in\mathcal{SM}_4 \, \mid \, f(\mathfrak{w})=0 \,\, \text{and} \,\,
    g(\mathfrak{w})=0\right\}
\end{equation}
The surface $\mathfrak{L}_2$  is lagrangian if the symplectic  $2$-form vanishes
when restricted to it, which is equivlent to say that  the Poisson bracket of the two embedding functions
is zero:
\begin{equation}\label{lagrangianita}
    \left. \omega \right.|_{\mathfrak{L}_2} \, = \, 0 \quad
    \Leftrightarrow \quad \left\{f\, , \, g\right\} \, = \, 0
\end{equation}
Substituting the equations of state (\ref{termiceos}-\ref{caloreos})
into the symplectic form (\ref{simpletformgen}) we obtain:
\begin{equation}\label{loscucheros}
  \left. \omega \right.|_{\mathfrak{L}_2} \, = \,  \frac{\mathrm{d}T\wedge \mathrm{d}V \, \left(\,T
   \partial_T\mathcal{A}(T,V)-\mathcal{A}(T,V)-\partial_V\mathcal{B}(T,V)\, \right)}{T^2}
\end{equation}
Hence the constraint to be satisfied by the two immersion functions
$\mathcal{A}(T,V),\mathcal{B}(T,V)$ in order, for
the  immersed submanifold to be  lagrangian is that the
image through the projection $\pi$ of a legendrian submanifold
$\mathcal{L}_2$  immersed in the  contact manifold
$\mathcal{M}_5$ should be  the following:
\begin{equation}\label{boacostrictor}
    \,T
   \partial_T\mathcal{A}(T,V)-\mathcal{A}(T,V)-\partial_V\mathcal{B}(T,V)\,
   \, = \,0
\end{equation}
Therefore the canonical Riemannian metric  on the  lagrangian submanifold
$\mathfrak{L}_2$  is the following:
\begin{equation}\label{canolagrometro}
    ds^2_{\mathfrak{L}_2} \, = \, \mathrm{d}\left[T^{-1}\right ]\otimes \mathrm{d}\mathcal{B}(T,V)\, +
    \, \mathrm{d}\left[T^{-1}\,\mathcal{A}(T,V)\right]\otimes \mathrm{d}V
\end{equation}
which makes sense if and only if the constraint (\ref{boacostrictor}) is satisfied. We can verify eq.(\ref{boacostrictor}) in the familiar case of Ideal Gases, recalling eq.s (\ref{tabelIG}-\ref{internIG}) from which we see that in the ideal gas case we have:
\begin{equation}\label{barramelone}
  \mathcal{A}(T,V)\, = \, \frac{k_B \,N \, T}{V} \quad ; \quad \mathcal{B}(T,V) \, = \, \ft 32 \, k_B \, T
\end{equation}
In this case the thermodynamical metric (\ref{canolagrometro}) becomes:
\begin{equation}\label{gattolagrometro}
    ds^2_{IG} \, = \,  - \,k_B \, \left( \ft 3 2 \, \left(\frac{dT}{T}\right)^2 \, + \, N \, \left(\frac{dV}{V}\right)^2 \right)
\end{equation}
which is obviously a flat metric. Indeed it suffices to change variables ($T=\sqrt(\ft 2 3) \,\log[x]$, $V=\frac{1}{\sqrt{N}} \, \log[y]$) and (\ref{gattolagrometro}) becomes proportional to the standard Euclidian metric on $\mathbb{R}^2$.
\par
In appendix \ref{vandervallus}, as an illustrative counterexample we briefly discuss the van der Waals model of a
real gas equation of state and we show that the immersion functions of the lagrangian equilibrium submanifold
(\ref{barramelone}) are substituted by the new immersion functions (\ref{AeBVDW}) which also satisfy the
lagrangian constraint (\ref{boacostrictor}), as they should. Correspondingly the flat thermodynamical metric
(\ref{gattolagrometro}) is replaced by its van der Waals equivalent shown in eq.(\ref{VDWmetraRiem}) which is not flat. Its curvature shown in fig.(\ref{curvoilvallo}) has a non trivial behavior and displays a singularity along the critical curve separating the gas from the liquid phase.
\subsection{General Conclusion of This Section}
What we have shown above is the use of the geometrical definition of equilibrium states as lagrangian submanifolds in the context of equation of states for conventional thermodynamical systems. Yet we should keep in mind that the identification of the thermodynamical metric with the hessian of the stochastic hamiltonian  displayed in eq.(\ref{cessiano}) is general and applies to any Gibbs state probability distribution
of whatever type. Indeed given any dynamical system we can define the partition function as in eq.(\ref{predone}) and we obtain the stochastic hamiltonian from eq.(\ref{tremoamillo}). This is true also for the geodesic dynamical system that we discuss in the next section and for K\"ahler thermodynamics of non-compact symmetric spaces discussed in later sections. The thermodynamical curvature as we are going to see exists also in these case and might display singular behaviors signaling critical phenomena.
\section{The Geodesic Dynamical System}
\label{geodynsys}
As recalled in the introduction and fully explained in
\cite{TSnaviga}, Cartan Neural Networks are based on the scheme:
\begin{equation}\label{formaggio}
 \mathbb{V}_{input} \, \underbrace{\stackrel{\iota_{[\mathcal{Q},\Lambda]}}{\hookrightarrow}}_{\text{injection}} \,\underbrace{\mathrm{U_1/H_1} \, \stackrel{\hat{\mathcal{K}}^1_{[\mathcal{W}_1,\Psi_2]}}{\longrightarrow} \, \mathrm{U_2/H_2}\, \stackrel{\hat{\mathcal{K}}^2_{[\mathcal{W}_2,\Psi_3]}}{\longrightarrow}  \,\dots \, \stackrel{\hat{\mathcal{K}}^{N-1}_{[\mathcal{W}_{N-1},\Psi_{N}]}}{\longrightarrow} \,\mathrm{U_N/H_N}}_{\text{hidden layers}} \, \underbrace{\boldsymbol{\Rrightarrow} \, \mathbb{V}_{output}}_{\text{output map}}
\end{equation}
where the hidden layers $\mathrm{U}_i/\mathrm{H}_i$ are non compact symmetric spaces, metrically equivalent to as many solvable Lie groups $\mathcal{S}_i \subset \mathrm{U}_i$ and the maps $K_i \, : \, \mathrm{U_i/H_i} \to \mathrm{U_{i+1}/H_{i+1}}$ are  isometric mapping endowed with general covariance with respect to the transformations of both the $\mathrm{U_{i}}$ and the $\mathrm{U_{i+1}}$ group.
The initial injection map and the final output map depend on the type of task for which the network architecture is designed and from its categorical type. For instance in the frequent  case of the classification task (\textit{e.g} for images) implemented with a simple CaNN, the \textit{initial datum} is regarded as a \textit{single vector} $\Xi_i$ (\textit{e.g.} the list of all pixels) and the \textit{injection map} is a linear relation between the \textit{solvable coordinates}\footnote{See \cite{pgtstheory} and \cite{TSnaviga} for the general definition of solvable coordinates on $\mathrm{U/H}$ manifolds} $\Upsilon_A$ of a point in the first hidden layer and  the components of the datum vector:
\begin{equation}\label{tangoliscio}
  \Upsilon_A \,= \,  Q_{A}^i \, \Xi_i \, + \, \Lambda_A
\end{equation}
where $Q_{A}^i$ is a $\mathrm{dim}[{\mathrm{U_1/H_1}}] \times
\mathrm{dim}[\mathbb{V}_{input}]$ matrix and $\Lambda_A$ a
$\mathrm{dim}[{\mathrm{U_1/H_1}}]$-dimensional vector, both being target of the learning process (see section 6.3 of \cite{TSnaviga}). In the same case of the classification task and in the same categorical type of a simple CaNN, the output map is schematically described below (see eq.(6.11) of \cite{TSnaviga}):
\begin{equation}\label{rimasuglio}
 \underbrace{\boldsymbol{\Rrightarrow} \, \mathbb{V}_{output}}_{\text{output map}} \, = \,  \underbrace{\stackrel{\mathfrak{S}}{\longrightarrow}\, \mathcal{M}_+\cup \mathcal{M}_-}_{\text{partition}} \underbrace{\, \stackrel{\sigma}{\longrightarrow}\,\left[0,1\right]_{out}}_{\text{log. regr.}}
\end{equation}
where $\stackrel{\mathfrak{S}}{\longrightarrow}$ is the partition map of the last layer  into two or more disjoined components, induced by one or more \textbf{separator submanifolds},  defined in Definition 6.1 of \cite{TSnaviga}, whose general constructive theory is  instead presented in \cite{tassellandum}. After separation, the final classification output is obtained with the probabilistic setup of the \textit{logistic regression} or of its multi-component generalization, \textit{i.e.} the \textit{softmax}.
\par
In the case of the Convolutional Cartan Neural Networks outlined in section 1.4 of \cite{tassellandum}, the hidden layers are chosen as  $\mathrm{U_i/H_i}=\mathcal{M}^{[r,q_i]}$ where:
\begin{equation}\label{titstot}
  \mathcal{M}^{[r,q]}\, \equiv \,\frac{\mathrm{SO(r,r+q)}}{\mathrm{SO(r) \times SO(r+q)}}
\end{equation}
Fixing $r$ once for all, the hidden layers have to be regarded as the total spaces of Tits Satake vector bundles sharing the same base manifold, namely the Tits Satake submanifold $\mathcal{M}^{[r,1]}$:
\begin{eqnarray}\label{carneadtwo}
\mathcal{M}^{[r,q]}& = & \,\text{tot}\left[\mathcal{E}^{[r(q-1)]}\right] \nonumber\\
\mathcal{E}^{[r(q-1)]} & \stackrel{\pi_{TS}}{{ \longrightarrow}}& \mathcal{M}^{[r,1]}
\end{eqnarray}
that have structural group
\begin{equation}\label{carneadstruc}
  \mathrm{G_{struc}} \, = \, \mathrm{SO(r)\times SO(q-1)}
\end{equation}
and having as standard fibre a vector space of dimension $r\times (q-1)$
\begin{equation}\label{carneadthree}
  \mathrm{F}  \, = \, \mathbb{V}^{(r\mid q-1)}
\end{equation}
in the mentioned direct product representation of $\mathrm{G_{struc}}$.
\par
In any case the important feature of the manifolds corresponding to all the inner layers of the network is what we already emphasized, namely that they are Cartan-Hadamard manifolds, hence diffeomorphic to $\mathbb{R}^n$ and each metrically equivalent to a suitable solvable Lie subgroup manifold $\mathcal{S} \subset \mathrm{U}$. As such all the $\mathrm{U_i/H_i}$ admit a uniquely defined distance function $\mathrm{d}(u,v)$ between any two points $u,v\in \mathrm{U_i/H_i}$ and such distance function is the length of the unique geodesic arc starting at $u$ and ending in $v$. The existence of a distance function is essential for all Machine Learning algorithms and it is for this reason that the properties of the geodesics, their general construction for $\mathrm{U_i/H_i}$ symmetric manifolds and the structure of the distance function were carefully analyzed in \cite{pgtstheory}.
\par
In the present section we reconsider the problem of geodesics from the point of view of hamiltonian mechanics. This is a necessary step in order to transform the space of geodesics into a symplectic manifold and introduce the thermodynamical structures discussed in sections \ref{brillantone} and
\ref{geostructa}. We begin with a general framework by recasting the geodesic problem on a generic Riemannian manifold $(\mathcal{M},g)$ into the form of a hamiltonian dynamical system. Then we specialize such general setup to the case of $\mathrm{U/H}$ non-compact symmetric spaces, metrically equivalent to appropriate solvable Lie groups, $\mathcal{S}_{\mathrm{U/H}}$ and we show how the general framework specializes to such a case displaying additional relevant properties.
\subsection{The Geodesic Dynamical System in General}
\label{gengeodynsys}
Let $(\mathcal{M},g)$ be a generic finite dimensional Riemannian space $\mathrm{dim}_\mathbb{R}(\mathcal{M}) \,=\, d < \infty$ and let us consider the problem of deriving the second order differential equations whose solutions are its geodesic curves:
\begin{equation}\label{geocurvina}
\gamma \quad : \quad  \mathbb{R}\, \longrightarrow \, \mathcal{M}
\end{equation}
In each coordinate patch $x^\alpha \, = \, \{x^1,\dots , x^d\}$ every geodesic is described by $d$ functions $x^\alpha(t)$, where $t\in \mathbb{R}$ is the affine parameter. As explained in \cite{pietrobook} (volume 1, page 145) the well known geodesic differential equations can be derived as the Euler Lagrange equations of a mechanical system whose Lagrangian is the following:
\begin{equation}\label{lagrangeod}
  \mathcal{L}(x,\dot{x}) \, = \, \ft 12 \, g_{\alpha\beta}(x) \,
  \dot{x}^\alpha \, \dot{x}^\beta
\end{equation}
having denoted by $\dot{x}^\alpha \equiv d{x}^\alpha/dt$ the generalized velocities and by $g_{\alpha\beta}(x)$ the metric tensor. We can easily convert the lagrangian system (\ref{lagrangeod}) into a hamiltonian one introducing as usual the canonical momenta:
\begin{equation}\label{canmoment}
  p_\alpha \, \equiv \, \frac{\partial \mathcal{L}}{\partial \dot{x}^\alpha} \, = \, g_{\alpha\beta}(x) \, \dot{x}^\beta
  \quad \Rightarrow \quad \dot{x}^\alpha \, = \, g^{\alpha\beta}(x) \, p_\beta
\end{equation}
where $g^{\alpha\beta}(x)$ is the inverse metric, and defining the Hamiltonian as the Legendre transform of the Lagrangian:
\begin{equation}\label{hamillaregia}
  \mathcal{H}(p,x) \, \equiv \, \dot{x}^\alpha \, p_\alpha \,  - \,
  \mathcal{L}(x,\dot{x}) \, = \, \ft 12 \, g^{\alpha\beta}(x) \,
  p_\alpha \, p_\beta
\end{equation}
According with the definitions and conventions of appendix \ref{pescione} we define the canonical symplectic manifold introducing the $2d$ canonical coordinates:
\begin{equation}\label{cannonicordi}
  Z^\Lambda \, = \,\left\{p_\alpha \, , \, x^\beta \right\} \quad; \quad \alpha,\beta\, = \,1, \dots,d
\end{equation}
and the symplectic $2$-form:
\begin{equation}\label{symp2form}
  \omega \, \equiv\, \omega_{\Lambda\Sigma} \, \mathrm{d}Z^\Lambda \wedge \mathrm{d}Z^\Sigma  \quad ; \quad
  \omega_{\Lambda\Sigma} \, = \, \ft 12 \, \left(
                                   \begin{array}{c|c}
                                     \mathbf{0}_{d\times d} & \mathbf{1}_{d\times d} \\
                                     \hline
                                    \mathbf{ -}\,\mathbf{1}_{d\times d} & \mathbf{0}_{d\times d} \\
                                   \end{array}
                                 \right)
\end{equation}
that agrees with definition \ref{simp2def}, being closed, non degenerate and of maximal rank as in eq.(\ref{simplicioformo}).
In terms of canonical momenta and canonical coordinates one has:
\begin{equation}\label{cromellito}
  \omega \, = \, \sum_{\alpha = 1}^d \, dp_\alpha \wedge dx^\alpha
\end{equation}
and according with the definition \ref{simp2def}, the total space $\mathrm{tot}\left[\mathcal{TM}\right]$ of the tangent bundle
\begin{equation}\label{ghino}
  \mathcal{TM}\, \stackrel{\pi}{\longrightarrow} \,\mathcal{M}
\end{equation}
to the Riemannian manifold $(\mathcal{M},g)$
equipped with the $2$-form $\omega$, locally defined in each coordinate patch $U$, by the expression (\ref{cromellito}), becomes a symplectic $2d$-dimensional manifold $\left(\mathrm{tot}\left[\mathcal{TM}\right],\omega\right)$ namely the phase-space of the geodesic dynamical system.
Since $\omega$ is non degenerate, the symplectic manifold $\left(\mathrm{tot}\left[\mathcal{TM}\right],\omega\right)$ is also a Poisson-manifold. Indeed in every coordinate patch it suffices to invert the matrix $\omega_{\Lambda\Sigma}$, so as to obtain the \textbf{Poissonian bivector}:
\begin{equation}\label{crisalide}
  \pi^{\Lambda\Sigma} \, = \, \left(\omega^{-1}\right)^{\Lambda\Sigma}
\end{equation}
and given any function $f(Y) \in \mathcal{C}^\infty\left(\mathrm{tot}\left[\mathcal{TM}\right] \right)$, according with eq.(\ref{carneadehamilt}) we can associate to it, the corresponding hamiltonian vector field:
\begin{equation}\label{carolinagalbani}
  \boldsymbol{X}_f \, \equiv \, \pi^{\Lambda\Sigma}\, \partial_\Lambda f \, \partial_\Sigma \quad ; \quad \partial_\Gamma \, \equiv \, \frac{\partial}{\partial Z^\Gamma}
\end{equation}
and we obtain the Poisson bracket fulfilling the properties listed in its definition \ref{poissone} by setting:
\begin{equation}\label{curmatore}
\forall \, f,g \, \in \mathcal{C}^\infty\left(\mathrm{tot}\left[\mathcal{TM}\right] \right)\quad : \quad \left\{f,g\right\} \, = \, \omega\,\left(\boldsymbol{X}_f
\, , \, \boldsymbol{X}_g\right)
\end{equation}
like in eq.(\ref{agniziono}).
\par
The standard geometric geodesic equation:
\begin{equation}\label{geodequate}
  \ddot{x}^\alpha \, + \, \Gamma^\alpha_{\beta\gamma}(x)\,
  \dot{x}^\beta \dot{x}^\gamma \, = \, 0
\end{equation}
where
\begin{equation}\label{cristoffelo}
  \Gamma^\alpha_{\beta\gamma}(x) \, = \, \ft 12 \, g^{\alpha\mu} \left(\partial_\beta g_{\gamma\mu} + \partial_\gamma g_{\beta\mu}\, - \, \partial_\mu g_{\beta\gamma} \right)
\end{equation}
denotes the Christoffel symbols, \textit{i.e.} the components of the Levi Civita connection on $(\mathcal{M},g)$ is retrieved
 from the hamiltonian equations:
\begin{alignat}{6}\label{hamilleque}
  \dot{x}^\alpha \quad & = &\quad \left\{\mathcal{H}\, , \, x^\alpha \right\} \quad& = &\quad &\phantom{-\, \ft 12\,\,} g^{\alpha\nu} \, p_\nu  \null \nonumber \\
 \dot{p}_\alpha \quad & = &\quad \left\{\mathcal{H}\, , \, p_\alpha \right\} \quad& = & \quad & - \, \ft 12 \, \partial_\alpha g^{\rho\sigma} \, p_\rho\,p_\sigma  \null
\end{alignat}

\subsection{The Geodesic Dynamical System for Non Compact Symmetric Spaces}
\label{zaklyuchenie}
In the case of non-compact symmetric spaces $\mathrm{U/H}$, thanks to their metric equivalence with a suitable solvable group manifold $\mathcal{S}_{\mathrm{U/H}}$, the structure of the geodesic dynamical system can be recast into a more algebraic and very convenient form. To this effect we observe that the unique Einstein $\mathrm{U}$-invariant metric on $\mathrm{U/H}$ can be written
as:
\begin{equation}\label{cordelia}
  ds^2 \, \equiv \, \mathbf{g} \, = \, \kappa_{AB} \, e^A \times e^B
\end{equation}
where $\kappa_{AB}\, = \, \kappa_{BA}$ is a \textbf{constant symmetric matrix} and
\begin{equation}\label{forminesabbia}
  e^A \, = \, e(\boldsymbol{\Upsilon})^{A}_{\phantom{A}\alpha} \, \mathrm{d}\Upsilon^\alpha
\end{equation}
are the \textbf{left-invariant Maurer Cartan $1$-forms} on the group manifold $\mathcal{S}_{\mathrm{U/H}}$ satisfying the Maurer Cartan equations of the solvable Lie algebra $Solv_{\mathrm{U/H}}$:
\begin{equation}\label{MCequesolv}
  \mathrm{d}e^A \, + \, \ft{1}{2} \, f^{\phantom{BC}A}_{BC} \, e^B \wedge e^C \, = \, 0
\end{equation}
having denoted by $\boldsymbol{\Upsilon}$ the solvable coordinates, namely the parameters of the solvable Lie Group
$\mathcal{S}_{\mathrm{U/H}} \subset \mathrm{U}$ and by $f^{\phantom{BC}A}_{BC}$ the solvable Lie algebra structure constants (see \cite{pgtstheory} and \cite{TSnaviga} for further details).
Metric equivalence is nothing more than  eq.s(\ref{cordelia},\ref{MCequesolv}). Indeed let
\begin{equation}\label{cartaccia}
  \mathbf{t}_A \, = \, e(\boldsymbol{\Upsilon})^{\phantom{A}\alpha}_A \,   \frac{\partial}{\partial \Upsilon^\alpha}
\end{equation}
be a basis of sections of the tangent bundle $\mathcal{T}(\mathrm{U/H})$ made of \textit{left-invariant vector fields dual to the left-invariant one-forms $e^A$}:
\begin{equation}\label{neanderthal}
  e^A\left(\mathbf{t}_B\right) \, \equiv \, e(\boldsymbol{\Upsilon})^{A}_{\phantom{A}\alpha}\, e(\boldsymbol{\Upsilon})^{\phantom{B}\alpha}_B \, = \, \delta^A_B \quad ; \quad \left[ \mathbf{t}_B \, , \, \mathbf{t}_C\right] \, = \, f^{\phantom{BC}A}_{BC} \,\mathbf{t}_A
\end{equation}
The linear combinations with constant coefficients of the $\mathbf{t}_A$ vector fields constitute a vector space endowed with a Lie bracket that is the very definition of the Lie algebra $Solv_{\mathrm{U/H}}$ of the solvable group $\mathcal{S}_{\mathrm{U/H}}$ (see for instance \cite{fre2023book}):
\begin{equation}\label{algsolvay}
  \mathbf{X} \in Solv_{\mathrm{U/H}} \, \Leftrightarrow \,
  \mathbf{X}\, = \,X^A \, \mathbf{t}_A \quad ; \quad X^A \in \mathbb{R}
\end{equation}
and we obtain:
\begin{equation}\label{solvanorma}
  \forall \mathbf{X},\mathbf{Y} \in Solv_{\mathrm{U/H}} \quad : \quad \mathbf{g}(\mathbf{X},\mathbf{Y}) \, = \, \kappa_{AB}\, X^A Y^B \, \equiv \, \boldsymbol{\kappa}(\mathbf{X},\mathbf{Y})
\end{equation}
In this way we see that once reduced to the \textbf{left-invariant vector fields} the metric is a scalar product on the solvable Lie Algebra, and the symmetric matrix $\kappa_{AB}$ provides the coefficients of the corresponding symmetric quadratic form. 
Conversely any quadratic form $\boldsymbol{k}(\,, \,)$ on $Solv_{\mathrm{U/H}}$  induces a metric on the solvable Lie group $\mathcal{S}_{\mathrm{U/H}}$:
\begin{equation}\label{altrametra}
  {ds}_k^2 \, = \, k_{AB} \, e^A \times e^B
\end{equation}
that is $\mathcal{S}_{\mathrm{U/H}}$-invariant but not necessarily $\mathrm{U}$-invariant, nor Einstein. So any { positive definite quadratic bilinear} $\boldsymbol{k}(\,, \,)$ equips $\mathcal{S}_{\mathrm{U/H}}$ with the structure of an \textit{Alekseevskian normal Riemannian space} \cite{Alekseevsky1975,Alekseevsky:2003vw,Cortes,TSnaviga}, yet there is only one quadratic form that corresponds to the unique Einstein metric\footnote{The Einstein metric is unique up to a homothety, namely up to a overall constant rescaling of the metric tensor or of the vielbein and such is the corresponding invariant quadratic form on the solvable Lie algebra.} of the symmetric space $\mathrm{U/H}$.
\par
The structure of eq.(\ref{cordelia}) being clarified we reconsider the generic form of the geodesic dynamical system lagrangian \ref{lagrangeod} and we rewrite it as:
\begin{equation}\label{uhlagrageo}
  \mathcal{L}_{geoUH} \, = \, \ft 12 \,\kappa_{AB} \,
  e(\boldsymbol{\Upsilon})^{A}_{\phantom{A}\alpha} \,\dot{\Upsilon}^\alpha \,e(\boldsymbol{\Upsilon})^{B}_{\phantom{B}\beta} \dot{\Upsilon}^\beta
\end{equation}
Next we introduce the \textit{anholonomic lagrangian velocities}\footnote{The anholonomic lagrangian velocities are defined as the velocities in an anholonomic basis and, as such, should not be intended as time derivatives of coordinates. By an abuse of notation we still denote them using an upper dot.}
\begin{equation}\label{anhovelo}
  \dot{\mathfrak{q}}^A \, \equiv \, e(\boldsymbol{\Upsilon})^{A}_{\phantom{A}\alpha} \,\dot{\Upsilon}^\alpha \, = \, i_{\partial_t} e^{A}
\end{equation}
which are just the contraction of the Maurer Cartan $1$-forms with the time derivative. Similarly we introduce the \textit{anholonomic hamiltonian momenta} as:
\begin{equation}\label{anhomome}
  {\mathfrak{p}}_A \, \equiv \, \frac{\partial \mathcal{L}_{geoUH} }{\partial \dot{\mathfrak{q}}^A}\, =\,   \kappa_{AB} \,  \dot{\mathfrak{q}}^B
\end{equation}
Then the hamiltonian is defined as usual by Legendre transform
\begin{equation}\label{anhohamilla}
  \mathcal{H}_{geoUH} \, = \, \dot{\mathfrak{q}}^A \,{\mathfrak{p}}_A \, - \, \mathcal{L}_{geoUH} \, = \, \, \ft 12 \,\kappa^{AB} \, {\mathfrak{p}}_A \, {\mathfrak{p}}_B
\end{equation}
where according with standard conventions $\kappa^{AB}$ is the inverse of the quadratic form matrix $\kappa_{AB}$:
\begin{equation}\label{grimaldello}
  \kappa^{AB} \, \kappa_{BC} \, = \, \delta^A_C
\end{equation}
\subsubsection{The Symplectic $2$-Form}
Next we have to convert the standard symplectic form of eq. (\ref{cromellito}) to the new basis of hamiltonian coordinates:
\begin{equation}\label{newhamilcord}
  \Phi^\Lambda \, = \, \left\{\mathfrak{p}_A \, , \, \Upsilon^\alpha\right\}
\end{equation}
We write the identification:
\begin{equation}\label{crulino}
  \omega \, = \, \omega_{\Lambda\Sigma}(Z)\, \mathrm{d}Z^\Lambda \wedge \mathrm{d}Z^\Sigma \, = \, \hat{\omega}_{\Lambda\Sigma}(\Phi) \,
  \mathrm{d}\Phi^\Lambda \wedge \mathrm{d}\Phi^\Sigma
\end{equation}
so that our convention is that we name $\omega_{\Lambda\Sigma}$
the components of $\omega$ in the old hamiltonian basis and
$\hat{\omega}_{\Lambda\Sigma}$ the components of the same 2-form
in the new hamiltonian basis. The derivation of $\hat{\omega}_{\Lambda\Sigma}$ is the simple and direct calculation sketched below:
\begin{eqnarray}
\label{cicorietta}
  -\,\omega &=& \mathrm{d}\Upsilon^\alpha \wedge \mathrm{d}p_\alpha \, = \,\mathrm{d}\Upsilon^\alpha \wedge \mathrm{d}\left[g_{\alpha\beta} \dot{\Upsilon}^\beta\right]\, = \,\mathrm{d}\Upsilon^\alpha \wedge \mathrm{d}\left[e^P_\alpha(\boldsymbol{\Upsilon})\,\kappa_{PQ}
  \,e^Q_\beta(\boldsymbol{\Upsilon}) \,   \dot{\Upsilon}^\beta\right] \nonumber\\
  \null &=& \mathrm{d}\Upsilon^\alpha \wedge \mathrm{d}\Upsilon^\mu \, \partial_\mu e^P_\alpha(\boldsymbol{\Upsilon}) \left (\kappa_{PQ}
  \,e^Q_\beta(\boldsymbol{\Upsilon})\,\dot{\Upsilon}^\beta)\right)+ \mathrm{d}\Upsilon^\alpha \, e^P_\alpha(\boldsymbol{\Upsilon}) \wedge \mathrm{d}\left[\kappa_{PQ}
  \,e^Q_\beta(\boldsymbol{\Upsilon})\,\dot{\Upsilon}^\beta\right]     \nonumber\\
  \null &=& - de^A \,\mathfrak{p}_A + e^A \wedge \mathrm{d}\mathfrak{p}_A \nonumber\\
  \null &=& \ft 12 \, f^{\phantom{BC}A}_{BC} \,e^B \wedge e^C \, + \, e^A \wedge \mathrm{d}\mathfrak{p}_A
\end{eqnarray}
Summarizing we have the simple and elegant formula:
\begin{equation}\label{crautibavaresi}
  \omega \, = \, - \, \ft 12 \, f^{\phantom{BC}A}_{BC} \,e^B \wedge e^C \,\mathfrak{p}_A \, - \, e^A \wedge \mathrm{d}\mathfrak{p}_A
\end{equation}
which is defined on the total space of the tangent bundle of the symmetric space, coinciding with the total space of the tangent bundle of the corresponding solvable Lie group manifold:
$\mathcal{T}(\mathrm{U/H})\simeq \mathcal{T}(\mathcal{S}_\mathrm{U/H})$. The $2$-form $\omega$ is closed and of maximal rank:
\begin{eqnarray}
\label{sgraffio}
  d\omega &=& 0 \nonumber\\
  \underbrace{\omega\wedge\omega\wedge\dots\wedge\omega}_{\text{$d$-times}} &\neq& 0 \quad ; \quad d\, \equiv \, \mathrm{dim}_\mathbb{R} \left[\frac{\mathrm{U}}{\mathrm{H}} \right] \, = \, \mathrm{dim}_\mathbb{R} \left[\mathcal{S}_{\mathrm{U/H}} \right]
\end{eqnarray}
The first line in eq.(\ref{sgraffio}) follows from the consistency of the Maurer Cartan equations (\ref{MCequesolv}) while the second is evident since we get:
\begin{equation}\label{volumetto}
  \omega\wedge\omega\wedge\dots\wedge\omega \, = \, \underbrace{\text{const}}_{\neq \, 0} \,\times \,  \underbrace{e^1\wedge e^2 \wedge \dots \wedge e^d \wedge \mathrm{d}\mathfrak{p}_1 \wedge \dots \wedge \mathrm{d}\mathfrak{p}_d}_{\mathrm{Vol}(\mathcal{M}_{2d})}
\end{equation}
where by $\mathrm{Vol}(\mathcal{M}_{2d})$ we have denoted the top $2d$-form on the manifold $\mathcal{M}_{2d}$ defined as the total space of the tangent bundle $\mathcal{T}(\mathcal{S}_\mathrm{U/H})$ :
\begin{equation}\label{totalbenzina}
  \mathcal{M}_{2d} \, = \, \mathrm{tot}\left[\mathcal{T}(\mathcal{S}_\mathrm{U/H})\right]
\end{equation}
Hence the pair $(\mathcal{M}_{2d},\omega)$ as defined by eq.s (\ref{totalbenzina},\ref{crautibavaresi}) is a bona-fide symplectic manifold as defined and illustrated in appendix \ref{pescione} and such a statement is true for the solvable Lie group $\mathcal{S}_{\mathrm{U/H}}$ singled out by any non-compact symmetric space $\mathrm{U/H}$ with $\mathrm{U}$ simple as thoroughly discussed in \cite{pgtstheory,TSnaviga}.
\subsubsection{The Poissonian Bi-Vector}
According with the general theory discussed in section \ref{pescione} and partially already recalled above in eq.s(\ref{crisalide}-\ref{curmatore}), every symplectic manifold
is also a Poissonian manifold, although the reverse is not true. Indeed the \textit{Poissonian bivector} $\pi^{\Lambda\Sigma}$ can be obtained from the symplectic $2$-form components as in
eq.(\ref{crisalide}). Therefore our next task is that of retrieving the Poissonian bivector starting from the symplectic $2$-form in eq.(\ref{crautibavaresi}). This is easily done. In the coordinate basis (\ref{newhamilcord}) the $2d\times 2d$ matrix $\hat{\omega}_{\Lambda\Sigma}$ has the following structure:
\begin{eqnarray}\label{castigliano}
  \hat{\omega}_{\Lambda\Sigma} & = &
  \left(
  \begin{array}{c|c}
  \mathbf{0}_{d\times d} & \hat{\omega}^M_{\phantom{M}\beta} \\
  \hline
  \hat{\omega}_{\alpha}^{\phantom{\alpha}N} & \hat{\omega}_{\alpha\beta} \\
  \end{array}
  \right) \nonumber\\
 \hat{\omega}_{\alpha\beta} & = & - \ft 12 \, f^{\phantom{BC}A}_{BC} \,\mathfrak{p}_A\,e^B_\alpha \, e^C_\beta \quad ; \quad
  \hat{\omega}^M_{\phantom{M}\beta}\, = \, \ft 12 \, e^M_\beta \quad ; \quad
  \hat{\omega}_{\alpha}^{\phantom{\alpha}N} \, = \, - \, \ft 12 \, e^N_\alpha
\end{eqnarray}
The Poissonian bivector is an antisymmetric matrix $\pi^{\Lambda\Sigma}$ such that:
\begin{equation}\label{farlocco}
  \omega_{\Lambda\Sigma} \,\pi^{\Sigma\Delta} \, = \, \delta^\Delta_\Lambda
\end{equation}
We immediately find :
\begin{eqnarray}\label{manchego}
  \pi_{\Lambda\Sigma} & = &
  \left(
  \begin{array}{c|c}
  \pi_{MN} & \pi_M^{\phantom{M}\beta} \\
  \hline
  \pi^{\alpha}_{\phantom{\alpha}N} & \mathbf{0}_{d\times d} \\
  \end{array}
  \right) \nonumber\\
 \pi^{MN} & = &  -\, 2  \, f^{\phantom{MN}A}_{MN} \, \mathfrak{p}_A \quad ; \quad
  \pi_M^{\phantom{M}\beta}\, = \, - 2 \, e_M^\beta \quad ; \quad
  \pi^{\alpha}_{\phantom{\alpha}N} \, = \, 2 \, e_N^\alpha
\end{eqnarray}
where $e^N_\alpha$ are the components of the \textbf{left-invariant Maurer Cartan forms} $e^M$ and $e_M^\beta$ the components of their dual \textbf{left-invariant vector fields} on the solvable Lie group manifold:
\begin{equation}\label{porcellino}
  e^A \, = \, e^A_\alpha \, \mathrm{d}\Upsilon^\alpha \quad ; \quad ; \quad \mathbf{t}_B \, = \,e_B^\beta \, \frac{\partial}{\partial \Upsilon^\beta}  \quad ;
  \quad e^A(\mathbf{t}_B) \, = \, e^A_\alpha \, \,e_B^\alpha \, = \, \delta^A_B
\end{equation}
that generate right-translations.
\subsubsection{Hamiltonian Vector Fields and the Poisson Bracket}
Having determined the Poissonian bivector, we can write the explicit form of the hamiltonian vector field associated to any
function $\varphi(\Phi) \in \mathbb{C}^\infty\left (\mathrm{tot}\left[\mathcal{T}(\mathcal{S}_\mathrm{U/H})\right] \right)$. Recalling eq.(\ref{carolinagalbani}) we set:
\begin{alignat}{5}\label{hamilfreccia}
  \varphi(\Phi) & \quad \rightarrow \quad & \mathbf{X}_\varphi & = & \quad\pi^{\Lambda\Sigma}\, \partial_\Lambda\varphi \, \frac{\partial}{\partial \Phi^\Sigma} \quad\quad\quad\quad\quad\quad\quad\quad\quad\quad\quad\quad\quad\quad & \null \nonumber\\
   \null & \null & \null & = & \quad - 2 \, \mathfrak{p}_A \, f^{\phantom{BC}A}_{BC} \, \frac{\partial \varphi}{\partial \mathfrak{p}_B} \, \frac{\partial }{\partial \mathfrak{p}_C}\,- 2 \, \frac{\partial \varphi}{\partial \mathfrak{p}_M} \, \mathbf{t}_M \, + 2 \, \mathbf{t}_M\varphi \, \frac{\partial }{\partial \mathfrak{p}_M} & \null
\end{alignat}
and we define the Poisson bracket as:
\begin{equation}\label{corbellus}
\begin{array}{rcccl}
  \forall \varphi(\Phi),\psi(\Phi) \in \mathbb{C}^\infty\left (\mathrm{tot}\left[\mathcal{T}(\mathcal{S}_\mathrm{U/H})\right] \right) & : & \left\{\varphi\, , \, \psi\right\} & \equiv & \omega\left(\mathbf{X}_\varphi \, ,\,\mathbf{X}_\psi \right) \\
  \null & \null & \null & = & -2 \left( \, \mathfrak{p}_A \,f^{\phantom{BC}A}_{BC} \, \frac{\partial \varphi}{\partial \mathfrak{p}_B} \, \frac{\partial \psi }{\partial \mathfrak{p}_C}\,+\, \frac{\partial \varphi}{\partial  \mathfrak{p}_M} \, \mathbf{t}_M \psi \, - \, \mathbf{t}_M\varphi \, \frac{\partial \psi}{\partial \mathfrak{p}_M} \right) \\
  \end{array}
\end{equation}
\subsubsection{Symplectic Moment Map}
Given a vector field $\mathbf{k} \in \Gamma\left[\mathcal{TM},\mathcal{M}\right]$, namely a section of the tangent bundle to\footnote{Here and in the following lines we are always talking about the solvable Lie group $\mathcal{S}$ metrically equivalent to the symmetric space $\mathrm{U/H}$ and for notation simplicity we drop the subscrit $\mathrm{U/H}$.}
\begin{equation}\label{matotto}
  \mathcal{M} = \mathrm{tot}[\mathcal{TS}]
\end{equation}
which has the symplectic structure specified by the $2$-form  (\ref{crautibavaresi}) we define the moment map:
\begin{equation}\label{momentimappo}
  \mu \quad : \quad \mathbf{k} \, \longrightarrow \, \mu_\mathbf{k}\left(\Phi \right) \in \mathbb{C}^{\infty} \left(\mathcal{M}\right)
\end{equation}
by imposing the condition:
\begin{equation}\label{obbligohamil}
\forall f(\Phi)\in \mathbb{C}^{\infty} \quad : \quad
\mathbf{k} f\, = \, \left\{\mu_\mathbf{k}\, , \, f\right\} \, = \, \omega\left(\mathbf{k}\, , \, \mathbf{X}_f \right)
\end{equation}
where $\mathbf{X}_f$ is the hamiltonian vector field associated with the function $f$ (see eq.(\ref{hamilfreccia})). Hence the moment map $\mu_\mathbf{k}$ is a solution to the following differential equation:
\begin{equation}\label{diffeque}
  \mathbf{k} \, = \,- 2 \, f^{\phantom{BC}A}_{BC} \, \mathfrak{p}_A \, \frac{\partial \mu_\mathbf{k}}{\partial \mathfrak{p}_B} \, \frac{\partial }{\partial \mathfrak{p}_C}\,- 2 \, \frac{\partial \mu_\mathbf{k}}{\partial \mathfrak{p}_M} \, \mathbf{t}_M \, + 2 \, \mathbf{t}_M\mu_\mathbf{k} \, \frac{\partial }{\partial \mathfrak{p}_M}
\end{equation}
Consider in particular the vector fields
\begin{equation}\label{cartieradiriso}
  \mathbf{k}_N \, \equiv \, \mathbf{t}_N \, + \, \mathbf{v}_N
\end{equation}
where $\mathbf{t}_N$ are  the purely \textbf{horizontal}, right-invariant vector fields  defined over the solvable group manifold $\mathcal{S}$ that generate its solvable Lie algebra $Solv$ since
\begin{equation}\label{ticommi}
  \left[\mathbf{t}_N\, , \,\mathbf{t}_R  \right] \, = \,
  f^{\phantom{NR}A}_{NR} \, \mathbf{t}_R
\end{equation}
while
\begin{equation}\label{cremacatalana}
  \mathbf{v}_N \, \equiv \, f^{\phantom{NB}A}_{NB}\,\mathfrak{p}_A \, \frac{\partial}{\partial \mathfrak{p}_B}
\end{equation}
are  purely \textbf{vertical} vector fields that, as a cosequence of the Jacobi identities satisfy the commutation relations of $Solv$:
\begin{equation}\label{vicommi}
  \left[\mathbf{v}_N\, , \,\mathbf{v}_R  \right] \, = \,
  f^{\phantom{NR}A}_{NR} \, \mathbf{v}_R
\end{equation}
and commute with the horizontal partners:
\begin{equation}\label{zerocommi}
  \left[\mathbf{v}_N\, , \,\mathbf{t}_R  \right]\, = \, 0
\end{equation}
Hence the vector fields $\mathbf{k}_N$ also satisfy the solvable Lie algebra commutation relations:
\begin{equation}\label{vicommi}
  \left[\mathbf{k}_N\, , \,\mathbf{k}_R  \right] \, = \,
  f^{\phantom{NR}A}_{NR} \, \mathbf{k}_R
\end{equation}
and are the infinitesimal generators for the action of the solvable Lie group $\mathcal{S}$ on the total space of its tangent bundle, namely the phase-space of the geodesic dynamical system.
 We claim that for the vector fields $\mathbf{k}_N$ the appropriate moment map is the following
\begin{equation}\label{aspettamo}
  \mu_N \, \equiv \, \mu_{\mathbf{t}_N} \, = \, \, - \, \ft 12 \, \mathfrak{p}_N
\end{equation}
\subsubsection{Relation with the Nomizu Operator}
\label{numiziato}
\par
Once the metric form is given, the construction of geometry and of
the associated geodesic equations follows uniquely. The issue is
just that of calculating the Levi--Civita connection of the metric
$g$ induced on the manifold by the form $< \, ,\, >$ defined on
the solvable Lie algebra $Solv$. One way of describing this Levi--Civita
connection is by means of the so called \textit{Nomizu operator}
acting on  $Solv$. The latter is defined as follows:
\begin{eqnarray}\label{nomizuAb}
 \mathbb{L}& : & Solv \, \otimes\,
Solv \rightarrow  Solv, \\
 \forall X, Y, Z \in
Solv & : & 2<\mathbb{L}_XY,Z> = <[X,Y],Z> - <X,[Y,Z]> - <Y,[X,Z]>\nonumber
\end{eqnarray}
The \textit{Riemann curvature operator} on $Solv$ can be expressed as
\begin{equation} \Riem(X,Y) =
[\mathbb{L}_X,\mathbb{L}_Y] - \mathbb{L}_{[X,Y]}~
\end{equation}
If we introduce a basis of abstract generators $\{T_A\}$ for $Solv$
and the corresponding structure constants:
\begin{equation}\label{strutteconste}
    \left [ T_A \, , \, T_B \right ] \, = \, f_{AB}^{\phantom{AB}C} \, T_C
\end{equation}
that are the same as those appearing in eq.(\ref{MCequesolv},\ref{neanderthal}) together with the metric tensor:
\begin{equation}\label{Gciccius}
    < T_A ,T_B> \, = \, \kappa_{AB}
\end{equation}
the connection defined by eq.(\ref{nomizuAb}) leads to the following  connection coefficients:
\begin{eqnarray}
  \mathbb{L}_A T_B\, &=& \Gamma_{AB}^C \, T_C \nonumber\\
  \Gamma_{AB}^C &=&  f_{AB}^{\phantom{AB}C} \, - \, \kappa_{AD} \kappa^{CE}\, f_{BE}^{\phantom{AB}D}
  - \, \kappa_{BD} \kappa^{CE}\, f_{AE}^{\phantom{AB}D}
\end{eqnarray}
which are constant numbers.
\par
Given the connection coefficients the differential geodesic equations can be immediately written. In the chosen basis
the tangent vector to the geodesic is described by $n$ fields
\begin{equation}\label{certosio}
  \Pi^A(t)\, \equiv \, \kappa^{AB}\, \mathfrak{p}_B(t)
\end{equation}
which depend on the affine parameter $t$ along
the curve. The geodesic equation is given by the following first order differential system:
\begin{equation}\label{geoeque}
    \frac{d}{dt} \,\Pi^{A} \, + \, \Gamma^{A}_{BC} \, \Pi^B \, \Pi^C \, = \,0~
\end{equation}
The above equation contains two data:
\begin{description}
  \item[1)] the structure constants of the solvable Lie algebra
  $f_{AB}^{\phantom{AB}C}$,
  \item[2)] the constant tensor $\kappa_{AB}$ defining the norm on the solvable Lie algebra.
\end{description}
Equation (\ref{geoeque}) can be obtained as hamiltonian equations from the definition (\ref{anhohamilla}) of the hamiltonian
$\mathcal{H}_{geoUH}$ and the explicit expression of the Poisson bracket (\ref{corbellus}):
\begin{eqnarray}
  \partial_t \, \Pi^A &=& \left\{\mathcal{H}_{geoUH}\, , \, \Pi^A\right\} \label{bonita}\\
  \partial_t \, \dot{\Upsilon}^\alpha &=& \left\{\mathcal{H}_{geoUH}\, , \,\Upsilon^\alpha \right\} \, =\, e^\alpha_M \, \Pi^M
  \label{dimenticata}
\end{eqnarray}
Note that the first equation (\ref{bonita}) yielding eq.(\ref{geoeque}) was obtained in \cite{sashaebog} using the definition (\ref{corbellus}) of Poisson brackets reduced to functions only of the momenta $\mathfrak{p}_A$, namely:
\begin{equation}\label{gongo}
  \left\{\varphi(\mathfrak{p})\, , \, \psi(\mathfrak{p}) \right\}_{red} \, = \, -2 \, \mathfrak{p}_A \, f^{\phantom{BC}A}_{BC} \, \frac{\partial \varphi}{\partial \mathfrak{p}_B} \, \frac{\partial \psi }{\partial \mathfrak{p}_C}
\end{equation}
It was observed in \cite{sashaebog} that eq.(\ref{gongo}) equips the dual vector space $Solv^\star$ to any solvable Lie algebra $Solv$ with the structure of a Poisson manifold, which is not a symplectic manifold since the bivector provided by the solvable structure constants $f^{\phantom{BC}A}_{BC}$ is not invertible. Yet, as recalled in \cite{sashaebog}, when the solvable Lie algebra is the Borel subalgebra of the special linear group
\begin{equation}\label{borelliana}
Solv \, = \, \mathbb{B}_N \, \equiv \, \mathbb{B}\left[\slal(N,\mathbb{R})\right]
\end{equation}
namely the space of $N\times N$ triangular traceless matrices, it was proved  by Arkhangel'skii in \cite{arcangelo} that the hamiltonian system based on the Poisson bracket (\ref{gongo}) is always Liouville integrable since it possesses the  required number of hamiltonians in involution. Specifically, distinguishing the even case $\mathbb{B}_{2\nu}$ from the odd one $\mathbb{B}_{2\nu+1}$ (where $\nu \in \mathbb{N}$) we have that for $\mathbb{B}_{2\nu}$ there are $\nu^2+\nu-1$ hamiltonians in involution, while for $\mathbb{B}_{2\nu+1}$  the number of such objects is $\nu^2 + 2\nu$. Furthermore, in both cases,  among the hamiltonians   in  involution there are respectively $r=\nu -1$ and $r=\nu$ Casimirs $\mathcal{C}_i(\mathfrak{p})$ ($i=1,\dots,r$), namely such functions of the momenta $\mathfrak{p}$ that have vanishing Poisson bracket with all of them:
\begin{equation}\label{casimirro}
  \forall i,A \quad : \quad  \left\{ \mathfrak{p}_A\, , \,\mathcal{C}_i(\mathfrak{p})\right\} \, = \, 0
\end{equation}
 According with the general theory of dynamical systems, one can define level surfaces of the Casimirs:
 \begin{equation}\label{livelle}
   \mathcal{L}_{k_1,\dots, k_r} \subset Solv^\star \quad ; \quad \forall \mathfrak{p} \in \mathcal{L}_{k_1,\dots, k_r} \quad
   :  \quad\mathcal{C}_i(\mathfrak{p}) \, = \, k_i \, = \, \text{const} \in \mathbb{R}
 \end{equation}
and, as explained in \cite{sashaebog}, such level surfaces have always even dimension and become symplectic manifolds since
 when restricted to them the poissonian bi-vector becomes invertible.
 \par
 Although this is very interesting, as one sees from the above discussion it is only one aspect of the full story. Indeed the complete space
 that includes not only the momenta $\mathfrak{p}_A$ but also the coordinates $\Upsilon^\alpha$ is always symplectic and the full definition of the Poisson bracket is that given in eq.(\ref{corbellus}).  Yet the nice point about Arkhangel'skii hamiltonians is that those functions of the momenta $\mathfrak{p}$ that are in involution with respect to the reduced Poisson bracket (\ref{gongo}) remain in involution also with respect to the full Poisson bracket (\ref{corbellus}), so that they are constant along any geodesic.

\section{A Master Example for the Geodesic Dynamical System:  \boldmath{$\mathrm{SL(3,\mathbb{R})/SO(3)}$}  }
\label{magister}
As a concrete illustration of the above discussed concepts and
constructions we choose the $5$-dimensional symmetric space:
\begin{equation}\label{masterexamp}
  \mathcal{M}_{5} \, \equiv \, \frac{\mathrm{SL(3,\mathbb{R})}}{\mathrm{SO(3)}}
\end{equation}
The reasons for such a choice are several:
\begin{enumerate}
  \item $\mathcal{M}_{5}$  is the smallest symmetric space with a non compact rank $r>1$ and a non trivial solvable Lie algebra $Solv_5$.
  \item $\mathcal{M}_{5}$ belongs to the mother series of non-compact symmetric spaces $\frac{\mathrm{SL(N,\mathbb{R})}}{\mathrm{SO(N)}}$ that, thanks to the triangular embedding, explained in \cite{pgtstheory} and \cite{TSnaviga}, contains all the members of the other series as submanifolds.
  \item $\mathcal{M}_{5}$ is not a K\"ahler manifold, yet its $10$-dimensional tangent bundle $\mathcal{TM}_{5}$ has the symplectic structure discussed in sections \ref{zaklyuchenie} and \ref{magister} as any other tangent bundle. This allows to illustrate  the distinction among the symplectic manifold of the GDS utilized here  with respect to what is done in paper \cite{barbaresco2}, where, as we extensively stressed in the introduction, the moment maps and the thermodynamical states  are defined with respect to the Souriau symplectic $2$-form (\ref{KKSdefi}) constructed on coadjoint orbits and also with respect to what was done in the above mentioned papers\cite{sashaebog,arcangelo},  where the Poisson structure is defined only on the standard fibre of the tangent bundle to $\mathcal{TU}/\mathcal{H}$.  As we stressed in the introduction, the Souriau case corresponds to thermodynamics on K\"aehler manifolds.
\end{enumerate}
\paragraph{The solvable Lie algebra generators}
An explicit basis for the solvable Lie algebra $Solv_{5}$ of traceless, upper triangular matrices in $3$-dimension is the following one:
\begin{alignat}{9}\label{barbagianni}
  T_1 & = & \left(
\begin{array}{ccc}
 1 & 0 & 0 \\
 0 & 0 & 0 \\
 0 & 0 & -1 \\
\end{array}
\right) & \quad ; \quad &  T_2 & = & \left(
\begin{array}{ccc}
 0 & 0 & 0 \\
 0 & 1 & 0 \\
 0 & 0 & -1 \\
\end{array}
\right) & \quad ; \quad
  T_3 & = & \left(
\begin{array}{ccc}
 0 & 1 & 0 \\
 0 & 0 & 0 \\
 0 & 0 & 0 \\
\end{array}
\right) & \quad ; \quad &  T_4 & = & \left(
\begin{array}{ccc}
 0 & 0 & 0 \\
 0 & 0 & 1 \\
 0 & 0 & 0 \\
\end{array}
\right) \nonumber\\
  T_5 & = & \left(
\begin{array}{ccc}
 0 & 0 & 1 \\
 0 & 0 & 0 \\
 0 & 0 & 0 \\
\end{array}
\right) & ; &  \null & \null & \null \nonumber\\
\end{alignat}
where the diagonal $T_{1,2}$ are Cartan generators, while $T_{3,4}$ correspond to the simple roots $\alpha_{1,2}$ and $T_5$ is associated with the highest root $\alpha_1+\alpha_2$.
\paragraph{The solvable Lie group generic element}
Next it is very simple to write the generic element of  the solvable Lie group $\mathcal{S}_5 \cong \exp[Solv_5]$ obtained according with the rules of the exponential map $\Sigma$ as defined in \cite{TSnaviga,pgtstheory}.
\begin{equation}\label{cromagnone}
 \mathcal{S}_5 \, \ni \, \mathbb{L}(\boldsymbol{\Upsilon})\, \equiv\, \Sigma[\Upsilon^A \, T_A ] \, = \, \prod_{A=1}^5 \exp\left[\Upsilon^A\, T_A \right] \, = \, \left(
\begin{array}{ccc}
 e^{\Upsilon^1} & e^{\Upsilon^1} \Upsilon^3 &
   e^{\Upsilon^1} \left(\Upsilon^3 \Upsilon^4+\Upsilon
   ^5\right) \\
 0 & e^{\Upsilon^2} & e^{\Upsilon^2} \Upsilon^4 \\
 0 & 0 & e^{-\Upsilon^1-\Upsilon^2} \\
\end{array}
\right)
\end{equation}
\paragraph{The left invariant Cartan $1$-form matrix on $\mathcal{S}_5$}
Given the generic solvable Lie group element $\mathbb{L}(\boldsymbol{\Upsilon})$ we easily compute the solvable Lie algebra valued
left-invariant $1$-form $\Theta$:
\begin{eqnarray}\label{carlinga}
\Theta & \equiv  &\mathbb{L}^{-1}(\boldsymbol{\Upsilon})\cdot \mathrm{d}\mathbb{L}(\boldsymbol{\Upsilon})\nonumber \\
  &=&\left(
\begin{array}{ccc}
 \mathrm{d}\Upsilon^1 & \mathrm{d}\Upsilon^3+\Upsilon^3 (\mathrm{d}\Upsilon
  ^1-\mathrm{d}\Upsilon^2) & \mathrm{d}\Upsilon^5+\Upsilon^4
   \mathrm{d}\Upsilon^3+\Upsilon^3 \Upsilon^4 \mathrm{d}\Upsilon
   ^1-\Upsilon^3 \Upsilon^4 \mathrm{d}\Upsilon^2+2
   \Upsilon^5 \mathrm{d}\Upsilon^1+\Upsilon^5 \mathrm{d}\Upsilon
   ^2 \\
 0 & \mathrm{d}\Upsilon^2 & \mathrm{d}\Upsilon^4+\Upsilon^4
   (\mathrm{d}\Upsilon^1+2 \mathrm{d}\Upsilon^2) \\
 0 & 0 & -\mathrm{d}\Upsilon^1-\mathrm{d}\Upsilon^2 \\
\end{array}
\right)
\end{eqnarray}
\paragraph{The Maurer Cartan forms $1$-forms on $\mathcal{S}_5$}
The left-invariant Maurer Cartan forms $e^A$ are obtained decomposing the upper triangular traceless matrix $\Theta$, whose matrix elements are $1$-forms on the solvable Lie group manifold $\mathcal{S}_5$, along the generator basis (\ref{barbagianni}) :
\begin{eqnarray}\label{mcformi}
  \Theta & \quad = \quad & e^A \, T_A \nonumber\\
  e^1 &\quad = \quad &  \mathrm{d}\Upsilon^1  \nonumber\\
  e^2 &\quad = \quad & \mathrm{d}\Upsilon^2  \nonumber\\
  e^3 &\quad = \quad &  \mathrm{d}\Upsilon^3 +\Upsilon^3 (\mathrm{d}\Upsilon^1 -\mathrm{d}\Upsilon
   _2 ) \nonumber\\
  e^4 &\quad = \quad&  \mathrm{d}\Upsilon^4 +\Upsilon^4 (\mathrm{d}\Upsilon^1 +2 \mathrm{d}\Upsilon
   _2 ) \nonumber\\
  e^5 &\quad = \quad & \mathrm{d}\Upsilon^5 +\Upsilon^4 \mathrm{d}\Upsilon^3 +\Upsilon^3
   \Upsilon^4 \mathrm{d}\Upsilon^1 -\Upsilon^3 \Upsilon^4
   \mathrm{d}\Upsilon^2 +2 \Upsilon^5 \mathrm{d}\Upsilon^1 +\Upsilon
   ^5 \mathrm{d}\Upsilon^2
\end{eqnarray}
By construction, the Maurer Cartan forms $e^A$ satisfy  the Maurer Cartan equations:
\begin{equation}\label{Maurocarto}
  \mathrm{d}e^A\,+\,\ft 12 f^{\phantom{BC}A}_{BC} \, e^B \wedge e^C \, = \, 0
\end{equation}
where $f^{\phantom{BC}A}_{BC}$ are the structure constants of the solvable Lie algebra $Solv_5$:
\begin{equation}\label{ciamblocca}
  \left[ T_B \, ,\, T_C\right] \, = \, f^{\phantom{BC}A}_{BC} \, T_A
\end{equation}
\paragraph{The vielbein of the unique  $\mathrm{SL(3,\mathbb{R})}$ invariant Einstein metric on the symmetric space $\mathcal{M}_5$}
As explained above and in \cite{pgtstheory}, the unique Einstein $\mathrm{U}$-invariant metric on the symmetric manifold is algebraically  obtained from an orthonormal basis of $\mathbb{K}$ generators in the Cartan decomposition of the $\mathbb{U}$-Lie algebra: $\mathbb{U}=\mathbb{H} \oplus \mathbb{K}$, where $\mathbb{K}$ constitutes an irreducible representation of the maximal compact subalgebra $\mathbb{H}$ under its adjoint action. In our case $\mathbb{H} = \so(3)$
and the 5-dimensional vector space $\mathbb{K}$ corresponds to the $J=2$ irrep of the three-dimensional rotation group. An orthonormal basis is provided by the following symmetric matrices:
\begin{alignat}{9}
  K^1 & = & \left(
\begin{array}{ccc}
 \frac{1}{\sqrt{2}} & 0 & 0 \\
 0 & 0 & 0 \\
 0 & 0 & -\frac{1}{\sqrt{2}} \\
\end{array}
\right) & \quad ; \quad &  K^2 & = & \left(
\begin{array}{ccc}
 -\frac{1}{\sqrt{6}} & 0 & 0 \\
 0 & \sqrt{\frac{2}{3}} & 0 \\
 0 & 0 & -\frac{1}{\sqrt{6}} \\
\end{array}
\right) & \quad ; \quad
  K^3 & = & \left(
\begin{array}{ccc}
 0 & \frac{1}{\sqrt{2}} & 0 \\
 \frac{1}{\sqrt{2}} & 0 & 0 \\
 0 & 0 & 0 \\
\end{array}
\right) & \quad ; \quad &  K^4 & = & \left(
\begin{array}{ccc}
 0 & 0 & 0 \\
 0 & 0 & \frac{1}{\sqrt{2}} \\
 0 & \frac{1}{\sqrt{2}} & 0 \\
\end{array}
\right) \nonumber\\
  K^5 & = & \left(
\begin{array}{ccc}
 0 & 0 & \frac{1}{\sqrt{2}} \\
 0 & 0 & 0 \\
 \frac{1}{\sqrt{2}} & 0 & 0 \\
\end{array}
\right) & ; &  \null & \null & \null
\end{alignat}
that satisfy the relation:
\begin{equation}\label{kapponi}
  \mathrm{Tr} \left( K^I\cdot K^J\right) \, = \, \delta^{IJ} \quad ; \quad I,J\, = \, 1,2\,\dots, 5
\end{equation}
The vielbein determining the metric is obtained from the Cartan
left invariant matrix $1$-form (\ref{carlinga}) by setting:
\begin{equation}\label{viabidone}
  V^I \, \equiv \, \mathrm{Tr} \left( K^I \cdot \Theta\right)
\end{equation}
and the metric is:
\begin{eqnarray}\label{maggiorana}
  ds^2 & \equiv & \delta_{IJ} V^I \times V^J \, = \, g_{\alpha\beta} \left(\boldsymbol{\Upsilon}\right) \,\mathrm{d}\Upsilon_\alpha \, \mathrm{d}\Upsilon_\beta \nonumber\\
  &=&\frac{1}{2} \left\{3 \mathrm{d}\Upsilon^2+(2 \mathrm{d}\Upsilon
   ^1+\mathrm{d}\Upsilon^2)^2+(\mathrm{d}\Upsilon^3+\Upsilon^3
   (\mathrm{d}\Upsilon^1-\mathrm{d}\Upsilon^2))^2+(\mathrm{d}\Upsilon
   ^4+\Upsilon^4 (\mathrm{d}\Upsilon^1+2 \mathrm{d}\Upsilon
   ^2))^2\right.\nonumber\\
   &&\left. +(\mathrm{d}\Upsilon^5+\Upsilon^4 \mathrm{d}\Upsilon
   ^3+\Upsilon^3 \Upsilon^4 \mathrm{d}\Upsilon^1-\Upsilon
   ^3 \Upsilon^4 \mathrm{d}\Upsilon^2+2 \Upsilon^5
   \mathrm{d}\Upsilon^1+\Upsilon^5 \mathrm{d}\Upsilon^2)^2\right\}
\end{eqnarray}
The vielbein $1$-forms $V^I$ are linear combinations of the Maurer Cartan $1$-forms $e^A$:
\begin{equation}\label{nucrata}
  V^I\, = \, \nu^I_A \, e^a \quad ; \quad \nu \, = \, \left(
\begin{array}{ccccc}
 \sqrt{2} & \frac{1}{\sqrt{2}} & 0 & 0 & 0 \\
 0 & \sqrt{\frac{3}{2}} & 0 & 0 & 0 \\
 0 & 0 & \frac{1}{\sqrt{2}} & 0 & 0 \\
 0 & 0 & 0 & \frac{1}{\sqrt{2}} & 0 \\
 0 & 0 & 0 & 0 & \frac{1}{\sqrt{2}} \\
\end{array}
\right)
\end{equation}
So that the metric can also be written as:
\begin{eqnarray}\label{crisippo}
  ds^2 & =& \kappa_{AB}\, \, e^A \times e^B \nonumber\\
  \kappa & \equiv & \nu^T\cdot\nu \, = \, \left(
\begin{array}{ccccc}
 2 & 1 & 0 & 0 & 0 \\
 1 & 2 & 0 & 0 & 0 \\
 0 & 0 & \frac{1}{2} & 0 & 0 \\
 0 & 0 & 0 & \frac{1}{2} & 0 \\
 0 & 0 & 0 & 0 & \frac{1}{2} \\
\end{array}
\right)
\end{eqnarray}
\subsection{Hamiltonians in Involution and Generalized Thermodynamics}
As we stressed in section \ref{numiziato} the general definition (\ref{corbellus}) of the Poisson bracket on the total space of the tangent bundle can be reduced to functions $f(\mathfrak{p})$ that depend only on the momenta, namely on the vertical fibre coordinates, and one obtains the reduced formula (\ref{gongo}) which still satisfies all the properties mentioned in definition \ref{poissone} of section
\ref{pescione} in order to define a Poisson bracket. In the case of the here considered master model eq.(\ref{gongo}) takes the following explicit form for any pair of functions $F(\mathfrak{p}),G(\mathfrak{p})$:
\begin{eqnarray}
   \left\{F(\mathfrak{p})\, , \, G(\mathfrak{p})\right\}_{red} &=& -2 \mathfrak{p}_3 \left(-\partial_3 F \partial_1G+\partial_3 F \partial_2G+\partial_1F \partial_3G-\partial_2F \partial_3G\right)\nonumber\\
  &&-2 \mathfrak{p}_4
   \left(-\partial_4 F \partial_1G-2 \partial_4 F \partial_2G+\partial_1F \partial_4G+2 \partial_2F \partial_4G\right)\nonumber\\
   &&-2 \mathfrak{p}_5 \left(-2 \partial_5 F
   \partial_1G-\partial_5 F \partial_2G-\partial_4 F \partial_3G+\partial_3 F \partial_4G+2 \partial_1F \partial_5G+\partial_2F \partial_5G\right)
\end{eqnarray}
where we have used the shorthand notation:
\begin{equation}\label{notazia}
  \partial_{a}f \, \equiv \, \frac{\partial f(\mathfrak{p})}{\partial \mathfrak{p}_a}
\end{equation}
Utilizing the constructive recipe introduced by Arkhangel'skii in \cite{arcangelo} and recalled in eq.(2.49) of \cite{sashaebog}) we obtain the following three hamiltonians in involution:
\begin{eqnarray}
  \mathfrak{H}_1 &=& \frac{1}{3} \left(\mathfrak{p}_1^2-\mathfrak{p}_2 \mathfrak{p}_1+\mathfrak{p}_2^2+3
   \left(\mathfrak{p}_3^2+\mathfrak{p}_4^2+\mathfrak{p}_5^2\right)\right)\label{marino}\\
  \mathfrak{H}_2 &=&\frac{1}{27} \left(-2 \mathfrak{p}_1^3+3 \mathfrak{p}_2 \mathfrak{p}_1^2+3
   \left(\mathfrak{p}_2^2-3 \left(\mathfrak{p}_3^2-2
   \mathfrak{p}_4^2+\mathfrak{p}_5^2\right)\right) \mathfrak{p}_1-2 \mathfrak{p}_2^3-54
   \mathfrak{p}_3 \mathfrak{p}_4 \mathfrak{p}_5-9 \mathfrak{p}_2
   \left(\mathfrak{p}_3^2+\mathfrak{p}_4^2-2 \mathfrak{p}_5^2\right)\right)\label{montano} \\
  \mathfrak{H}_3 &=& \frac{1}{3} \left(\mathfrak{p}_1-2 \mathfrak{p}_2+\frac{3 \mathfrak{p}_3
   \mathfrak{p}_4}{\mathfrak{p}_5}\right)\label{desertico}\\
0&=& \left\{\mathfrak{H}_i\, , \, \mathfrak{H}_j\right\}_{red} \quad \quad i,j\, =\, 1,2,3\label{tundra}
\end{eqnarray}
The first crucial observation is the that the  quadratic first hamiltonian $\mathfrak{H}_1=\mathcal{H}_{geoUH}$ coincides with the
quadratic hamiltonian (\ref{anhohamilla}) obtained from the Legendre transform of the geodesic dynamical system Lagrangian. Indeed in our case we have that $\kappa$ is given by the second of equations (\ref{crisippo}) and one immediately verifies that
\begin{eqnarray}\label{verificone}
  \mathfrak{H}_1  &= &\ft 12 \left(\kappa^{-1}\right)^{AB} \, \mathfrak{p}_A \, \mathfrak{p}_B \, \equiv \, \ft 12 \kappa^{AB} \, \, \mathfrak{p}_A \, \mathfrak{p}_B \, = \, \mathcal{H}_{geoUH}\nonumber\\
  \kappa^{-1} & = & \left(
\begin{array}{ccccc}
 \frac{2}{3} & -\frac{1}{3} & 0 & 0 & 0 \\
 -\frac{1}{3} & \frac{2}{3} & 0 & 0 & 0 \\
 0 & 0 & 2 & 0 & 0 \\
 0 & 0 & 0 & 2 & 0 \\
 0 & 0 & 0 & 0 & 2 \\
\end{array}
\right)
\end{eqnarray}
This implies that the method of hamiltonians in involution selects the unique invariant norm on the solvable Lie algebra that corresponds to the unique $\mathrm{U}$-symmetric Einstein metric on the equivalent symmetric space $\mathrm{SL(N,\mathbb{R})/SO(N)}$. This fact was already pointed out in \cite{sashaebog}.
\par
The second important observation is that the functions depending only on the momenta $\mathfrak{p}_A$ that are in involution with respect to the reduced Poisson bracket, are in involution with respect to the complete Poisson bracket so that they are conserved along all trajectories, namely all geodesic lines.
\par
Different is the case of the Casimir $ \mathfrak{H}_3$. It commutes with all the momenta, yet it does not commute with all the coordinates $\Upsilon^\alpha$. Indeed we have:
\begin{equation}\label{rattus}
  \left\{\mathfrak{H}_3\, , \, \mathfrak{p}_A \right\} \, = \, 0 \quad ; \quad
  \left\{\mathfrak{H}_3\, , \, \Upsilon^\alpha \right\} \, = \,  \frac{\partial \mathfrak{H}_3}{\partial \mathfrak{p}_A} \, e_A^\alpha(\boldsymbol{\Upsilon})
\end{equation}
\subsection{Generalized Thermodynamics for a Geodesic Dynamical System on $\mathrm{U/H}$} As we see from eq.(\ref{rattus}) there is no way of fixing consistently the hamiltonian $\mathfrak{H}_3$ to a  constant value on the whole ten dimensional space with canonical coordinates $\{\mathfrak{p}_A,\Upsilon^\alpha\}$ obtaining in this way  a reduced symplectic manifold of dimension eight. This conclusion is illustrated explicitly in the present master example, yet it is true for all manifolds $\mathrm{SL(N,\mathbb{R})/SO(N)}$ since in all cases the Casimirs $\mathcal{C}_i(\mathfrak{p})$ that have a vanishing reduced Poisson bracket with the momenta $\mathfrak{p}$, have, with the manifold coordinates $\Upsilon^\alpha$, a non vanishing Poisson bracket:
\begin{equation}\label{ruminante}
   \left\{\mathcal{C}_i(\mathfrak{p})\, , \, \Upsilon^\alpha \right\} \, = \,  \frac{\partial \mathcal{C}_i(\mathfrak{p})}{\partial \mathfrak{p}_A} \, e_A^\alpha(\boldsymbol{\Upsilon})
 \end{equation}
 Nevertheless one can introduce, just as in the classical statistical mechanics of free gases, a Gibbs state probability distribution that minimizes the Shannon functional as we discussed in section \ref{geostructa} and we announced in section \ref{cromatillo}. Indeed, given the set of conserved hamiltonians in involution, that depend only on the momenta $\mathfrak{p}$ and that we denote $\mathfrak{H}_i(\mathfrak{p})$ we can construct the Gibbs state probability distribution defined by eq.s (\ref{predone},\ref{larvatarlo}).
 For any symmetric space $\mathrm{U/H}$ of non compact type we have:
  \begin{eqnarray}
   \mathrm{G}(\boldsymbol{\lambda},V) &=& \frac{\exp\left[\, -\, \boldsymbol{\lambda} \cdot \boldsymbol{\mathfrak{H}}\left(
   \boldsymbol{\mathfrak{p}}\right)\right]}{Z(\boldsymbol{\lambda},V)} \label{gipsyUH} \\
   Z(\boldsymbol{\lambda},V)&=& \int_{\mathcal{M}_{2d}} \, \exp\left[\, -\, \boldsymbol{\lambda} \cdot \boldsymbol{\mathfrak{H}}\left(
   \boldsymbol{\mathfrak{p}}\right)\right] \mathrm{d}\lambda(\boldsymbol{\mathfrak{p}},\boldsymbol{\Upsilon}) \label{partfunUH}\\
   \mathrm{d}\lambda(\boldsymbol{\mathfrak{p}},\boldsymbol{\Upsilon})  &=&  \underbrace{e^1\wedge e^2 \wedge \dots \wedge e^d \wedge \mathrm{d}\mathfrak{p}_1 \wedge \dots \wedge \mathrm{d}\mathfrak{p}_d}_{\mathrm{Vol}(\mathcal{M}_{2d})} \\
 \end{eqnarray}
 where $\mathcal{M}_{2d}$ is the total space of the tangent bundle as specified in eq.(\ref{totalbenzina}) and
 $\mathrm{d}\lambda(\boldsymbol{\mathfrak{p}},\boldsymbol{\Upsilon}) $ is the Liouville integration measure on such a space
 presented in eq.(\ref{volumetto}). Since all the hamiltonians depend only on the momenta $\mathfrak{p}_A$ and not on the coordinates
 $\Upsilon^\alpha$, we are in a situation similar to that of Ideal Gases (see appendix \ref{msidgas}) and the partition function (\ref{partfunUH}) factorizes into the product of two integrals (or summations):
 \begin{equation}\label{fattorino}
   Z(\boldsymbol{\lambda},V) \, = \, \underbrace{\int_{\mathcal{T}} \,\exp\left[\, -\, \boldsymbol{\lambda} \cdot \boldsymbol{\mathfrak{H}}\left(
   \boldsymbol{\mathfrak{p}}\right)\right] \, \mathrm{d}\mathfrak{p}_1 \wedge \dots \wedge \mathrm{d}\mathfrak{p}_d }_{\zeta(\boldsymbol{\lambda})}\,\times \underbrace{\int_{\text{Box}\subset \mathcal{S}} \, e^1\wedge e^2 \wedge \dots \wedge e^d}_{V=\text{volume}}
 \end{equation}
 \paragraph{Discussion on the Box conception.} In relation with the factorization in eq.(\ref{fattorino}) we have to pause for a moment and analyse the comparison with Ideal Gas Statistical Mechanics reviewed in section \ref{msidgas}. In both cases the basic hamiltonian is a quadratic  form in the momenta and the integral over $\boldsymbol{\mathfrak{p}}$ is a multiple gaussian integral, the multiplicity being the number of degrees of freedom $n_f$. In the Ideal Gas case this number is $n_f\, = \, 3\times N$ where $N$ is the number of molecules composing the gas, namely of the order of the Avogadro number. In the geodesic dynamical system case the number of degrees of freedom is $n_f \, = \, d$, namely a figure of the dimension of the symmetric space $\mathrm{U/H}$ or, if you prefer, of the corresponding solvable Lie group $\mathcal{S}_{\mathrm{U/H}}$. In the Ideal Gas case the integral $V$ is the volume of the portion of
 physical space $\mathbb{R}^3$ in which the sample of gas is confined and the integration on it occurs $N$ times, one for each molecule composing the gas. Hence the factor that multiplies $\zeta(\boldsymbol{\lambda})$ is $V^N$. For the gases the Gibbs state
 distribution expresses the probability that the $N$-particles be in a state where each of them has a given momentum and it is in a given point. The independence of the hamiltonian from the coordinates implies that such probability is insensitive to the position, namely all positions have the same probability, while the momenta follow a gaussian distribution. In the case of the geodesic dynamical system,
 the Gibbs state describes the probability that a geodesic starts at a given point with a given initial tangent vector (the momentum). The coordinate independence of the hamiltonians implies that all points of the base manifold have the same probability, while the initial tangent vectors follow, as far as we consider only the quadratic hamiltonian, a gaussian distribution.  Introducing more hamiltonians the distribution becomes more complicated and more structured in momentum space, yet it remains flat in coordinate space. In both cases this flatness is the consequence of translation invariance, with respect to $\mathbb{R}^3$ in the Ideal Gas case, with respect to the solvable Lie group $\mathcal{S}_{\mathrm{U/H}}$ in the case of the geodesic dynamical system on $\mathrm{U/H}$. Just as in the Ideal Gas case the \textit{Box} is the portion of physical space in which the gas is confined, in the same way for the Geodesic Dynamical System case the \textit{Box} is the portion of base manifold to which we confine the possible initial points of the geodesics of our interest: indeed the statistical distribution described by our thermodynamics is a statistical distribution in the space of geodesics.
 \subsubsection{Generalized Thermodynamics for the Chosen Master Example}
 In order to study the generalized thermodynamics of our master example, where the conserved hamiltonians are given by eq.s ( \ref{marino}-\ref{desertico}) we have to calculate the nontrivial part of the partition function:
 \begin{equation}\label{notrivialo}
   \zeta(\boldsymbol{\lambda}) \, = \, \int\exp\left[-\sum_{i=1}^3\lambda_i \, \mathfrak{H}_i (\boldsymbol{\mathfrak{p}})\right]
   \, \mathrm{d}^5\boldsymbol{\mathfrak{p}}
 \end{equation}
 To this effect it is convenient to diagonalize the quadratic form in $\mathfrak{p}_A$ that corresponds to the first hamiltonian. Since we have:
 \begin{equation}\label{diagahamilla}
   -\,\lambda_1 \, \mathfrak{H}_1 \, = \, \lambda_1 \, \underbrace{\left( -\ft 12 \, \kappa^{-1}\right)^{AB}}_{=\,\,\boldsymbol{k}^{AB}} \, \mathfrak{p}_A\, \mathfrak{p}_B
 \end{equation}
 we need to diagonalize the above defined matrix $\boldsymbol{k}$. An easy calculation shows that the following matrix:
 \begin{equation}\label{carnenonvale}
   \mathcal{U} \, \equiv \, \left(
\begin{array}{ccccc}
 0 & 0 & 0 & -1 & \sqrt{3} \\
 0 & 0 & 0 & 1 & \sqrt{3} \\
 0 & 0 & 1 & 0 & 0 \\
 0 & 1 & 0 & 0 & 0 \\
 1 & 0 & 0 & 0 & 0 \\
\end{array}
\right)
 \end{equation}
 satisfies the condition:
 \begin{equation}\label{ufesta}
   \mathcal{U}^T\cdot \boldsymbol{k} \cdot \mathcal{U} \, = \, - \, \mathbf{1}_{5\times 5}
 \end{equation}
 Hence defining new variables:
 \begin{equation}\label{candillo}
   \boldsymbol{\mathfrak{w}} \, \equiv \, \mathcal{U}\, \boldsymbol{\mathfrak{p}}
 \end{equation}
 we obtain:
 \begin{equation}\label{frescovile}
   - \, \boldsymbol{\mathfrak{w}}^T \, \boldsymbol{\mathfrak{w}} \, = \, - \, \ft 12 \, \mathfrak{H}_1(\boldsymbol{\mathfrak{p}})
 \end{equation}
 Explicitly the change of variables is as follows
 \begin{equation}\label{farfullo}
   \begin{array}{ccc}
      \mathfrak{w}_1 & = & \mathfrak{p}_5 \\
      \mathfrak{w}_2 & = & \mathfrak{p}_4 \\
      \mathfrak{w}_3 & = & \mathfrak{p}_3 \\
      \mathfrak{w}_4 & = & \frac{1}{2} \left(\mathfrak{p}_2-\mathfrak{p}_1\right) \\
      \mathfrak{w}_5 & = &  \frac{\mathfrak{p}_1+\mathfrak{p}_2}{2 \sqrt{3}} \\
      \end{array} \quad ; \quad
      \begin{array}{ccc}
      \mathfrak{p}_1 & = & \sqrt{3} \mathfrak{w}_5-\mathfrak{w}_4 \\
      \mathfrak{p}_2  & = & \mathfrak{w}_4+\sqrt{3} \mathfrak{w}_5 \\
      \mathfrak{p}_3  & = & \mathfrak{w}_3 \\
      \mathfrak{p}_4  & = & \mathfrak{w}_2 \\
      \mathfrak{p}_5  & = & \mathfrak{w}_1
    \end{array}
 \end{equation}
 In terms of new variables $\mathfrak{w}$ the three hamiltonians have the following expression:
\begin{eqnarray}\label{whamil}
   \mathfrak{H}_1(\boldsymbol{\mathfrak{w}}) &=& \mathfrak{w}_1^2+\mathfrak{w}_2^2+\mathfrak{w}_3^2+\mathfrak{w}_4^2+\mathfrak{w}_5^2 \nonumber\\
   \mathfrak{H}_2(\boldsymbol{\mathfrak{w}}) &=& \left(\mathfrak{w}_4+\frac{\mathfrak{w}_5}{\sqrt{3}}\right) \mathfrak{w}_1^2-2
   \mathfrak{w}_2 \mathfrak{w}_3 \mathfrak{w}_1+\frac{1}{9} \left(\left(3 \sqrt{3}
   \mathfrak{w}_5-9 \mathfrak{w}_4\right) \mathfrak{w}_2^2+2 \sqrt{3} \mathfrak{w}_5
   \left(\mathfrak{w}_5^2-3
   \left(\mathfrak{w}_3^2+\mathfrak{w}_4^2\right)\right)\right) \nonumber\\
   \mathfrak{H}_3(\boldsymbol{\mathfrak{w}}) &=& \frac{\mathfrak{w}_2
   \mathfrak{w}_3}{\mathfrak{w}_1}-\mathfrak{w}_4-\frac{\mathfrak{w}_5}{\sqrt{3}}
 \end{eqnarray}
 Hence the partition function reduces to the following multiple integral
 \begin{equation}\label{moltiplico}
   \zeta(\lambda_1,\lambda_2,\lambda_3) \, = \, \int_{-\infty}^{\infty} d\mathfrak{w}_1\,\int_{-\infty}^{\infty} d\mathfrak{w}_2
   \,\int_{-\infty}^{\infty} d\mathfrak{w}_3\,\int_{-\infty}^{\infty} d\mathfrak{w}_4\,\int_{-\infty}^{\infty} d\mathfrak{w}_5 \,
   \exp\left[-\sum_{i=1}^3 \,\lambda_i \,\mathfrak{H}_i(\boldsymbol{\mathfrak{w}}) \right]
 \end{equation}
 Although we have not explored all the possible strategies to calculate the integral in the case that the three generalized temperatures
 $\lambda_i$ are all different from zero such calculation appears though. On the other hand the calculation is rather easy if we put
 $\lambda_2=0$, while keeping $\lambda_1\neq 0$ and $\lambda_3 \neq 0$. We present such explicit result that provides an illustrative example of what the geodesic thermodynamics on a non-compact symmetric space can be.
 \par
 With an iterative gaussian integration we obtain:
 \begin{equation}\label{zerlina2zero}
   \zeta(\lambda_1,0,\lambda_3) \, = \, \frac{2 \pi ^{5/2} e^{\frac{\lambda _3^2}{12 \lambda _1}}}{\lambda _1^{5/2}} \quad \Rightarrow \quad Z(\lambda_1,0,\lambda_3,V) \, = \, \frac{2 \pi ^{5/2} e^{\frac{\lambda _3^2}{12 \lambda _1}}}{\lambda _1^{5/2}} \, V
 \end{equation}
 and consequently the stochastic hamiltonian (see eq. (\ref{tremoamillo})) reads as follows:
 \begin{equation}\label{maialotto}
   \mathcal{H}\mid_{\lambda_2=0} \, = \, -\log \left(\frac{2 \pi ^{5/2} e^{\frac{\lambda _3^2}{12 \lambda _1}}}{\lambda
   _1^{5/2}}\right)-\log (V) \, = \, -\frac{\lambda _3^2}{12 \lambda _1}-\frac{5}{2} \log \left(\frac{\pi }{\lambda
   _1}\right)-\log (V)-\log (2)
 \end{equation}
 Recalling eq.(\ref{leggendoleggo}) the Shannon Information Entropy (which coincides with minus the thermodynamical entropy $S$) is the following:
 \begin{equation}\label{lucidapavimenti}
   \mathcal{I} \, = \,  \mathcal{H} \, - \, \lambda_1 \, \frac{\partial \mathcal{H}}{\partial\lambda_1} \, - \,
   \lambda_3 \, \frac{\partial \mathcal{H}}{\partial\lambda_3} \, = \, \frac{5 \log \left(\lambda _1\right)}{2}-\log (2 V)-\frac{5}{2}-\frac{5 \log (\pi )}{2}
 \end{equation}
 Finally the Gibbs state probability distribution is
\begin{equation}\label{cartilagine}
   \mathrm{G}\left(\lambda_1,\lambda_3, V,\boldsymbol{\mathfrak{p}},\boldsymbol{\Upsilon}\right) \, = \,
   \frac{\lambda _1^{5/2} \exp \left(\frac{1}{12} \left(-\frac{\lambda _3^2}{\lambda _1}-4
   \lambda _3 \left(\mathfrak{p}_1-2 \mathfrak{p}_2+\frac{3 \mathfrak{p}_3
   \mathfrak{p}_4}{\mathfrak{p}_5}\right)-4 \lambda _1
   \left(\mathfrak{p}_1^2-\mathfrak{p}_2 \mathfrak{p}_1+\mathfrak{p}_2^2+3
   \left(\mathfrak{p}_3^2+\mathfrak{p}_4^2+\mathfrak{p}_5^2\right)\right)\right)\right)
   }{2 \pi ^{5/2} V}
 \end{equation}
 %
 \subsubsection{Final Remarks on the GDS Generalized Thermodynamics of the Master Model $\mathrm{SL(3,\mathbb{R})/SO(3)}$}
 Let us observe that if we put $\lambda_3=0$ and we interpret $\lambda_1=1/(k_B\, T)$, the partition function
 (\ref{zerlina2zero})  takes apart from constant factors  the form of that of an ideal gas (compare the following with eq.(\ref{idgaspartfundevel3})):
  \begin{equation}\label{kastriula}
    Z(T,0,V) \, = \,2 \,(\pi \,k_B \, T)^{5/2} \, V
  \end{equation}
 where $3N$ is replaced by $5$ and $N$ by $1$. The meaning of this is quite transparent. It is like we were dealing with a gas made by just one particle (hence $N=1$) that moves in a space with $5$ dimensions instead of $3$. Correspondingly the thermodynamical metric at $\lambda_3=0$ is flat as for ideal gases. On the other hand if we switch on the generalized temperature $\lambda_3$ we have a surprise. Considering the three thermodynamical coordinates:
 \begin{equation}\label{tcord}
   \mathbf{t}\, = \, \{\lambda_1,\lambda_3,V\}
 \end{equation}
 and working out the thermodynamical metric from the stochastic hamiltonian in equation (\ref{maialotto}) we obtain:
 \begin{eqnarray}
 \label{pernacchiageo}
   ds^2_{therm} &\equiv& \frac{\partial^2 \mathcal{H}}{\partial t^i \partial t^j} dt^i \, dt^j \nonumber \\
   \null &=& \frac{ {dV}^2}{V^2}-\frac{ {d \lambda_  1}^2 \left(15  { \lambda_
    1}+ { \lambda_  3}^2\right)-2  {d \lambda_  1}  {d \lambda_  3}
    { \lambda_  1}  { \lambda_  3}+ {d \lambda_  3}^2  { \lambda_  1}^2}{6
    { \lambda_  1}^3}
 \end{eqnarray}
Clearly the metric (\ref{pernacchiageo}) is a metric on a direct product manifold, that spanned by the two temperatures $\lambda_{1,3}$ and that spanned by the volume. Furthermore the metric has a Lorentzian signature
$\{-1,-1,1\}$. Finally the two-dimensional space associated with the two temperatures is not flat as in the ideal gas case rather it is a constant negative curvature manifold. Indeed introducing the following dreibein:
\begin{eqnarray}\label{ciromania}
  \mathbf{E}& = & \left\{E^1,\, E^2, \, E^3 \right\}\nonumber\\
  E^1& = & \frac{  \sqrt{15  { \lambda_  1}+ { \lambda_  3}^2}}{\sqrt{6}
    { \lambda_  1}^{3/2}}\, {d \lambda_  1} \, -\, \frac{  { \lambda_  3}}{\sqrt{6}
   \sqrt{ { \lambda_  1}} \sqrt{15  { \lambda_  1}+ { \lambda_  3}^2}}\, {d \lambda_  3} \nonumber\\
   E^2 & = & \frac{\sqrt{\frac{5}{2}}  }{\sqrt{15  { \lambda_
    1}+ { \lambda_  3}^2}}\, {d \lambda_  3} \nonumber\\
   E^3 & = & \frac{1 }{V}\, {dV}
\end{eqnarray}
we can write:
\begin{equation}\label{crollante}
   ds^2_{therm}\, = \, - \, (E^1)^2\,   - \, (E^2)^2\,   + \, (E^3)^2
\end{equation}
and we derive the following spin-connection
\begin{equation}\label{spinnoconno}
  \omega^{ij} \, = \,\left(
                       \begin{array}{ccc}
                         0 & -\frac{15 \sqrt{\frac{3}{2}} \lambda _1^{3/2}}{\left(\lambda _3^2+15 \lambda
   _1\right){}^{3/2}}\,E^2-\frac{\lambda _3^3}{\sqrt{10} \left(\lambda _3^2+15
   \lambda _1\right){}^{3/2}}\,E^1 & 0 \\
                         \frac{15 \sqrt{\frac{3}{2}} \lambda _1^{3/2}}{\left(\lambda _3^2+15 \lambda
   _1\right){}^{3/2}}\,E^2+\frac{\lambda _3^3}{\sqrt{10} \left(\lambda _3^2+15
   \lambda _1\right){}^{3/2}}\,E^1 & 0 & 0 \\
                         0 & 0 & 0 \\
                       \end{array}
                     \right)
\end{equation}
which yields the following curvature $2$-form:
\begin{equation}\label{krugavoi}
  \mathrm{R}^{ij} \, = \, d\omega^{ij} \, + \, \omega^{ik} \wedge \omega^{pj} \eta_{kp} \, = \,
  \ft {1}{10} \, \left(
    \begin{array}{ccc}
      0 & E^1 \wedge E^2 & 0 \\
     - \,  E^1 \wedge E^2 & 0 & 0 \\
      0 & 0 & 0 \\
    \end{array}
  \right)
\end{equation}
which clearly demonstrates what we just stated. The two dimensional thermodynamical subspace spanned by the generalized temperatures $\lambda_1,\lambda_3$ is a portion of a constant curvature manifold namely a portion of a
a hyperbolic plane since the negative signature of the metric implies a negative value of the constant curvature. In any case the constancy of the thermodynamical curvature  signals the absence of any critical point and phase transition.
 \section{Generalized Thermodynamics {\`a} la Souriau on K\"ahler Non Compact $\mathrm{U/H}$.s}
 \label{kellomastero}
 Having clarified the status of generalized thermodynamics associated with the Geodesic Dynamical System and in general with Integrable Dynamical Systems it clearly appears  that the corresponding Gibbs Probability Distributions are of little use for Machine Learning algorithms, since they are non trivial distributions only in momentum space and not on the very manifolds $\mathrm{U/H}$ that constitute the hidden layers of a neural network. Although nothing can be excluded a priori, what one needs in the ML context are  non trivial probability distributions on the very manifolds $\mathrm{U/H}$ that constitute the layers of the network. Such distributions are provided by
 generalized thermodynamics {\`a} la Souriau as schematically defined in eq.s(\ref{partofungo}-\ref{momentico}). Such kind of generalized thermodynamics replaces the hamiltonians in involution of integrable systems with the moment-maps $\mathfrak{P}_A(\boldsymbol{\Upsilon})$ of a typically non abelian, actually semisimple or simple, Lie algebra $\mathbb{U}$ that has a Poissonian realization:
\begin{equation}\label{larvadigatto}
  \left\{\mathfrak{P}_A\, , \, \mathfrak{P}_B \right\} \, = \, f_{AB}^{\phantom{AB}C} \, \mathfrak{P}_C
\end{equation}
and exists only on K\"ahler manifolds, as we extensively discussed in the introduction, analysing the original conception of Barberesco et al based on the notion of coadjoint orbits. As one sees generalized thermodynamics {\`a} la Souriau is almost the opposite of the generalized thermodynamics related with integrable systems and with the corresponding Liouville hamiltonians in involution. Its relevance precisely resides in the non-abelian character of the algebra satisfied by the hamiltonians that is the algebra of Killing vector fields of the corresponding Riemannian manifold.  The principle underlying the construction of generalized thermodynamics {\`a} la Souriau is another one, totally different from Liouville integrability: it is the convergence of the partition function integral (\ref{partofungo}), namely
\begin{equation}
  Z_K(\boldsymbol{\beta}) \, \equiv\, \int_{\mathrm{U/H}}\,  \exp\left[-\boldsymbol{\beta}\cdot \boldsymbol{\mathfrak{P}}(\Upsilon)\right]\,  \underbrace{\boldsymbol{\mathcal{K}}\wedge\boldsymbol{\mathcal{K}}\wedge
  \dots\wedge\boldsymbol{\mathcal{K}}}_{\text{$n$-times}} \, < \infty\label{convpartofungo}
\end{equation}
that poses constraints on the vector $\boldsymbol{\beta}$ of generalized temperatures which, generically, identifies an element of the Lie algebra $\mathbb{U}$ in the chosen basis $\mathfrak{t}_A$ of its generators:
\begin{equation}\label{temperone}
  \boldsymbol{\beta}^A \,\mathfrak{t}_A \in   \mathbb{U}
\end{equation}
The determination of the subset $\Omega \subset \mathbb{U}$ (typically  not a subalgebra), for which the
convergence constraint (\ref{convpartofungo}) is satisfied, encodes the new quality of this peculiar K\"ahlerian generalized thermodynamics whose use in Machine Learning appears to be much more perspective than generalized thermodynamics based on integrable dynamical systems.
\par
We will demonstrate how the conditions defining  the subspace $\Omega$ of generalized temperatures  is distinctively  handy when one uses solvable coordinates, namely while relying on the metric equivalence of non compact $\mathrm{U/H}$ symmetric spaces with their corresponding solvable Lie Group Manifolds $\mathcal{S}_{\mathrm{U/H}}$.
Indeed the inequalities defining the range of $\boldsymbol{\beta}$ are successively extracted from the negativity requirement of the quadratic term in gaussian integrals over the real line:
\begin{equation}\label{positivityconstraint}
  \int_{-\infty}^{\infty}e^{-\alpha _2 x^2-\alpha _1 x} dx \, = \, \frac{\sqrt{\pi } e^{\frac{\alpha _1^2}{4 \alpha _2}}}{\sqrt{\alpha _2}} \quad  \text{iff} \quad  \alpha _2 >0
\end{equation}
In order to arrive at this we first have to single out among the non-compact symmetric manifolds $\mathrm{U/H}$ the series of those that are K\"ahlerian. There are essentially two series, since, as we explained in the introduction, the K\"ahler character of a coset manifold $\mathrm{U/H}$ is uniquely signaled by the presence of a $\uu(1) \simeq\so(2)$ addend in the compact subalgebra $\mathbb{H}$. Recalling the results and the discussion of the foundational paper \cite{pgtstheory} we see that the two series are:
\begin{equation}\label{supercaronte}
  \mathcal{M}^{[2,q]} \, \equiv \,  \frac{\mathrm{SO(2,2+q)}}{\mathrm{SO(2)}\times \mathrm{SO(2+q)}}
  \quad ; \quad \mathbb{SH}_{n}\, \equiv \, \frac{\mathrm{Sp(2n,\mathbb{R})}}{\mathrm{U(1)}\times \mathrm{SU(n)}}
  \underbrace{\, = \,}_{\text{by Cayley
  map}}\frac{\mathrm{USp(n,n)}}{\mathrm{U(1)}\times \mathrm{SU(n)}}
\end{equation}
For the generalized Cayley map see for instance \cite{advancio} (page 356, formulae (7.2.12)-(7.2.13)) and more generally \cite{mylecture}.
The first series was already mentioned in eq.(\ref{r2class}) and corresponds to an entire Tits Satake Universality class, the common Tits Satake submanifold of the entire class being:
\begin{equation}\label{tittusatakr2}
  \mathcal{M}^{[2,q]}_{TS} \, = \, \mathcal{M}^{[2,1]} \, = \, \frac{\mathrm{SO(2,3)}}{\mathrm{SO(2)}\times \mathrm{SO(3)}} \, \underbrace{\simeq}_{\begin{array}{c}
   \text{spinor double}\\
   \text{covering}\\
   \end{array}}\, \frac{\mathrm{Sp(4,\mathbb{R})}}{\mathrm{U(1)}\times \mathrm{SU(2)}}
\end{equation}
As we see from eq.(\ref{tittusatakr2}) the Tits Satake submanifold of the first series is equivalent, due to the low dimension Lie algebra isomorphisms to the second manifold in the second series. Indeed the second series is the series of Siegel half spaces of genus $n$ and $\frac{\mathrm{Sp(4,\mathbb{R})}}{\mathrm{U(1)}\times \mathrm{SU(2)}}$ is a double covering of $\frac{\mathrm{SO(2,3)}}{\mathrm{SO(2)}\times \mathrm{SO(3)}}$ obtained through the use of the $4$-dimensional spinor representation of $\mathrm{SO(2,3)}$ rather then its $5$-dimensional vector one (see
\cite{pgtstheory} and also \cite{fre2023book}).
The first manifold $n=1$ of the second series is also the Tits Satake submanifold of the Hyperbolic Space universality class:
\begin{equation}\label{cannoniconpanna}
  \mathcal{M}^{[1,1+q]} \equiv \,  \frac{\mathrm{SO(1,1+q)}}{\mathrm{SO(1+q)}}
\end{equation}
In the above series the manifolds are not K\"ahlerian except for the case $q=1$ which also corresponds to the Tits Satake submanifold of the entire class.
\begin{equation}\label{tittusatakr2}
  \mathcal{M}^{[1,q]}_{TS} \, = \, \mathcal{M}^{[1,1]} \, = \, \frac{\mathrm{SO(1,2)}}{ \mathrm{SO(2)}} \, \underbrace{\simeq}_{\begin{array}{c}
   \text{spinor double}\\
   \text{covering}\\
   \end{array}}\, \frac{\mathrm{Sp(2,\mathbb{R})}}{\mathrm{U(1)}}\, = \, \frac{\mathrm{SL(2,\mathbb{R})}}{\mathrm{SO(2)}}\underbrace{\, = \,}_{\text{by Cayley
  map}} \frac{\mathrm{SU(1,1)}}{\mathrm{U(1)}}
\end{equation}
Finally, in view of the visions of Paint Group invariance and of its relevance in ML algorithms discussed in
\cite{pgtstheory,tassellandum} one realizes that the most interesting  K\"ahler manifolds are those of the series $\mathcal{M}^{[2,q]}$ in (\ref{supercaronte}) that are also named Calabi-Vesentini manifolds. An interesting point, whose implication for ML are all to be
studied, is that the Calabi-Vesentini manifolds times a hyperbolic plane constitute Special K\"ahler manifolds (see
\cite{mylecture} and \cite{pgtstheory})  and can be described by a suitable section of the corresponding flat holomorphic symplectic bundle, as recalled in \cite{pgtstheory}.
\subsection{The General Setup}
\label{settonasale}
Having singled out the series of non-compact symmetric spaces $\mathrm{U/H}$ that are K\"ahlerian and correspondingly apt to support generalized thermodynamics {\`a} la Souriau let us develop the general setup for the construction of this latter.
\par
For all $\mathrm{U/H}$ we have the two sinergic Lie algebra decompositions:
\begin{eqnarray}
\label{sinergico}
  \mathbb{U} &=& \mathbb{H} \oplus \mathbb{K} \label{cane1}\\
  \mathbb{U} &=& \mathbb{H} + Solv_{\mathrm{U/H}} \label{cane2}
\end{eqnarray}
where $\mathbb{H}$ is the maximal compact subalgebra of $\mathbb{U}$ and $\mathbb{K}$ constitutes a linear representation of $\mathbb{\mathbb{H}}$ under its adjoint action but it is not a closed subalgebra:
\begin{equation}\label{grullo1}
  \left[\mathbb{H} \, , \, \mathbb{K} \right] \subset \mathbb{K} \quad ; \quad \left[\mathbb{K} \, , \, \mathbb{K} \right]\nsubseteq \mathbb{K} \quad \text{rather} \quad \left[\mathbb{K} \, , \, \mathbb{K} \right] \subset \mathbb{H}
\end{equation}
while $Solv_{\mathrm{U/H}}$, that has the same dimension as $\mathbb{K}$, is a closed Lie subalgebra (a solvable one) but it is not a linear representation of $\mathbb{H}$ under its adjoint action:
\begin{equation}\label{grullo2}
  \left[\mathbb{H} \, , \, Solv_{\mathrm{U/H}} \right] \nsubseteq Solv_{\mathrm{U/H}} \quad ; \quad \left[Solv_{\mathrm{U/H}} \, , \, Solv_{\mathrm{U/H}} \right] \subset Solv_{\mathrm{U/H}}
\end{equation}
\par
In the case when $\mathrm{U/H}$ is K\"ahlerian we have the additional essential property:
\begin{equation}\label{essenziando}
  \mathbb{H} \, = \, \mathbb{H}_0 \oplus \so_c(2) \quad ; \quad \left[\mathbb{H}_0 \, , \, \mathbb{H}_0 \right] \subset \mathbb{H}_0 \quad ; \quad \left[\mathbb{H}_0 \, , \, \so_c(2) \right] \, = \, \mathbf{0}
\end{equation}
In all cases (compare with section \ref{zaklyuchenie}) we construct the $\mathrm{U}$-invariant metric on $\mathrm{U/H}$ utilizing the vielbein extracted from the left-invariant matrix $1$-form $\Theta$ of the metric equivalent solvable Lie group $\mathcal{S}_{\mathrm{U/H}} \subset \mathrm{U}$:
\begin{equation}\label{tettona}
  \boldsymbol{\Theta} \, \equiv \, \mathbb{L}^{-1}(\boldsymbol{\Upsilon}) \, \mathrm{d}\mathbb{L}(\boldsymbol{\Upsilon})
\end{equation}
by projecting it onto an orthonormal basis of $\mathbb{K}$ generators:
\begin{eqnarray}\label{congelato}
  \mathbf{V}^A & \equiv & \text{Tr}\left(\boldsymbol{\Theta}\cdot K^A\right) \quad ; \quad \text{Tr}(K^A\cdot K^B) \, = \, \delta^{AB}\nonumber\\
  ds_{\mathrm{U/H}}^2 &=& \mathbf{V}^A \times \mathbf{V}^B \, \delta_{AB}
\end{eqnarray}
and we find that we can equivalently write (compare with eq.(\ref{cordelia}):
\begin{equation}\label{critta}
  ds_{\mathrm{U/H}}^2 \, = \, \underbrace{\kappa_{AB}}_{\text{const. symm.}} \, \mathbf{e}^A \times \mathbf{e}^B
\end{equation}
where the $1$-forms $\mathbf{e}^A$ are defined by the expansion of $\boldsymbol{\Theta}$ along a basis of generators
$T_A$ of the solvable Lie algebra $Solv_{U/H}$:
\begin{equation}\label{crucifige}
  \boldsymbol{\Theta} \, = \, \mathbf{e}^A \, T_A
\end{equation}
since the relation between the two synergic decompositions (\ref{cane1},\ref{cane2}) of the same  Lie algebra $\mathbb{U}$ implies that there always exists a constant matrix $\nu$ such that (compare with eq.(\ref{crisippo})):
\begin{equation}\label{ermenegildo}
  V^A \, = \, \nu^A_{\phantom{A}B} \, e^B \quad \Rightarrow \quad \kappa = \nu^T \cdot \nu
\end{equation}
\par
In the K\"ahlerian case we have the additional item of the K\"ahler $2$-form whose form is general in terms of the matrix representing the adjoint action of the $\so_c(2)$ generator on the space $\mathbb{K}$. Let us name $\mathbf{X}^c$ such generator and construct its $d$-dimensional matrix representation on $\mathbb{K}$:
\begin{equation}\label{cratello}
  \mathbf{K}^{c}_{AB} \, = \, \delta_{AC} \, \delta_{BD}\, \text{Tr}\left( \left[\mathbf{X}^c\, ,\,K^C\right]\cdot K^D\right)
\end{equation}
Since $\so_c(2)$ is a compact subalgebra, the matrix $\mathbf{K}^{c}$ is necessarily antisymmetric and the
$\mathrm{H}$ invariance of the metric guarantees that the K\"ahler $2$-form defined by:
\begin{equation}\label{kalledue}
  \boldsymbol{\mathcal{K}} \, \equiv \, \mathbf{K}^{c}_{AB} \, \mathbf{V}^A \wedge \mathbf{V}^B
\end{equation}
is necessarily closed. The proof is simple. By definition of the Levi Civita connection in vielbein/spin-connection formalism we have:
\begin{equation}\label{crinolo}
  \mathrm{d}\mathbf{V}^A \, = \, \omega^{AC}\wedge \mathbf{V}^C
\end{equation}
where $\omega^{IJ}\, = \, -\, \omega^{JI}$ is valued in the $d$-dimensional representation $\mathbb{K}$ of the
$\mathbb{H}$ Lie algebra. Using (\ref{crinolo}) we obtain
\begin{equation}\label{barlacco}
  \mathrm{d}\boldsymbol{\mathcal{K}} \, = \, - \underbrace{\left[\mathbf{K}^{c}, \omega\right]_{IJ}}_{=0} \, V^I \wedge V^J
\end{equation}
Indeed $\mathbf{K}^{c}$ is the $\so_c(2)$ subalgebra and it commutes with the whole $\mathbb{H}$-algebra.
\par
The next question concerns the Killing vector fields and their hamiltonian representation. It is important to stress another clearcut distinction. The symmetric space $\mathrm{U/H}$ is diffeomorphic to the solvable Lie group
$\mathcal{S}_{\mathrm{U/H}}$ and the latter, as any Lie group manifold $\mathrm{G}$, possesses two commuting sets of vector fields $\boldsymbol{\mathfrak{t}}^{[L/R]}_A$ satisfying the Lie algebra $\mathbb{G}$ of the group:
\begin{equation}\label{tournedorossini}
  \left[\boldsymbol{\mathfrak{t}}^{[L/R]}_A\, , \, \boldsymbol{\mathfrak{t}}^{[L/R]}_B\right] \, = \,
  f_{AB}^{\phantom{AB}C} \,\, \boldsymbol{\mathfrak{t}}^{[L/R]}_B C \quad ; \quad \left[\boldsymbol{\mathfrak{t}}^{[L]}_A\, , \, \boldsymbol{\mathfrak{t}}^{[R]}_B\right]\, = \, 0
\end{equation}
the left-invariant ones $\boldsymbol{\mathfrak{t}}_A\, \equiv \,\boldsymbol{\mathfrak{t}}^{[L]}_A$, dual to the left-invariant $1$-forms $\mathbf{e}^B$, according with eq.s(\ref{cartaccia},\ref{neanderthal}), generate right-translations, while the right-invariant ones $\boldsymbol{\mathfrak{t}}^{[R]}_A$, dual to the right-invariant 1-forms $\mathbf{e}_{[R]}^A$ defined by the expansion along a generator basis $T_A$ of $Solv_{U/H}$ of the right-invariant matrix $1$-form:
\begin{equation}\label{contaminazia}
  \boldsymbol{\Theta}_{[R]} \, \equiv \,  \mathrm{d}\mathbb{L}\cdot \mathbb{L}^{-1} \, = \, \mathbf{e}_{[R]}^A \, T_A \quad ; \quad \mathbf{e}_{[R]}^A\left(\boldsymbol{\mathfrak{t}}^{[R]}_A\right) \, = \, \delta^A_B
 \end{equation}
 generate the left-translations. For this reason only the right-invariant vector fields $\boldsymbol{\mathfrak{t}}^{[R]}_A$ are \textbf{Killing vector fields} of the symmetric space metric (\ref{congelato})
 and, as such, they are also \textbf{symplectic Killing vector fields} for the symplectic structure provided by the K\"aehler
 $2$-form in eq.(\ref{kalledue}):
 \begin{equation}\label{liuvillico}
   \ell_{\mathfrak{t}^{[R]}_A} \boldsymbol{\mathcal{K}} \, = \,0
 \end{equation}
where $\ell_\mathbf{t} $ denotes the Lie derivative along the vector field $\mathbf{t}$. The right-invariant vector fields $\mathfrak{t}^{[R]}_A$ are not the only Killing vectors and hence symplectic Killing vectors: indeed
the symmetric space metric is invariant with respect to the whole group $\mathrm{U}$ and we need
a set of Killing vector fields satisfying the whole $\mathbb{U}$ Lie algebra. According with the decomposition (\ref{cane2}) we just have to add the Killing vector fields spanning the compact subalgebra $\mathbb{H}$. The question is how to obtain their explicit expression.
\subsubsection{General Construction Method of the Killing Vector Fields}
\label{generalecuster}
In order to construct the expression in solvable coordinate $\boldsymbol{\Upsilon}$ of the Killing vector fields associated with the compact generators we utilize the following procedure which is general and applies to any
Killing vector field.
\par
From the general theory of coset manifolds (see for instance \cite{pietrobook}, second volume, section 5.2.3 page 114 and following ones) we know that when we act on the left on a coset representative $\mathbb{L}(y)$ with any element $g\in \mathrm{U}$ we have:
\begin{equation}\label{barengo}
  g\, \mathbb{L}(y)\, = \, \mathbb{L}\left(g\left(y\right)\right)\cdot h(y,g) \quad ; \quad h(y,g)\in \mathrm{H} \subset \mathrm{U}
\end{equation}
where $y^\prime=g(y)$ is the coordinate of the new point in the coset reached by the $g$ transformation and
$h(y,g)$ which lies in the subgroup is named the compensator. Typically the determination of the compensator is a cumbersome task, yet in the solvable parameterization there is a well defined universal algorithm for the determination of $g(y)$ that bypasses  the determination of the compensator. Our coset representative is
an element of the solvable Lie group and as demonstrated in \cite{pgtstheory,TSnaviga} we can always use for any
$\mathrm{U/H}$ the so called triangular basis where the solvable coset representative is an upper triangular matrix.
Hence the solvable coset representative $\mathbb{L}(\boldsymbol{\Upsilon})$ is upper triangular and the matrices $h\in \mathrm{H}$ of the compact subgroup, in particular the compensators are all orthogonal $h\cdot h^T \, =\, \mathbf{1}$. It follows that defining the symmetric matrix:
\begin{equation}\label{barnabeo}
  \mathcal{M}(\boldsymbol{\Upsilon}) \, \equiv \, \mathbb{L}(\boldsymbol{\Upsilon})\cdot\mathbb{L}^T(\boldsymbol{\Upsilon})
\end{equation}
we have:
\begin{equation}\label{fioridizucca}
  \forall g \in \mathrm{U} \quad : \quad g\cdot \mathcal{M}\left(\boldsymbol{\Upsilon}\right) \cdot g^{T} \, = \,
  \mathcal{M}\left(g\left(\boldsymbol{\Upsilon}\right)\right)
\end{equation}
In order to obtain $g\left(\boldsymbol{\Upsilon}\right)$, which is our goal, it suffices to utilize the finite recursive Cholewski-Crout algebraic algorithm (see \cite{pgtstheory,TSnaviga}) that, given $\mathcal{M}\left(g\left(\boldsymbol{\Upsilon}\right)\right)$, uniquely determines the upper triangular matrix
$\mathbb{L}\left(g\left(y\right)\right)$ such that:
\begin{equation}\label{fioridizucca}
  \mathcal{M}\left(g\left(\boldsymbol{\Upsilon}\right)\right)\, = \, \mathbb{L}\left(g\left(\boldsymbol{\Upsilon}\right)\right)\cdot\mathbb{L}^T\left(g\left(\boldsymbol{\Upsilon}\right)\right)
\end{equation}
Then applying the inverse of the $\Sigma$ exponential map according with the definitions and conventions of
\cite{pgtstheory,TSnaviga} we get:
\begin{equation}\label{quittus}
  \Sigma^{-1}\left[\mathbb{L}\left(g\left(\boldsymbol{\Upsilon}\right)\right)\right] \, = g\left(\boldsymbol{\Upsilon}\right)^\alpha\, T_\alpha
\end{equation}
In equation (\ref{quittus}) consider the compact group subgroup elements of the form:
\begin{equation}\label{pergolatorotto}
  g_i[\theta]=\exp\left[\theta \, J_i \right]
\end{equation}
where $J_i$ ($i=1,\dots, m=\text{dim}\mathbb{H}$) is a basis of generators of the compact subalgebra $\mathbb{H}$.
The $g_i[\theta]$ are elements of the $m$ one-parameter subgroups of $\mathrm{H}$.
Expanding in power series of $\theta$ we get:
\begin{equation}\label{caramellina}
  g_i[\theta]\left(\boldsymbol{\Upsilon}\right)^\alpha\, = \, \boldsymbol{\Upsilon}^\alpha \, + \, \mathfrak{f}_i^\alpha (\boldsymbol{\Upsilon}) \, + \,\mathcal{ O}(\theta^2)
\end{equation}
The searched for Killing vector fields are:
\begin{equation}\label{transeunte}
  \boldsymbol{\mathfrak{k}}_i \, \equiv \, \mathfrak{f}_i^\alpha (\boldsymbol{\Upsilon}) \, \frac{\partial}{\partial \Upsilon^\alpha}
\end{equation}
and note that by construction  they satisfy the Lie algebra $\mathbb{H}$ and are symplectic Killing vector fields, according with the definition (\ref{liuvillico}). Note also that the above construction of the Killing vector fields that was indispensable for the compact subalgebra $\mathbb{H}$ could also be applied to the determination of the Killing vector fields associated with the solvable Lie subalgebra, if we did note know them independently, or to those associated with the $\mathbb{K}$-generators, corresponding to the decomposition (\ref{cane1}). Indeed the above procedure is completely general for any set of generators $J_\Lambda$ of the entire Lie algebra $\mathbb{U}$.
\subsubsection{The General form of the Moment-Maps}
\label{deagostini}
Given any set $\boldsymbol{\mathfrak{k}}_\Lambda$ of Killing vector fields, each uniquely associated with a generator $J_\Lambda$ of the Lie algebra $\mathbb{U}$ via the construction mentioned above in eq.s
(\ref{pergolatorotto},\ref{transeunte}), one obtains the corresponding moment map via another fully general formula
which is extremely simple:
\begin{eqnarray}\label{gelindoelapecora}
  \mathfrak{P} & : & \mathbb{U} \, \longrightarrow \, \mathbb{C}^{\infty}\left(\frac{\mathrm{U}}{\mathrm{H}}\right)\nonumber\\
  \mathfrak{P}_\Lambda\left(\boldsymbol{\Upsilon}\right) &=& \ft 12 \, \text{Tr} \left[ X_c\cdot \mathbb{L}^{-1}
  \left(\boldsymbol{\Upsilon}\right)\cdot J_\Lambda \cdot \mathbb{L}
  \left(\boldsymbol{\Upsilon}\right)\right]
\end{eqnarray}
The functions $\mathfrak{P}_\Lambda\left(\boldsymbol{\Upsilon}\right)$  satisfy the necessary condition with respect to the Killing vector fields $\boldsymbol{\mathfrak{k}}_\Lambda$:
\begin{equation}\label{lescontamines}
  i_{\boldsymbol{\mathfrak{k}}_\Lambda}\boldsymbol{\mathcal{K}}  \, = \, \mathrm{d}\mathfrak{P}_\Lambda\left(\boldsymbol{\Upsilon}\right)
\end{equation}
and their definition (\ref{gelindoelapecora}) has a very simple interpretation; they are the projection onto the $\so(2)_c$ central subalgebra defining the K\"ahler structure of the adjoint transformation of the generator $J_\lambda$. An important comment is the following. Since all the non-compact symmetric spaces $\mathrm{U/H}$ are
Hadamard-Cartan manifolds diffeomorphic to $\mathbb{R}^n$ (see \cite{TSnaviga}), they can be covered by just one open chart and the solvable coordinates $\boldsymbol{\Upsilon}$ provide such chart. For this reason the moment maps in solvable coordinates $\mathfrak{P}_\Lambda\left(\boldsymbol{\Upsilon}\right)$ are globally defined functions over the whole manifold and eq. (\ref{lescontamines}) holds true globally.
\subsubsection{The Partition Function and the Gibbs Probability Distribution}
Equipped with the above general weapons one can address the conditions on the temperature vector $\boldsymbol{\beta}$ for the convergence of the integral (\ref{convpartofungo}) defining the partition function
$Z(\boldsymbol{\beta})$. The first observation is that in the privileged solvable coordinate chart the volume form reduces to:
\begin{equation}\label{garibaldino}
  \underbrace{\boldsymbol{\mathcal{K}}\wedge\boldsymbol{\mathcal{K}} \wedge \dots \wedge \boldsymbol{\mathcal{K}}}_{n \text{ times}}  \, = \, \text{const} \times \underbrace{\mathrm{d}\Upsilon^1 \wedge\dots\wedge \mathrm{d}\Upsilon^{r_{n.c.}}}_{\text{Cartan coordinates}} \underbrace{\mathrm{d}\Upsilon^{r_{n.c.}+1}\wedge\dots \wedge \mathrm{d}\Upsilon^{2n}}_{\text{nilpotent coordinates}}
\end{equation}
where $r_{n.c.}$ is the non compact rank of the considered K\"ahlerian symmetric space $\mathrm{U/H}$. If we choose the first series in eq.(\ref{supercaronte}), then the non-compact rank is always $r_{n.c.}$ and we just have two solvable coordinates associated with the unique two Cartan generators, while all the other coordinates $\Upsilon^{r_{n.c.}+1},\dots,\Upsilon^{2n}$ are associated with nilpotent generators corresponding to long and short roots, the latter organized in two multiplets assigned to the fundamental representation of the Paint Group $\mathrm{G_{Paint}} \, = \, \mathrm{SO(q)}$.  On the contrary all the K\"ahler manifolds in the second series
are maximally split and the non-compact rank increases linearly in $n$. In any case the strategy to calculate the partition function is that of starting with the integration one-by-one on the nilpotent coordinates  which leads each time to a gaussian integral  of the form (\ref{positivityconstraint} and to a corresponding positivity constraint on the $\boldsymbol{\beta}$ vector. As we show in the two explicitly constructed examples the integration on the nilpotents for $r_{n.c.}\, = \,2$ can be explicitly performed analytically until we reach a function to be integrated on the Cartan coordinates which can be verified to be explicitly positive definite and exponetially decreasing at infinity so as to guaranteed partition function convergence. In the simplest case of the hyperbolic plane we are able to perform also the last integration analytically and we obtain the partition function in closed form and also the thermodynamical metric. For the Siegel $n=2$ plane the last integration on the Cartan fields has to be done numerically, yet as compiled functions all the thermodynamical items are accessible. The extension of the
results for the Siegel plane to all the manifolds of the first series in (\ref{supercaronte}) via a convenient use of Paint Group invariance is an issue of further research. In appendix \ref{belohorizonte} we present the preliminary setup calculation for the case $q=2$.
\subsection{Generalized Thermodynamics {\`a} la Souriau of the Poincar\'e-Lobachevsky Hyperbolic Plane $\mathbb{H}_2$}
 Let us begin with the simplest example, namely with the hyperbolic plane $\mathbb{H}_2$. Utilizing the representation:
 \begin{equation}\label{hyperplane}
   \mathbb{H}_{2}\, = \, \frac{\mathrm{SL(2,\mathbb{R})}}{\mathrm{SO(2)}}
 \end{equation}
 and following all the conventions of \cite{pgtstheory}, we write the generators of the full
  $\mathrm{SL(2,\mathbb{R})}$ Lie algebra  in two ways, namely in the orthogonal decomposition:
 \begin{equation}\label{banderuola1}
   \slal(2,\mathbb{R})\, = \, \so(2) \oplus \mathbb{K}
 \end{equation}
 and in the solvable subalgebra decomposition:
 \begin{equation}\label{banderuola2}
   \slal(2,\mathbb{R})\, = \, \so(2) + Solv_2
 \end{equation}
  For the $1$-dimensional compact subalgebra $\mathbb{H}\, = \, \so(2)$, both in eq.(\ref{banderuola1}) and in eq.(\ref{banderuola2}), we choose the same generator:
 \begin{equation}\label{Xcompact}
   X_c \, = \, \left(
\begin{array}{cc}
 0 & 1 \\
 -1 & 0 \\
\end{array}
\right)
 \end{equation}
The two generators of the orthogonal non-compact space $\mathbb{K}$ are given by:
 \begin{equation}\label{tafferuglio}
   K_1 \, = \, \left(
\begin{array}{cc}
 \frac{1}{\sqrt{2}} & 0 \\
 0 & -\frac{1}{\sqrt{2}} \\
\end{array}
\right) \quad , \quad K_2 \, = \, \left(
\begin{array}{cc}
 0 & \frac{1}{\sqrt{2}} \\
 \frac{1}{\sqrt{2}} & 0 \\
\end{array}
\right)
 \end{equation}
 while our chosen basis of generators for the solvable Lie algebra is:
 \begin{equation}\label{sodacaustica}
   Solv_2\, = \, \text{span}\left\{T_1\,T_2\right\}\quad ; \quad T_1 \, = \, \left(
\begin{array}{cc}
 1 & 0 \\
 0 & -1 \\
\end{array}
\right) \quad ; \quad T_2 \, = \, \left(
\begin{array}{cc}
 0 & 1 \\
 0 & 0 \\
\end{array}
\right)
 \end{equation}
 As explained in \cite{pgtstheory}, the general form of a solvable Lie group element is:
 \begin{equation}\label{trullo}
   \mathbb{L}(\boldsymbol{\Upsilon}) \,= \, \left(
\begin{array}{cc}
 e^{\Upsilon _1} & e^{\Upsilon _1} \Upsilon _2 \\
 0 & e^{-\Upsilon _1} \\
\end{array}
\right)
 \end{equation}
and the left-invariant matrix $1$-form decomposes as follows in the solvable Lie algebra basis:
 \begin{equation}\label{invarform}
   \Theta \, \equiv \, \mathbb{L}^{-1}\, \mathrm{d}\mathbb{L} \, = \, \mathbf{e}^1 \, T_1 \, + \, \mathbf{e}^2 \,  T_2
 \end{equation}
where the two left-invariant $1$-forms $\mathbf{e}^{1,2}$ have the following appearance:
\begin{eqnarray}\label{epissili}
   \mathbf{e}^1 &=& \text{d$\Upsilon $}_1\nonumber\\
   \mathbf{e}^2&=& 2 \Upsilon_ 2 \,\text{d$\Upsilon $}_1 +\text{d$\Upsilon
   $}_2
\end{eqnarray}
The zweibein of the $2$-dimensional space is instead defined by the projection of the left-invariant $1$-form along the $\mathbb{K}$ generators:
\begin{equation}\label{kurchatov}
  \mathbf{V}^i \, = \, Tr[\Theta\cdot K_i]
\end{equation}
and one obtains the relation:
\begin{equation}\label{krumito}
     \mathbf{V} \, = \, \nu \, \mathbf{e} \quad ; \quad \nu \, = \, \left(
\begin{array}{cc}
 \sqrt{2} & 0 \\
 0 & \frac{1}{\sqrt{2}} \\
\end{array}
\right)
\end{equation}
so that the norm form on the solvable Lie algebra mentioned in eq.(\ref{cordelia}) is as follows:
\begin{equation}\label{corneliogracco}
   \kappa \, = \, \nu^T\cdot\nu \, = \, \left(
\begin{array}{cc}
 2 & 0 \\
 0 & \frac{1}{2} \\
\end{array}
\right)
\end{equation}
 Defining the adjoint action of the unique $\so(2)$ generator on the $K_i$ generators and hence on the zweibein $\mathbf{V}$:
 \begin{equation}\label{aggiungisale}
   K^{c} \, = \, \text{adj}_{X_c}[\mathbb{K}] \, = \,\text{Tr}\left(\left[X_c \, , \, K_i\right].K_j\right) \, = \, \left(
\begin{array}{cc}
 0 & -1 \\
 1 & 0 \\
\end{array}
\right)
 \end{equation}
 we obtain the K\"ahler $2$-form:
 \begin{equation}\label{kallerform}
   \mathcal{K} \, = \, K^{c}_{ij} \, \mathbf{V}^i \wedge \mathbf{V}^j \, = \, -2 \, \mathbf{V}^1 \wedge \mathbf{V}^2 \, = \, -2\, d\Upsilon_1 \wedge d\Upsilon_2
 \end{equation}
 whose closure $\mathrm{d}\mathcal{K}\,= \,0$ is immediately verified.
 \par
 The K\"ahler metric and its inverse are immediately calculated from the above data:
 \begin{equation}\label{metricHyp}
   \mathbf{g}_{ij} \, = \, \left(
\begin{array}{cc}
 2 \left(\Upsilon _2^2+1\right) & \Upsilon _2 \\
 \Upsilon _2 & \frac{1}{2} \\
\end{array}
\right) \quad ; \quad \mathbf{g}^{-1|ij} \, = \, \left(
\begin{array}{cc}
 \frac{1}{2} & -\Upsilon _2 \\
 -\Upsilon _2 & 2 \left(\Upsilon _2^2+1\right) \\
\end{array}
\right)
 \end{equation}
 and the complex structure tensor in the solvable coordinate basis is obtained from the explicit form of the K\"ahler $2$-form provided in eq.(\ref{kallerform}):
 \begin{equation}\label{formicolio}
   J_c \, = \, K^{c}\cdot g^{-1} \, = \, \left(
\begin{array}{cc}
 \Upsilon _2 & -2 \left(\Upsilon _2^2+1\right) \\
 \frac{1}{2} & -\Upsilon _2 \\
\end{array}
\right) \quad ; \quad  J_c^{\phantom{c}2} \, = \, - \, \mathbf{1}
 \end{equation}
 The metric (\ref{metricHyp}) is explicitly hermitian with respect to the complex structure (\ref{formicolio}):
 \begin{equation}\label{saturnali}
   J_c\cdot \mathbf{g} \cdot J_c^T \, = \, \mathbf{g}
 \end{equation}
 The left-invariant $1$-forms $\mathbf{e}$ on the solvable Lie group manifold $\mathcal{S}_2$, in terms of which we have constructed the K\"ahler metric satisfy the Maurer Cartan equation written below that define the structure constant of the solvable Lie algebra $Solv_2$
 \begin{equation}\label{sospirodelmoro}
   \text{d$\mathbf{e}$}^1\, = \, 0 \quad ; \quad \text{d$\mathbf{e}$}^2+2 \mathbf{e} ^1 \wedge\mathbf{e} ^2
   \, = \,  0
 \end{equation}
The same Maurer Cartan equations (\ref{sospirodelmoro}) are satisfied also by the right-invariant $1$-forms obtained decomposing the right-invariant matrix $1$-form $\mathrm{d}\mathbb{L}\cdot \mathbb{L}$.
\par
Following all the procedures detailed in the previous section \ref{settonasale} we obtain the explicit form of
the three Killing vector fields in solvable coordinates:
\begin{eqnarray}
\label{stendivettore}
  \boldsymbol{\mathfrak{k}}_0 &=&  e^{2 \Upsilon _1} \Upsilon _2\, \frac{\partial}{\partial \Upsilon^1}+ \left(e^{-2 \Upsilon _1}-e^{2
   \Upsilon _1} \left(\Upsilon _2^2+1\right)\right)\, \frac{\partial}{\partial \Upsilon^2} \nonumber \\
  \boldsymbol{\mathfrak{k}}_1 &=& \frac{\partial}{\partial \Upsilon^1} \nonumber \\
  \boldsymbol{\mathfrak{k}}_2 &=& \exp\left[-2 \, \Upsilon^1\right]\, \frac{\partial}{\partial \Upsilon^2}
\end{eqnarray}
where $\boldsymbol{\mathfrak{k}}_0$ is the Killing vector field associated with compact generator $X_c$,
while $\boldsymbol{\mathfrak{k}}_{1,2}$ are the Killing vector fields associated with the solvable Lie algebra generators $T_{1,2}$. The corresponding moment maps calculated by means of eq.(\ref{gelindoelapecora}) are the following ones:
\begin{eqnarray}
\label{natalepomeriggio}
  \mathfrak{P}_0(\boldsymbol{\Upsilon}) &=& \frac{1}{2} e^{-2 \Upsilon _1} \left(-e^{4 \Upsilon _1} \left(\Upsilon
   _2^2+1\right)-1\right) \nonumber\\
  \mathfrak{P}_1(\boldsymbol{\Upsilon}) &=& -\Upsilon _2 \nonumber\\
  \mathfrak{P}_2(\boldsymbol{\Upsilon}) &=& -\frac{1}{2} e^{-2 \Upsilon _1}
\end{eqnarray}
\subsubsection{Calculation of the Partition Function}
\begin{figure}
\begin{center}
\includegraphics[width=9cm]{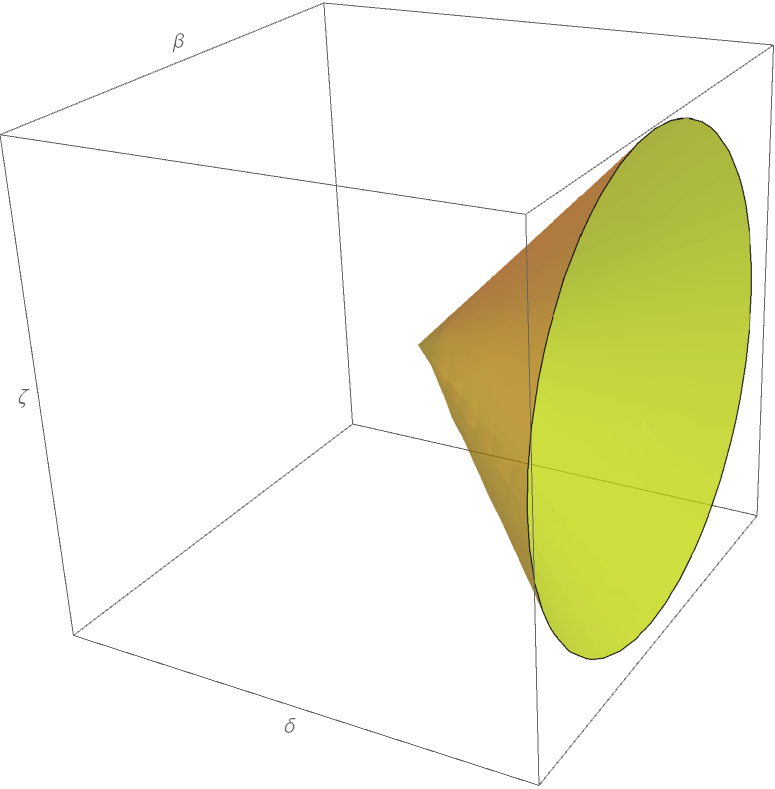}
 \caption{\label{balzano1} The cone $\Omega$ of Souriau allowed temperature vectors in the $\slal(2,\mathbb{R})$ Lie algebra space as defined in eq.(\ref{kunegonda}). }
\end{center}
\end{figure}

For simplicity of writing, naming $\alpha,\beta,\gamma$ the three component $\beta^0, \beta^1,\beta^2$ of the temperature vector $\boldsymbol{\beta}$, and  $x=\Upsilon_1,y=\Upsilon_2$ the two solvable coordinates, the
partition function to be computed is the following:
\begin{equation}\label{partitona}
  Z(\alpha,\beta,\gamma) \, = \, \int_{-\infty}^\infty \, dx \, \int_{-\infty}^\infty dy \,
  \exp \left[-\frac{1}{2} e^{-2 x} \left(\alpha +\gamma +\alpha  e^{4 x}
   \left(y^2+1\right)+2 \beta  e^{2 x} y\right)\right]
\end{equation}
and using the convergence condition of the gaussian integrals recalled in eq.(\ref{positivityconstraint}) we get the following constraints:
\begin{equation}\label{su11costretti}
  \alpha \, > \, 0 \quad ; \quad \alpha  (\alpha +\gamma )-\beta ^2\, > \, 0
\end{equation}
When the above conditions are satisfied the integrals are easily calculated and we obtain:
\begin{equation}\label{zetofono}
  Z(\alpha,\beta,\gamma)\, = \, \frac{\pi  e^{-\sqrt{\alpha  (\alpha +\gamma )-\beta ^2}}}{\sqrt{\alpha  (\alpha
   +\gamma )-\beta ^2}}
\end{equation}
The corresponding \textbf{Gibbs probability distribution} takes the following appearance:
\begin{equation}\label{gibbuto1}
  G(\alpha,\beta,\gamma,\Upsilon^1,\Upsilon^2)\, = \, \frac{\sqrt{\alpha  (\alpha +\gamma )-\beta ^2} \exp \left[\sqrt{\alpha  (\alpha +\gamma
   )-\beta ^2}-\frac{1}{2} e^{-2 \Upsilon _1} \left(\alpha  e^{4 \Upsilon _1}
   \left(\Upsilon _2^2+1\right)+\alpha +2 \beta  e^{2 \Upsilon _1} \Upsilon _2+\gamma
   \right)\right]}{\pi }
\end{equation}
Eq.s (\ref{su11costretti}-\ref{gibbuto1}) take a nicer form by a linear redefinition of the  temperatures:
\begin{equation}\label{crisantemo}
  \alpha \, = \, \delta + \zeta \quad ; \quad \gamma \, = \, -2 \, \zeta \quad ; \quad \beta \, = \, \beta
\end{equation}
where $\delta,\zeta$ are the new temperature parameters. We get:
\begin{eqnarray}
  \delta &>& 0  \quad ; \quad \delta^2 - \, \beta^2 \, - \, \zeta^2 \, > \, 0 \label{omegaregion} \\
  {Z}(\delta,\beta,\zeta) &=& \frac{\pi  e^{-\sqrt{\delta ^2-\beta ^2-\zeta ^2}}}{\sqrt{\delta ^2-\beta ^2-\zeta ^2}}\label{Zfinta} \\
   {G}(\delta,\beta,\zeta,\Upsilon^1,\Upsilon^2) &=& \frac{\sqrt{-\beta ^2+\delta ^2-\zeta ^2} \exp \left(\sqrt{-\beta ^2+\delta ^2-\zeta
   ^2}-\frac{1}{2} e^{-2 \Upsilon _1} \left(2 \beta  e^{2 \Upsilon _1} \Upsilon _2+e^{4
   \Upsilon _1} \left(\Upsilon _2^2+1\right) (\delta +\zeta )+\delta -\zeta
   \right)\right)}{\pi }\nonumber\\
   \label{gibbuti2}
\end{eqnarray}
It appears from eq.(\ref{omegaregion}) that the Souriau $\Omega \subset \slal(2,\mathbb{R})$ subspace of allowed temperatures is a cone in the three dimensional Lie algebra space:
\begin{equation}\label{kunegonda}
   \, = \, \left\{ \left(
\begin{array}{cc}
 \beta  & \delta -\zeta  \\
 -\delta -\zeta  & -\beta  \\
\end{array}
\right) \in \slal(2,\mathbb{R}) \quad \mid \quad \delta > 0 \, , \, \delta^2 - \, \beta^2 \, - \, \zeta^2 \, > \, 0\right \}
\end{equation}
as it is shown in fig.\ref{balzano1}
\subsubsection{Visualization of the Gibbs Probability Distributions}
In the perspective of Data Science applications the temperature vectors $\boldsymbol{\beta}=\{\delta,\beta,\zeta\}\in \Omega$ define, as the Gibbs states (\ref{gibbuti2}), probability distributions over the symmetric space $\mathrm{U/H} \, = \, \mathrm{SL(2,\mathbb{R})/O(2)}$, namely the Poincar\'e hyperbolic plane. Such Gibbs states are the appropriate generalizations to not-flat Cartan-Hadamard spaces of the
familiar gaussian distributions pertaining to flat space. The temperature vector models such distributions. It is very much helpful to visualize the Gibbs states (\ref{gibbuti2}) utilizing the disk model of the Poincar\'e plane:
\begin{equation}\label{diskmod}
  \text{Disk} \, = \, \left\{ \left\{x,y\right\} \in \mathbb{R}^2\, \mid \, x^2+y^2 \, < \, 1\right\}
\end{equation}
The relation between the solvable coordinates $\Upsilon_1,\Upsilon_2$ and the coordinates $x,y$ is provided by the following formula (see \cite{pgtstheory,tassellandum}):
\begin{equation}\label{frugiferentis}
  \Upsilon _2\, = \,  \frac{4 y}{x^2+y^2-1}\quad , \quad \Upsilon _1\, = \,  \log
   \left(-\frac{x^2+y^2-1}{x^2-2 x+y^2+1}\right)
\end{equation}
Substituting eq.(\ref{frugiferentis}) in eq.(\ref{gibbuti2}) and furthermore utilizing, polar coordinates in the
$\beta,\zeta$ plane of the temperature space:
\begin{equation}\label{fintereo}
  \beta \, = \, \mu \, \cos(\theta) \quad ; \quad  \zeta \, = \, \mu \, \sin(\theta)\quad ; \quad 0 < \mu < \delta
\end{equation}
we obtain the following three parameter family of probability distributions over the disk (\ref{diskmod}):
\begin{eqnarray}
\label{perigord}
 &&G(\delta,\mu,\theta,x,y) \,=\, \frac{\sqrt{\delta ^2-\mu ^2} }{\pi } \, \times \nonumber\\
   && \times \exp \left[\sqrt{\delta ^2-\mu ^2}-\frac{\left(x^2-2
   x+y^2+1\right)^2 \left(\delta -\mu  \sin (\theta )+\frac{\left(\frac{16
   y^2}{\left(x^2+y^2-1\right)^2}+1\right) \left(x^2+y^2-1\right)^4 (\delta +\mu  \sin
   (\theta ))}{\left(x^2-2 x+y^2+1\right)^4}+\frac{8 \mu  y \cos (\theta )
   \left(x^2+y^2-1\right)}{\left(x^2-2 x+y^2+1\right)^2}\right)}{2
   \left(x^2+y^2-1\right)^2}\right] \nonumber\\
\end{eqnarray}

\begin{figure}
\begin{center}
\includegraphics[width=12cm]{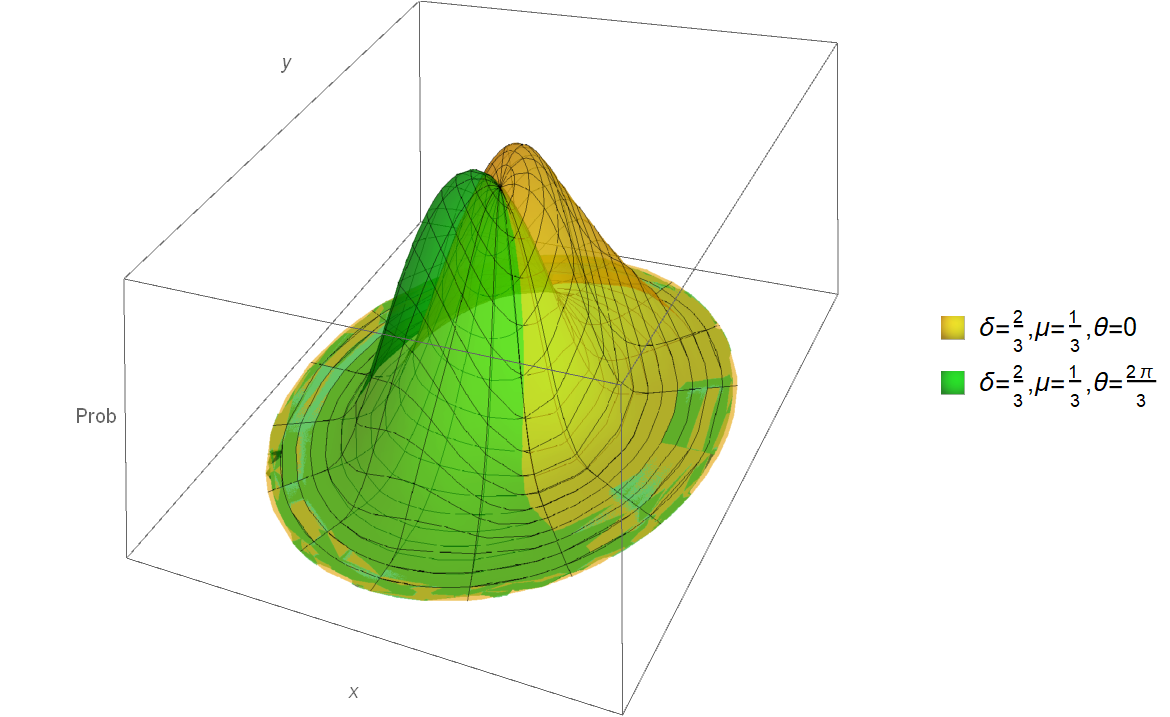}
\includegraphics[width=12cm]{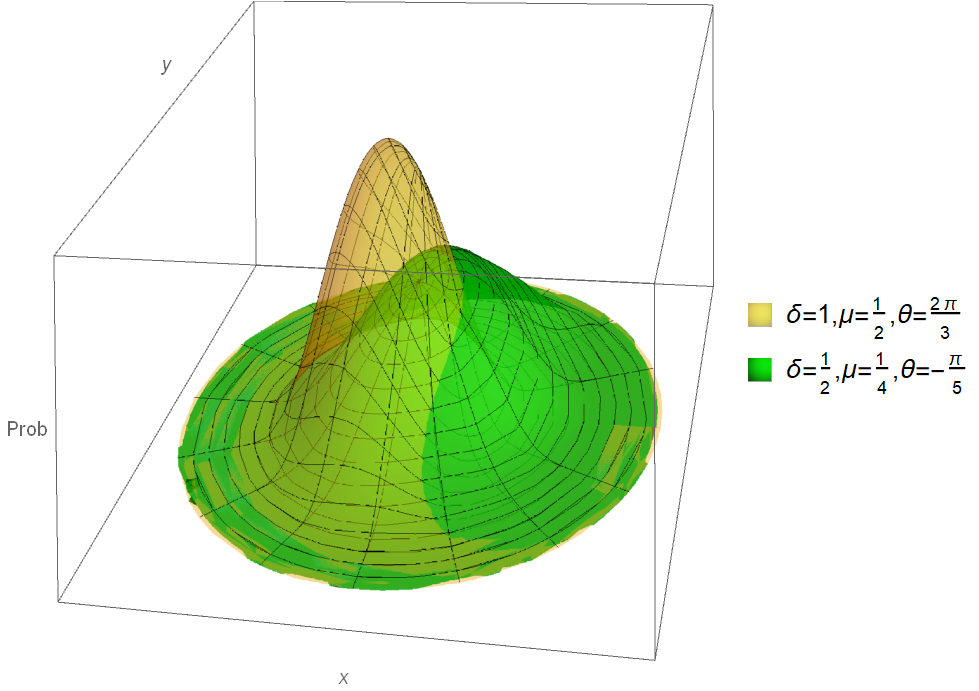}
 \caption{\label{balzano2} Examples of plots of the Gibbs probability distributions (\ref{perigord}) over the Poincar\'e disk, labeled by different set of temperatures. The exponential gaussian decay toward infinity is visually evident, as much as the deformed bell shape. In the first image we compare two distribution with the same values of $\delta\mu$ but with a different angle $\theta$. In the second image we compare two distributions that
 differ in all parameters.}
\end{center}
\end{figure}

We present in fig.\ref{balzano2} a few examples of plots of such probability distributions.

\subsubsection{The K\"ahler Geothermodynamic Metric and Curvature}
The convex cone conditions (\ref{omegaregion}) defining the Souriau temperature space in the case
$\so(2,1)\simeq \su(1,1)\simeq\slal(2,\mathbb{R})$ were found, in different notations and setups, also by the authors of \cite{marlentropia,caldobarbaresco,barbaresco2,barbaresco3,marlegibbs}. These authors interpreted the cone as the future directed light-cone of three-dimensional Minkowski space, yet this is just a special feature of the low dimensional case of $\mathrm{U/H}$ K\"ahler manifolds  under consideration. What is general is that the temperature associated with the $\so(2)_c$ subalgebra defining the K\"ahler structure must be strictly positive and then the temperatures associated with the other generators receive constraints in terms of the latter in order to maintain convergence of the other gaussian integrals.
\par
Independently from such observations, Gibbs states are parameterizd probability distributions that can be used to interpolate data by fitting their parameters, namely their temperatures.  Generalized thermodynamics provides a metric on the space of temperatures and therefore yields a distance between two distributions, each corresponding to an equilibrium state.
\par
\begin{figure}
\begin{center}
\includegraphics[width=12cm]{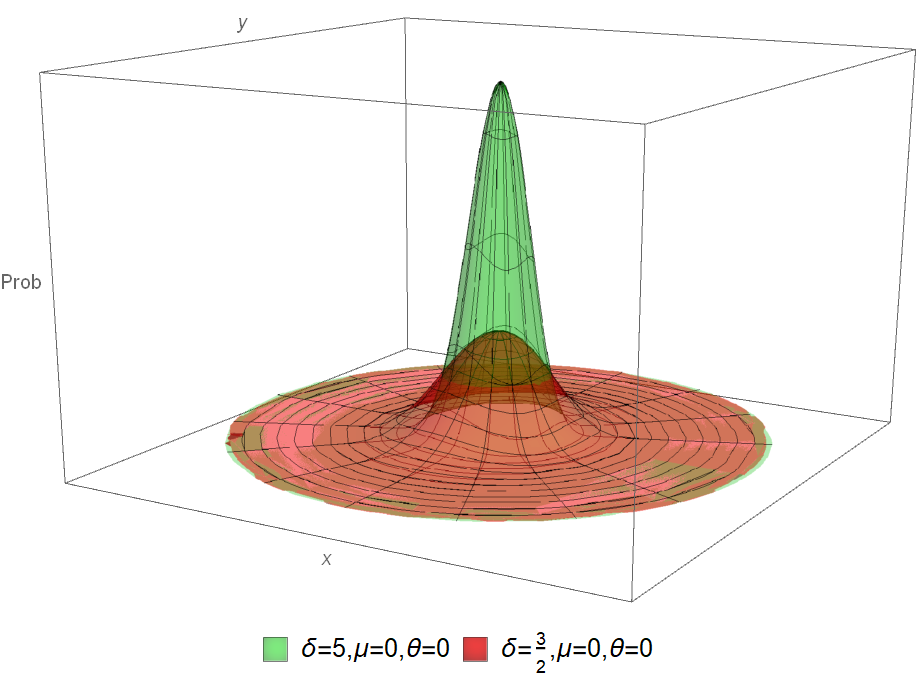}
 \caption{\label{pinnacoli}In this figure we compare two Gibbs distributions on the Poincar\'e plane corresponding to a lower and higher value of the norm (\ref{normaltale}). As one sees for a high value of the norm the distribution is very sharply shaped around its maximal value, while for lower norm it is much broader. For high norm we know with much more precision the actual location of the stochastic variable in the plane. }
\end{center}
\end{figure}

According with the general theory discussed in previous sections we have the stochastic hamiltonian:
\begin{equation}\label{cartolina}
  \mathcal{H}^{stoch} \, \equiv \, - \, \log\left[Z(\delta,\beta,\zeta)\right] \, = \, \sqrt{-\beta ^2+\delta ^2-\zeta ^2}+\frac{1}{2} \log \left[-\beta ^2+\delta ^2-\zeta
   ^2\right]-\log (\pi )
\end{equation}
and according with eq.(\ref{leggendoleggo}) we calculate  Shannon Information Functional:
\begin{equation}\label{cruciverba}
  \mathcal{I}\, = \, \mathcal{H}^{stoch} \, - \,\left(\delta \frac{\partial}{\partial\delta}\,+ \beta \frac{\partial}{\partial\beta} + \zeta\frac{\partial}{\partial\zeta}\right)\mathcal{H}^{stoch} \, = \, \frac{1}{2} \left(\log \left[\frac{\delta ^2-\beta ^2-\zeta ^2}{\pi^2}\right]-2\right)
\end{equation}
As we see the Information Functional tends to $-\infty$ on the boundary of the cone of fig.\ref{balzano1}, confirming that the norm  $\mathbb{N}(\boldsymbol{\beta})$ of the temperature vector is the analogue of the inverse thermodynamical temperature
\begin{equation}\label{normaltale}
  \mathbb{N}(\boldsymbol{\beta})\, = \, \sqrt{\delta ^2-\beta ^2-\zeta ^2} \sim \frac{1}{T}
\end{equation}
and that Shannon Information Functional $\mathcal{I}$ is just minus the Thermodynamical Entropy $S$:
\begin{equation}\label{cardinalegreco}
  \mathcal{I} \, \sim \, - \, S
\end{equation}
When  temperature $T$ goes to $\infty$ the Thermodynamical Entropy $S$ goes to infinity and the information content of the Gibbs probability distribution is largely negative since we have maximal disorder. On the contrary when $T\to 0$ which means that $\mathbf{N}(\boldsymbol{\beta}) \to \infty$ the Information Entropy tends to $\infty$ as well, logarithmically slowly, since we have a lot of information. As an illustration of this basic feature the reader is referred to fig.\ref{pinnacoli}

\par
Recalling next the general form of the thermodynamical metric expressed in terms of the Hessian of the stochastic  hamiltonian (see eq.(\ref{cessiano})) we write
\begin{eqnarray}
\label{curdaro}
  ds^2_{geotherm} &=& \frac{\partial^2 \mathcal{H}^{sto}}{\partial\boldsymbol{\beta}^i \partial\boldsymbol{\beta}^j }\, {\mathbf{d}}\boldsymbol{\beta}^i \times {\mathbf{d}}\boldsymbol{\beta}^j\nonumber\\
  &=& \frac{1}{\mathbb{N}^4}\, \times \left\{ {\mathbf{d} \beta  }^2 \left(-\left(\beta ^2+(\mathbb{N}+1) (\delta -\zeta ) (\delta
   +\zeta )\right)\right)+2 \, (\mathbb{N}+2)\, \beta \,  {\mathbf{d} \beta  } \times (\delta \,
    {\mathbf{d} \delta  }-\zeta\, {\mathbf{d} \zeta  }  )\right. \nonumber\\
   &&\left.- {\mathbf{d} \delta  }^2 \left(\delta ^2+\beta
   ^2 (\mathbb{N}+1)+\zeta ^2 (\mathbb{N}+1)\right)+2\, (\mathbb{N}+2)\, \delta \times \zeta\, \times  {\mathbf{d} \delta }\,
  \times  {\mathbf{d} \zeta  }   + {\mathbf{d} \zeta  }^2 \left((\mathbb{N}+1)
   (\beta -\delta ) (\beta +\delta )-\zeta ^2\right)\right\} \nonumber\\
\end{eqnarray}
where $\mathbb{N}$ is a shorthand for $\mathbb{N}(\boldsymbol{\beta})$ as defined in eq.(\ref{normaltale}) and
$\boldsymbol{\beta} \equiv \{\delta,\beta,\zeta\}$.
\par
\begin{figure}
\begin{center}
\includegraphics[width=12cm]{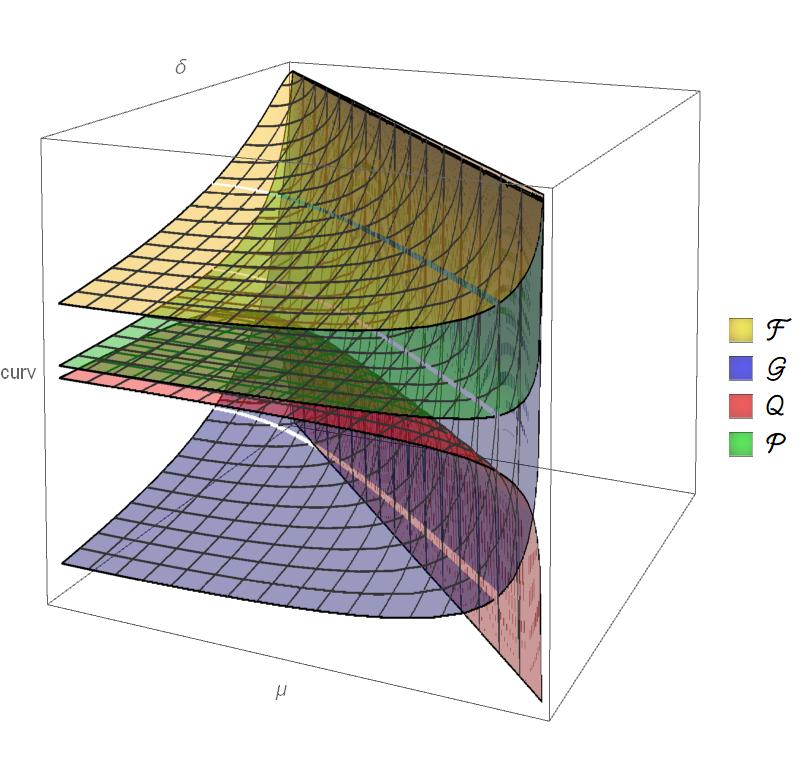}
\includegraphics[width=12cm]{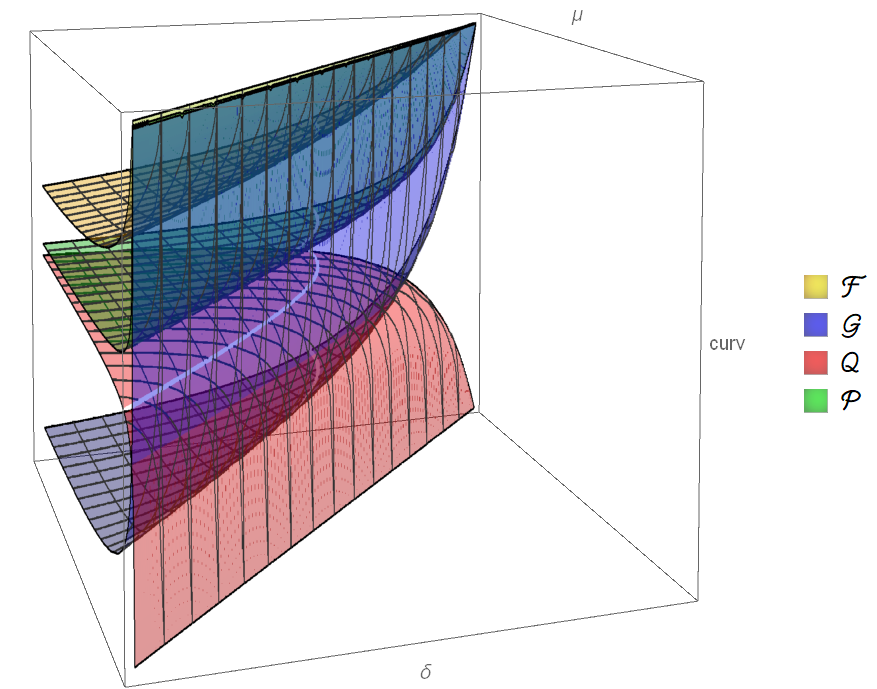}
\end{center}
\caption{\label{pignattoni} In the first and second picture of this figure we display the behavior of the four
 intrinsic curvature components $\mathcal{F},\mathcal{G},\mathcal{Q},\mathcal{P}$ as respectively seen from  the two sides of the vertical plane that has the  the line $\delta=\mu$ as base. Such line corresponds to the boundary of the cone in fig. \ref{balzano1}. All the curvature components become singular on such line.}
\end{figure}

 We are interested in calculating the Riemannian curvature of the metric (\ref{curdaro}) but in order to better understand the intrinsic properties of the curvature $2$-form we prefer to study it in the anholonomic basis provided by the vielbein formalism. For this reason we need to construct a suitable \textbf{dreibein} that reproduces (\ref{curdaro}) as the sum of squares of appropriate $1$-forms:
\begin{equation}\label{culargiones}
  ds^2_{geotherm} \, = \, - \,\sum^{3}_{i=1} \, V^i \times V^i
\end{equation}
With some work we have found that the following dreibein fulfils its own job:
\begin{eqnarray}
\label{perculatore}
  V^1 &=& \frac{1}{\mathbb{N}^2\, \sqrt{\mathbb{N}+2} \,  (\delta
   -\zeta )} \times \left\{-\beta \, \mathbf{d}\beta \, (\mathbb{N}+2) (\delta -\zeta )+\mathbf{d}\delta
   \left(\beta ^2 (\mathbb{N}+1)+(\delta -\zeta ) (\delta -\zeta
   (\mathbb{N}+1))\right)\right. \nonumber\\
   &&\left.+\,\mathbf{d}\zeta  \left((\delta -\zeta ) (\delta -\zeta +\delta
    \mathbb{N})-\beta ^2 (\mathbb{N}+1)\right)\right\} \nonumber \\
  V^2 &=& \frac{\sqrt{\mathbb{N}+1} (\mathbf{d}\beta  (\delta -\zeta )+\beta  (\mathbf{d}\zeta
   -\mathbf{d}\delta ))}{\mathbb{N} (\delta -\zeta )} \nonumber \\
  V^3 &=& (\mathbf{d}\delta -\mathbf{d}\zeta )
   \sqrt{\frac{-\mathbb{N}^2+\mathbb{N}+2}{\left(4-\mathbb{N}^2\right) (\delta -\zeta
   )^2}}
\end{eqnarray}
and using the MATHEMATICA code \textbf{Vielbgrav23} developed by one of us (see \cite{fre2023book}) we have found
the explicit form of the spin connection and of the curvature $2$-form:
\begin{equation}\label{definitori}
  \mathbf{d}V^i \, + \, \omega^{ij}\wedge V^k\,\eta_{jk} \, = \, 0 \quad ; \quad \mathfrak{R}^{ij} \, \equiv \,
   \mathbf{d}\omega^{ij} \, + \omega^{ik}\wedge \omega^{\ell j} \, \eta_{k\ell} \quad ; \quad \eta= \text{diag}(-1,-1,-1)
\end{equation}
We got:
\begin{equation}\label{curvasinistra}
  \mathfrak{R}^{ij} \, = \, \left(
\begin{array}{ccc}
 0 & \mathcal{G}(\boldsymbol{\beta})\, V^2\wedge V^3 +\mathcal{F}(\boldsymbol{\beta}) \, V^1\wedge V^2 & \mathcal{Q}(\boldsymbol{\beta})\, V^1\wedge
   V^3 \\
 -\mathcal{G}(\boldsymbol{\beta})\, V^2\wedge V^3-\mathcal{F}(\boldsymbol{\beta})\, V^1\wedge V^2 & 0 & \mathcal{P}(\boldsymbol{\beta})\, V^2\wedge
   V^3+\mathcal{G}(\boldsymbol{\beta})\, V^1\wedge V^2 \\
 -\mathcal{Q}(\boldsymbol{\beta})\, V^2\wedge V^3 & -\mathcal{P}(\boldsymbol{\beta})\, V^2\wedge V^3-\mathcal{G}(\boldsymbol{\beta})\, V^1\wedge V^2
   & 0 \\
\end{array}
\right)
\end{equation}
where the four coefficients $\mathcal{F}(\boldsymbol{\beta}),\mathcal{G}(\boldsymbol{\beta}),\mathcal{Q}(\boldsymbol{\beta}),
\mathcal{P}(\boldsymbol{\beta})$ are not constants rather they have a non trivial dependence on the temperature vector components. However they  display the $\mathrm{SO(2)}_c$ invariance of the geothermic metric. Indeed using the polar parameterization (\ref{fintereo}) of the  two non-compact temperatures $\beta,\zeta$, we find that the intrinsic components of the curvature $2$-form, $\mathcal{F}(\boldsymbol{\beta}),\mathcal{G}(\boldsymbol{\beta}),\mathcal{Q}(\boldsymbol{\beta}),
\mathcal{P}(\boldsymbol{\beta})$,  depend only on $\delta,\mu$ and do not depend on the angle $\theta$.
Explicitly we found:
\begin{eqnarray}\label{Ffun}
  \mathcal{F} & = & \frac{N_F}{D_F} \nonumber\\
  N_F & = &
 -\left(\delta ^4-2 \delta ^2 \mu ^2-4 \delta ^2+\mu ^4+4 \mu ^2\right) \left(\delta ^8-4
   \delta ^6 \mu ^2+71 \delta ^6+6 \delta ^4 \mu ^4-213 \delta ^4 \mu ^2+384 \delta ^4-4
   \delta ^2 \mu ^6 \right.\nonumber\\
   &&\left. +213 \delta ^2 \mu ^4-768 \delta ^2 \mu ^2-426 \delta ^2 \mu ^2
   \sqrt{\delta ^2-\mu ^2}+426 \delta ^2 \sqrt{\delta ^2-\mu ^2}-426 \mu ^2 \sqrt{\delta
   ^2-\mu ^2}+104 \sqrt{\delta ^2-\mu ^2}\right.\nonumber\\
   &&\left.-13 \mu ^6 \sqrt{\delta ^2-\mu ^2}+39 \delta ^2
   \mu ^4 \sqrt{\delta ^2-\mu ^2}+213 \mu ^4 \sqrt{\delta ^2-\mu ^2}+284 \delta ^2+13
   \delta ^6 \sqrt{\delta ^2-\mu ^2}-39 \delta ^4 \mu ^2 \sqrt{\delta ^2-\mu ^2}\right.\nonumber\\
   &&\left.+213
   \delta ^4 \sqrt{\delta ^2-\mu ^2}+\mu ^8-71 \mu ^6+384 \mu ^4-284 \mu ^2+16\right)\nonumber\\
   D_F & = & 4 \left(\sqrt{\delta ^2-\mu ^2}+1\right) \left(\sqrt{\delta ^2-\mu ^2}+2\right)^3
   \left(\sqrt{\delta ^2-\mu ^2}+\delta ^2-\mu ^2\right)^2 \times \nonumber\\
   &&\times  \left(\sqrt{\delta ^2-\mu
   ^2}-\delta ^2+\mu ^2+2\right) \left(3 \sqrt{\delta ^2-\mu ^2}+\delta ^2-\mu
   ^2+2\right)
\end{eqnarray}
\begin{eqnarray}
\label{Gfun}
  \mathcal{G} &=& \frac{N_G}{D_G} \nonumber\\
  N_G & = & \left(\delta ^2-\mu ^2\right) \left(\frac{-\delta ^2+\mu ^2+4}{\sqrt{\delta ^2-\mu
   ^2}-\delta ^2+\mu ^2+2}\right)^{3/2} \left(\delta ^6-3 \delta ^4 \mu ^2+26 \delta
   ^4+3 \delta ^2 \mu ^4 \right.\nonumber\\
   &&\left.-52 \delta ^2 \mu ^2-16 \delta ^2 \mu ^2 \sqrt{\delta ^2-\mu
   ^2}+44 \delta ^2 \sqrt{\delta ^2-\mu ^2}-44 \mu ^2 \sqrt{\delta ^2-\mu ^2}+20
   \sqrt{\delta ^2-\mu ^2}+8 \mu ^4 \sqrt{\delta ^2-\mu ^2}\right.\nonumber\\
   &&\left. +41 \delta ^2+8 \delta ^4
   \sqrt{\delta ^2-\mu ^2}-\mu ^6+26 \mu ^4-41 \mu ^2+4\right)\nonumber\\
  D_G &=& 2 \left(\sqrt{\delta ^2-\mu ^2}+1\right) \left(\sqrt{\delta ^2-\mu ^2}+2\right)^{5/2}
   \left(\sqrt{\delta ^2-\mu ^2}+\delta ^2-\mu ^2\right)^2 \left(3 \sqrt{\delta ^2-\mu
   ^2}+\delta ^2-\mu ^2+2\right)
\end{eqnarray}
\begin{eqnarray}
\label{Qfun}
  \mathcal{Q} &=& \frac{N_Q}{D_Q} \nonumber \\
  N_Q &=& \left(\delta ^2-\mu ^2-4\right)^4 \left(\delta ^6-3 \delta ^4 \mu ^2+25 \delta ^4+3
   \delta ^2 \mu ^4-50 \delta ^2 \mu ^2-16 \delta ^2 \mu ^2 \sqrt{\delta ^2-\mu ^2}+38
   \delta ^2 \sqrt{\delta ^2-\mu ^2}\right.\nonumber\\
   &&\left.-38 \mu ^2 \sqrt{\delta ^2-\mu ^2}+8 \sqrt{\delta
   ^2-\mu ^2}+8 \mu ^4 \sqrt{\delta ^2-\mu ^2}+28 \delta ^2+8 \delta ^4 \sqrt{\delta
   ^2-\mu ^2}-\mu ^6+25 \mu ^4-28 \mu ^2\right)\nonumber \\
  D_Q &=&4 \left(\sqrt{\delta ^2-\mu ^2}+2\right)^6 \left(\sqrt{\delta ^2-\mu ^2}-\delta ^2+\mu
   ^2+2\right)^4
\end{eqnarray}
\begin{eqnarray}
\label{Pfun}
  \mathcal{P} &=& \frac{N_P}{D_P} \nonumber \\
  N_P &=& \left(\delta ^2-\mu ^2-4\right)^2 \left(\delta ^2-\mu ^2\right) \left(14 \delta ^{10}-70
   \delta ^8 \mu ^2+340 \delta ^8+140 \delta ^6 \mu ^4-1360 \delta ^6 \mu ^2+1562 \delta
   ^6-140 \delta ^4 \mu ^6\right.\nonumber\\
   &&\left.+2040 \delta ^4 \mu ^4-4686 \delta ^4 \mu ^2+1864 \delta ^4+70
   \delta ^2 \mu ^8-1360 \delta ^2 \mu ^6+4686 \delta ^2 \mu ^4-3728 \delta ^2 \mu
   ^2\right.\nonumber\\
   &&\left.-4030 \delta ^2 \mu ^2 \sqrt{\delta ^2-\mu ^2}+1210 \delta ^2 \sqrt{\delta ^2-\mu
   ^2}-1210 \mu ^2 \sqrt{\delta ^2-\mu ^2}+136 \sqrt{\delta ^2-\mu ^2}-\mu ^{10}
   \sqrt{\delta ^2-\mu ^2}\right.\nonumber\\
   &&\left.+5 \delta ^2 \mu ^8 \sqrt{\delta ^2-\mu ^2}+89 \mu ^8
   \sqrt{\delta ^2-\mu ^2}-356 \delta ^2 \mu ^6 \sqrt{\delta ^2-\mu ^2}-869 \mu ^6
   \sqrt{\delta ^2-\mu ^2}+2607 \delta ^2 \mu ^4 \sqrt{\delta ^2-\mu ^2}\right.\nonumber\\
   &&\left. +2015 \mu ^4
   \sqrt{\delta ^2-\mu ^2}+524 \delta ^2+\delta ^{10} \sqrt{\delta ^2-\mu ^2}\right.\nonumber\\
   &&\left.-5 \delta
   ^8 \mu ^2 \sqrt{\delta ^2-\mu ^2}+89 \delta ^8 \sqrt{\delta ^2-\mu ^2}-356 \delta ^6
   \mu ^2 \sqrt{\delta ^2-\mu ^2}+869 \delta ^6 \sqrt{\delta ^2-\mu ^2}+10 \delta ^6 \mu
   ^4 \sqrt{\delta ^2-\mu ^2}\right.\nonumber\\
   &&\left.-2607 \delta ^4 \mu ^2 \sqrt{\delta ^2-\mu ^2}+2015 \delta
   ^4 \sqrt{\delta ^2-\mu ^2}-10 \delta ^4 \mu ^6 \sqrt{\delta ^2-\mu ^2}+534 \delta ^4
   \mu ^4 \sqrt{\delta ^2-\mu ^2}-14 \mu ^{10}+340 \mu ^8\right.\nonumber\\
   &&\left.-1562 \mu ^6+1864 \mu ^4-524
   \mu ^2+16\right) \nonumber\\
  D_P &=&4 \left(\sqrt{\delta ^2-\mu ^2}+1\right)^2 \left(\sqrt{\delta ^2-\mu ^2}+2\right)^3
   \left(\sqrt{\delta ^2-\mu ^2}+\delta ^2-\mu ^2\right)^2 \times \nonumber\\
   &&\times\left(\sqrt{\delta ^2-\mu
   ^2}-\delta ^2+\mu ^2+2\right)^2 \left(3 \sqrt{\delta ^2-\mu ^2}+\delta ^2-\mu
   ^2+2\right)^2
\end{eqnarray}
The behavior of these four functions is displayed in two pictures in fig.\ref{pignattoni}.
As it becomes clear from the previous detailed discussion, the generalized thermodynamics \`a la Souriau yields a $3$-dimensional temperature space equipped with a completely non trivial metric which describes the distance between different Gibbs state probability distributions. The richness and non triviality of this generalized thermodynamics is to be contrasted with the essentially trivial thermodynamics (Ideal-Gas like) associated with the geodesic dynamical system and with any other conceivable integrable dynamical system. What one needs in Machine Learning algorithms are probability distributions on the very base manifold, constituting the mathematical model of the hidden layers. Probability distributions on the fibres of the tangent bundle are not that much useful in this context.
\subsection{Generalized Thermodynamics {\`a} la Souriau of the Siegel Half Plane $\mathbb{SH}_2$}
The solvable coordinate description of the Siegel half plane $\mathbb{SH}_2$ and its theory are presented in section 7.2 of the foundational paper \cite{pgtstheory} written by two of us together with Ugo Bruzzo. To that paper and to that section we refer the reader for all the items we use here to derive the generalized thermodynamics {\`a} la Souriau of this manifold. First of all we recall from the introduction to section 7.2 of  \cite{pgtstheory} two  conceptual points that are relevant in the perspective of Machine Learning applications.
\par
In the four papers \cite{pgtstheory,TSnaviga,naviga,tassellandum} by means of which the new paradigm of \textbf{Cartan Neural Networks} was elaborated and presented to the scientific community, we mainly focused on the
series of $\mathrm{U/H}$ manifolds of the type:
\begin{equation}\label{manirq}
  \mathcal{M}^{[r,q]} \, \equiv \, \frac{\mathrm{SO(r,r+q)}}{\mathrm{SO(r) \times SO(r+q)}}
\end{equation}
in order to enlighten the role of non maximal split manifolds having a non trivial Tits Satake projection conjectured to be a universal mechanism for clustering of data. In particular we analyzed the cases of non compact rank $r=1,2$ and in \cite{pgtstheory}, at the beginning of section 7.2, we wrote:
\par
\textit{
We focus on the case $r=2$ of the series of symmetric manifolds (\ref{manirq}) in a completely synoptic setup with respect to our previous treatment of the case $r=1$.The motivation for this synopsis is twofold:
\begin{enumerate}
  \item On one hand we want to stress that the $r=1$ case is completely aligned with all the subsequent $r>1$ ones and that the Tits Satake projection, which went unnoticed and unexploited by the authors in \cite{francesi1,francesi2,francesi3}, is actually the conceptual back-bone for all the members of the considered series of manifolds.
  \item On the other hand we want  to emphasize that the $r=2$ and $r=1$ cases are twins inside the entire series since their respective Tits Satake submanifolds $\mathcal{M}^{[1,1]},\mathcal{M}^{[2,1]}$ are  just the first and the second instance of a \textit{Siegel upper complex plane}, which is the appropriate
      generalization of the Lobachevsky-Poincar\'e hypebolic plane. Instead, for values $r>2$, the Tits Satake submanifold, that, by definition, is always  a maximally split symmetric space, is not a further instance of a Siegel upper complex plane. Indeed the appearance of the first two Siegel planes is strictly linked with the low rank sporadic isomorphisms of simple Lie algebras.
\end{enumerate}
}
As we already noticed above from the point of view of Souriau generalized thermodynamics the relevant $\mathrm{U/H}$ are the K\"ahlerian ones which essentially means the two infinite series mentioned in eq.(\ref{supercaronte}). The second series (the Siegel planes) is made of maximally split symmetric manifolds, while the first series displays, for
$q>1$ the Tits Satake projection mechanism and gives rise to non trivial Paint Group invariances. Hence if we want to join the Tits Satake projection with generalized thermodynamics {\`a} la Souriau we have to choose the first series
$\mathcal{M}^{[2,q]}$. Yet, due to low-dimensional Lie algebra isomorphism, the Tits Satake submanifold $\mathcal{M}^{[2,1]}$ for the TS universality class $\mathcal{M}^{[2,q]}$ is locally isomorphic (by double covering) to the second element of the second series, namely the Siegel plane $\mathbb{SH}_2$, as already emphasized in eq.(
\ref{tittusatakr2}). These preliminary observations are instrumental to understand the reason behind our following exposition which, summarizing and importing the results of section 7.2 of \cite{pgtstheory}, emphasizes the double description of $\mathbb{SH}_2$ in terms of the $4$-dimensional spinor representation of $\mathrm{SO(2,3)}$, identified with the $\mathrm{Sp(4,\mathbb{R})}$ fundamental representation, and in terms of the $5$-dimensional vector, defining, representation of $\mathrm{SO(2,3)}$. This is due to our interest in extending the results inherent to the Tits Satake submanifold to the full TS-universality class by relying on Paint Group covariance. Indeed in appendix \ref{belohorizonte} we provide a preliminary study of the generalized thermodynamics setup for the  case $\mathcal{M}^{[2,2]}$ in order to emphasize the role of Paint Group in this context.
\par
After these preliminary clarifications we start the analysis of $\mathbb{SH}_2$ following section 7.2 of our
foundational paper. The relation between the spinor and the vector representation that provides the local isomorphism of $\mathrm{SO(2,3)}$ with $\mathrm{Sp(4,\mathbb{R})}$ is given by the gamma-matrices $\Gamma_i$, $i=1,\dots,5$. Those well adapted to the triangular basis which is that where the preserved metric with signature $(2,3)$ is the $\eta_t$-matrix in eq.(7.34) of \cite{pgtstheory} are displayed in eq.(7.41) of the same paper.
The charge conjugation matrix $\mathrm{C}_s$ that becomes the symplectic invariant form of $\mathrm{Sp(4,\mathbb{R})}$ is displayed in eq.(7.37) of the same paper. The $10$ generators $J_{ij}$ of the
$\so(2,3) \simeq \sym(4,\mathbb{R})$ Lie algebra are defined in eq.(7.39) of \cite{pgtstheory} as commutators of gamma matrices. The explicit expression of the double covering of $\mathrm{SO(2,3)}$ group by means of
$\mathrm{Sp(4,\mathbb{R})}$ group is provided by eq.(7.44) of the same reference   that we repeat here for reader's convenience:
\begin{equation}\label{cogrutto}
\forall \mathcal{S} \in \mathrm{Sp(4,\mathbb{R})} \quad : \quad    \mathcal{O}^i_{\phantom{i}j}[\mathcal{S}] \, \equiv \, \ft 14 \, \text{Tr} \, \left( \Gamma^T_{i} \, \mathcal{S}^{-1} \, \Gamma_j \, \mathcal{S}\right) \, \in \, \mathrm{SO(2,3)}
\end{equation}
The fundamental next step is the construction of the solvable coset representative both in the vector and in the spinor representation. We utilize the notations and the results of \cite{pgtstheory} sect. 7.2. Hence we use the letter \textbf{W} instead of $\boldsymbol{\Upsilon}$ for the vector of solvable coordinates, namely of parameters of the solvable Lie group metrically equivalent to $\mathrm{U/H}$. Thinking of $\mathrm{SO(2,3)}$ as the Tits Satake subgroup of $\mathrm{SO(2,2+2s)}$ the solvable coordinate vector is the one given in eq.(7.52) of \cite{pgtstheory}, namely:
\begin{equation}\label{craniato}
  \boldsymbol{W}=\left\{w_1,w_2,w_3,w_4,w_5,\underbrace{0,\dots,0}_{(2s-1)},w_6,\underbrace{0,\dots,0}_{(2s-1)}\right\}
\end{equation}
the zeros  corresponding to the Tits Satake projection. The corresponding solvable group element constructed with the $\Sigma$-exponential map (see a discussion of its definition in \cite{TSnaviga}) in terms of the solvable coordinates is the following one:
\begin{equation}\label{LBmat}
  \mathbb{L}\left(\mathbf{W}\right)\, =\, \left(
\begin{array}{ccccc}
 e^{ {w_1}} & \frac{e^{ {w_1}}  {w_3}}{\sqrt{2}} & \frac{1}{2}
   e^{ {w_1}} \left( {w_3}  {w_6}+\sqrt{2}  {w_5}\right) &
   \frac{1}{8} e^{ {w_1}} \left(-\sqrt{2}  {w_3}  {w_6}^2+4 \sqrt{2}
    {w_4}-4  {w_5}  {w_6}\right) & -\frac{1}{4} e^{ {w_1}} \left(2
    {w_3}  {w_4}+ {w_5}^2\right) \\
 0 & e^{ {w_2}} & \frac{e^{ {w_2}}  {w_6}}{\sqrt{2}} & -\frac{1}{4}
   e^{ {w_2}}  {w_6}^2 & -\frac{e^{ {w_2}}  {w_4}}{\sqrt{2}} \\
 0 & 0 & 1 & -\frac{ {w_6}}{\sqrt{2}} & -\frac{ {w_5}}{\sqrt{2}} \\
 0 & 0 & 0 & e^{- {w_2}} & -\frac{e^{- {w_2}}  {w_3}}{\sqrt{2}} \\
 0 & 0 & 0 & 0 & e^{- {w_1}} \\
\end{array}
\right)
\end{equation}
which is eq.(7.54) of \cite{pgtstheory}. Imposing the identification:
\begin{equation}\label{lanerossi}
  \mathcal{O}\left[\mathcal{S}_{s}(\mathbf{W}) \right] \, = \, \mathbb{L}\left(\mathbf{W}\right)
\end{equation}
one finds a unique solution for the $\mathcal{S}_{s}$ that is the following:
\begin{equation}\label{cartoon}
  \mathcal{S}_{s}[\mathbf{W}] \, =\,\left(
\begin{array}{cccc}
 e^{\frac{1}{2} \left(-w_1-w_2\right)} & \frac{1}{2} e^{\frac{1}{2}
   \left(w_2-w_1\right)} w_6 & \frac{1}{4} e^{\frac{1}{2} \left(w_1+w_2\right)}
   \left(w_5 w_6-2 \sqrt{2} w_4\right) & \frac{1}{4} e^{\frac{1}{2}
   \left(w_1-w_2\right)} \left(2 w_5+\sqrt{2} w_3 w_6\right) \\
 0 & e^{\frac{1}{2} \left(w_2-w_1\right)} & \frac{1}{2} e^{\frac{1}{2}
   \left(w_1+w_2\right)} w_5 & \frac{e^{\frac{1}{2} \left(w_1-w_2\right)}
   w_3}{\sqrt{2}} \\
 0 & 0 & e^{\frac{1}{2} \left(w_1+w_2\right)} & 0 \\
 0 & 0 & -\frac{1}{2} e^{\frac{1}{2} \left(w_1+w_2\right)} w_6 & e^{\frac{1}{2}
   \left(w_1-w_2\right)} \\
\end{array}
\right)
\end{equation}
\subsubsection{The Siegel Upper Plane}
For completeness we import from \cite{pgtstheory} the necessary concepts and formulae for the Siegel upper plane. This is particularly important in order to compare our results with those of Barbaresco et al \cite{caldobarbaresco,barbaresco2,barbaresco3,marlegibbs,marlentropia}.
The \textit{Siegel upper complex plane} of degree (or genus) $g$ is the generalization to higher dimensions of the Lobachevsky-Poincar\'e
hyperbolic plane.
\par
Just as the standard hyperbolic plane with Poincar\'e metric is a complex analytic realization of a maximally split
symmetric space, namely $\mathrm{SL(2,\mathbb{R})/SO(2)}$, in the same way, the upper Siegel plane of degree $g$ is the complex analytic
realization of the symmetric space:
\begin{equation}\label{siegelsymspac}
  \mathcal{M}_{Siegel} \, = \, \frac{\Sp\mathrm{(2\,g, \mathbb{R})}}{\mathrm{S[U(1)\times U(g)]}}
\end{equation}
The key observation is the following. Just in the same way as the \textit{fractional linear transformation}:
\begin{equation}\label{fraclin}
  z \, \rightarrow \, \tilde{z} \, \equiv \, \frac{a \,z +b }{c\, z + d } \quad ; \quad \left(
  \begin{array}{cc}
     a & b \\
     c & d \\
     \end{array}
 \right) \, \in \, \mathrm{PSL(2,\mathbb{R})}
\end{equation}
maps complex numbers $z$ with strictly positive imaginary part into complex numbers $\tilde{z}$ with the same property, \textbf{the fractional
linear matrix transformation}:
\begin{equation}\label{fraclinmat}
  Z_{g\times g} \, \rightarrow \, \tilde{Z}_{g\times g} \, \equiv \, (A_{g\times g} \,Z_{g\times g}
  +B_{g\times g} )\cdot(C_{g\times g}\,Z_{g\times g} + D_{g\times g})^{-1} \quad ; \quad \left(
  \begin{array}{c|c}
   A_{g\times g} & B_{g\times g} \\
   \hline
   C_{g\times g} &  D_{g\times g} \\
   \end{array}
\right) \, \in \, \Sp(2\,g,\mathbb{R})
\end{equation}
maps \textbf{complex symmetric matrices}:
\begin{equation}\label{Zetone}
  Z_{g\times g} \, = \,  X_{g\times g} \, + \, \mathit{i} \, Y_{g\times g} \quad ; \quad  Z_{g\times g}^T \, = \,  Z_{g\times g}
\end{equation}
whose imaginary part $Y_{g\times g}$ is positive definite (namely has strictly positive eigenvalues) into \textbf{complex symmetric matrices}
$\tilde{Z}_{g\times g}$ with the same property. The relations among the $g \times g$ blocks:
\begin{equation}\label{blocrela}
  A^T \, C \, = \, C^T \, A \quad ; \quad B^T \, D \, = \, D^T \, B \quad ; \quad A^T D\, - \, C^T \, B \, = \, \mathbf{1 }
\end{equation}
following from the very definition of the $\Sp(2\,g,\mathbb{R})$ group, are instrumental in the lenghty yet straighforward proof of what was
stated above.
\par
The number of real components of $Z_{g\times g}$ exactly matches the dimension of the \textit{maximally split symmetric space} defined in
eq.(\ref{siegelsymspac}) so that the upper Siegel plane constitutes its holomorphic realization. Furthermore the choice of the Borel solvable
subgroup inside $\Sp(2\,g,\mathbb{R})$ provides a convenient parameterization of the matrix $Z_{g\times g}$. Indeed this latter is the orbit under
the fractional linear action of the Borel subgroup of the special matrix $Z_0 \, = \,\mathit{i} \, \mathbf{1}_{g\times g}$.
\par
Applying this idea to the case $g=r=2$ which is ours and utilizing the parameterization of the Borel solvable subgroup provided in
eq.(\ref{cartoon}), we obtain:
\begin{eqnarray}\label{muradilegno}
  Z & = & \, X \, + \, \mathit{i} \, Y \nonumber\\
  X & = & \left(
\begin{array}{cc}
 \frac{1}{8} \left(w_6 \left(4 w_5+\sqrt{2} w_3 w_6\right)-4 \sqrt{2} w_4\right)
   & \frac{1}{4} \left(2 w_5+\sqrt{2} w_3 w_6\right) \\
 \frac{1}{4} \left(2 w_5+\sqrt{2} w_3 w_6\right) & \frac{w_3}{\sqrt{2}} \\
\end{array}
\right) \nonumber\\
  Y & = & \left(
\begin{array}{cc}
 \frac{1}{4} e^{-w_1-w_2} \left(e^{2 w_2} w_6^2+4\right) & \frac{1}{2}
   e^{w_2-w_1} w_6 \\
 \frac{1}{2} e^{w_2-w_1} w_6 & e^{w_2-w_1} \\
\end{array}
\right)
\end{eqnarray}
According with equation \ref{fraclinmat} the action of any element $g\in \mathrm{Sp(4,\mathbb{R})}$ on the coset manifold is
the  fractional linear transformation of the symmetric complex matrix:
\begin{equation}\label{zetone}
  Z \, = \,\left(
\begin{array}{cc}
 z & \omega  \\
 \omega  & \zeta  \\
\end{array}
\right) \quad ; \quad z,\omega,\zeta \in \mathbb{C}
\end{equation}
which represents  the entire manifold. When $Z$ is diagonal, namely when $\omega = 0$ the two remaining complex entries $z,\zeta$ represent the
coordinates of two hyperbolic upper planes. The subgroup $\Gamma \subset \mathrm{Sp(4,\mathbb{R})}$ which respects diagonality, namely the
condition $\omega =0$ is $\Gamma \, = \, \mathrm{SL(2,\mathbb{R})} \times \mathrm{SL(2,\mathbb{R}) }$.
\par
It is convenient to recall eq.(\ref{muradilegno}) which provides the parameterization of the complex
matrix (\ref{zetone}) in terms of the solvable coordinates $w_1,w_2,w_3, w_4,w_5,w_6$ that can be summarized by stating:
\begin{eqnarray}\label{lisciva}
  z &=& \frac{1}{8} \left(-4 \sqrt{2} w_4+w_6 \left(4 w_5+\sqrt{2} w_3 w_6\right)+2 i
   e^{-w_1-w_2} \left(e^{2 w_2} w_6^2+4\right)\right) \nonumber\\
  \zeta &=& \frac{w_3}{\sqrt{2}}+i e^{w_2-w_1} \nonumber\\
 \omega &=& \frac{1}{4} \left(2 w_5+2 i e^{w_2-w_1} w_6+\sqrt{2} w_3 w_6\right)
\end{eqnarray}
As one sees from eq.(\ref{lisciva}) the diagonalization condition of the matrix $Z$ corresponds to setting $w_5 = w_6 =0$ which implies that the
other two solvable coordinates $w_3,w_4$ obtain the interpretation of real parts of the complex coordinates $\zeta$ and $z$, respectively.
\par
The three complex numbers $z,\zeta,\omega$ can be utilized as complex coordinates of the symmetric space and the
K\"ahler metric can be derived from a suitable K\"ahler potential. Similarly the K\"ahlerian moment maps for all the Killing vector fields can be obtained from the K\"ahler potential, just as the K\"ahler 2-form. This however is not the best approach for our goals. In order to construct the generalized thermodynamics {\`a} la Souriau it is much more convenient to utilize real solvable coordinates and obtain the K\"ahler 2-form just as we did in the Poincar\'e case from the unique $\mathrm{U(1)}$ generator. This is what we do in next subsection.
\subsubsection{The K\"ahler $2$-Form, the Killing Vector Fields and the Moment Maps}
In Order to construct all the items of generalized thermodynamics we need the vielbein, the K\"ahler $2$-form and the moment
maps of all Killing vectors. To this effect we need a well-normalized basis of generators of the full $\mathbb{U}$ Lie algebra. Such basis is presented in two versions in the spinor representation in table \ref{s10genni} and in the vector representation in table \ref{v10genni}
\begin{table}[htb]
\begin{center}
\caption{\it In this table we display the complete list of the 10 generators  of the $\sym(4,R)$ Lie algebra. \label{s10genni}}
$$
\begin{array}{||ccc||ccc||}
\hline\hline
 T^s_1 & = & \left(
\begin{array}{cccc}
 -\frac{1}{2} & 0 & 0 & 0 \\
 0 & -\frac{1}{2} & 0 & 0 \\
 0 & 0 & \frac{1}{2} & 0 \\
 0 & 0 & 0 & \frac{1}{2} \\
\end{array}
\right) & T^s_2 & = & \left(
\begin{array}{cccc}
 -\frac{1}{2} & 0 & 0 & 0 \\
 0 & \frac{1}{2} & 0 & 0 \\
 0 & 0 & \frac{1}{2} & 0 \\
 0 & 0 & 0 & -\frac{1}{2} \\
\end{array}
\right)\\
\hline
 T^s_3 & = & \left(
\begin{array}{cccc}
 0 & 0 & 0 & 0 \\
 0 & 0 & 0 & \frac{1}{\sqrt{2}} \\
 0 & 0 & 0 & 0 \\
 0 & \frac{1}{\sqrt{2}} & 0 & 0 \\
\end{array}
\right) & T^s_4 & = & \left(
\begin{array}{cccc}
 0 & 0 & -\frac{1}{\sqrt{2}} & 0 \\
 0 & 0 & 0 & 0 \\
 -\frac{1}{\sqrt{2}} & 0 & 0 & 0 \\
 0 & 0 & 0 & 0 \\
\end{array}
\right) \\
\hline
 T^s_5 & = & \left(
\begin{array}{cccc}
 0 & 0 & 0 & \frac{1}{2} \\
 0 & 0 & \frac{1}{2} & 0 \\
 0 & \frac{1}{2} & 0 & 0 \\
 \frac{1}{2} & 0 & 0 & 0 \\
\end{array}
\right) & T^s_6 & = & \left(
\begin{array}{cccc}
 0 & \frac{1}{2} & 0 & 0 \\
 \frac{1}{2} & 0 & 0 & 0 \\
 0 & 0 & 0 & -\frac{1}{2} \\
 0 & 0 & -\frac{1}{2} & 0 \\
\end{array}
\right)\\
\hline
 T^s_7 & = & \left(
\begin{array}{cccc}
 0 & \frac{1}{2} & 0 & 0 \\
 -\frac{1}{2} & 0 & 0 & 0 \\
 0 & 0 & 0 & \frac{1}{2} \\
 0 & 0 & -\frac{1}{2} & 0 \\
\end{array}
\right) & T^s_8 & = & \left(
\begin{array}{cccc}
 0 & 0 & 0 & \frac{1}{2} \\
 0 & 0 & \frac{1}{2} & 0 \\
 0 & -\frac{1}{2} & 0 & 0 \\
 -\frac{1}{2} & 0 & 0 & 0 \\
\end{array}
\right)\\
\hline
 T^s_9 & = & \left(
\begin{array}{cccc}
 0 & 0 & -\frac{1}{2} & 0 \\
 0 & 0 & 0 & \frac{1}{2} \\
 \frac{1}{2} & 0 & 0 & 0 \\
 0 & -\frac{1}{2} & 0 & 0 \\
\end{array}
\right) & T^s_{10} & = & \left(
\begin{array}{cccc}
 0 & 0 & \frac{1}{2} & 0 \\
 0 & 0 & 0 & \frac{1}{2} \\
 -\frac{1}{2} & 0 & 0 & 0 \\
 0 & -\frac{1}{2} & 0 & 0 \\
\end{array}
\right)\\
\hline\hline
\end{array}
$$
\end{center}
\end{table}
The generators in the two lists are in one-to-one correspondence and satisfy the Lie algebra commutation relations with the very same structure constants. The $10$ generators, irrespectively of their label $s$ or $v$ are ordered in the following way:
\begin{equation}\label{cosettico}
  T^{s/v}_{1,\dots,6} \,=\,K_{i}
\end{equation}
are the $6$ non-compact coset generators spanning the vector subspace $\mathbb{K}$ in the orthogonal decomposition
\begin{equation}\label{trappolatura}
  \mathbb{U} \, = \, \mathbb{H} \, \oplus \, \mathbb{K}
\end{equation}
Furthermore the first two generators $T^{s/v}_{1,2}$ are the two non-compact Cartan generators.
\par
The generators:
\begin{equation}\label{hattico}
  T^{s/v}_{7,8,9} \,=\,H_{1,2,3}
\end{equation}
are the generators of the $\su(2)\simeq\so(3)$ subalgebra of $\mathbb{H}\, = \, \su(2)\oplus \uu(1)$.
\begin{table}[htb]
\caption{\it In this table we display the complete list of the 10 generators  of the $\so(2,3)$ Lie algebra, that are in one-to-one correspondence with and in the same order as the generators of the $\sym(4,\mathbb{R})$ Lie algebra listed
in table \ref{s10genni}. \label{v10genni}}
$$
\begin{array}{||lclcl||lclcl||}
\hline\hline
 T^v_1 & = & \sqrt{2}\,K_1 & = &  \left(
\begin{array}{ccccc}
 1 & 0 & 0 & 0 & 0 \\
 0 & 0 & 0 & 0 & 0 \\
 0 & 0 & 0 & 0 & 0 \\
 0 & 0 & 0 & 0 & 0 \\
 0 & 0 & 0 & 0 & -1 \\
\end{array}
\right) & T^v_2 & = & \sqrt{2}\,K_2 & = & \left(
\begin{array}{ccccc}
 0 & 0 & 0 & 0 & 0 \\
 0 & 1 & 0 & 0 & 0 \\
 0 & 0 & 0 & 0 & 0 \\
 0 & 0 & 0 & -1 & 0 \\
 0 & 0 & 0 & 0 & 0 \\
\end{array}
\right)\\
\hline
 T^v_3 & = & \sqrt{2}\,K_3 & = & \left(
\begin{array}{ccccc}
 0 & \frac{1}{\sqrt{2}} & 0 & 0 & 0 \\
 \frac{1}{\sqrt{2}} & 0 & 0 & 0 & 0 \\
 0 & 0 & 0 & 0 & 0 \\
 0 & 0 & 0 & 0 & -\frac{1}{\sqrt{2}} \\
 0 & 0 & 0 & -\frac{1}{\sqrt{2}} & 0 \\
\end{array}
\right) & T^v_4 & = & \sqrt{2}\,K_4 & = & \left(
\begin{array}{ccccc}
 0 & 0 & 0 & \frac{1}{\sqrt{2}} & 0 \\
 0 & 0 & 0 & 0 & -\frac{1}{\sqrt{2}} \\
 0 & 0 & 0 & 0 & 0 \\
 \frac{1}{\sqrt{2}} & 0 & 0 & 0 & 0 \\
 0 & -\frac{1}{\sqrt{2}} & 0 & 0 & 0 \\
\end{array}
\right) \\
\hline
 T^v_5 & = & \sqrt{2}\,K_5 & = & \left(
\begin{array}{ccccc}
 0 & 0 & \frac{1}{\sqrt{2}} & 0 & 0 \\
 0 & 0 & 0 & 0 & 0 \\
 \frac{1}{\sqrt{2}} & 0 & 0 & 0 & -\frac{1}{\sqrt{2}} \\
 0 & 0 & 0 & 0 & 0 \\
 0 & 0 & -\frac{1}{\sqrt{2}} & 0 & 0 \\
\end{array}
\right) & T^v_6 & = & \sqrt{2}\,K_6 & = & \left(
\begin{array}{ccccc}
 0 & 0 & 0 & 0 & 0 \\
 0 & 0 & \frac{1}{\sqrt{2}} & 0 & 0 \\
 0 & \frac{1}{\sqrt{2}} & 0 & -\frac{1}{\sqrt{2}} & 0 \\
 0 & 0 & -\frac{1}{\sqrt{2}} & 0 & 0 \\
 0 & 0 & 0 & 0 & 0 \\
\end{array}
\right)\\
\hline
 T^v_7 & = & H_1 & = & \left(
\begin{array}{ccccc}
 0 & 0 & 0 & 0 & 0 \\
 0 & 0 & \frac{1}{\sqrt{2}} & 0 & 0 \\
 0 & -\frac{1}{\sqrt{2}} & 0 & -\frac{1}{\sqrt{2}} & 0 \\
 0 & 0 & \frac{1}{\sqrt{2}} & 0 & 0 \\
 0 & 0 & 0 & 0 & 0 \\
\end{array}
\right) & T^v_8 & = & H_2 & = &  \left(
\begin{array}{ccccc}
 0 & 0 & \frac{1}{\sqrt{2}} & 0 & 0 \\
 0 & 0 & 0 & 0 & 0 \\
 -\frac{1}{\sqrt{2}} & 0 & 0 & 0 & -\frac{1}{\sqrt{2}} \\
 0 & 0 & 0 & 0 & 0 \\
 0 & 0 & \frac{1}{\sqrt{2}} & 0 & 0 \\
\end{array}
\right)\\
\hline
 T^v_9 & = & H_3 & = &  \left(
\begin{array}{ccccc}
 0 & \frac{1}{2} & 0 & \frac{1}{2} & 0 \\
 -\frac{1}{2} & 0 & 0 & 0 & -\frac{1}{2} \\
 0 & 0 & 0 & 0 & 0 \\
 -\frac{1}{2} & 0 & 0 & 0 & -\frac{1}{2} \\
 0 & \frac{1}{2} & 0 & \frac{1}{2} & 0 \\
\end{array}
\right) & T^v_{10} & = & H_0 & = &  \left(
\begin{array}{ccccc}
 0 & \frac{1}{2} & 0 & -\frac{1}{2} & 0 \\
 -\frac{1}{2} & 0 & 0 & 0 & \frac{1}{2} \\
 0 & 0 & 0 & 0 & 0 \\
 \frac{1}{2} & 0 & 0 & 0 & -\frac{1}{2} \\
 0 & -\frac{1}{2} & 0 & \frac{1}{2} & 0 \\
\end{array}
\right)\\
\hline\hline
\end{array}
$$
\end{table}
Finally
\begin{equation}\label{konditorei}
  T^{s/v}_{10} \, = \, H_0
\end{equation}
is the $\uu(1)\simeq \so(2)$ generator responsible for the K\"ahler structure.
\par
Introducing the left-invariant $1$-form in either the spinor or the vector form and projecting it onto the $\mathbb{K}$-subspace we find the sechsbein of the $6$-dimensional space which turns out to be the same
in the two cases, as it should. We choose the vector form and we write:
\begin{eqnarray}
\label{carnevaletroiano}
  \Theta^v &\equiv& \mathbb{L}^{-1}(\mathbf{W})\cdot \mathbf{d}\mathbb{L}(\mathbf{W}) \nonumber \\
  \mathbf{e}^i &=& \text{Tr}\left(\Theta^v\cdot K^\dagger_i\right) \nonumber \\
  \delta_{ij} &=& \text{Tr}\left(K_i\cdot K^\dagger_j \right) \, ; \, \text{Tr}(H_i\cdot K^\dagger_j) \, = \,0 \quad \quad \text{normalization of the adjoints $K_j^\dagger$}
\end{eqnarray}
and explicitly we have:
\begin{eqnarray}
\label{sexbein}
  \mathbf{e}^1 &=& \text{$\mathbf{d}$w}_1  \nonumber \\
  \mathbf{e}^2 &=& \text{$\mathbf{d}$w}_2  \nonumber \\
  \mathbf{e}^3 &=& \frac{1}{2} \left(\text{$\mathbf{d}$w}_3+\text{$\mathbf{d}$w}_1
   w_3-\text{$\mathbf{d}$w}_2 w_3\right)  \nonumber \\
  \mathbf{e}^4 &=& \frac{1}{8} \left(w_6^2 \left(-\left(\text{$\mathbf{d}$w}_3+\text{$\mathbf{d}$w}_1
   w_3-\text{$\mathbf{d}$w}_2 w_3\right)\right)-2 \sqrt{2} w_6
   \left(\text{$\mathbf{d}$w}_5+\text{$\mathbf{d}$w}_1 w_5\right)+4
   \left(\text{$\mathbf{d}$w}_4+\left(\text{$\mathbf{d}$w}_1+\text{$\mathbf{d}$w}_2\right) w_4\right)\right)  \nonumber \\
  \mathbf{e}^5 &=& \frac{1}{4} \left(2 \text{$\mathbf{d}$w}_5+2 \text{$\mathbf{d}$w}_1 w_5+\sqrt{2} w_6
   \left(\text{$\mathbf{d}$w}_3+\text{$\mathbf{d}$w}_1 w_3-\text{$\mathbf{d}$w}_2
   w_3\right)\right)  \nonumber \\
  \mathbf{e}^6 &=& \frac{1}{2} \left(\text{$\mathbf{d}$w}_6+\text{$\mathbf{d}$w}_2 w_6\right)
\end{eqnarray}
In this case the sechsbein coincide with the left-invariant $1$-forms and satisfy the Maurer Cartan equations of the solvable Lie group, this however is not particulary relevant for our goals. The explicit form of the metric is given by:
\begin{equation}\label{dsqSiegel}
  ds^2_{Siegel} \, = \, \sum_{i=1}^{6} \mathbf{e}^i \times \mathbf{e}^i
\end{equation}
The K\"ahler $2$-form is obtained by constructing the adjoint representation of the generator $H_0$ on the subspace $\mathbb{K}$. Working in the vector representation and defining the normalized generators:
\begin{equation}\label{loffo}
  K_i \, =\, \frac{1}{\sqrt{2}}\, T^v_{i} \quad (i,1,\dots, 6) \quad; \quad \text{Tr}\left(K_i\cdot K_j\right) \, = \, \delta_{ij}
\end{equation}
we obtain
\begin{equation}\label{Cosroe}
  \left[H_0\, , \, K_i\right] \, = \, \left(\text{AdjH}_0\right)_{ij} \, K_j
\end{equation}
where:
\begin{equation}\label{cortanze}
  \text{AdjH}_0 \, = \,  \left(
\begin{array}{cccccc}
 0 & 0 & -\frac{1}{\sqrt{2}} & \frac{1}{\sqrt{2}} & 0 & 0 \\
 0 & 0 & \frac{1}{\sqrt{2}} & \frac{1}{\sqrt{2}} & 0 & 0 \\
 \frac{1}{\sqrt{2}} & -\frac{1}{\sqrt{2}} & 0 & 0 & 0 & 0 \\
 -\frac{1}{\sqrt{2}} & -\frac{1}{\sqrt{2}} & 0 & 0 & 0 & 0 \\
 0 & 0 & 0 & 0 & 0 & -1 \\
 0 & 0 & 0 & 0 & 1 & 0 \\
\end{array}
\right)
\end{equation}
and the K\"ahler $2$-form takes the form:
\begin{eqnarray}\label{siegelkaller}
  \boldsymbol{\mathcal{K}} & = &\left(\text{AdjH}_0\right)_{ij} \mathbf{e}^i\wedge \mathbf{e}^j \nonumber\\
  & = & -\sqrt{2} \, \mathbf{e}^1\wedge \mathbf{e}^3+\sqrt{2} \,\mathbf{e}^1\wedge \mathbf{e}^4+\sqrt{2} \, \mathbf{e}^2\wedge \mathbf{e}^3+\sqrt{2}\,
   \mathbf{e}^2\wedge \mathbf{e}^4-2 \, \mathbf{e}^5\wedge \mathbf{e}^6
\end{eqnarray}
The explicit expression of the K\"ahler $2$ in the solvable coordinate basis is obtained from eq.(\ref{siegelkaller}) by substituting the explicit form of the sechsbein (\ref{sexbein}); so doing  one immediately verifies that
the $2$-form $\boldsymbol{\mathcal{K}}$ is closed, namely $\mathbf{d}\boldsymbol{\mathcal{K}}\, = \, 0$.
\par
Then, utilizing the vector representation and utilizing  the method described in section \ref{generalecuster} we derive the explicit form of the $10$ Killing vector fields expressed in terms of the solvable coordinate basis associated with the Lie algebra generators, as ordered and displayed in table \ref{v10genni}. We obtain the following result. The $6$ Killing vector fields, generating the coset translations associated with the $\mathbb{K}$-generators, have the following explicit form:
\begin{eqnarray}
\label{KillvetK}
  \boldsymbol{\mathfrak{k}}_1 &=& \, \boldsymbol{\partial}_1 \nonumber \\
  \boldsymbol{\mathfrak{k}}_2 &=& \, \boldsymbol{\partial}_2  \nonumber \\
  \boldsymbol{\mathfrak{k}}_3 &=& -\frac{1}{2} e^{w_1-w_2} w_3 \, \boldsymbol{\partial}_1+\frac{1}{2} e^{w_1-w_2} w_3
   \, \boldsymbol{\partial}_2+\left(\frac{1}{2} e^{w_1-w_2} \left(w_3^2+2\right)+e^{w_2-w_1}\right)
   \, \boldsymbol{\partial}_3+\frac{1}{4} e^{w_1-w_2} \left(w_5-w_6\right) \left(w_5+w_6\right)
   \, \boldsymbol{\partial}_4\nonumber \\
   &&-\frac{e^{w_1-w_2} w_6 }{\sqrt{2}}\, \boldsymbol{\partial}_5 +\frac{e^{w_1-w_2} w_5
   }{\sqrt{2}} \, \boldsymbol{\partial}_6 \nonumber \\
  \boldsymbol{\mathfrak{k}}_4 &=& -\frac{1}{2} e^{w_1+w_2} w_4 \, \boldsymbol{\partial}_1+\frac{1}{4} e^{w_1+w_2} \left(\sqrt{2} w_5
   w_6-2 w_4\right) \, \boldsymbol{\partial}_2+\frac{1}{4} e^{w_1+w_2} \left(w_5^2+\sqrt{2} w_3 w_6
   w_5-w_6^2\right) \, \boldsymbol{\partial}_3 \nonumber\\
   &&+\left(\frac{1}{16} e^{w_1+w_2} \left(8 w_4^2-4 \sqrt{2}
   w_5 w_6 w_4+\left(w_6^2+4\right){}^2\right)+e^{-w_1-w_2}\right)
   \, \boldsymbol{\partial}_4 \nonumber \\
   &&+\frac{e^{w_1+w_2} w_6 \left(w_6^2+4\right) }{4
   \sqrt{2}}\, \boldsymbol{\partial}_5-\frac{e^{w_1+w_2} w_5 \left(w_6^2+4\right) }{4 \sqrt{2}}\, \boldsymbol{\partial}_6 \nonumber\\
  \boldsymbol{\mathfrak{k}}_5 &=& -\frac{1}{2} e^{w_1} w_5 \, \boldsymbol{\partial}_1
  -\frac{e^{w_1} w_3 w_6 }{2
   \sqrt{2}}\, \boldsymbol{\partial}_2-\frac{e^{w_1} \left(w_3^2+2\right) w_6 }{2
   \sqrt{2}}\, \boldsymbol{\partial}_3+\frac{e^{w_1} w_6 \left(w_6^2+2 w_3 w_4+4\right) }{4
   \sqrt{2}}\, \boldsymbol{\partial}_4\nonumber\\
   &&+\left(\frac{1}{4} e^{w_1} \left(w_5^2+2 w_6^2+2 w_3
   w_4+4\right)+e^{-w_1}\right) \, \boldsymbol{\partial}_5+\frac{e^{w_1} \left(w_3
   \left(w_6^2+4\right)-4 w_4\right) }{4 \sqrt{2}} \, \boldsymbol{\partial}_6 \nonumber \\
   \boldsymbol{\mathfrak{k}}_6 &=& -\frac{1}{2} e^{w_2} w_6 \, \boldsymbol{\partial}_2-\frac{1}{2} e^{w_2} \left(\sqrt{2} w_5+w_3
   w_6\right) \, \boldsymbol{\partial}_3+\left(\frac{e^{-w_2} w_5}{\sqrt{2}}+\frac{1}{2} e^{w_2} w_4
   w_6\right) \, \boldsymbol{\partial}_4\nonumber\\
   &&+\frac{e^{-w_2} \left(e^{2 w_2} w_4-w_3\right)
   }{\sqrt{2}}\, \boldsymbol{\partial}_5+\left(\frac{1}{4} e^{w_2}
   \left(w_6^2+4\right)+e^{-w_2}\right) \, \boldsymbol{\partial}_6
\end{eqnarray}
The $3$ Killing vector fields closing the $\su(2)$ Lie subalgebra of the isotropy algebra $\mathbb{H}$
have the following explicit form:
\begin{eqnarray}
\label{Killvetsu2}
  \boldsymbol{\mathfrak{k}}_7 &=& \frac{1}{2} e^{w_2} w_6 \, \boldsymbol{\partial}_2+\frac{1}{2} e^{w_2} \left(\sqrt{2} w_5+w_3
   w_6\right) \, \boldsymbol{\partial}_3+\left(\frac{e^{-w_2} w_5}{\sqrt{2}}-\frac{1}{2} e^{w_2} w_4
   w_6\right) \, \boldsymbol{\partial}_4\nonumber\\
   &&-\frac{e^{-w_2} \left(w_3+e^{2 w_2} w_4\right)
   }{\sqrt{2}}\, \boldsymbol{\partial}_5+\left(e^{-w_2}-\frac{1}{4} e^{w_2}
   \left(w_6^2+4\right)\right) \, \boldsymbol{\partial}_6  \nonumber \\
  \boldsymbol{\mathfrak{k}}_8 &=& \frac{1}{2} e^{w_1} w_5 \, \boldsymbol{\partial}_1+\frac{e^{w_1} w_3 w_6 }{2
   \sqrt{2}}\, \boldsymbol{\partial}_2+\frac{e^{w_1} \left(w_3^2+2\right) w_6 }{2
   \sqrt{2}}\, \boldsymbol{\partial}_3-\frac{e^{w_1} w_6 \left(w_6^2+2 w_3 w_4+4\right) }{4
   \sqrt{2}}\, \boldsymbol{\partial}_4\nonumber \\
   &&+\left(e^{-w_1}-\frac{1}{4} e^{w_1} \left(w_5^2+2 w_6^2+2 w_3
   w_4+4\right)\right) \, \boldsymbol{\partial}_5+\frac{e^{w_1} \left(4 w_4-w_3
   \left(w_6^2+4\right)\right) }{4 \sqrt{2}}\, \boldsymbol{\partial}_6  \nonumber \\
  \boldsymbol{\mathfrak{k}}_9 &=& \frac{e^{w_1-w_2} \left(w_3+e^{2 w_2} w_4\right) }{2 \sqrt{2}}\, \boldsymbol{\partial}_1+\frac{1}{4}
   e^{w_1-w_2} \left(e^{2 w_2} \left(\sqrt{2} w_4-w_5 w_6\right)-\sqrt{2} w_3\right)
   \, \boldsymbol{\partial}_2\nonumber\\
   &&+\frac{e^{-w_1-w_2} \left(e^{2 w_2} \left(4-e^{2 w_1}
   \left(w_5^2+\sqrt{2} w_3 w_6 w_5-w_6^2\right)\right)-2 e^{2 w_1}
   \left(w_3^2+2\right)\right) }{4 \sqrt{2}}\, \boldsymbol{\partial}_3\nonumber\\
   &&+\frac{e^{-w_1-w_2} \left(e^{2
   w_1} \left(-4 w_5^2+4 w_6^2-e^{2 w_2} \left(8 w_4^2-4 \sqrt{2} w_5 w_6
   w_4+\left(w_6^2+4\right){}^2\right)\right)+16\right) }{16
   \sqrt{2}}\, \boldsymbol{\partial}_4 \nonumber\\
   &&+\frac{1}{8} e^{w_1-w_2} w_6 \left(4-e^{2 w_2} \left(w_6^2+4\right)\right)
   \, \boldsymbol{\partial}_5+\frac{1}{8} e^{w_1-w_2} w_5 \left(e^{2 w_2}
   \left(w_6^2+4\right)-4\right) \, \boldsymbol{\partial}_6
\end{eqnarray}
Finally, the Killing vector field associated with the  $\uu(1)$ subalgebra of the $\mathbb{H}$ isotropy algebra
has the following explicit form:
\begin{eqnarray}
 \label{Kilvecu1}
  \boldsymbol{\mathfrak{k}}_{10} &=& \frac{\left(e^{w_1-w_2} w_3-e^{w_1+w_2} w_4\right) }{2 \sqrt{2}}\, \boldsymbol{\partial}_1-\frac{1}{4}
   e^{w_1-w_2} \left(\sqrt{2} w_3+e^{2 w_2} \left(\sqrt{2} w_4-w_5 w_6\right)\right)
   \, \boldsymbol{\partial}_2\nonumber\\
   &&+\frac{e^{-w_1-w_2} \left(e^{2 w_2} \left(e^{2 w_1} \left(w_5^2+\sqrt{2}
   w_3 w_6 w_5-w_6^2\right)+4\right)-2 e^{2 w_1} \left(w_3^2+2\right)\right)
   }{4 \sqrt{2}}\, \boldsymbol{\partial}_3\nonumber\\
   &&+\frac{e^{-w_1-w_2} \left(e^{2 w_1} \left(-4 w_5^2+4
   w_6^2+e^{2 w_2} \left(8 w_4^2-4 \sqrt{2} w_5 w_6
   w_4+\left(w_6^2+4\right){}^2\right)\right)-16\right) }{16
   \sqrt{2}}\, \boldsymbol{\partial}_4\nonumber\\
   &&+\frac{1}{8} e^{w_1-w_2} w_6 \left(e^{2 w_2} \left(w_6^2+4\right)+4\right)
   \, \boldsymbol{\partial}_5-\frac{1}{8} e^{w_1-w_2} w_5 \left(e^{2 w_2}
   \left(w_6^2+4\right)+4\right) \, \boldsymbol{\partial}_6
\end{eqnarray}
Last but not least we need the moment maps associated with each of the above Killing vector fields. To this effect we utilize the general method described in sect.\ref{deagostini} and we apply formula(\ref{gelindoelapecora}). Hence we write:
\begin{equation}\label{carnedivenerdi}
  \boldsymbol{\mathfrak{P}}_\Lambda\left(\mathbf{W}\right) \, = \, \ft 12 \, \text{Tr} \left(T^v_{10} \cdot
  \mathbb{L}^{-1}(\mathbf{W})\cdot T^v_\Lambda \cdot \mathbb{L}(\mathbf{W}) \right) \quad ; \quad \Lambda \, = \, 1,\dots , 10
\end{equation}
and the explicit result that we obtain is displayed below:
\begin{eqnarray}\label{carambola}
&&  \begin{array}{|lcl|}
  \hline
 \mathfrak{P}_1 & = & \frac{1}{16} \left(4 \sqrt{2} w_4-4 w_5 w_6-\sqrt{2}
   w_3 \left(w_6^2+4\right)\right)  \\
 \mathfrak{P}_2 & = & \frac{4 w_4+w_3 \left(w_6^2+4\right)}{8 \sqrt{2}} \\
 \mathfrak{P}_3 & = & \frac{1}{32} \left(e^{w_1-w_2} \left(\sqrt{2}
   \left(w_6^2+4\right) w_3^2+4 w_5 w_6 w_3+2 \sqrt{2} \left(w_5^2+4\right)\right)-2
   \sqrt{2} e^{w_2-w_1} \left(w_6^2+4\right)\right) \nonumber \\
 \mathfrak{P}_4 & = & \frac{1}{64} e^{-w_1-w_2} \left(16 \sqrt{2}-e^{2
   \left(w_1+w_2\right)} \left(8 \sqrt{2} w_4^2-8 w_5 w_6 w_4+\sqrt{2}
   \left(w_5^2+4\right) \left(w_6^2+4\right)\right)\right) \nonumber \\
 \mathfrak{P}_5 & = & \frac{1}{32} e^{w_1} \left(-4 \sqrt{2} w_4 w_5+\sqrt{2}
   w_3 \left(w_6^2+4\right) w_5-4 w_3 w_4 w_6+2 \left(w_5^2-4\right)
   w_6\right)-\frac{1}{4} e^{-w_1} w_6 \nonumber \\
 \mathfrak{P}_6 & = & \frac{1}{16} e^{-w_2} \left(2 \sqrt{2} \left(w_3-e^{2
   w_2} w_4\right) w_6+w_5 \left(e^{2 w_2} \left(w_6^2+4\right)+4\right)\right)  \\
   \hline
 \mathfrak{P}_7 & = & \frac{1}{16} e^{-w_2} \left(2 \sqrt{2} \left(w_3+e^{2
   w_2} w_4\right) w_6+w_5 \left(4-e^{2 w_2} \left(w_6^2+4\right)\right)\right)  \\
 \mathfrak{P}_8 & = & -\frac{1}{4} e^{-w_1} w_6-\frac{1}{32} e^{w_1} \left(-4
   \sqrt{2} w_4 w_5+\sqrt{2} w_3 \left(w_6^2+4\right) w_5-4 w_3 w_4 w_6+2
   \left(w_5^2-4\right) w_6\right)  \\
 \mathfrak{P}_9 & = & \frac{1}{64} e^{-w_1-w_2} \left[-4 e^{2 w_2}
   \left(w_6^2+4\right)-2 e^{2 w_1} \left(\left(w_6^2+4\right) w_3^2+2 \sqrt{2} w_5 w_6
   w_3+2 \left(w_5^2+4\right)\right)\right. \\
   \null & \null & \left.+e^{2 \left(w_1+w_2\right)} \left(8 w_4^2-4 \sqrt{2}
   w_5 w_6 w_4+\left(w_5^2+4\right) \left(w_6^2+4\right)\right)+16\right]  \\
   \hline
 \mathfrak{P}_{10} & = & \frac{1}{64} e^{-w_1-w_2} \left[-4 e^{2 w_2}
   \left(w_6^2+4\right)-2 e^{2 w_1} \left(\left(w_6^2+4\right) w_3^2+2 \sqrt{2} w_5 w_6
   w_3+2 \left(w_5^2+4\right)\right)\right.\\
   \null & \null & \left.-e^{2 \left(w_1+w_2\right)} \left(8 w_4^2-4 \sqrt{2}
   w_5 w_6 w_4+\left(w_5^2+4\right) \left(w_6^2+4\right)\right)-16\right] \\
   \hline
\end{array}\nonumber\\
&&
\end{eqnarray}
In eq.(\ref{carambola}) we have separated the moment-maps in the three groups. The first group of six are the moment maps of the $\mathbb{K}$ translations. The subsequent group of three yields the moment maps of the $\su(2)$ compact generators, while the last group of just one is the moment map of the $\uu(1)$ generator associated with the K\"ahler structure.
\section{On the Partition Function and Gibbs Distributions in General and for $\mathbb{SH}_2$ in Particular}
\label{ciurlacco}
Before addressing the determination of the partition function and the Gibbs distributions for the case of the Siegel half plane, that is now accessible, since we have prepared all the necessary instruments, we pose for a moment to consider a very important general property of the temperature vectors $\boldsymbol{\beta}$. We might have anticipated the forthcoming discussion to previous sections, yet we chose to postpone it to the present junction since we are now in a favorable position to illustrate it with a concrete and non trivial example. Once again the
metric equivalence of the non-compact symmetric spaces with a solvable Lie group manifold $\mathcal{S}_{\mathrm{U/H}}$ and the double synergic decompositions (\ref{cane1},\ref{cane2}) play an essential role.
\par
The solvable Lie algebra generators can always be reexpressed as suitable linear combinations of the $K_i$ coset generators and of the $H_\alpha$ compact subalgebra generators. Let us name $\mathcal{T}_i$ ($i=1,\dots,6$) the solvable Lie algebra generators in our $\mathbb{SH}_2$ case. With reference to table \ref{v10genni} we have:
\begin{equation}\label{orosubile}
 \mathcal{T}_i \, = \, \underbrace{\mathcal{Q}_{ij}}_{6\times 6} \, K_j \, + \,\underbrace{ \mathcal{Q}_{i\alpha}}_{6\times 4} \, H_\alpha
 \quad : \quad \left\{\begin{array}{rcl}
 \mathcal{T}_1 & = & K_1 \\
 \mathcal{T}_2 & = & K_2 \\
 \mathcal{T}_3 & = & \frac{1}{\sqrt{2}}\,H_0+\frac{1}{\sqrt{2}}\, H_3+K_3 \\
 \mathcal{T}_4 & = & -\frac{1}{\sqrt{2}}\,H_0+\frac{1}{\sqrt{2}}\, H_3+K_4
   \\
 \mathcal{T}_5 & = & H_2+K_5 \\
 \mathcal{T}_6 & = & H_1+K_6 \\
\end{array} \right.
\end{equation}
The same linear transformation applies to the moment maps and allows to find the moment maps of the Killing vectors associated to the solvable Lie algebra generators. Hence instead of using the basis $(\mathbb{K},\mathbb{H})$ for the moment-maps and the temperature vector $\boldsymbol{\beta}$, we can use the basis $(Solv,\mathbb{H})$ and write the argument of the exponential in the partition function as
\begin{equation}\label{merovingio}
  \widehat{\boldsymbol{\beta}}\cdot \widehat{\boldsymbol{\mathfrak{P}}}(\mathbf{W})\, = \, \widehat{\boldsymbol{\beta}}^\Lambda \, \widehat{\boldsymbol{\mathfrak{P}}}_\Lambda(\mathbf{W})
\end{equation}
where now:
\begin{eqnarray}\label{polifonico}
  \widehat{\boldsymbol{\mathfrak{P}}}_\Lambda(\mathbf{W})& = &\ft 12 \, \text{Tr} \left(H_{0} \cdot
  \mathbb{L}^{-1}(\mathbf{W})\cdot \widehat{T}_\Lambda \cdot \mathbb{L}(\mathbf{W}) \right) \quad ; \quad \Lambda\, = \, 1,\dots,10 \nonumber\\
  \widehat{T}_\Lambda & = & \left\{ \mathcal{T}_1 \, \dots , \mathcal{T}_6\, , \, H_1,\dots, H_4\right\}
\end{eqnarray}
Consider next the formal definition of the partition function:
\begin{eqnarray}
  Z(\widehat{\boldsymbol{\beta}}) & = & \int \exp\left[ -\,\widehat{\boldsymbol{\beta}}\cdot \widehat{\boldsymbol{\mathfrak{P}}}(\mathbf{W})\right]\, \mu(\mathbf{W})\label{spartito}\\
   \mu(\mathbf{W}) & \equiv & \boldsymbol{\mathcal{K}} \wedge \boldsymbol{\mathcal{K}} \wedge \boldsymbol{\mathcal{K}} \simeq \underbrace{\sqrt{\text{det}\left[g(\mathbf{\mathbf{W}})\right]}}_{\text{just a constant}} \, \mathrm{d}^6 \mathbf{W} \label{misura}
\end{eqnarray}
where $\mu(\mathbf{W}) $ in eq.(\ref{misura}) is the integration measure that is invariant with respect to isometries of the coset manifold $\mathrm{U/H}$. Recalling next the metric equivalence of the coset manifold with the solvable Lie
group $\mathcal{S}_{\mathrm{U/H}}$ that has free transitive action on the base manifold we consider any group element $\mathfrak{s}_\mathbf{u}\in \mathcal{S}_{\mathrm{U/H}}$ whose corresponding parameters we name $\mathbf{U}$.
By definition the abstract group element is represented in $5$-dimension by the matrix $\mathbb{L}(\mathbf{U})$ and we have:
\begin{eqnarray}\label{corellus}
  \mathfrak{s}_\mathbf{u} \, : \, \mathbb{L}(\mathbf{W}) \, \longrightarrow \, \mathbb{L}(\mathbf{U})\cdot \mathbb{L}(\mathbf{W}) \, = \, \mathbb{L}\left(\mathfrak{s}_\mathbf{u}(\mathbf{W})\right)
\end{eqnarray}
Since the integration mesure is invariant under the solvable Lie group translations that are isometries we can change integration variables and write:
\begin{equation}\label{canellone}
  Z(\widehat{\boldsymbol{\beta}}) \, = \, \int \exp\left[ -\,\widehat{\boldsymbol{\beta}}\cdot \widehat{\boldsymbol{\mathfrak{P}}}(\mathfrak{s}_\mathbf{u}(\mathbf{W}))\right] \mu(\mathfrak{s}_\mathbf{u}(\mathbf{W})) \, = \, \int \exp\left[ -\,\widehat{\boldsymbol{\beta}}\cdot \widehat{\boldsymbol{\mathfrak{P}}}(\mathfrak{s}_\mathbf{u}(\mathbf{W}))\right]\, \mu(\mathbf{W})
\end{equation}
Focusing on the argument of the exponential integrand we have:
\begin{eqnarray}\label{pignattone}
  \widehat{\boldsymbol{\beta}}\cdot \widehat{\boldsymbol{\mathfrak{P}}}(\mathfrak{s}_\mathbf{u}(\mathbf{W}))
  & = &\widehat{\boldsymbol{\beta}}^\Lambda \, \times \, \ft 12 \, \text{Tr} \left(H_{0} \cdot
  \mathbb{L}^{-1}(\mathbf{W}) \cdot\mathbb{L}^{-1}(\mathbf{U})\cdot \widehat{T}_\Lambda
   \cdot\mathbb{L}(\mathbf{U})\cdot\mathbb{L}(\mathbf{W}) \right) \nonumber \\
   & = & \widehat{\boldsymbol{\beta}}^\Lambda \, \times \, \mathrm{Adj}\left(\mathfrak{s}_\mathbf{u}\right)_\Lambda^{\phantom{\Lambda}\Sigma} \times
   \ft 12 \, \text{Tr} \left(H_{0} \cdot
  \mathbb{L}^{-1}(\mathbf{W})\cdot \widehat{T}_\Sigma \cdot \mathbb{L}(\mathbf{W}) \right) \nonumber\\
  &=& \widehat{\boldsymbol{\beta}}^\Lambda \, \times \, \mathrm{Adj}\left(\mathfrak{s}_\mathbf{u}\right)_\Lambda^{\phantom{\Lambda}\Sigma} \, \widehat{\boldsymbol{\mathfrak{P}}}_\Sigma(\mathbf{W})
\end{eqnarray}
The conclusion of the above formal calculation is that the partition function of generalized thermodynamics {\`a} la Souriau on K\"ahler non compact symmetric spaces $\mathrm{U/H}$ has the following extremely important symmetry:
\begin{eqnarray}\label{invariasolv}
  Z(\widehat{\boldsymbol{\beta}}) & = & Z(\mathrm{Adj}^T(\mathfrak{s})\cdot \widehat{\boldsymbol{\beta}}) \quad ; \quad \forall \mathfrak{s}\in \mathcal{S}_{\mathrm{U/H}} \nonumber\\
  \left(\mathrm{Adj}^T(\mathfrak{s})\cdot \widehat{\boldsymbol{\beta}}\right)^\Sigma & \equiv & \widehat{\boldsymbol{\beta}}^\Lambda \,  \mathrm{Adj}\left(\mathfrak{s}\right)_\Lambda^{\phantom{\Lambda}\Sigma}
\end{eqnarray}
The relevance of the above symmetry is that the co-adjoint action (transpose) of the solvable Lie group can always
rotate a generic temperature vector $\boldsymbol{\beta}$ into a new one $\boldsymbol{\beta}_c$ that has non vanishing components only along the compact subalgebra generators $H_\alpha$, as we might explicitly illustrate for the Siegel case $\mathbb{SH}_2$, but we skip the somewhat lengthy calculations. This is not the end of the story. There is still  the symmetry with respect
to the compact isotropy subgroup that we can utilize. Consider the partition function reduced to a compact temperature $\boldsymbol{\beta}_c$, namely:
\begin{eqnarray}\label{minnesota}
  Z(\boldsymbol{\beta}_c)& = &\int \exp\left[- \boldsymbol{\beta}_c^\alpha \, \boldsymbol{\mathfrak{P}}_\alpha(\mathbf{W})\right] \, \mu(\mathbf{W}) \nonumber\\
  \boldsymbol{\beta}_c^\alpha \, \boldsymbol{\mathfrak{P}}_\alpha(\mathbf{W})& = &\sum_{\alpha=1}^{\text{dim} \mathbb{H}}\, \boldsymbol{\beta}_c^\alpha \, \times \,\ft 12 \, \text{Tr} \left(H_{0} \cdot
  \mathbb{L}^{-1}(\mathbf{W})\cdot H_\alpha \cdot \mathbb{L}(\mathbf{W}) \right)
\end{eqnarray}
The transformation of the compact isotropy subgroup $\mathrm{H}$ are just isometries as all other transformations of $\mathrm{U}$, hence they leave the integration measure $\mu(\mathbf{W})$ invariant and act on the solvable coset representative $\mathbb{L}(\mathbf{W})$ in the canonical way as follows:
\begin{equation}\label{cumpensami}
  \forall \mathrm{h}\in \mathrm{H} \quad ; \quad \mathrm{h}\cdot \mathbb{L}(\mathbf{W}) \, = \, \mathbb{L}(\mathfrak{h}(\mathbf{W}) ) \cdot \mathfrak{H}(\mathrm{h},\mathbf{W}) \quad ; \quad \mathfrak{H}(\mathrm{h},\mathbf{W}) \, \in \, \mathrm{H}
\end{equation}
where $\mathfrak{h}(\mathbf{W})$ are the new solvable coordinates after the $\mathrm{h}$-isometry transformation
and $\mathfrak{H}(\mathrm{h},\mathbf{W})$ is the H-compensator, which, by definition  also lies in $\mathrm{H}$ and depends both on the point $\mathbf{W}$ and on the chosen $\mathrm{h}$ group element.
\par
With the same strategy utilized above, we change the integration variable $\mathbf{W}\to\mathfrak{h}(\mathbf{W}) $ and we rewrite:
\begin{eqnarray}\label{dakota}
  Z(\boldsymbol{\beta}_c)& = &\int \exp\left[- \boldsymbol{\beta}_c^\alpha \, \boldsymbol{\mathfrak{P}}_\alpha(\mathfrak{h}(\mathbf{W}))\right] \, \mu(\mathbf{\mathfrak{h}(\mathbf{W})}) \nonumber\\
  &=& \int \exp\left[- \boldsymbol{\beta}_c^\alpha \, \boldsymbol{\mathfrak{P}}_\alpha(\mathfrak{h}(\mathbf{W}))\right] \, \mu(\mathbf{W})
\end{eqnarray}
Next using eq.s(\ref{minnesota},\ref{cumpensami}) we obtain:
\begin{eqnarray}
 \boldsymbol{\beta}_c^\alpha \, \boldsymbol{\mathfrak{P}}_\alpha(\mathfrak{h}(\mathbf{W})) &=&
 \sum_{\alpha=1}^{\text{dim} \mathbb{H}}\, \boldsymbol{\beta}_c^\alpha \, \times \,\ft 12 \, \text{Tr}
 \left(H_{0} \cdot \mathfrak{H}(\mathrm{h},\mathbf{W}) \cdot
  \mathbb{L}^{-1}(\mathbf{W})\cdot \mathrm{h}^{-1} \cdot H_\alpha \cdot \mathrm{h} \cdot \mathbb{L}(\mathbf{W})\cdot\mathfrak{H}^{-1}(\mathrm{h},\mathbf{W}) \right)  \label{paesibaschi}\\
  & = & \boldsymbol{\beta}_c^\alpha \, \mathrm{Adj}(h)_\alpha^{\phantom{\alpha}\beta} \, \boldsymbol{\mathfrak{P}}_\beta(\mathbf{W})  \label{superganzo}
\end{eqnarray}
In order to establish the equality of the r.h.s in (\ref{paesibaschi}) with that in (\ref{superganzo}) we utilized
three properties:
\begin{enumerate}
  \item The cyclic invariance of the trace.
  \item The crucial fact that $H_0$ is the center of the compact Lie algebra $\mathbb{H}$, which is the very reason why the considered manifold is K\"ahlerian, so that $H_0$ is invariant against any adjoint transformation of $\mathrm{H}$ group.
  \item {The adjoint $H$ representation in the space of its Lie algebra $\mathbb{H}$:
  \begin{equation}\label{aggiungiunposto}
   \forall \mathrm{h}\in \mathrm{H} \quad : \quad \mathrm{h}^{-1} \cdot H_\alpha \cdot \mathrm{h} \, =
  \, \mathrm{Adj}(h)_\alpha^{\phantom{\alpha}\beta} \, H_\beta
  \end{equation}
  }
  \end{enumerate}
The final crucial fact is that by means of a suitable $\mathrm{\mathrm{h}}$-transformation we can always bring any
$\mathbb{\mathbb{H}}$ Lie algebra element into the Cartan subalgebra $\mathcal{C}\subset \mathbb{H}$ and then reduce the $\boldsymbol{\beta}_c$ to $\boldsymbol{\beta}_c^0$ that has non vanishing component only along the Cartan generators, one being $H_0$ the remaining ones being the Cartan generators of $H^\prime$ in the decomposition
(\ref{crunacammello}).
\subsection{Canonical form of the Partition Functions and of the Gibbs Probability Distributions, in General}
In this way the space of allowed temperatures turns out to be the $\mathrm{U}$
(co)adjoint orbit of a proper subset $\Omega_c \mathcal{C}\subset \mathbb{H}$ of the compact Cartan subalgebra $\mathcal{C}$. $\Omega_c$ is provided by those compact Cartan temperatures that have the correct sign in order to guarantee convergence of the Gaussian integrals. As we advocate further on, for the non maximally split manifolds, where the  solvable Lie algebra admits the Paint-Group automorphism group, which is a proper subgroup $\mathrm{G^{Paint}} \subset \mathrm{H}\subset \mathrm{U}$, the essential Cartan temperatures might be further reduced. We postpone this study to a next publication and confine ourselves to the preliminary observations of appendix
\ref{belohorizonte} relative to the case $\mathcal{M}^{[2,2]}$.
\par
Learning the lesson that the temperature vector $\boldsymbol{\beta}$ can be reduced to its minimal Cartan form
$\boldsymbol{\beta}_0$ by means of $\mathrm{U}$-isometry transformations, the general form of the Gibbs probability distributions becomes very simple and clear. It contains always, as many parameters as the dimension of the $\mathrm{U}$ Lie group but it can be written in the following very compact and useful way:
\begin{eqnarray}\label{carbonero}
  \mathrm{G}\left(\boldsymbol{\beta}_0\, ; \,  g \, \mid \, \mathbf{W} \right) & \equiv & \frac{\exp\left[ - \sum_{i=0}^{\ell=\text{rank }\mathbb{H}^\prime} \,\boldsymbol{\beta}_0^{i} \,\boldsymbol{\mathfrak{P}}_i \left(\mathfrak{g}[\mathbf{W}]\right) \right]} {Z(\boldsymbol{\beta}^0)} \nonumber\\
  \boldsymbol{\beta}_0^{i} & = & \text{temperatures associated with the compact Cartan generators $H^c_{0,1,\dots, \ell}$ of $\mathbb{H}$} \nonumber\\
  \boldsymbol{\mathfrak{P}}_i \left(\mathbf{W}\right) &=& \text{moment maps of the compact Cartan generators $H^c_{0,1,\dots, \ell}$ of $\mathbb{H}$} \nonumber\\
  g & = & \text{any group element in $\mathrm{U}$} \nonumber\\
  \mathfrak{g}[\mathbf{W}] & = & \text{new solvable parameters after a $g$-isometry transformation}
  \end{eqnarray}
  In other words the Cartan temperatures define the Gibbs probability distribution centered around the origin of the coset manifold, namely the identity of the metrically equivalent solvable group. The other temperatures simply
  would rotate such a distribution (H-transformations) or translate it to be centered around any other point of the manifold (solvable Lie group transformations). Hence instead of introducing such parameters we can simply evaluate
  the original Gibbs distribution in a transformed point. Both for analytic calculations and for practical applications in Machine Learning this view-point is extremely useful.  {Stated differently, when reducing the temperature vector to the compact Cartan subalgebra, the suppressed  $\beta$-parameters are replaced by the parameters of a generic ${\rm U}$-transformation $g$ of the point, appearing in the argument of the exponential distribution.}
\subsection{Calculation of the Partition Function for the Siegel Plane in Canonical Form}
Having clarified that we jus need two compact temperatures associated with two Cartan generators we choose as second Cartan generator the one listed as $H_3\, = \, T^v_9$ in table \ref{v10genni}. This means that using eq.(\ref{carambola}) for the explicit form of the moment maps, the argument of the exponential in the partition function integrand is the following:
\begin{eqnarray}\label{planetario}
 \mathfrak{A}\, \equiv \, \boldsymbol{\beta}_0\cdot \boldsymbol{\mathfrak{P}}(\mathbf{W}) & = & \mu \, \boldsymbol{\mathfrak{P}}_9(\mathbf{W}) \, + \,  \lambda
  \boldsymbol{\mathfrak{P}}_{10}(\mathbf{W}) \nonumber\\
  &=&\frac{1}{64} e^{-w_1-w_2} \left(\lambda
   \left(-4 e^{2 w_2} \left(w_6^2+4\right)-2
   e^{2 w_1} \left(\left(w_6^2+4\right)
   w_3^2+2 \sqrt{2} w_5 w_6 w_3+2
   \left(w_5^2+4\right)\right)\right.\right. \nonumber\\
   &&\left.\left.-e^{2
   \left(w_1+w_2\right)} \left(8 w_4^2-4
   \sqrt{2} w_5 w_6 w_4+\left(w_5^2+4\right)
   \left(w_6^2+4\right)\right)-16\right)\right.\nonumber\\
   &&\left. +\mu
   \left(-4 e^{2 w_2} \left(w_6^2+4\right)-2
   e^{2 w_1} \left(\left(w_6^2+4\right)
   w_3^2+2 \sqrt{2} w_5 w_6 w_3+2
   \left(w_5^2+4\right)\right)\right.\right.\nonumber\\
   &&\left.\left.  +e^{2
   \left(w_1+w_2\right)} \left(8 w_4^2-4
   \sqrt{2} w_5 w_6 w_4+\left(w_5^2+4\right)
   \left(w_6^2+4\right)\right)+16\right)\right)
\end{eqnarray}
where we have named $\beta_9 \, = \, \mu$ and $\beta_{10} \, = \lambda$. For calculation convenience it is
useful to redefine $w_{1,2} \, = \, \log[\rho_{1,2}$. In this way we get:
\begin{eqnarray}
\label{acco}
  \mathfrak{A} &=&\frac{N_A}{D_A} \nonumber\\
  N_A&=& \rho _1^2 \left(-\left(\rho _2^2 \left(8
   w_4^2-4 \sqrt{2} w_5 w_6
   w_4+\left(w_5^2+4\right)
   \left(w_6^2+4\right)\right) (\lambda -\mu
   )\right.\right. \nonumber\\
  &&\left.\left.  +2 \left(\left(w_6^2+4\right) w_3^2+2
   \sqrt{2} w_5 w_6 w_3+2
   \left(w_5^2+4\right)\right) (\lambda +\mu
   )\right)\right)-4 \left(4 (\lambda -\mu
   )+\rho _2^2 \left(w_6^2+4\right) (\lambda
   +\mu )\right)  \nonumber\\
   D_A &= & 64 \rho _1 \rho _2 \nonumber\\
\end{eqnarray}
We have to calculate the 6-integrals of $\exp[\mathfrak{A}]$ on the nilpotent coordinate $w_3,w_4,w_5,w_6$ and finally on $\rho_1,\rho2$. We begin with the $w_3$ integral that is perfectly gaussian and imposes the convergence condition:
\begin{equation}\label{condit1}
  \lambda+\mu >0
\end{equation}
Next we calculate also the integrand on $w_4$ which is again gaussian and imposes the second convergence condition:
\begin{equation}\label{condit2}
  \lambda-\mu >0
\end{equation}
We get
\begin{eqnarray}
\label{lianaorfei}
  \int_{-\infty}^{\infty}\, dw_4 \int_{-\infty}^{\infty} dw_3 \, \exp\left[-\mathfrak{A}\right]&=& \frac{16 \pi  e^{-\mathfrak{B}}}{\rho _1 \sqrt{\lambda -\mu } \sqrt{\left(w_6^2+4\right)
   (\lambda +\mu )}} \nonumber\\
 \mathfrak{B} &=& \frac{N_B}{64 \rho _1 \rho _2}\nonumber \\
 N_B &=& 16 \lambda -16 \mu +\rho _2^2 \rho _1^2 w_5^2 w_6^2 (-(\lambda -\mu ))+4 \rho _2^2
   \left(w_6^2+4\right) (\lambda +\mu ) \nonumber\\
 &&+\frac{\rho _1^2
   \left(\rho _2^2 \left(w_5^2+4\right) \left(w_6^2+4\right){}^2 (\lambda -\mu )+16
   \left(w_5^2+w_6^2+4\right) (\lambda +\mu )\right)}{w_6^2+4} \nonumber\\
\end{eqnarray}
The next integration on the nilpotent coordinate $w_5$ is just gaussian and it does not impose further contraints on the two temperatures $\mu$ and $\lambda$. We get:
\begin{eqnarray}\label{paramento}
  \int_{-\infty}^{\infty} dw_5 \, \frac{16 \pi  e^{-\mathfrak{B}}}{\rho _1 \sqrt{\lambda -\mu } \sqrt{\left(w_6^2+4\right)
   (\lambda +\mu )}}& = & \mathfrak{C}\nonumber\\
   \mathfrak{C} & = & \frac{64 \pi ^{3/2} \exp \left(-\frac{4 \left(\rho _1^2 (\lambda +\mu )+\lambda -\mu
   \right)+\rho _2^2 \left(w_6^2+4\right) \left(\rho _1^2 (\lambda -\mu )+\lambda +\mu
   \right)}{16 \rho _1 \rho _2}\right)}{\rho _1 \sqrt{\left(w_6^2+4\right) (\lambda -\mu
   ) (\lambda +\mu )} \sqrt{\rho _2 \rho _1 (\lambda -\mu )+\frac{4 \rho _1 (\lambda
   +\mu )}{\rho _2 \left(w_6^2+4\right)}}}\nonumber\\
\end{eqnarray}
The fourth integration on the nilpotent variable $w_6$ can also be performed and yields an analytical result:
\begin{eqnarray}\label{cartamura}
 &&\mathfrak{F}(\rho_1,\rho_2,\lambda,\mu) \,\equiv \, \int_{-\infty}^{\infty} dw_6 \, \mathfrak{C}\nonumber\\
 &=& \frac{64 \pi ^{3/2} \sqrt{\lambda -\mu } \exp \left(-\frac{\lambda ^2+(\lambda -\mu )
   \left(\rho _1^2 \left(\rho _2^2 (\lambda -\mu )+\lambda +\mu \right)+\rho _2^2
   (\lambda +\mu )\right)-6 \lambda  \mu +\mu ^2}{8 \rho _1 \rho _2 (\lambda -\mu
   )}\right) K_0\left(\frac{\left((\lambda -\mu ) \rho _1^2+\lambda +\mu \right)
   \left((\lambda -\mu ) \rho _2^2+\lambda +\mu \right)}{8 (\lambda -\mu ) \rho _1 \rho
   _2}\right)}{\sqrt{\rho _2 (\lambda +\mu )} \left(\frac{\rho _1 (\lambda -\mu )}{\rho
   _2^2 (\lambda -\mu )+\lambda +\mu }\right){}^{3/2} \left(\rho _2^2 (\lambda -\mu
   )+\lambda +\mu \right){}^{3/2}}\nonumber\\
\end{eqnarray}
where $K_0(x)$ is the Bessel function of type $K$ and index $0$.
\par
\begin{figure}
\begin{center}
\includegraphics[width=8cm]{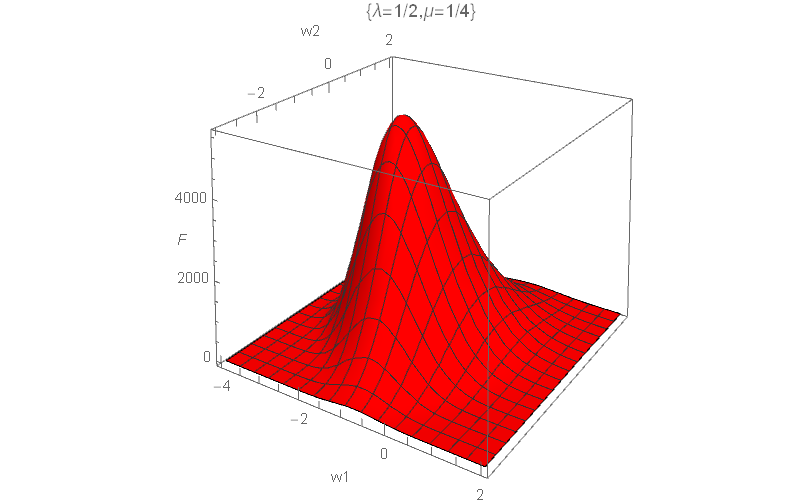}
\includegraphics[width=8cm]{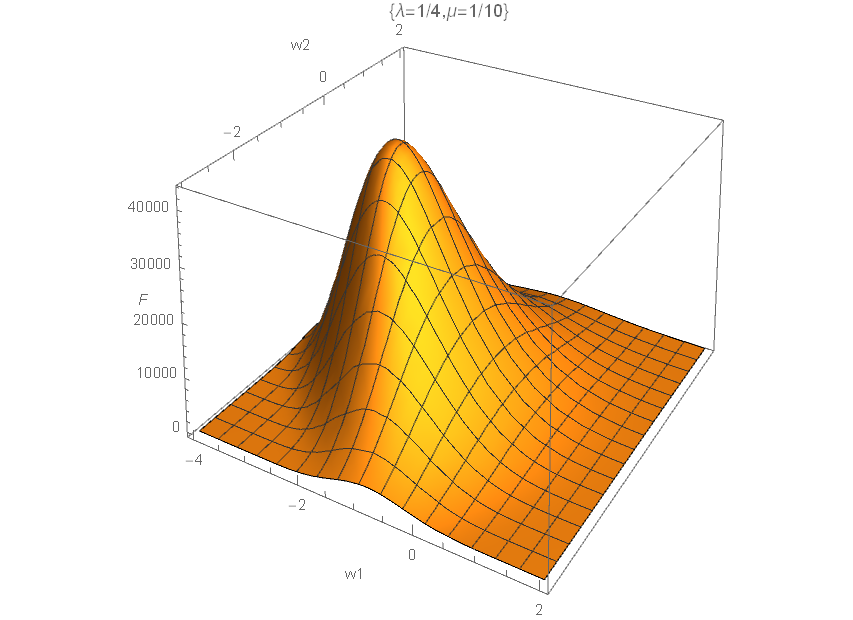}\\
\includegraphics[width=13cm]{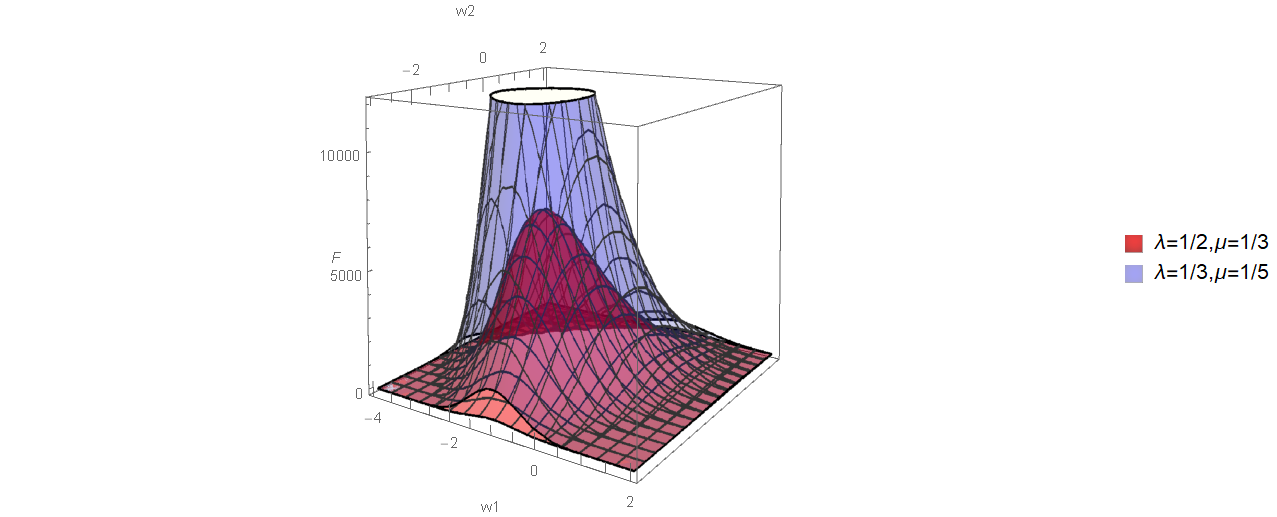}
 \caption{\label{convergenza} In this figure we show for a few pairs of values of $\lambda$ and $\mu$ the integrand
 $\mathfrak{F}$ in the integration variables $w_{1,2}\, = \, \log{\rho_{1,2}}$. The bell shape and the uniform exponential decay to zero at infinity in all directions guarantees the convergence of the two remaining integrals on $w_{1,2}$.}
\end{center}
\end{figure}
\begin{figure}
\begin{center}
\includegraphics[width=8cm]{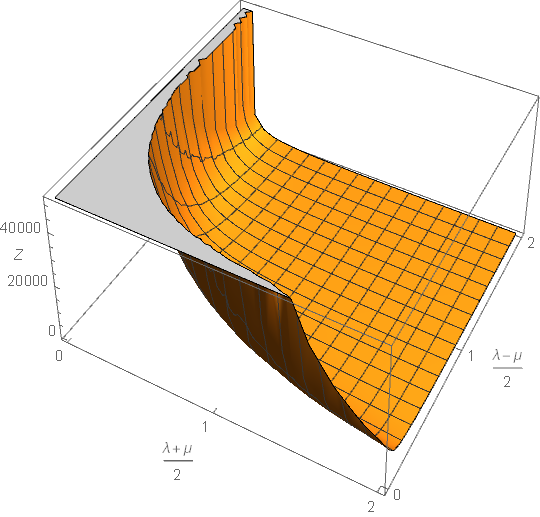}\\
\includegraphics[width=8cm]{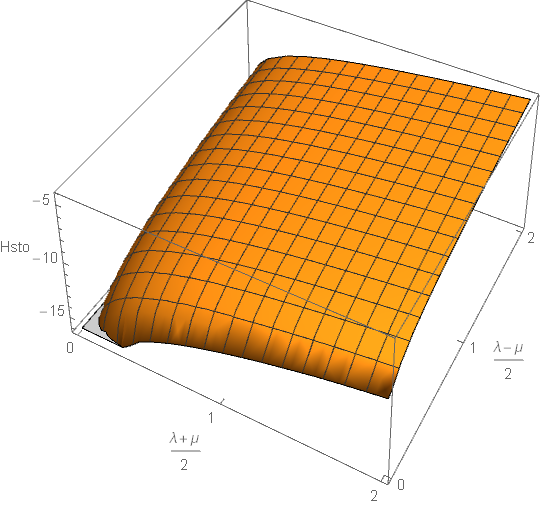}
 \caption{\label{compilatus} Plots of the numerical evaluations of the Geothermodynamical partition function and stochastic Hamiltonian \`a la Souriau for the Siegel plane.}
\end{center}
\end{figure}

Unfortunately the last two  integrals on the remaining two variables $\rho_{1,2}$ or better on their logarithms
$w_{1,2}$ cannot be done analytically and one has to perform them numerically introducing in this way compiled functions. By plotting the integrand we can however very easily verify that it always dacays exponentially to zero in all directions so that the integral is  always convergent (see fig.\ref{convergenza}).
For this reason one can define the partition function as compiled function by performing numerically on a
computer the last two integrals:
\begin{equation}\label{graccus}
  Z(\lambda,\mu) \, = \,\int_{-\infty}^{\infty}\, dw_1 \int_{-\infty}^{\infty} dw_2 \,
  \mathfrak{F}(\exp[w_1],\exp[w_2],\lambda,\mu)
\end{equation}
In fig.\ref{compilatus} we display a plot of the partition function and of minus its logarithm, namely of the stochastic hamiltonian.
\section{Conclusions}
\label{concludo}
Machine Learning and so named Artificial Intelligence algorithms rely on two mathematical pillars: Differential Geometry on one-side and Probability Theory on the other. The first is needed to model the spaces whereto data are to be encoded, the second to classify and elaborate the data by assigning a probability to their actual location in such spaces and to their correlations. Hence the two mathematical pillars have to be reconciled to each other and to be solidly entangled. A statistical viewpoint on whatever set of objects always leads to thermodynamics, in a generalized sense, and,  already since some decades, an abstract geometrical formulation of thermodynamics has been developed that starts from Shannon's information entropy and leads to the identification of thermodynamical equilibrium states with lagrangian submanifolds of symplectic manifolds also endowing them with a canonical Riemannian structure. In separate data science literature the same Riemannian structure has been developed under the name of Fisher's information geometry. In such a variegated conceptual landscape and in strong correlation with the newly introduced paradigm of \textbf{Cartan neural networks}, where all the hidden layers of neural networks are modeled as non compact symmetric spaces $\mathrm{U/H}$, that are all Cartan Hadamard manifolds and,  consequently, are equipped with a canonical distance function, an important issue for Machine Learning is that of gaussian-like
probability distributions on the encoding spaces. Starting from the hints provided by the work of a group of French authors \cite{marlentropia,caldobarbaresco,barbaresco2,barbaresco3,marlegibbs} who suggested the use of Gibbs states related with the Lie Group Thermodynamics proposed long ago by Souriau, in the present paper we took on ourselves the following task:
\begin{enumerate}
  \item Clarify the relation of  Geometrical Thermodynamics {\`a} la Ruppeiner \cite{Ruppeiner_2010,Ruppeiner_2012,Ruppeiner_2012b,Ruppeiner_2013,ruppoRdiag,Ruppeiner_2020} and Lychagin  \cite{lychaginlecture} with Souriau's proposals\cite{marlentropia,marlegibbs} and with Information Geometry
      \cite{Nielsen_2020}.
  \item Distinguish among Souriau non abelian thermodynamics and the geometrical thermodynamics associated with Integrable Dynamical Systems, in particular the Geodesic Dynamical System associated with the calculation of geodesics on the same $\mathrm{U/H}$ symmetric spaces that enter the Machine Learning game as hidden layer models.
  \item Investigate the basic principle of Souriau's thermodynamics that is the characterization of the locus $\Omega \subset \mathbb{U}$ in the relevant Lie algebra, whose elements are possible generalized temperatures in the sense that for them the partition function integral is convergent.
  \item Clarify the role of the coadjoint orbit conception, Souriau's favorite one, that turns out to be equivalent to the more practical and algorithmic conception based on coset manifolds.
\end{enumerate}
We think that we have attained  all the goals we aimed at. Indeed our results can be summarized as follows.
\begin{description}
  \item[1)] We have established the identity of Fisher's Information metric, given as the Hessian of a certain matrix with the  metric obtained as the Hessian of the stochastic hamiltonian $\mathcal{H}^{sto}(\boldsymbol{\lambda})$,  derived in Lychagin's approach as the canonical Riemannian metric on Lagrangian submanifolds of a symplectic manifold where, by definition, the symplectic $2$-form $\omega$ vanishes identically. Such lagrangian submanifolds are the thermodynamical equilibrium states and the $1$-st and $2$nd Principle of Thermodynamics are incorporated in their very definition. These notions are fully general and equally apply to any generalized thermodynamics, irrespectively whether the generalized temperatures $\boldsymbol{\lambda}$ are dual to a set of hamiltonians in involution, as it happens with integrable dynamical systems, or the moment-maps of a non-abelian algebra as it happens in the thermodynamics {\`a} la Souriau.
  \item[2)] With respect to the Poissonian structures on the dual $Solv^\star$ of solvable Lie algebras $Solv$ utilized also by two of us (P.F. and A.S.) in their 2009 paper \cite{sashaebog} on the integrability of the geodesic equations on non compact symmetric spaces $\mathrm{U/H}$ and investigated by Arkhangelsky in
      \cite{arcangelo} where he derived their hamiltonians in involution, we show here that such Poissonian structure is only half of the full story, since it is defined only on the momentum subspace of phase space. Introducing also the coordinates, which is what one should always do, there is a complete symplectic manifold with a symplectic $2$-form of maximal rank and what one describes is just the geodesic dynamical system in hamiltonian formalism. Arkhangelsky hamiltonians depend only on the momenta but they are hamiltonians in involution also with respect to the complete symplectic structure. The geometric thermodynamics associated with such integrable dynamical systems can be constructed but it is essentially
      uninteresting for three reasons:
      \subitem[a] The dependence of the partition function on volume is factorized and the equation of state resembles the trivial one of Ideal gases.
      \subitem[b] The degrees of freedom are few and a statistical description seem unappropriate.
      \subitem[c] Last but not least the Gibbs probability distributions have a non trivial structure only in momentum space, namely along the fibres of the tangent bundle $\mathcal{T}\mathrm{U/H}$ not on the very base manifold $\mathrm{U/H}$. All that is of little appeal for Machine Learning applications where one looks for probability distributions (Gibbs states) on $\mathrm{U/H}$.
  \item[3)] The searched for gaussian-like probability distributions on $\mathrm{U/H}$ are instead provided by the construction of Gibbs states {\`a} la Souriau. This requires a symplectic structure on the very manifold $\mathrm{U/H}$ and not on its tangent bundle. After demonstrating that a coadjoint $\mathrm{U}$-orbit of some element $\mathbf{b} \in \mathbb{U}$ is always diffeomorphic and algebraically equivalent to a coset manifold $\mathrm{U/H}$ where $\mathrm{H}\subset \mathrm{U}$ is the stabilizer of the element $\mathbf{b}$ in $\mathrm{U}$, we abandon the coadjoint orbit conception and we focus on non-compact symmetric spaces
      $\mathrm{U/H}$. In order to have the symplectic structure and construct Souriau thermodynamics, $\mathrm{H}$
      must be the stabilizer of some Lie algebra element and this, as we show, implies that $\mathbb{H}$ has a $\uu(1)$ which endows the symmetric space with a K\"ahler structure and the K\"ahler $2$-form $\boldsymbol{\mathcal{K}}$ is the required symplectic $2$-form. In this way we come to the conclusion that the relevant non compact symmetric spaces are the K\"ahler ones corresponding only to two infinite series, the Siegel half-planes $\mathbb{SH}_n$ and the Calabi-Vesentini manifolds $\mathcal{M}^{[2,q]}$ mentioned in eq.(\ref{supercaronte}). The first series is composed of maximally split manifolds of increasing non-compact rank, while the second constitute a Tits Satake universality class having the Siegel $\mathbb{SH}_2$ manifold as universal Tits Satake submanifold. In application to Machine Learning if one wants to take advantage of the Paint Group symmetry (see \cite{pgtstheory}) and its potentiality in data clustering, the Calabi-Vesentini choice is the preferred one.
  \item[4)] The central point of Souriau generalized thermodynamics, namely the determination of the subspace
   $\Omega \subset \mathbb{U}$ of allowed temperature vectors was also solved by us in a simple and elegant way.
   $\Omega$ is just the \textbf{the adjoint $\mathrm{U}$ orbit} of a \textbf{positivity chamber in the Cartan subalgebra $\mathcal{C}\subset {H}$ of the compact subalgebra $\mathbb{H}\subset \mathbb{U}$}. One fixes the sign of the $\ell$ independent temperatures associated with $\ell$ generators of $\mathcal{C}$ and by an adjoint transformation of $\mathrm{U}$ generates all the other possible temperature vectors that respect convergence of the partition function integral. This property apart from solving the convergence issue is also of practical relevance. Indeed the true temperatures are just the compact Cartan ones, all the others are an effect of translation of the central point of a Gibbs probability distribution to any other in the manifold by means of an isometry. Hence we can always utilize the same partition function depending on a very small number of temperature and change the point in the Gibbs distribution from a given one to its image under any element of the isometry group $\mathrm{U}$.
   \item[5)] For the case of the Poincar\'e plane we have explicitly constructed the partition function depending on three temperatures and even studied the $3$-dimensional thermodynamical Riemannian metric, showing that it is non trivial and not that of a space. For the case of the Siegel half-plane $\mathbb{SH}_2$
        we have reduced the partition function to an integral in two variables, whose integrand is a combination of exponentials, roots and Bessel functions. The integral is convergent and one can construct a compiled function of which we have shown a plot. The only problem is the computing velocity of the utilized computer.
\end{description}
\paragraph{What to do next.}
As we have shown in appendix \ref{belohorizonte} the here obtained results for the Tits Satake submanifold $\mathbb{SH}_2$ are liable to be extended to the entire universality class of Calabi-Vesentini manifolds by careful use of
Paint Group invariance. This is certainly the most urgent task in the program. Succeeding in that we will have Gibbs probability distributions for each of the candidate hidden layers of a Cartan Neural Network based on Calabi Vesentini manifolds. As Barbaresco et al have shown \cite{marlentropia,caldobarbaresco,barbaresco2,barbaresco3,marlegibbs} geometric thermodynamics {\`a} la Souriau can be utilized to study time/sequences, in particular those provided by radar data. This is just the tip of an iceberg. On one side there are many other possible sequential data, on the other side the use of Gibbs probability distributions on the hidden layers or anyhow on the manifold whereto various types of data can be mapped, introduces a new powerful weapon in designing algorithm architectures. Furthermore, recalling results of
\cite{pgtstheory} so far not yet used, one activity direction might be that of restricting the Gibbs probability distributions to infinite discrete subsets of the Calabi Vesentini manifold, corresponding to orbits of the origin under the action of any of the large class of discrete subgroups of $\mathrm{SO(2,2+q)}$ that were classified and constructed in the foundational paper \cite{pgtstheory}.
\paragraph{Final Comment} When the present article was ready for posting on ArXiv and for submission to a Journal we learned about the beautiful and inspiring paper by
Laurent Bonnasse-Gahot and Jean Pierre Nadal \cite{nadalcategor} focused on  \textit{geometry of the internal representation}, that, mathematically, means  Fisher's information geometry, namely what we have here shown to be identical with the thermodynamical geometry of
Lychagin, Roop and Ruppeiner, in whose general scheme we could also fit the Gibbs states à la Souriau on the
K\"ahler non compact symmetric spaces $\mathrm{U/H}$. It follows that an additional challenging direction of further investigation is the
possible application of
\cite{nadalcategor} general methods to data mapped to
K\"ahler non compact symmetric spaces and to the Gibbs states defined over them.
\newpage
\appendix
\section{Basic  Structures of Contact and Symplectic Geometry}
\label{geoprobo}
In this appendix we recall the basic structures of symplectic and contact geometry that are the two tightly connected arenas both for the geometric reformulation of thermodynamics and for dynamical systems, \textit{i.e.} the underlying mathematical basis of all phenomena studied in stochastic processes, data science and (geometric) deep learning. Indeed the prototype of a symplectic manifold is the phase space of a dynamical system.
\subsection{Contact Geometry}
\label{contattiamoci} Contact Geometry is at the same time a new and ancient chapter in Mathematics, since it originates in classical results dating back to Goursat, Darboux, Lie and other Masters of the 19th century, but has been vigorously developed in the last two-three decades by a relatively small community of mathematicians. Regarding such structures, there is an extensive literature provided by the following bibliographical references and the additional articles they cite
\cite{Gilbert},\cite{Etnyre2000},\cite{Ghrist2007},\cite{contatoregeiges}
,\cite{cardone2019},\cite{miranda2021},
\cite{trasversozero},\cite{unochiuseinteg},\cite{guglielminoconmiranda},
\cite{pollosingolare}. To summarize, it can be said that \textit{Contact Geometry} is a mathematical theory that aims at establishing an intrinsic geometric-topological characterization of \textit{non-integrability} and, in a sense, is formulated in an inverted perspective from that usual in cohomological theories and integrable systems theory.
\par
Not surprisingly, Contact Geometry comes into play whenever one intends to study \textit{chaos and disorder}, rather than order as is the case with \textit{integrable systems}. Disorder and lack of information are characteristics of turbulent regimes in \textit{Fluid Dynamics} and is the essential attribute of statistical thermodynamic systems: so it is quite natural that Contact Geometry has relations to both Fluid Dynamics and Thermodynamics.
\par
Contact Geometry deals exclusively with
\textit{Differentiable Manifolds of odd dimension
$\mathcal{M}_{2n+1}$ }  and on the other hand it has a symbiotic relationship with
 \textit{ Simplectic Manifolds  $\mathcal{S}_{2n+2}$ and
$\mathcal{S}_{2n}$} in the two adjacent even dimensions, the upper and the lower one.
\par
The fundamental notions of Contact Geometry are easily stated and require, to be assimilated, nothing more than the most basic and elementary concepts of Differential Geometry. In the concise summary we present in this section, we closely follow the excellent review article \cite{contatoregeiges}.
\subsection{Contact Structures}
Let us consider an odd-dimensional differentiable manifolds
$\mathcal{M}_{2n+1}$ and itstangent bundle:
\begin{equation}\label{tangentobundolo}
    \mathcal{TM}_{2n+1} \, \stackrel{\pi}{\longrightarrow} \,
    \mathcal{M}_{2n+1} \quad ; \quad \forall \, p \,
    \in \, \mathcal{M}_{2n+1} \quad , \quad \pi^{-1}(p) \, \sim \,
    \mathbb{R}^{2n+1}
\end{equation}
By definition, the transition function between any pair of local trivializations of the tangent vector bundle is provided by the \textit{inverse Jacobian} of the transition function
$\psi_{ij}(x)$  between the two corresponding open charts
$(U_i,\varphi_i)$ e $(U_j,\varphi_j)$  in any atlas  that covers the entire manifold $\mathcal{M}_{2n+1}$ (see, for instance, \cite{fre2023book}).
\par
The space of the sections of the tangent bundle $\Gamma\left[\mathcal{TM}_{2n+1},\mathcal{M}_{2n+1}\right]$ is made by all the vector fields, whose local description is in terms of first-order differential operators :
\begin{eqnarray}\label{vettorifildi}
    &&\mathbf{X} \, \in \, \Gamma\left[\mathcal{TM}_{2n+1}\right] \quad
    \Downarrow \nonumber\\
    &&  \mathbf{X} \, = \, X^\mu \left(x\right)
    \,\frac{\partial}{\partial x^\mu} \quad : \quad \text{In any open chart $U$
    with coordinates $x^1\dots x^{2n+1}$}
\end{eqnarray}
The cotangent bundle
\begin{equation}\label{cotangentobundolo}
    \mathcal{T}^\star\mathcal{M}_{2n+1} \, \stackrel{\pi_\star}{\longrightarrow} \,
    \mathcal{M}_{2n+1} \quad ; \quad \forall \, p \,
    \in \, \mathcal{M}_{2n+1} \quad , \quad \pi_\star^{-1}(p) \, \sim \,
    \left(\mathbb{R}^{2n+1}\right)^\star
\end{equation}
is the dual of the tangent one, in the sense that its fibre over any point $p\in\mathcal{M}_{2n+1}$  of the basis space is the dual of the fibre vector space over the same point as the tangent fibre, that is, the space of linear functionals over the latter:
\begin{equation}\label{cricco}
    \forall p \in \, \mathcal{M}_{2n+1} \quad : \quad
    \pi_\star^{-1}(p) \, = \, \text{space of linear functionals on}
    \, \pi^{-1}(p)
    \,
\end{equation}
By construction, the transition function between two local trivializations of the cotangent bundle is provided by the \textit{direct Jacobian} of the two corresponding open charts $(U_i,\varphi_i)$ and
$(U_j,\varphi_j)$  in any atlas that covers the entire base-manifold
 $\mathcal{M}_{2n+1}$.
\par
The space of sections of the cotangent bundle
$\Gamma\left[\mathcal{T}^\star\mathcal{M}_{2n+1},\mathcal{M}_{2n+1}\right]$
is made by differential 1-forms whose local description is recalled here below
\begin{eqnarray}\label{vettorifildi}
    &&\omega \, \in \, \Gamma\left[\mathcal{T}^\star\mathcal{M}_{2n+1},\mathcal{M}_{2n+1}\right] \quad
    \Downarrow \nonumber\\
    &&  \omega \, = \, \omega_\mu \left(x\right)
    \,dx^\mu \quad \quad : \quad \text{In any open chart $U$
    with coordinates $x^1\dots x^{2n+1}$}
\end{eqnarray}
\par
In terms of these fundamental concepts, which apply to differentiable manifolds of any dimension, whether even or odd, the concept of hyperplane bundles can be introduced.
A hyperplane bundle is a reduction of the tangent bundle where the fibres above each point constitute a codimension one subspace of the tangent space above the same point, the transition functions being derived accordingly:
\begin{eqnarray}\label{iperpiano}
    \mathcal{HY}\,
    \stackrel{\mathcal{P}}{\longrightarrow} \, \mathcal{M} & ;
    & \forall p \in \mathcal{M} \, , \,
    \mathcal{P}^{-1}(p)
    \subset\pi^{-1}(p) \quad \text{where} \quad \mathcal{T}\mathcal{M} \, \stackrel{\pi}{\longrightarrow} \,
    \mathcal{M}\nonumber\\
    \mathrm{dim}_\mathbb{R}\mathcal{M}\, = \, m & ;&\mathrm{dim}_\mathbb{R}\pi^{-1}(p) \, = \, m \quad ; \quad
    \mathrm{dim}_\mathbb{R}\mathcal{P}^{-1}(p) \, = \, m-1
\end{eqnarray}
A simple way to construct a hyperplane bundle is by means of a section of the cotangent bundle, namely by means of a 1-form  $\omega \in \Gamma\left[\mathcal{T}^\star
\mathcal{M},\mathcal{M}\right]$.
The desired hyperplane sub-bundle
$\mathcal{HY}^\omega \subset \mathcal{TM}$ of the tangent bundle
is implictly defined by specifying  what is the space of all of its sections
$\Gamma\left[\mathcal{HY}^\omega,\mathcal{M}\right]$,  namely by specifying all vector fields that are sections of
$\mathcal{HY}^\omega$. In a precise mathematical language, let
$\mathbf{X}\in \Gamma\left[\mathcal{TM},\mathcal{M}\right]$ be a vector field
then we write:
\begin{equation}\label{kernello}
    \mathbf{X} \, \in \, \Gamma\left[\mathcal{HY}^\omega,\mathcal{M}\right] \quad
    \text{iff} \quad \mathbf{X} \, \in \,\mathrm{ker}\, \omega \quad
    \textit{i.e.} \quad \omega\left(\mathbf{X}\right) \equiv 0 \quad
    (\text{everywhere})
\end{equation}
\begin{definizione}
\label{struttamorfa} Given an odd-dimensional manifold $\mathcal{M}_{2n+1}$,
a \textbf{contact structure} on
$\mathcal{M}_{2n+1}$  is a rank $2n$  sub-bundle  $\xi \,
\stackrel{\mathcal{P}}{\longrightarrow}\, \mathcal{M}_{2n+1}$  of
 tangent bundle $\mathcal{TM}_{2n+1} \,
\stackrel{\pi}{\longrightarrow}\, \mathcal{M}_{2n+1}$  that can be identified with the hyperplane bundle
 $\mathcal{HY}^\alpha$
where the  1-form $\alpha$  satisfies the following condition:
\begin{equation}\label{latisanaditiglio}
    \alpha \, \wedge \, \underbrace{\mathrm{d}\alpha \, \wedge \,\dots \, \, \wedge
    \,\mathrm{d}\alpha}_{n-\text{volte}} \, \neq \, 0 \quad
    (\text{everywhere on $\mathcal{M}_{2n+1}$})
\end{equation}
The $1$-forma $\alpha$  is named the \textbf{contact form} of the contact structure.
\end{definizione}
\begin{definizione}
\label{cuntavaria} A \textbf{contact manifold} is a pair
 $\left(\mathcal{ M}_{2n+1},\xi\right)$  made by an odd-dimensional manifold and a
contact structure $\xi \,
\stackrel{\mathcal{P}}{\longrightarrow}\, \mathcal{M}_{2n+1}$.
\end{definizione}
Some observations are obligatory in connection with the two definitions above. The first is that the same contact structure can be defined by different contact forms
$\alpha$, $\alpha^\prime$,$\dots$. In fact all multiples of a
given contact form by means  of a scalar function that does not vanish at any point $\lambda\, : \, \mathcal{ M}_{2n+1}\, \to
\, \mathbb{R}$  define the same contact structure.
Secondly it is perfectly possible that the same odd-dimensional manifold $\mathcal{ M}_{2n+1}$ admits more than one contact structure.
The classification of such contact structures modulo diffeomorphism connected to the identity map is an
interesting and relevant mathematical problem for odd-dimensional manifolds just as it is interesting and relevant the classification of  complex structures that  exist on an even-dimensional manifold.
It is therefore obligatory to single out the concept of
\textbf{contactomorphism}
\begin{definizione}
\label{cuntamorfa} Let $\left(\mathcal{ M},\xi\right)$ and
$\left(\mathcal{ N},\chi\right)$ be two contact manifolds and let:
\begin{equation}\label{diffeomorfus}
    \varphi \, : \, \mathcal{M} \, \longrightarrow \,\mathcal{N}
\end{equation}
be a diffeomorphism of the former  onto the latter manifold
(obviously  $\mathcal{M}$ and $\mathcal{N}$ must have the same dimensions
in order for $\varphi$ to possibly exist). Let
$\alpha$ be a contact form that defines $\xi$ and let $\beta$ be a contact form
that definies $\chi$. The considered  diffeomorphism
$\varphi $ is named a \textbf{contactomorphism}  if and only if:
\begin{equation}\label{pullabacco}
    \varphi^\star\left(\beta\right) \, = \, \lambda \, \, \alpha
\end{equation}
where $\varphi^\star$  is the pull-back map and
\begin{equation}\label{nowhere}
    \lambda \, : \, \mathcal{M} \, \longrightarrow \, \mathbb{R}
\end{equation}
is a nowhere vanishing function on $\mathcal{M}$ . If  a  contactomorphism between them does exist,
the two considered manifolds are named
\textbf{contactomorphic}.
\end{definizione}
In the definition \ref{cuntamorfa} the two manifolds $\mathcal{M}$ and
$\mathcal{N}$ might be the same manifold. In this case
what we are considering is the transformation of a contact structure into another one by means of a contactomorphism.
\begin{definizione}
\label{agignasco} Given two contact structures $\xi$ and $\chi$
on the same manifold $\mathcal{M}_{2n+1}$  they must be identified as the same contact structure   if
a  \textit{contactomorphism} does exist that maps one into the other.
\end{definizione}
In conclusion the relevant mathematical problem is that of
classifying contact structures on a manifold
$\mathcal{M}_{2n+1}$ modulo contactomorphisms.
\subsection{Integrability and Frobenius Theorem}
We do not go into the details of Frobenius theorem (see for example
esempio \cite{arnoldometodica} e \cite{contatoregeiges}) but we merely summarize the concepts underlying its formulation.
We begin by noting that any vector field
$\mathbf{X}$  on a differentiable manifold $\mathcal{M}$ of any dimension defines its own integral curves
$\mathcal{I}_\mathbf{X}$, i.e., those curves that at each of their points admit the local value of the vector field $\mathbf{X}$  as a tangent vector.
Since any $p\in
\mathcal{M}$ lies on some integral curve
$\mathcal{I}_\mathbf{X}$, we are guaranteed that a single vector field induces a \textit{foliation} of the manifold
$\mathcal{M}$ into  one-dimensional submanifolds.
\par
On the other hand, it is a more complicated matter to determine whether or not a rank $r > 1$ sub-bundle
$\mathcal{E} \longrightarrow \mathcal{M}$  of the tangent bundle does or does not  induce a \textit{foliation} of $\mathcal{M}$. To this effect, utilizing a not completely rigorous, yet intuitive and qualitatively correct way of speaking, by means of the wording
\textit{foliation} one means the covering of the manifold with a family of
\textit{leaves}, i.e., submanifolds all diffeomorphic to each other, $\mathcal{F}_{\pmb{\nu}} \subset
\mathcal{M}$ having dimension equal to the rank $r$ of the sub-bundle
$\mathcal{E}$, each of which can be regarded as the level hypersurface
of $r$ functions $u_i(p)$ ($i=1,\dots,r$)
that originate from the integration of a basis of sections
$\mathbf{X}_i$ of the sub-bundle $\mathcal{E} \longrightarrow
\mathcal{M}$:
\begin{eqnarray}\label{sfogliatellanapoletana}
   \mathcal{F}_{\pmb{\nu}} & = & \left\{p \, \in \mathcal{M} \,
    \mid \, u_i(p) \, = \,\nu_i \right\} \quad ; \quad \pmb{\nu}\,
    \equiv\, \{\nu_1, \dots \nu_r\} \quad ; \quad \nu_i\, = \,
    \textit{ real constants}\nonumber\\
    \nabla u_i(p) &=& \mathbf{X}_i\mid_p
\end{eqnarray}
When this situation is realized one says that the sub-bundle
$\mathcal{E} \longrightarrow \mathcal{M}$  is integrable.
\textbf{ Frobenius Theorem} establishes the necessary conditions
for such an integrability.
\begin{teorema}\label{frobenico}Let $\mathcal{M}$ be a differentiable manifold  and let $\mathcal{E}
\longrightarrow \mathcal{M}$ be a rank $r> 1$ sub-bundle of the tangent bundle $\mathcal{TM}$.
The necessary and sufficient condition in order for
$\mathcal{E}$ to be  integrable is the following one:
\begin{equation}\label{frobeniata}
    \forall\, \mathbf{X},\mathbf{Y}  \in  \Gamma[\mathcal{E},\mathcal{M}] \quad :
    \quad
    \, \left[\mathbf{X},\mathbf{Y} \right] \, \in \, \Gamma[\mathcal{E},\mathcal{M}]
\end{equation}
\end{teorema}
In the case where $\mathcal{E} \longrightarrow \mathcal{M}$  is a
hyperplane bundle defined by a  1-form $\omega$,  Frobenius integrability condition
can be formulated as follows
\begin{equation}\label{frombolino}
    \omega \, \wedge \, d\omega \, = \, 0
\end{equation}
In order to appreciate the equivalence of the  condition (\ref{frombolino}) with
condition (\ref{frobeniata}) it suffices to recall Cartan's
formula for the value of $\mathrm{d}\omega$ on two arbitrary vector fields
 $\mathbf{X},\mathbf{Y}$:
\begin{equation}\label{cornicione}
    \mathrm{d}\omega\left(\mathbf{X}\, ,\,\mathbf{Y}\right) \, = \, \frac{1}{2} \, \left(\mathbf{X} \, \omega\left(\mathbf{Y}\right) -
    \mathbf{Y }\, \omega \left(\mathbf{X}\right)
    - \omega\left( \left[\mathbf{X} \, , \, \mathbf{Y}\right]\right)\right)
\end{equation}
Let $\mathbf{Z} \notin \Gamma[\mathcal{E},\mathcal{M}]$ and
$\mathbf{X},\mathbf{Y} \in \Gamma[\mathcal{E},\mathcal{M}]$;  next evaluate
the  3-forma $\omega \, \wedge \, d\omega $ on the  triplet
 $\mathbf{Z},\mathbf{X},\mathbf{Y}$ of vector fields. We find:
\begin{equation}\label{grugliasco}
    \omega \, \wedge \, d\omega\left(\mathbf{Z},\mathbf{X},\mathbf{Y}\right) \, \propto \,
    \underbrace{\omega\left(Z\right)}_{\neq 0} \, \omega\left(\left[\mathbf{X}\, ,
    \,\mathbf{Y}\right]\right)
\end{equation}
Hence equation (\ref{frombolino}) implies
$\omega\left(\left[\mathbf{X}\, ,
    \,\mathbf{Y}\right]\right)\, = 0$ that on its turn means $\left[\mathbf{X}\, ,
    \,\mathbf{Y}\right] \in \Gamma[\mathcal{E},\mathcal{M}]$.
\par
In this way we see that a contact structure defined by a contact 1-form is the exact opposite of an integrable sub-bundle. Indeed, one can show that it corresponds to maximum nonintegrability.
\par In section \ref{zaklyuchenie}, considering the integrable dynamical systems constructed on \textit{normed solvable Lie algebras} and solvable groups,
we get deeper into the meaning of Frobenius theorem. For integrable systems the foliation of the differentiable manifold is provided by the level set surfaces of a maximal set of hamiltonian functions in involution whose corresponding vector fields span the integrable sub-bundle of the tangent bundle.
\subsection{Isotropic Submanifolds of a Contact Manifold and Non Integrability}
Let us introduce the following:
\begin{definizione}
Let $\left(\mathcal{M}_{2n+1}\, , \, \xi\right)$  be a contact manifold and  $\mathcal{L}\subset\mathcal{M}_{2n+1}$  be one of its
submanifolds. Let us consider the tangent bundle of such a submanifold
$\mathcal{TL}\,\stackrel{\pi_{\tau}}{\longrightarrow}\, \mathcal{L}$
and the contact structure $\xi
\stackrel{\pi_\xi}{\longrightarrow}\, \mathcal{M}$. The
submanifold $\mathcal{L}$  is named \textbf{isotropic} if and only if
\begin{equation}\label{isotropicamente}
    \forall p \in \mathcal{L} \quad  : \quad \pi_{\tau}^{-1}(p) \,
    \subset \, \pi_{\xi}^{-1}(p)
\end{equation}
Equivalently, if the contact structure  $\xi$  is defined by contact 1-form
$\alpha$, the submanifold
$\mathcal{L}$  is \textbf{isotropic} if
any vector field $\mathbf{X}$ that is tangent to $\mathcal{L}$,  is contained
in $\text{ker}\,\alpha$:
\begin{equation}\label{cundoiso}
\mathbf{X}\in \Gamma\left[\mathcal{TL},\mathcal{L}\right] \quad
\Rightarrow \quad \alpha(\mathbf{X}) \, = \, 0
\end{equation}
\end{definizione}
Let us introduce the further:
\begin{definizione}
Let $\left(\mathcal{M}_{2n+1}\, , \, \xi\right)$ be a contact manifold
and $\tilde{\mathcal{M}}_{2m+1} \subset \mathcal{M}_{2n+1}$
be an odd-dimensional submanifold with codimension
$2(n-m)\geq 0$. Let $\alpha$ be the contact 1-form that defines the contact structure
$\xi$ and let  $\iota$  be the  inclusion map:
\begin{equation}\label{includendoiota}
    \iota \quad : \quad \tilde{\mathcal{M}}_{2m+1} \,
    \longrightarrow \, \mathcal{M}_{2n+1}
\end{equation}
In this case  $\left(\tilde{\mathcal{M}}_{2m+1},\chi\right)$  is
named \textbf{a contact sbmanifold} of
$\left(\mathcal{M}_{2n+1}\, , \, \xi\right)$  if the contact structure
 $\chi$ on $\tilde{\mathcal{M}}_{2m+1}$  is defined
by the   \textbf{contact form} $\iota^\star \, \alpha$, namely if
\begin{equation}\label{cariota}
    \chi \, = \, {\mathrm{ker}} \, \iota^\star\,\alpha
\end{equation}
\end{definizione}
A result of the highest relevance both for applications in Fluid Dynamics and for the Geometrization of Thermodynamics
is the following theorem due to  Arnol'd:
\begin{teorema}
\label{barbagianni} Let $\left(\mathcal{M}_{2n+1},\xi\right)$ be
a contact manifold  in $2n+1$-dimensions and
$\mathcal{L}\subset\mathcal{M}_{2n+1}$ an isotropic submanifold.
Then $\mathrm{dim}\, \mathcal{L} \leq n$.
\end{teorema}
In order to prove the theorem \ref{barbagianni} the following Lemma is needed
\begin{lemma}
Let $\left(\mathcal{M}_{2n+1},\xi\right)$  be a contact manifold whose
contact sctruture$\xi$  is defined as
$\mathrm{ker} \,\alpha$, in terms of a contact 1-form
$\alpha$. As a consequence of the condition
$\ref{latisanaditiglio}$ included in the definition, it follows that $\mathrm{d}\alpha\mid_\xi \neq 0$
and for each point $p\in \mathcal{M}_{2n+1}$  the
$2n$-dimensional fibre $\xi_p \subset T_p \,\mathcal{M}_{2n+1}$ is a vector
space endowed with an antisymmetric  2-form of maximal rank
massimale  (namely without vanishing eigenvalues) which is  exactly provided by the restriction
to $\xi_p$  of $\mathrm{d}\alpha$
\textit{i.e.} $\mathrm{d}\alpha\mid_{\xi_p}$. Hence the contact structure is
a \textit{symplectic bundle} with respect to the
2-form $\mathrm{d}\alpha\mid_\xi$.
\end{lemma}
The proof of the lemma is almost evident from its own formulation.
\begin{proof}
{\rm In order to prove the theorem we consider the inclusion map:
\begin{equation}\label{legendro}
    \iota \quad : \quad \mathcal{L} \, \longrightarrow \, \mathcal{M}_{2n+1}
\end{equation}
and also the\textit{pull-back} of the  contact 1-form on the isotropic submanifold
 By definition
$\iota^\star\alpha = 0$. Hence we have $\iota^\star
\,\mathrm{d}\alpha = 0$. In any point $p\in \mathcal{L}$, the tangent space
 $\mathcal{T}_p \mathcal{L}$ is a subspace of the symplectic space
  $\xi_p$ on which the symplectic  2-form  vanishes $\mathrm{d}\alpha\mid_{\xi_p}$. Using elementary linear algebra
we conclude that such a subspace has maximal dimension one-half of the dimension of  $\xi_p$. Indeed it suffices to put, by means of change of basis the antisymmetric 2-form in canonical form:
\begin{equation}
    \left(\begin{array}{c|c}
    \mathbf{0}_{n\times n}& \mathbf{1}_{n\times n}\\
    \hline
    -\mathbf{1}_{n\times n} &\mathbf{0}_{n\times n} \\
    \end{array}\right)\nonumber\\
\end{equation}
and the statement becomes evident. This proves the theorem
 $\blacksquare$.}
\end{proof}
\par
What are the consequences of this theorem? It states that
if we have a contact structure $\xi$, induced by a contact 1-form
 $\alpha$, then we can exclude  a foliation of the contact manifold
into hypersufaces $\Sigma_h \subset
\mathcal{M}_{2n+1}$  codimension one:
\begin{equation}\label{sfoliazione}
  \mathcal{M}_{2n+1}\, \backsimeq \, \Sigma_h \times \mathbb{R}_h
\end{equation}
such that for each $h\in \mathbb{R}$ the tangent bundle of
$\Sigma_h$ be contained in the contact structure. Indeed if this were true se
any leave  $\Sigma_h$ of the foliation would be an
 isotropica submanifold of dimension $2\times n$ which is exactly what is ruled out by the theorem.
\begin{definizione}
\label{leggiadro} An isotropic  submanifold $\mathcal{L}
\subset \mathcal{M}_{2n+1}$ of a $(2n+1)$-dimensional  contact manifold that has the maximal allowed dimension
 $n$ is named  a
\textbf{Legendrian submanifold}.
\end{definizione}
\paragraph{\sc An example relevant for Fluid Dynamics} In Fluid Dynamics
the relevant manifold $\mathcal{M}_3$,
 is locally isomorphic to our three dimensional euclidian space $\mathbb{E}^3 \cong \mathbb{R}^3$  since fluids of interest flow in a portion
$\mathcal{M}_3 \subset \mathbb{R}^3$ of such a space
that can be compact
(the fluid is  confined in some container) or partially non
compact (river, lake, ocean, atmosphere). In any case we have $n=1$.
Hence if the flow, which is a vector field
$\mathbf{U}\in \Gamma\left[\mathcal{TM}_3,\mathcal{M}_3\right]$
induces a contact structure on  $\mathcal{M}_3$ (in the next subsection
we will see that this happens when  $\mathbf{U}$
is proportional to Reeb vector  of a contact form
$\alpha$), then the unique  Legendrian submanifolds are one-dimensional
so that the ambient space does not admit a foliation into 2-dimensional submanifolds
that contain the flows or that are transverse to them. This is the geometrical foundation
 of turbulence: the flows are chaotic.
\paragraph{\sc An example relevant for Thermodynamics}
As we are going to see later on the relevant contact manifold for thermodynamics is not the euclidian
physical space rather the space of thermodynamical variables (both extensive and intensive)
that are in number of: $2n+1\, \equiv \, 2m+3$
($n=m+1$).
The simplest minimal case corresponds to $m=1$
and the $5$ thermodynamic variables are
$\left\{U,S,V,T,P\right\}$, namely \textit{ internal energy, entropy,
volume, temperature and pressure}. As we will explain the thermodynamical space
 $\mathcal{M}_{2m+3}$ \`{e} is always endowed with a canonical contact $1$-form that summarizes
the principles of classical thermodynamics
and available equilibrium thermodynamical states
correspond to all possible Legendrian submanifolds
$\mathcal{L}_{E} \subset \mathcal{M}_{2m+3}$ that have, as a consequence of theorem
\ref{barbagianni} and of definition  \ref{leggiadro}
the following dimension:
\begin{equation}\label{dimlegendro}
    \mathrm{dim}\mathcal{L}_{E}\, = \, m+1
\end{equation}
In the simplest case $m+1 \, = \,2$  so that the Legendrian subvarieties are portions of an $\mathbb{R}^2$-plane.
It is in such planes that one studies the phase diagrams of chemicals. In the case of multiphase, multicomponent mixtures we have  $m>1$ and phase diagrams develops in spaces of higher dimensions.
\paragraph{\sc Big Data spaces}
It is to be investigated whether certain Big Data spaces might be endowed with a hidden contact structure responsible for chaotic motions and phase transitions.
\subsection{The Reeb Vector}
Let us now come to the definition of  the Reeb vector which shifts the definition of a contact structure from the cotangent to the tangent bundle of the considered  $\mathcal{M}_{2n+1}$  manifold.
\begin{definizione}\label{ribatriestina} Associated wih a contact form $\alpha$
we always have the so named  \textbf{Reeb vector field}
$\mathbf{R}_\alpha$, defined  by two conditions:
\begin{eqnarray}\label{Ribbo}
&&\alpha\left(\mathbf{R}_\alpha\right) \, = \, \lambda(\mathbf{x})
\quad =
 \quad
 \text{nowhere vanishing  function on $\mathcal{M}_{2n+1}$}\nonumber\\
 &&\forall \mathbf{X}\in \Gamma\left[\mathcal{TM}_{2n+1},\mathcal{M}_{2n+1}\right] \quad :
 \quad
 \mathrm{d}\alpha\left(\mathbf{R}_\alpha,\mathbf{X}\right)\, = \,
 0
\end{eqnarray}
\end{definizione}
If the contact manifold $\mathcal{M}_{2n+1}$  is endowed with a
Riemanniana metric $g$ (as it is the case of the
euclidian space$\mathbb{R}^{2n+1}$), then the contact form  $\alpha$ and its Reeb vector field
$\mathbf{R}_\alpha$  are related to each other  by the operation of raising and lowering of world indices.
 Suppose we start from the  Reeb vector field:
\begin{equation}\label{ribbone}
    \mathbf{R} \, = \, R^\mu \,
    \frac{\partial}{\partial x^\mu}
\end{equation}
The corresponding $\alpha$ is retrieved by setting:
\begin{equation}\label{contactribbo}
    \alpha \, = \, \Omega^{[\mathbf{R}]} \,  \equiv \, g_{\mu\nu} R^\mu \,
    dx^\nu
\end{equation}
and the condition (\ref{latisanaditiglio}) that such a 1-form should be a contact form
 becomes the following equation on the Reeb field components:
\begin{equation}\label{gridolino}
    \epsilon^{\lambda \mu_1\nu_1\mu_2\nu_2\dots\mu_n\nu_m} \,
    R_\lambda \, \partial_{\mu_1}\,R_{\nu_1}
    \,\partial_{\mu_2}\,R_{\nu_2}\, \dots\,
    \partial_{\mu_n}\,R_{\nu_n}\, \neq \, 0 \quad \text{nowhere vanishes}
\end{equation}
In the opposite direction, if one start from the contact form $\alpha$,
le componens of the  Reeb vector field are obtained by setting:
\begin{equation}\label{crinolina}
    R_\alpha^\mu \, = \, g^{\mu\nu} \, \alpha_\nu
\end{equation}
One should note that the nowhere vanishing function $\lambda$ mentioned
in the definition \ref{ribatriestina} is simply the squared norm of the Reeb vector field or of the contact form  that make sense only in a Riemannian space and there they coincide:
\begin{equation}\label{normaquadrata}
    \lambda \, =\, \parallel \mathbf{R}\parallel^2 \, = \, \parallel
    \Omega^{[\mathbf{R}]}\parallel^2 \, \equiv \, g_{\mu\nu} \, R^\mu \,
    R^\nu
\end{equation}
\paragraph{\sc Contact structures for $n=1$ and  hydrodynamical flows}
Let us now consider the case relevant to the hydrodynamics of three-dimensional contact manifolds
$\left(\mathcal{M}_3\,\xi_\alpha\right)$, where, by the notation
$\xi_\alpha$, we refer to the contact form $\alpha$
that defines the contact structure $\xi$. The consequence of theorem\ref{barbagianni},
as we have already pointed out, is that in these contact varieties, the Legendrian subvarieties are all one-dimensional, that is, they are \textit{curves} or, as it is usual to refer to them in this context \textit{knots}.
\par
Therefore in three dimensions, there are two types of knots the \textbf{Legendrian knots}, whose tangent vector belongs to
$\text{ker}\, \alpha$  and the \textbf{transverse knots} whose tangent vector is parallel to the Reeb field vector at every point in their trajectory.
\par
\begin{figure}
\begin{center}
\includegraphics[height=150mm]{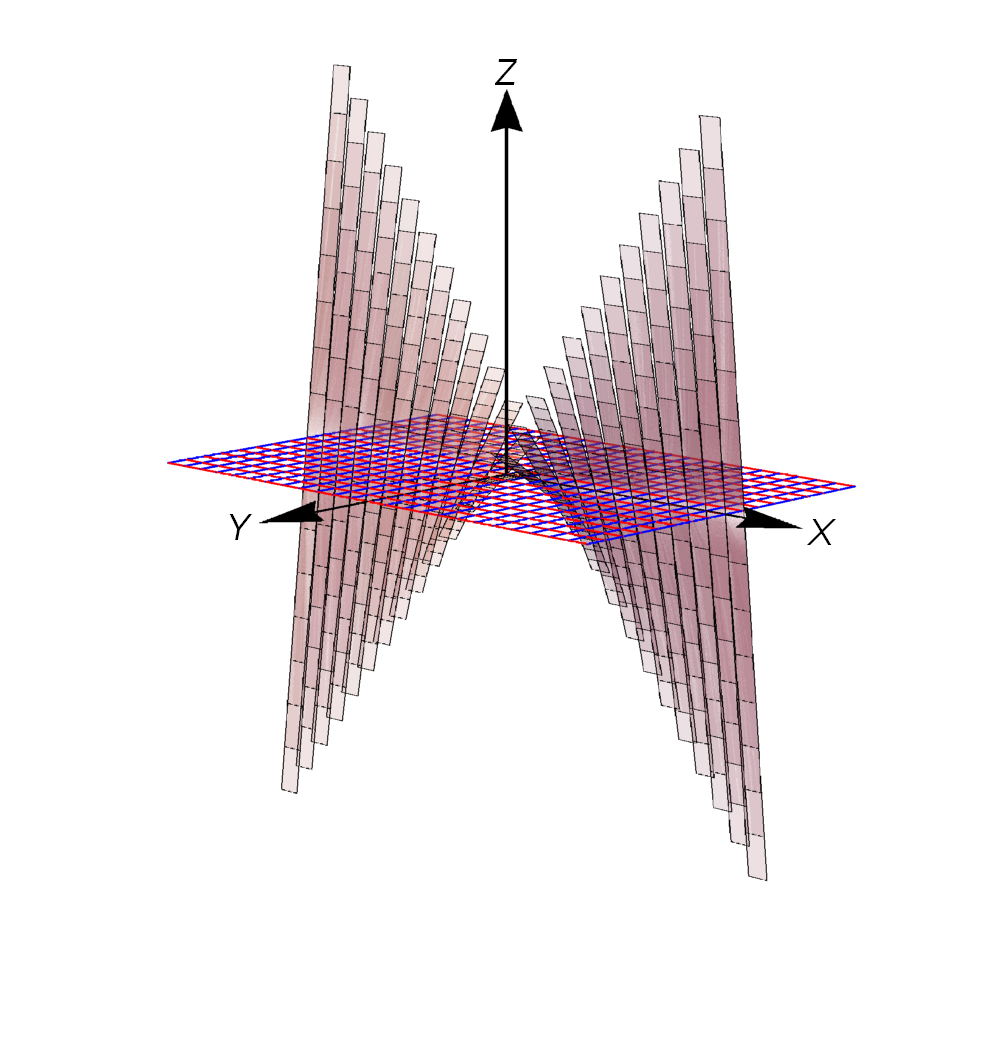}
\end{center}
\caption{\it Schematic vision of the standard contact structure
in $\mathbb{R}^3$.} \label{berlucchino}
 \hskip 1cm \unitlength=1.1mm
\end{figure}

Furthermore in D=3 condition(\ref{gridolino}) becomes:
\begin{equation}\label{urlonotturno}
    \epsilon^{\lambda\mu\nu}  \,R_\lambda \, \partial_\mu \, R_\nu
    \, \neq \, 0
\end{equation}
\paragraph{\sc The standard contact structure on $\mathbb{R}^3$.}
Three-dimensional flat space, whose coordinates we denote
$x,y,z$ is equipped with a standard contact structure that admits the following
contact $1$-form
\begin{equation}\label{romildino}
    \alpha_s \, = \, \mathrm{d}z+ x \,\mathrm{d}y
\end{equation}
A picture of the local planes defining the contact structure (\ref{romildino}) is shown in fig.\ref{berlucchino}.
In hydrodynamics, a vector field $\mathbf{U}$ that we identify with the velocity field of a fiuid is potentially interesting for the study of chaotic regimes if it is the Reeb field of a contact $1$-form $\alpha$.
If our work arena is a Riemanianna variety endowed with a metric $g$ cio\`{e} $(\mathcal{M}_3,g)$  we can always invert the procedure
la procedura and define the contact form $\alpha$  by writing:
\begin{equation}\label{gominato}
    \alpha \, = \, \Omega^{[\mathbf{U}]}\
  \, \equiv \, g_{ij} \, U^i(x,t) \, dx^j
\end{equation}
where $U^i(x,t)$ are the three components of the velocity field at time $t$ and the Latin indices $i,j=1,2,3$ identify the three
coordinates $x^i \, = \, \{x,y,z\}$. In this way  the first of the two
conditions (\ref{Ribbo}) is automatically satisfied:
${\rm i}_\mathbf{U}\Omega^{[\mathbf{U}]} \, = \,
\parallel \mathbf{U}\parallel^2 \, > 0$. It remains to be seen if
 $\Omega^{[\mathbf{U}]}$ is a true contact form,
namely if $\Omega^{[\mathbf{U}]} \, \wedge \, \mathrm{d}
\Omega^{[\mathbf{U}]} \, \ne \,0$. It is at this level that one finds  the relation
with  Beltrami equations which makes sense only  in
dimension $D=3$ and hence for $n=1$. In the thermodynamic case  $n
\geq 2$: hence we stop the discussion here and we refer the interested  reader to \cite{mariopietroinpress}.
\subsection{Darboux Theorem and the Case of Thermodynamics}
\label{darbone}
A classical theorem due to Darboux, the proof of which we omit by referring the reader to \cite{contatoregeiges} where it is presented, highlights the important fact that the standard contact structure of $\mathbb{R}^3$ mostrata nella
eq.(\ref{romildino}) and illustrated graphically in
fig.\ref{berlucchino} is not an arbitrary choice, bensl corresponds to the canonical local form
of any contact structure on any contact manifold.
\begin{teorema}
\label{darbuso} Let  $\left(\mathcal{M}_{2n+1},\xi\right)$  be
$(2n+1)$-dimensional contact manifold  and let $\alpha$ be
 a contact $1$-forma that defines $\xi = \text{ker}\,\alpha$. Let
$p\in \mathcal{M}_{2n+1}$ be any point of the manifold and
$U \subset \mathcal{M}_{2n+1}$ an open neighborhood of  $p$. Then we
can always find a local homomorphism : $\varphi \, : \, U
\, \to \, \mathbb{R}^{2n+1}$ such that, naming
$\{x_0,x_i,y_i\}$,$(i=1,\dots,n)$ the coordinates of $\varphi(U)
\subset \mathbb{R}^{2n+1}$ we get:
\begin{equation}\label{localU}
    \alpha\mid_{U} \, = \, dx_0 \, + \,\sum_{i=1}^n \,y^i\,dx_i
\end{equation}
\end{teorema}
In the case $n=1$ eq.(\ref{localU}) reproduces eq.(\ref{romildino}).
\par
Darboux's canonical form is the most inspiring one to suggest a connection between contact geometry and thermodynamics. As we shall see in more detail later, in classical thermodynamics it is natural to distinguish two types of variables,
the extensive ones such as $U,S,V,N$ \textit{ ( internal energy,
entropy, volume, number of particles or of moles)} and the intensive ones such as $T,P,\mu$ \textit{(temperature, pressure, potenziale chemical potential)}.
We can identify the coordinates $x_i=\{S,V,N,\dots\}$ with the extensive
variables and the  coordinates $y_i=\{T,P,\mu\,\dots\}$ with the intensive ones,
further  idenfying the priviledged extensive
variable $x_0=U$ with internal energy. For a reason that will become clear later, after the discussion of probability measures, it is convenient to generally rename the intensive variables as $\lambda_i$ since in the general view of information/probability theory they play the role of Lagrange multipliers. Thus the contact differential $1$-form
\begin{equation}\label{formalpt}
    \alpha_{termo} \, \equiv \, dx_0 \, + \, \sum_{\mu=1}^{3} \lambda^i \,
    dx_i
\end{equation}
can be used to formulate the first and second Principles of Thermodynamics stating that $\alpha_{termo}$ should vanish.
Such vanishing must be understood in a correct mathematical way: it vanishes on  Legendrian submanifolds
$\mathcal{L}_E\subset \mathcal{M}_{2m+3}$ which represent equilibrium thermodynamical states whose embedding functions in the ambient space $\mathcal{M}_{2m+3}$ are the physical laws defining such states.
\subsection{Symplectic and Poisson Manifolds}
\label{pescione}
In the further study of the immersion of legendrian submanifolds in the thermodynamic contact manifold $\mathcal{M}_{2m+3}$
an important role is played by the symbiotic relationship of all contact manifolds $\mathcal{M}_{2n+1}$ with the adiacent
symplectic manifolds that are also Poissonian.
\par
Hence let us begin with the following:
\begin{definizione}
\label{simp2def}
A symplectic manifold  is a pair
$\left(\mathcal{SM}_{2n+2},\omega\right)$ of a smooth manifold
$\mathcal{SM}_{2n+2}$ in dimension $2n+2$ and one differential $2$-form
$\omega$ that is closed, non degenerate (namely admitting no vanishing eigenvalue)
and of maximal  rank:
\begin{equation}\label{simplicioformo}
    \mathrm{d}\omega \, = \, 0 \quad ; \quad
    \omega\wedge\omega\wedge \dots \wedge \omega \, \neq 0 \quad
    \text{everywhere on $\mathcal{SM}_{2n+2}$}
\end{equation}
\end{definizione}
On a symplectic manifold we have a naturally defined antisymmetric 2-form on the space of sections of
the tangent bundle, \textit{i.e.} on vector fields:
\begin{eqnarray}\label{tardicchio}
    \omega &: & \Gamma\left[\mathcal{TSM}_{2n+2}\, , \,
    \mathcal{SM}_{2n+2}\right] \times \Gamma\left[\mathcal{TSM}_{2n+2}\, , \,
    \mathcal{SM}_{2n+2}\right] \, \longrightarrow \,
    C^{(\infty)}\left(\mathcal{SM}_{2n+2}\right)\nonumber\\
   \forall X,Y\, &\in & \Gamma\left[\mathcal{TSM}_{2n+2}\, , \,
    \mathcal{SM}_{2n+2}\right] \quad , \quad  \omega\left(X,Y\right) \,
    \in\,
    C^{(\infty)}\left(\mathcal{SM}_{2n+2}\right)
\end{eqnarray}
Poisson manifolds are instead defined as it follows:
\begin{definizione}
\label{poissone} A Poisson manifold
$\left(\mathcal{PM}_\ell,\{,\}\right)$  is a pair of a smooth
manifold $\mathcal{PM}_\ell$ of dimension $\ell$ and a Poisson bracket
 $\{,\}$  which  is a binary operation
on the space of smooth functions defined over the manifold:
\begin{equation}\label{pesceparente}
    \{,\} \quad : \quad C^{(\infty)} \left(\mathcal{PM}_\ell\right)
    \, \times \, C^{(\infty)} \left(\mathcal{PM}_\ell\right)\,
    \longrightarrow \, C^{(\infty)} \left(\mathcal{PM}_\ell\right)
\end{equation}
endowed with the following three properties:
\begin{description}
  \item[1)] Antisymmetry $\quad \{f\, ,\, g\} \, = \, - \, \{g\, ,\, f\}
  $, $\quad\forall f,g \in C^{(\infty)} \left(\mathcal{PM}_\ell\right)$
  \item[2)] Jacobi Identity $\quad \{f\, , \, \{g \, ,\, h\}\}+\{g\, , \, \{h \, ,\, f\}\}+\{h\, , \, \{f \, ,\, g\}\}\, = \,
  0$, $\quad \forall f,g,h \in C^{(\infty)} \left(\mathcal{PM}_\ell\right)$
  \item[3)] Leibniz rule $\quad \{f\, , \, g.h\} \, = \, \{f\, , \,
  g\} \, h \, + \, g \,\{f\, , \,
  h\}$,  $\quad \forall f,g,h \in C^{(\infty)} \left(\mathcal{PM}_\ell\right)$
\end{description}
\end{definizione}
The first two properties mentioned in Definition
\ref{poissone} guarantee that the space of functions on a Poisson variety becomes a Lie algebra when it is equipped with a Poisson bracket. On the other hand, the third property implies that every function $f \in
C^{(\infty)}$ is associated by the Poisson bracket with a derivation of the commutative algebra of functions on the manifold, which is, by definition, a vector field
$\mathbf{X}_f$; this latter is named  \textbf{the hamiltonian vector field} of the function $f$.
\par
Locally, in each open chart $\{x_1,\dots,x_j\}$, the Poisson bracket takes the following form::
\begin{equation}\label{localedelpesce}
    \{f,g\} \, = \,\pi^{ij}(x)\,\frac{\partial f}{\partial x^i}\, \frac{\partial g}{\partial
    x^j} \quad ; \quad \pi^{i,j}(x) \, = \, - \, \pi^{ji}(x)
\end{equation}
where the antisymmetric countervariant vector $\pi^{ij}(x)$ is usually called a \textbf{bivector}.
Thus the Hamiltonian vector field $\mathbf{X}_f$ is easily identified as:
\begin{equation}\label{carneadehamilt}
    \mathbf{X}_f \, = \, \pi^{ij}\partial_i f \, \partial_j
\end{equation}
Suppose now that the dimension of the Poisson variety is $\ell=2n+2$ and that the bivector $\pi^{ji}(x)$ is an everywhere invertible matrix. Posing $ \omega \, = \, \pi^{-1}_{k\ell} dx^k
\wedge dx^\ell $ we obtain a symplectic $2$-form of maximal rank that is closed as a consequence of the Jacobi identities satisfied by the bivector. In this way we see that such a Poisson manifold is a symplectic manifold and in particular we can write:
\begin{equation}\label{agniziono}
    \{f,g\} \, = \, \omega\left(\mathbf{X}_f,\mathbf{X}_g\right)
\end{equation}
\begin{definizione}
\label{liuvillocampillo} Let
$\left(\mathcal{SM}_{2n+2},\omega\right)$ be a symplectic manifold. A   \textbf{ Liouville vector field} $X$ is
a vector field for which the following condition holds:
\begin{equation}\label{taleggio}
   {\mathcal{L}_X\omega \,=\,\omega}
\end{equation}
where $\mathcal{L}_X\omega\, \equiv \, i_X\,
  \mathrm{d}\omega \, + \, \mathrm{d}\,(i_X\omega)\,$  denotes the Lie  derivative of the 2-form $\omega$ along the  vector field $X$.
\end{definizione}
Note that, being $\omega$ closed, if $X$ is a Liouvile vector field, we have: $\mathrm{d}\,(i_X\omega)=\mathcal{L}_X\omega\, =\,\omega$.
\subsection{The Relation Between Contact Manifolds and Symplectic Manifolds}
\label{liasone} Let us consider a symplectic manifold $\left(
\mathcal{SM}_{2n+2},\omega\right)$  and let us assume that it admits at least one Liouville vector field
 $\mathbf{L}$. Furthermore let
$\Sigma_{\mathbf{L}}\subset \mathcal{SM}_{2n+2} $ be the hypersurface which is transverse
to the  Liouville field $\mathbf{L}$. Then we realize that
 $\Sigma_{\mathbf{L}}$ is a contact manifold
with contact $1$-form  $\alpha \, =
\,i_{\mathbf{L}}\omega$. Since $\Sigma_{\mathbf{L}}$ is
transverse to $\mathbf{L}$, the $1$-form $\alpha$ vanishes along
$\mathbf{L}$ and, on the contrary,  it never vanishes on the tangent bundle
$T\Sigma_{\mathbf{L}}$. In order to verify that  $\alpha$ is a \textit{bona fide}
contact form we just have to perform the following calculation:
\begin{eqnarray}\label{gelso}
 \hspace{-72pt}   \alpha\wedge \underbrace{\mathrm{d}\alpha \wedge \dots \wedge
    \mathrm{d}\alpha}_{n-\text{times}} & = & i_\mathbf{L}\omega\wedge\underbrace{\mathrm{d} i_\mathbf{L}\omega \wedge \dots \wedge
    \mathrm{d} i_\mathbf{L}\omega}_{n-\text{times}}\,=\,i_\mathbf{L}\omega\wedge\underbrace{\omega \wedge \dots \wedge
    \omega}_{n-\text{times}}\, = \, \frac{1}{n+1} i_\mathbf{L}\left(\underbrace{\omega \wedge \dots \wedge
    \omega}_{(n+1)-\text{times}}
    \right)\nonumber\\
    & = & \text{Vol}_{\Sigma_\mathbf{L}}\nonumber\\
\end{eqnarray}
The last equality holds because  $\omega^{n+1}$ is the volume form of the ambient  symplectic manifold
forma di volume  della variet\`{a} simplettica ambiente la
and the hypersurface  $\Sigma_\mathbf{L}$ is, by hypothesis, transverse to the Liouville vector field.
\par
Reversely, given a contact manifold
$\left(\mathcal{M}_{2n+1},\xi\right)$  with contact $1$-form $\alpha$
and Reeb vector field $\mathbf{R}$, any hypersurface $\Sigma
\subset \mathcal{M}_{2n+1}$ that is  transverse to the Reeb vector,
automatically acquires the structure of a symplectic manifold
with symplectic  $2$-form$\tilde{\omega}\, = \,
\mathrm{d}\alpha\mid_\Sigma$.
\par
Therefore, we can have odd-dimensional contact manifolds lying in between two adjacent even-dimensional symplectic manifolds as it is shown in the following diagram:
\begin{equation}\label{diagrammus}
\begin{array}{ccccc}
    \left(\mathcal{SM}_{2n},\tilde{\omega}=d\alpha\right)
    &\stackrel{\iota}{\hookrightarrow}
    &\left(\mathcal{M}_{2n+1},\alpha=i_\mathbf{L}\omega\right)
    &\stackrel{\iota}{\hookrightarrow}
    & \left(\mathcal{SM}_{2n+2},\omega\right)\\
    \Downarrow &\null& \Downarrow &\null & \Downarrow\\
    \text{symplectic}&\null& \text{contact}& \null &
    \text{symplectic}\\
\text{transverse to Reeb field}&\null& \text{transverse to Liouville
field}& \null &
    \null\\
    \end{array}
\end{equation}
The pattern described in equation (\ref{diagrammus}) is reminescent
of what happens with sasakian manifolds
that lye in between two K\"ahler manifolds  of adiacent even dimensions.
This is not surprising  since
K\"ahler manifolds are particular instances of  symplectic manifolds where the
symplectic form is simply the   K\"ahler $2$-form.
\par
We stop here, for the moment with the exposition of fundamental geometric concepts, since we have to turn to the other
leg of our constructions, namely, Probability Theory. We shall come  back to geometry in later chapter about diffusion theory
that establishes a solid link between Markov random processes and Riemannian Geometry.
\section{Fundaments of  Probability Theory}
\label{fondiprobi}
In this appendix we recall the fundamental concepts, which will be indispensable to us, with regard to measurement and probability theory.
\subsection{$\sigma$-Algebras and  Probability Measures}
\begin{definizione}
\label{sigmalgeb} Given a set $\Omega$ one defines
$\sigma$-algebra on $\Omega$ a family $\mathcal{A}$ of
subsets $A_i \subset \Omega$ such that:
\begin{enumerate}
  \item $\emptyset \in \mathcal{A}$ and $\Omega \in \mathcal{A}$.
  \item If $A \in \mathcal{A}$ then its complement
  $A^c \equiv \Omega - A$ also belongs to the same family
  $A^c\in \mathcal{A}$.
  \item If the elements of a denumerable family  of sets
  $\left\{ A_i\right\}_{i\in \mathbb{N}}$ belong to
  $\mathcal{A}$ then also their union belongs to it:
  $$ A \, = \, \bigcup_{i=1}^{\infty} A_i \, \in  \mathcal{A}$$
\end{enumerate}
\end{definizione}
\begin{definizione}
\label{spazmisur} The pair $\left(\Omega\, , \, \mathcal{A}\right)$
where $\Omega$ is a set and $\mathcal{A}$ is a $\sigma$-algebra
on $\Omega$ is named a measurable space.
\end{definizione}
In Probability Theory the elements of the
$\sigma$-algebra, $X\in \mathcal{A}$ are named \textbf{events}
while the points  $p\in \Omega$ are named
\textbf{experiments}.
\begin{definizione}
\label{poweralg} One names \textbf{Part Algebra} of a set
 $\Omega$  and denotes it with the symbol $\mathfrak{P}(\Omega)$ the set of all subsets
 equipped with the boolean algebraic operations of  \textit{union,
intersection and   complement}.
\end{definizione}
Let us now consider two sets $\Omega_1$ ed $\Omega_2$; any map:
\begin{equation}\label{mappatrainsiemi}
    \phi \, : \,\Omega_1 \, \rightarrow \,\Omega_2
\end{equation}
induces a \textit{pullback map} $\phi_\star^{-1}$ on the
corresponding Part Algebras:
\begin{equation}\label{pullobacco}
    \phi_\star^{-1} \, : \,\mathfrak{P}(\Omega_2) \,\rightarrow \,\mathfrak{P}(\Omega_1)
\end{equation}
One can verify that the map $\phi_\star^{-1}$ satisfies the following properties
\begin{eqnarray}
\label{morfisema}
  \phi_\star^{-1}\left(X\bigcup Y \right) &=& \phi_\star^{-1}\left(X\right)\bigcup \phi_\star^{-1}\left(Y\right)\nonumber \\
  \phi_\star^{-1}\left(X\bigcap Y \right) &=& \phi_\star^{-1}\left(X\right)\bigcap \phi_\star^{-1}\left(Y\right)\nonumber \\
  \phi_\star^{-1}\left(X^c \right) &=& \phi_\star^{-1}\left(X \right)^c
\end{eqnarray}
Hence $\phi_\star^{-1}$ \`{e} is a morphism of boolean algebras.
\par
Once this general concepts have been established one can introduce the following  notion of
\textit{probability measure} by means of the following:
\begin{definizione}
\label{misaproba} Let $\left(\Omega\, , \, \mathcal{A}\right)$ be a measurable space. A probability measure on $\left(\Omega\,
, \, \mathcal{A}\right)$ is a map
\begin{equation}\label{cuccusprobrpus}
    \mathfrak{p} \, : \, \mathcal{A} \, \rightarrow \, [0,1] \subset \mathbb{R}
\end{equation}
that satisfies the following properties:
\begin{itemize}
  \item $\mathfrak{p}(\emptyset) \, = \, 0$ and $\mathfrak{p}(\Omega)=1$
  \item $\mathfrak{p}\left ( \bigcup_{i} X_i \right)\, = \, \sum_i
  \mathfrak{p}\left(X_i\right)$  for all denumerable unions of disjoint parts, i.e. such that
  $X_i\bigcap X_j \, =\, \emptyset$ se
  $i \neq j$.
\end{itemize}
\end{definizione}
When we have a triplet
$\left(\Omega,\mathcal{A},\mathfrak{p}\right)$ we say that we have
a stochastic  space and the value $\mathfrak{p}(X)$ is the probability of the event $X$.
\subsection{Stochastic Functions, Stochastic Vectors  and  Distributions}
Let us begin with:
\begin{definizione} If $\mathcal{T}$ is a separable topological space,  one names \textbf{Borel Algebra} of $\mathcal{T}$, denoted
$\mathcal{B}(\mathcal{T})$  the $\sigma$-algebra made by all
denumerable unions,  intersections and complements of open subsets
$U\subset\mathcal{T}$.
\end{definizione}
In particular, on all varieties $\mathbb{R}^n$ we have the ball-topology
which, for the real line $\mathbb{R}$, reduces to the
topology of open intervals $]x,y[ \subset \mathbb{R}$, and the correspondent
Borel algebra is clearly defined.
Hence, using as $\sigma$-algebra the natural
Borel algebra $\mathcal{B}(\mathbb{R})$,  we have that the pair$\left(
\mathbb{R}\, ,\,\mathcal{B}(\mathbb{R})\right)$, makes a measurable space
\par
Let us then consider a stochastic space
$\left(\Omega,\mathcal{A},\mathfrak{p}\right)$ and a map:
\begin{equation}\label{coronafine}
    \psi \, : \, \Omega \, \rightarrow \, \mathbb{R}
\end{equation}
which to any point   $p \in \Omega$  of the set $\Omega$ associates a real number (its coordinate).
Because of what we discussed above, the pullback map
\begin{equation}\label{polloalladiavola}
    \psi^{-1}_\star \, : \, \mathcal{B}(\mathbb{R}) \, \rightarrow
    \, \mathcal{A}
\end{equation}
is a morphism of  boolean algebras that  explicitly associates an
element $X\in \mathcal{A}$ to every element of the Borel
algebra of the real line. Thus, composing the maps we define:
\begin{equation}\label{cornutella}
    \mathfrak{p}_{\psi} \, \equiv \, \mathfrak{p}\circ\psi^{-1}_\star
\end{equation}
which is a map from the  Borel algebra of the real line to the interval
$[0,1]$:
\begin{equation}\label{cromatogno}
    \mathfrak{p}_\psi \, : \, \mathcal{B}\left(\mathbb{R}\right) \, \rightarrow \,
    [0,1]
\end{equation}
This is what we name a \textbf{stochastic function}. In practice  to every open interval $]x,y[$
the stochastic function associates a number between $0$ ed $1$  which is the probability
that while doing a measuring experiment the measured value happens to be in the considered interval.
In this way one can consider stochastic functions that are discontinuous, step-wise and the like, yet they are Lebesgue integrable
thanks to the measurability of the support space.
\paragraph{\sc Probability Density}
An interesting case is when the stochastic function can be described in terms of a probability density given by an integrable function $\rho_\psi(q)$ on the real line such that:
\begin{equation}\label{coriaceo}
    \mathfrak{p}_\psi\left(\left[a,b\right]\right) \, =
    \,\int_{a}^{b} \rho_\psi(q) \,dq
\end{equation}
For the probability density to be well defined, it is necessary that the probability density
$\rho_\psi(q)$ be properly normalized:
\begin{equation}\label{ranierus}
    \int_{-\infty}^{+\infty} \rho_\psi(q) \,dq \, = \, 1
\end{equation}
Under these conditions, one can define the average value of any function $f(q)$ of the  stochastic variable $q \in
\mathbb{R}$ writing
\begin{equation}\label{romildo}
    \langle f\rangle \, \equiv \,  \int_{-\infty}^{+\infty}\, f(q)\, \rho_\psi(q) \,dq \,
\end{equation}
\paragraph{\sc Stochastic Vector}
In a similar way we can define stochastic vectors.
\par
Consider a finite dimensional vector space $\mathbb{V}$:
\begin{equation}\label{lollobrigga}
    \mathrm{dim}_\mathbb{R} \,\mathbb{V}\, = \, r < \infty
\end{equation}
and a basis $\mathbf{e}_i$ ($i=1\dots,r$) of vectors such that
\begin{equation}\label{contellus}
    \forall \mathbf{X} \in \mathbb{V} \, : \, \mathbf{X} \, =\,
    \sum_{i=1}^r X^i(\chi)\mathbf{e}_i
\end{equation}
Where the components of the vector are thought of as functions of $\chi
\in \Omega$,  the space of events over which we defined the probability measure.  By reasoning entirely analogous to that above, each component $X^i(\chi)$ can be thought of as a probability density $X^i_\psi\left(
\mathbf{q}\right)$ on a space $\mathbb{R}^n$ where $n$ is the number of
coordinates necessary to identify a point
$\chi\in\Omega$, namely the dimension of the set $\Omega$, if this latter can be thought of as a differentiable manifold.
As in the previous case what we are constructing is, for each compenent $w^i$ of the stochastic vector, a map:
\begin{equation}\label{coronafine}
    \psi^i \, : \, \Omega \, \rightarrow \, \mathbb{R}^n
\end{equation}
which, by pullback, induces a map:
\begin{equation}\label{polloalladiavola}
    \psi^{-1|i}_{\star} \, : \, \mathcal{B}(\mathbb{R}^n) \, \rightarrow
    \, \mathcal{A}
\end{equation}
By composition of maps we obtain
\begin{equation}\label{scortabella}
    \mathfrak{p}_{\psi^i} \, \equiv \, \mathfrak{p}\circ\psi^{-1|i}_\star
\end{equation}
which is a map from the Borel algebra  of
$\mathbb{R}^n$ to the interval $[0,1]$:
\begin{equation}\label{cromatogno}
    \mathfrak{p}_{\psi^i} \, : \, \mathcal{B}\left(\mathbb{R}^n\right) \, \rightarrow \,
    [0,1]
\end{equation}
In this way we have defined  aa stochastic vector, namely a map:
\begin{equation}\label{coronaspessa}
    \pmb{\Psi} \, : \, \Omega \, \rightarrow \, \mathbb{V}
\end{equation}
Also for  stochastic vectors the most favorable and smooth situation occurs when
each of the vector components
is substituted by an integrable probability density:
\begin{equation}\label{cherasco}
    \mathbf{X}(\mathbf{q}) \, = \, \sum_{i=1}^r \,
    \rho_\Psi^i(\mathbf{q})\,  \mathbf{e}_i \quad ; \quad
    \int\int\dots\int_{\mathbb{R}^n}\, \rho_\Psi^i(\mathbf{q}) \,
    \underbrace{d^n\mathbf{q} }_{\equiv d\mu(\mathbf{q})}\, = \, 1
\end{equation}
where with  $\mathrm{d}\mu(\mathbf{q})$ we have denoted  the integration
measure on the space $\Omega$ which might be more elaborate and contain a factor
 $\sqrt{\text{det}g}$ when $\Omega$ is a Riemannian manifold endowed with a metric.
\section{A Summary of Classical Thermodynamics and Statistical Mechanics}
\label{richiamatermo} In this appendix in order to establish the notations and systematically organize the  geometrical treatments of both classical thermodynamics and statistical mechanics, we introduce in a very synthetic way the basic concepts of both disciplines
\cite{kersono,ulenbeccus} making when necessary reference to the general conceptual frameworks discussed in the previous appendix and in the main text. Indeed the main motivation of the present summary is to show how Shannon Information entropy and classical thermodynamical entropy do indeed coincide emphasizing that a thermodynamical, geometrical view is closely inherent to any Big Data system.
\subsection{Thermodynamical Potentials and State Functions}
\label{pottherm} In macroscopic thermodynamics one utilizes the  following
\textbf{extensive quantities}:
\begin{enumerate}
  \item $U$ = \textit{internal energy} of a thermodynamical system
  \item $S$ = \textit{entropy}
  \item $V$ = \textit{volume} occupied by the system under consideration, for instance a mixture of gases or a certain quantity of a liquid or of a solid.
  \item $N$ = \textit{the total number of particles} composing the system,
   ifor instance the number of molecules of a gas that can be measured in various units,
among which the most frequently utilized  is the \textit{number of moles} of the chemical compound under
investigation. Alternatively when the system is a mixture of more than one component
one utilizes:
\item  $N_i$ = \textit{the total number of particles of the
$i$-th component} of the mixture, typically measured in \textit{number or fractions
of moles.}
\end{enumerate}
Extensive quantities means that partitioning the  system into subsystems
the considered quantity is subdivided into percentual fractions. In other words, for instance the entropy of a system
composed  of two subsystems $A$ and $B$ is the entropy of
$A$ plus the  entropy of $B$:
\begin{equation}\label{ciucca}
    S_{A \cup B} \, = \, S_A + S_B
\end{equation}
Similarly can be said of the internal energy $U$, of the volume $V$
and of the numbers of particles $N_i$ or fractions of moles.
\par
The  \textbf{intensive quantities} of classical thermodynamics do not have the same property
and they are instead characterized by the fact that in the equilibrium states
they \textbf{assume the same value in every part of the system}, large or small. They are
\begin{enumerate}
  \item $T$ = \textit{ temperature} that determines the average energy per particle.
  \item $P$ = \textit{pressure} which, as in mechanics, is the force per unit area..
  \item $\mu$ = \textit{chemical potential} which is the intensive variable canonically conjugate to the number of particles $N$.
  \item $\mu_i$ = \textit{chemical potentials of the various components in the various phases}
 as happens in multicomponent and multiphase mixtures.
\end{enumerate}
\subsection{Thermodynamical Constants}
Before we begin our summary of classical thermodynamics and statistical mechanics, it is appropriate to recall the definition and numerical value of universal constants, both fundamental and empirical.
\begin{definizione}
\label{mole} One \textbf{mole} of a chemical substance $\mathcal{X}$ is
the quantity of  atoms or of molecules of that substance $\mathcal{X}$
necessary to form  a mass $\mathit{M}$  numerically equal in grams
to the weight  $w_\mathcal{X}$  of an atom or a  molecule of that substance
$\mathcal{X}$ expressed in atomic mass units.
\end{definizione}
Thanks to such a  definition one mole of any chemical substance
$\mathcal{X}$ always contains the same number of atoms or of
molecules that is the Avogadro number:
\begin{equation}\label{numeroagrado}
    N_A \, = \, 6,002214076 \times 10^{23}
\end{equation}
In view of the definition \ref{mole}  Avogadro number can be seen as the conversion factor  from the standard  mass unit, namely the gram and the atomic mass unit $\mathbf{u}$
\begin{equation}\label{conversiaconavo}
    1 g \, = \, N_A \, \mathbf{u}
\end{equation}
The Equation of State  (\ref{idealeqsta}) that  shortly after we  rigorously derive
from the partition function for a classical system of free identical particles,
coincides with the Equation of State of diluted gases, empirically known since long time
in the following form\footnote{It was experimentally determined  by \'{E}mile
Clapeyron in 1834}:
\begin{equation}\label{rudinosco}
    P\;V \, = \, n \; R \; T
\end{equation}
where $P$ is the pressure, $n$ denotes the \textbf{number of moles} of the gas that are present in the considered volume $V$,
and  $R$ is a universal physical constant with the following value:
\begin{equation}\label{valorediR}
    R \, = \, 8,31446261815324 \frac{J}{mol \times K}
\end{equation}
\begin{definizione}
\label{bolzmanno} Boltzmann constant, that appears in all formulae of
statistical mechanics, is  defined as
the ratio of the ideal gas constant $R$  and Avogadro number
(\ref{numeroagrado}):
\begin{equation}\label{kBdefi}
    k_B \, \equiv \, \frac{R}{N_A} \, = \, 1,380649 \times 10^{-16}
    \, erg \, K^{-1}
\end{equation}
\end{definizione}
Thanks to definition \ref{bolzmanno}, the empirical form of the equation of state
(\ref{rudinosco}) and that derived from Statistical Mechanics of free particles
do coincide since by substituting (\ref{kBdefi}) into eq.(\ref{idealeqsta})
we obtain the ratio $N/N_A$ between the number of particles and Avogadro number
which is by definition the number of moles of the considered chemical substance:
\begin{equation}\label{numolosso}
    n \, = \, \frac{N}{N_A} \, = \,  \text{\# di moli}
\end{equation}
\subsubsection{The First and Second Principles of  Thermodynamics}
\label{principiale} Classical thermodynamics is axiomatically formulated through two principles that are expressed in differential form and concern the infinitesimal changes in the quantities introduced in the previous paragraph when, by means of heat supply or subtraction, $\pm dQ$  an infinitesimal transformation of the thermodynamic system is carried out from its equilibrium state.
\paragraph{\sc The First Principle} The first principle asserts that if we call $\mathrm{d}W$
the mechanical work absorbed or given up by the system, in an infinitesimal transformation, the following relationship holds
\begin{equation}\label{principiumprimum}
    \mathrm{d}Q\,=\, \mathrm{d}U \, - \, \mathrm{d}W
\end{equation}
Typically mechanical work produces a change in the volume $V$ of the system and since pressure is force per unit area we have
  $\mathrm{d}W=P\mathrm{d}V$ so that the canonical formulation of the first principle is the following:
\begin{equation}\label{principiumprimumcan}
    \mathrm{d}Q\,=\, \mathrm{d}U \, - \, P\, \mathrm{d}V
\end{equation}
\paragraph{\sc The Second Principle} The second principle of thermodynamics can be formulated in several equivalent ways. The most concise is as follows:
\begin{equation}\label{principiumsecundum}
    \oint_C \, \frac{dQ}{T} \, =\,0
\end{equation}
which translates to saying that if we integrate the differential form
$\frac{dQ}{T}$ along a closed path in the space of state variables, that is, by performing through exchanges of work and heat a transformation that takes the system from the initial state to a final state equal to the initial one, then we obtain zero.  This implies that unlike heat differential $dQ$ the combination
$\frac{dQ}{T}$  is an exact differential namely it is the differential of a new state function that
we have already anticipated and which takes the name of entropy $S$. Hence we can write:
\begin{equation}\label{principiumsecundumcan}
    dQ=T \, dS
\end{equation}
where $S$ is  a function of state, for instance $S=S(T,V,N)$.
\par
Thus one introduces the following thermodynamic potentials, which are all functions of state
\begin{description}
  \item[1)] \textbf{Internal Energy}
  \begin{equation}\label{internalE}
    U \, = \, U(T,V,N)
  \end{equation}
\item[2)] \textbf{Helmholtz Free Energy}
\begin{equation}\label{Helmolzo}
    F \, \equiv \, U \, - \, T\,S
\end{equation}
  \item[3)] \textbf{Entalpy}
\begin{equation}\label{entalpo}
    H \, \equiv \, U \, + \, P\,V
\end{equation}
  \item[4)] \textbf{Gibbs potential}
\begin{equation}\label{ghibbo}
    G \, \equiv \, H \, - \, T\,S \, = \, U \, + \, P\, V \, - \,
    T\, S \, = \, F \, + \, PV
\end{equation}
\end{description}
The various thermodynamic potentials are all related to each other by relationships and are useful in describing thermodynamic transformations of various kinds, but their most important justification is the relationship of three of them,  with the partition function, respectively, of the three possible ensembles in the microscopic, statistical mechanic description of macroscopic thermodynamic systems.
\subsection{The Three Ensembles of Statistical Mechanics}
\label{tuttinsieme} Let us begin with the first of the three ensembles, which is perhaps the most fundamental of the three because it aims to directly explain disorder in terms of the enormous number of microscopic configurations corresponding  to the same macroscopic state.
\subsubsection{The Microcanonical Ensemble}
\label{insiememicro}Considering a number $N\gg 1$ of particles
(molecules) that are collected in a volume $V$ and have an overall energy $E$, we identify the internal energy of the system with the said energy:
\begin{equation}\label{energhia}
    U \, = \, E
\end{equation}
and the entropy of the same system with:
\begin{equation}\label{entropizzo}
    S\left(U,V\right) \, = \, k_B \, \log \left(\mathcal{N}_E\right)
\end{equation}
where $k_B$  is Boltzmann constant  and by definition
\begin{equation}\label{numerologo}
    \mathcal{N}_E \, = \, \# \text{microscopic states that have an overall energy $E$}
\end{equation}
From the first principle of thermodynamics, if we know the entropy function
$S\left(U,V\right)$ we retrieve the  temperature since:
\begin{equation}
\label{tempdaentrop}
    \left.\frac{\partial S}{\partial U}\right| _{V=cost} \, = \, \frac{1}{T}
\end{equation}
Using the same logical procedure we instead obtain:
\begin{equation}
\label{presdaentrop} T \,\left.\frac{\partial S}{\partial V}\right|
_{U=cost} \, = \, P
\end{equation}
and all the equations of  thermodynamics can be reconstructed
from the knowledge of the function $S\left(U,V\right)$  defined by means
of the indentification in eq.(\ref{entropizzo}).
\subsubsection{The Canonical Ensemble}
\label{insiememedio} In the microcanonical ensemble, the total energy $U=E$, the volume $V$  and the number of particles $N$ are kept constant.  In the canonical ensemble, on the other hand, the fixed energy condition is relaxed and only the volume $V$ and the number of particles $N$ constituting the system are kept fixed. Instead of energy, the temperature parameter $T$ is introduced.  With these elements one then constructs the so-called \textbf{canonical partition function}. In the following way.
Let $\Sigma(N,V)$  be the set of all possible microscopic states (at the level of classical mechanics we can say configurations in phase space) that can be constructed with $N\ggg 1$ particles (typically molecules) in volume $V$. Each state, i.e., each element $\sigma \in \Sigma(N,V)$ is endowed with a specific energy that we will denote $E_\sigma$. The canonical partition function is then written as the following sum:
\begin{equation}\label{canpartfun}
    Z_N\left(T,V\right)\, \equiv \, \sum_{\sigma\in\Sigma(N,V)} \,
    \exp \left[ - \,\frac{E_\sigma}{k_B \, T}\right]
\end{equation}
The connection with classical thermodynamics is made through the following identification:
\begin{equation}\label{interpretocanonico}
  F\left(T,V,N\right) \, =\,  A\left(T,V,N\right) \, \equiv \, -\, k_B \, T \, \log
    \left[Z_N\left(T,V\right)\right]
\end{equation}
where $F\left(T,V,N\right)$  is  Helmholtz free energy introduced in equation (\ref{Helmolzo}) and $A\left(T,V,N\right)$
is the name we give to the combination to its left.
The rationale for this identification becomes clear if we reason in terms of probabilities and refer to the general scheme described in section \ref{fondiprobi} in particular
recalling eq.s (\ref{distribuzionepesci},\ref{partifungo}).
The set of all states $\Sigma(N,V)$ is, in the present case, the measurable space $\Omega$ and the energy $E_\sigma$ of a state
$\sigma \in \Omega$ is  the relevant stochastic vector $\mathbf{X}$.
In this case the vector space $\mathbb{V}$ is
one-dimensional because the energy $E$ is a scalar quantity
at least in classical mechanics. Finally we can identify:
\begin{equation}\label{agniscolambdo}
    \lambda \, = \, \frac{1}{k_B T}
\end{equation}
and we see that the probability of the system being in state
$\sigma$, given by:
\begin{equation}\label{probabener}
    \mathfrak{p}(\sigma) \, = \, \frac{\exp\left[-\frac{E_\sigma}{k_B
    T}\right]}{Z_N\left(T,V\right)}
\end{equation}
coincides with that defined in the general formula (\ref{distribuzionepesci}). For the same valid reason in the general case this probability density is correctly normalized to $1$  if we sum over all states.
\par
Note that the measurable set $\Sigma(N,V)$ is typically a variety of large dimensions and therefore the variables
$\mathbf{q}$ that identify its points are many. For example, when the constituent particles of the system can be considered classical particles, $\Sigma(N,V)$ is the phase space for a system of
$N$ particles with $6^N$ dimensions. Each
configurations $\sigma$ in the phase space has  a given energy $E_\sigma$.
\par
We can now estimate the average energy of the system in the ensemble by writing:
\begin{equation}\label{Emediata}
    \langle E \rangle \, = \, \sum_{\sigma\in\Sigma(N,V)}\, E_\sigma
    \,\mathfrak{p}(\sigma) \, \equiv \, k_B \, T^2 \partial_T
    \,\log\left[ Z_N\left(T,V\right)\right]
\end{equation}
The fundamental conceptual identification is that between the thermodynamic internal energy and the average energy of the canonical ensemble calculated in equation (\ref{Emediata}):
\begin{equation}\label{agnusco}
    U \, = \, \langle E \rangle
\end{equation}
Referring again to the general formulas in section
\ref{fondiprobi} we can compare equation (\ref{Emediata})
with the general equation (\ref{variabiliestensive}). Setting
\begin{equation}\label{carisco}
    \mathcal{H}(\lambda,N) \, = \, - \, \log\left[Z(T,N)\right]
\end{equation}
and considering  relation (\ref{agniscolambdo}) according to
(\ref{variabiliestensive}) we find:
\begin{equation}\label{ruminante}
    \langle E \rangle \, = \, \frac{\partial}{\partial \lambda}\mathcal{H}(\lambda,N)
    \, = \, - \,\frac{\partial T}{\partial \lambda} \, \partial_T
    \log\left[Z(T,N)\right] \, = \, k_B \, T^2 \partial_T
    \,\log\left[ Z_N\left(T,V\right)\right]
\end{equation}
which exactly reproduces eq. (\ref{Emediata}).
\par
Let us compare eq. (\ref{interpretocanonico}) with
the general eq. (\ref{leggendoleggo}). Using the relation
(\ref{agniscolambdo}) we get:
\begin{equation}\label{parfumo}
    \mathcal{I}\left[\mathfrak{p}\right] \, = \, - \,
    \log\left[Z(T,N)\right] \, - \, \frac{1}{k_B T} \, U
\end{equation}
and multiplying both the left and the right hand side by $k_B T$
we find:
\begin{equation}\label{corbezzoli}
    U \, + \, T \left(k_B \, \mathcal{I}\left[\mathfrak{p}\right]\right )
    \, = \, A(T,V,N) \, \equiv \, - \, k_B \, T \, \log
    \left[Z_N\left(T,V\right)\right]
\end{equation}
Hence recalling  the definition of Helmholtz free energy
(\ref{Helmolzo}), we realize that the identification
(\ref{interpretocanonico}) is the correct one if Shannon
information entropy  $\mathcal{I}\left[\mathfrak{p}\right]$
is identified with thermodynamical  entropy modulo the factor
$-k_B$:
\begin{equation}\label{ramificato}
    S(T,V,N) \, = \, - \, k_B \, \mathcal{I}\left[\mathfrak{p}\right]
\end{equation}
In classical thermodynamics the identification  between the function
$A\left(T,V,N\right)$ and Helmholtz free energy
$F\left(T,V,N\right)$ follows from the fact that we  can  show that
$F$ and $A$ satisfy the same differential relation with internal energy.
\par
Let us begin with the thermodynamical definition (\ref{Helmolzo}). Utilizing both the first and second principles of thermodynamics
we get:
\begin{eqnarray}
\label{derivaF}
  dF &=& dU\, -\, dT  S \, - \, T dS \\
  \null &=& T dS -P\,dV - dT S -TdS \, =\, -P\, dV \, - \, S \,
  dT
\end{eqnarray}
from which we work out the relations:
\begin{equation}\label{pressiaentropica}
    P \, = \, - \, \left.\frac{\partial F}{\partial
    V}\right|_{T=cost} \quad ; \quad  S \, = \, - \, \left.\frac{\partial F}{\partial
    T}\right|_{V=cost}
\end{equation}
Equations(\ref{pressiaentropica}) imply that the definition
(\ref{Helmolzo}) can be rewritten  as the following
differential equation:
\begin{equation}\label{diffequoF}
    F \, = \, U \, + \, T \, \left.\frac{\partial F}{\partial
    T}\right|_{V=cost}
\end{equation}
Riconsidering now equation (\ref{Emediata}) that provides the expression for
internal energy, with obvious manipulations we can write what follows:
\begin{equation}\label{diffequoA}
   \langle E \rangle \, = \, U \, = \,A \, -  \, T \, \left.\frac{\partial A}{\partial
    T}\right|_{V=cost}
\end{equation}
which coincides with (\ref{diffequoF}) if $F=A$. Thus  the interpretation (\ref{interpretocanonico}) is fully justified within the set up
of classical thermodynamics.
\par In the previous discussion we emphasized  the deeper sense of the classical identification in relation with Information Theory.
In any case having established   identification
(\ref{interpretocanonico}),  the thermodynamic variables are all identified in the canonical ensemble as well.
Summarizing we have:
\begin{eqnarray}
\label{resumannoCan}
P &=&  \partial_V \left(k_B \, T \, \log \left[Z_N\left(T,V\right)\right]\right) \nonumber\\
S  &=& \partial_T \left(k_B \, T \, \log \left[Z_N\left(T,V\right)\right]\right) \nonumber\\
U  &=& k_B \, T^2 \partial_T
    \,\log\left[ Z_N\left(T,V\right)\right]\nonumber\\
F(T,V,N) &=& -\, k_B \, T \, \log
    \left[Z_N\left(T,V\right)\right]
\end{eqnarray}
Keeping the number $N$ of particles  or of moles fixed, the thermodynamic state state of the system is always a point in a two-dimensional variety for which we can use either the native variables $(T,V)$ which are the arguments of the partition function or any other pair of independent variables whose relations to $(T,V)$ we can derive by inverting the relations
(\ref{resumannoCan}), whenever this is analytically possible.
\subsubsection{The Grand Canonical Ensemble and the Gibbs Potential}
\label{insiemegrande} In the formulation of statistical mechanics through the canonical grand ensemble, not only is the energy of states not fixed, but neither is the total number of particles or moles of the substance. Thus the following grand partition function is written:
\begin{equation}\label{grandecane}
    \mathcal{Z}\left(T,V,\mu\right) \, = \, \sum_{N=0}^{\infty}
    \,z^N \,
    Z_N\left(T,V\right)
\end{equation}
where the variable $z=\exp\left[\frac{\mu}{k_B T}\right]$ is named
the fugacity and the symbol  $\mu$ is called the chemical potential.
The relationship between the statistical description by the canonical grand partition function and classical thermodynamics is encoded in the identification:
\begin{equation}\label{agnisco}
    \Phi(T,V,\mu) \, = \,-\, k_B \,T\,
    \log\left[\mathcal{Z}\left(T,V,\mu\right)\right]\, = \, U \, -
    \, T\, S \, - \, \mu \, N
\end{equation}
In the grand canonical description of thermodynamics as well as in the canonical description, the internal energy of the system is not an a priori fixed datum, but rather it takes an average value that we calculate with the exact analogue of the formula (\ref{Emediata})
\begin{equation}\label{Ugrancan}
    U \, = \, k_B \, T^2 \partial_T \log \left[
    \mathcal{Z}\left(T,V,\mu\right)\right]
\end{equation}
The same thing happens with the number of particles (or moles), which is not fixed but just assumes an average value calculated below:
\begin{equation}\label{Nmediano}
    \langle N \rangle \, = \, - \, k_B \, T \, \partial_\mu  \log \left[
    \mathcal{Z}\left(T,V,\mu\right)\right]
\end{equation}
Identification(\ref{agnisco}) also allows us to write the other analogs of equations (\ref{resumannoCan})
\begin{eqnarray}
\label{rastrellato}
  P &=& \partial_V \left(k_B \, T \, \log \left[\mathcal{Z}\left(T,V,\mu\right)\right]\right) \nonumber \\
  U &=& k_B \, T^2 \partial_T \log \left[
    \mathcal{Z}\left(T,V,\mu\right)\right]\nonumber \\
  S &=& \partial_T \left(k_B \, T \, \log \left[\mathcal{Z}\left(T,V,\mu\right)\right] \right)\nonumber\\
  N &=&  - \, k_B \, T \, \partial_\mu  \log \left[
    \mathcal{Z}\left(T,V,\mu\right)\right]
\end{eqnarray}
and suggests the introduction of a new thermodynamic potential, called the Gibbs potential, which has the following definition
\begin{equation}\label{gibbone}
    G(T,V,\mu) \, = \, PV \, + \, U \, - \, TS
\end{equation}
\subsection{Statistical Mechanics of Ideal Gases}
\label{msidgas} After the general exposition of the previous section, we briefly present  the fundamental example
of the ideal gas of identical particles with masses all equal and not interacting with each other.  As we recalled above
this is one of the rare cases where the partition function can be explicitly computedd in close form.
\par
Our system consists of $N$  non relativistic particles,
each with mass $\mathit{m}$, that are free to move inside a
volume $V$, which, for simplicity,  we regard as a cube  with
side $\ell = \sqrt[3]{V}$.
\par The phase space, in the Hamiltonian approach, is a space of dimension $2\times 3 \times N$ whose coordinates are the
$N$ pairs $(\mathbf{p}_i,\mathbf{q}_i)$ where
$\mathbf{p}_i=\{p_{i,x},p_{i,y},p_{i,z}\}$ is the momentum vector
and $\mathbf{q}_i=\{q_{i,x},q_{i,y},q_{i,z}\}$ is the position vector
of the $i$-th particle ($i=1,\dots, N$).
\par
The classical Hamiltonian of the system is very simple and is as follows:
\begin{equation}\label{idgasham}
    \mathfrak{H}(\boldsymbol{\mathfrak{p}})= \sum _{i=1}^N
   \frac{\mathfrak{p}_{x,i}^2+\mathfrak{p}_{y,i}^2+\mathfrak{p}_{z,i}^2}{2 \mathit{m}}
\end{equation}
In accordance with the general principles stated in previous sections, since the Hamiltonian represents the
energy of the system, introducing Planck's constant $h$ as the unit of measurement, the \textbf{canonical partition
function} is written as follows:
\begin{equation}\label{idgaspartfundef}
    Z_N(T,V) \, = \, \frac{1}{N!h^{3N}} \, \int \,
    \exp\left [- \frac{\mathfrak{H}(\boldsymbol{\mathfrak{p}})}{k_B T}\right]
    \,\mathrm{d}^{3N} \boldsymbol{\mathfrak{p}}\, \times \int \,\mathrm{d}^{3N} \mathbf{q}
\end{equation}
Since the Hamiltonian does not depend on the coordinates $\mathbf{q}$,
the last integral in $\mathrm{d}^{3N} \mathbf{q}$ is calculated immediately
and we get $V^N$.  Since the Hamiltonian is a sum of independent quadratic terms, the integral over moments is
transformed into a product of integrals and we have:
\begin{equation}\label{idgaspartfundevel1}
     Z_N(T,V) \, = \, \frac{V^N}{N!} \, \times \,\prod_{i=1}^N  \,
     \mathfrak{P}_{i,x} \, \mathfrak{P}_{i,y} \, \mathfrak{P}_{i,z}
\end{equation}
where:
\begin{equation}\label{idgaspartfundevel2}
     \mathfrak{P}_{i,x} \, = \, \int_{-\infty}^{+\infty} \frac{1}{h} \, \exp\left[ - \frac{p_{i,x}^2}{2
     \mathit{m}\,k_B\, T}\right]dp_{i,x}
\end{equation}
and similarly for $x\to y$ ed $x\to z$. Everything then reduces to the calculation of a single Gaussian integral. Through
the change of variable $t=p/\sqrt{2\mathit{m}k_B T}$ we obtain the final result:
\begin{equation}\label{idgaspartfundevel3}
     Z_N(T,V) \, = \, \frac{ V^N}{N!}
   \left(\frac{2 \pi k_B T \,
   \mathit{m}}{h^2}\right)^{3 N/2}
\end{equation}
At this point we are in a position to write the Helmholtz free energy as a function of temperature, volume and
number of particles (or number of moles). It suffices to use the general formula(\ref{interpretocanonico}) and apply it to the
case of the free gas partition function calculated in
(\ref{idgaspartfundevel3}). We get:
\begin{eqnarray}\label{IdGasHelm}
\hspace{-36pt}    F_{IG}\left(T,V,N\right) & = &-k_B T \left(N \left(\frac{3}{2} \log \left(\frac{2 \pi
    k_B
   \mathit{m}}{h^2}\right)+\frac{3 \log (T)}{2}+\log (V)\right)-\log (\Gamma
   (N+1))\right)\nonumber\\
   &\approx& \frac{1}{2} k_B N T \left(-3 \log \left(\frac{2
   \pi  k_B \mathit{m}}{h^2}\right)+2 \log (N)-3
   \log (T)-2 \log (V)-2\right)
\end{eqnarray}
where $\Gamma$ denotes the Euler Gamma function e $\Gamma (N+1)=N!$
and where the second line is obtained from the first using the Stirling approximation $\log[\Gamma[N+1]]\approx N\log[N] -N$
which is very accurate for large values of $N$  as happens in our case.
\par
Starting from equation (\ref{IdGasHelm}) and using the general definitions (\ref{resumannoCan}) and relation
(\ref{Helmolzo}) we obtain all thermodynamic state functions for ideal gases.
\begin{description}\label{tabelIG}
  \item[{\sc Pressure)}]
  \begin{equation}\label{pressIG}
    P_{IG}(T,V,N) \, = \, - \partial_V \,F_{IG} \, = \,\frac{k_B N T}{V}
  \end{equation}
  \item[{\sc Entropy)}]
\begin{equation}\label{entropIG}
 \hspace{-72pt}   S_{IG}(T,V,N) \, = \, - \partial_T \,F_{IG} \, = \,\frac{1}{2} k N \left(3 \log \left(\frac{2 \pi
    k_B \mathit{m}}{h^2}\right)-2 \log (N)+3
   \log (T)+2 \log (V)+5\right)
  \end{equation}
  \item[{\sc Internal Energy)}]
\begin{equation}\label{internIG}
    U_{IG}(T,V,N) \, = \, F_{IG}\left(T,V,N\right)\, + \, T \, S_{IG}(T,V,N)\,
    = \, \frac{3 k_B N T}{2}
  \end{equation}
\end{description}
\subsubsection{The Equation of State of Ideal Gases}
\label{Osserinfero}  Eq. (\ref{pressIG}) gives us the equation of state of ideal gases as a relation between pressure, volume and temperature at fixed number of moles $N$.
Leaving out the subscript $IG$ since it is clear that we are talking about ideal gases we have:
\begin{equation}\label{idealeqsta}
    P\;V \, = \, k_B \; N \; T
\end{equation}
\subsection{The Van Der Waals Model of a Real Gas}
\label{vandervallus}
The first and still important example of modification of the gas equation of state is due to the Dutch physicist-mathematician van der Waals  who replaced equation (\ref{idealeqsta}) with the following one
where two phenomenological parameters were introduced, $(a,b)$:
\begin{equation}\label{VDWEQstat}
     \left(P+\frac{a n^2}{V^2}\right)\, (V-b n)\,= \, n R T \quad;
\end{equation}
In (\ref{VDWEQstat}) $n$ denotes the number of moles of the chemical component of the gas, while, as before, $P$ is the pressure, $V$ geometrical volume in which the gas is enclosed. Furthermore  $R$ is a universal physical constant with the following value:
\begin{equation}\label{valorediR}
    R \, = \, 8,31446261815324 \frac{J}{mol \times K}
\end{equation}
\begin{definizione}
\label{bolzmanno} The Boltzmann constant, already utilized before and  that appears in all formulae of
statistical mechanics, is  defined as
the ratio of the ideal gas constant $R$  and Avogadro number
(\ref{numeroagrado}):
\begin{equation}\label{kBdefi}
    k_B \, \equiv \, \frac{R}{N_A} \, = \, 1,380649 \times 10^{-16}
    \, erg \, K^{-1}
\end{equation}
\end{definizione}
\begin{definizione}
\label{mole} One \textbf{mole} of a chemical substance $\mathcal{X}$ is
the quantity of  atoms or of molecules of that substance $\mathcal{X}$
necessary to form  a mass $\mathit{M}$  numerically equal in grams
to the weight  $w_\mathcal{X}$  of an atom or a  molecule of that substance
$\mathcal{X}$ expressed in atomic mass units.
\end{definizione}
Thanks to such a  definition one mole of any chemical substance
$\mathcal{X}$ always contains the same number of atoms or of
molecules that is the Avogadro number:
\begin{equation}\label{numeroagrado}
    N_A \, = \, 6,002214076 \times 10^{23}
\end{equation}
In view of the definition \ref{mole}  Avogadro number can be seen as the conversion factor  from the standard  mass unit, namely the gram and the atomic mass unit $\mathbf{u}$
\begin{equation}\label{conversiaconavo}
    1 g \, = \, N_A \, \mathbf{u}
\end{equation}
The two phenomenological parameters $a>0$ and  $b>0$ were introduced to account for two effects. The first effect
encrypted in parameter  $b$ takes into account the fact that each molecule does not have the entire geometric
volume $V$ at its disposal because the repulsive forces of the other molecules that are short-range create interdiction zones whose total volume is the greater, the larger is the number of molecules present. The second effect modeled through the introduction of the parameter  $a$  is the reduction in the effective force exerted by
molecular collisions on the walls of the container because of the attractive force, at longer radius, exerted on
each molecule by all the first ones nearby. The physical dimensions of the introduced parameters can be read
directly from the equation. Considering the number of moles dimensionless  we have:
\begin{equation}\label{dimemparam}
    \left[b\right]\, =\, \ell^3 \, ; \,  \left[a\right]\, =
   \,m  \,\ell^7\, t^{-2}
\end{equation}
where $\ell$ denotes length, $m$ denotes mass and $t$ denotes time.
\par
Skipping a lot of classical thermodynamic manipulation we arrive at the conclusion that for the van der Waals model of a real gas eq.(\ref{barramelone}) is replaced by
\begin{eqnarray}
\label{AeBVDW}
  \mathcal{A}_W(T,V) &=& \frac{a b n^3-a n^2 V+n \,R\,T\, V^2}{V^2 (V-b n)}\nonumber\\
  \mathcal{B}_W(T,V) &=& \,n \, \frac{3}{2}R \, T \, -\, \frac{a n^2}{V}
\end{eqnarray}
Functions (\ref{AeBVDW}) satisfy
the general constraint (\ref{boacostrictor}) of lagrangian  immersion. Correspondingly, the Riemannian metric that is induced on the Lagrangian variety as defined in eq.(\ref{canolagrometro}) takes, in
the van der Waals case, the following explicit form:
\begin{equation}\label{VDWmetraRiem}
   ds^2_{VDW}\, = \,  \frac{n \left(-\frac{2 T dV^2 \left(R T V^3-2 a n (V-b n)^2\right)}{V^3 (V-b
   n)^2}-3 R dT^2\right)}{2 T^2}
\end{equation}
The metric (\ref{VDWmetraRiem}) can be immediately rewritten
in terms of  zweibein $1$-forms :
\begin{equation}\label{dsinzweibainVDW}
    ds^2_{VDW}\, = \, - \, \mathbf{e}^1 \otimes \mathbf{e}^1 \, - \,
    \mathbf{e}^2 \otimes \mathbf{e}^2
\end{equation}
where:
\begin{eqnarray}\label{zweibainVDW}
  \mathbf{e}^1 &=& \frac{\sqrt{\frac{3}{2}} dT \sqrt{n R}}{T} \nonumber\\
  \mathbf{e}^2 &=& dV \sqrt{\frac{n \left(R T V^3-2 \alpha  n (V-\beta  n)^2\right)}{T V^3
   (V-\beta  n)^2}}
\end{eqnarray}
It is very easy to calculate the unique component of the spin connection $\omega^{12}$ and the curvature $2$-form
that happens to be the following:
\begin{eqnarray}
\label{Rvandervallo}
  \mathfrak{R}_{VDW} &=& R_{VDW}(T,V)\, \left(
\begin{array}{cc}
 0 & \mathbf{e}^1 \wedge \mathbf{e}^2 \\
 -\mathbf{e}^1 \wedge \mathbf{e}^2 & 0 \\
\end{array}
\right)\nonumber \\
  R_{VDW}(T,V) &=& \frac{2 a (V-b n)^2 \left(a n (V-b n)^2-R T V^3\right)}{3 R \left(R T V^3-2 a n
   (V-b n)^2\right)^2}
\end{eqnarray}
One immediately sees that if the two parameters  $a,b$ are set to zero the curvature $2$-form vanishes.
Hence the equation of state of Ideal Gases corresponds to a \textit{flat lagrangian variety}
deprived of phase transitions and critical phenomena. It follows that \textit{thermodynamic curvature},
as it was first proposed  by  Ruppeiner \cite{Ruppeiner_2010}  measures the interaction among molecules
and defines critical phenomena through its own critical surfaces and critical curves.
\par
What is essentially new compared with the equation of state for ideal gases is that the  function $\mathcal{A}_W\left(V,n,T,a,b\right)$ possesses a critical point at which both its first and second derivatives with respect to the volume cancel. The two equations $\partial_V\mathcal{A}_W\, =\,0$ and $\partial_V^2\mathcal{A}_W\, =\,0$ have a single solution for the temperature variable $\mathcal{T}$  and the volume variable $V$ such that we find a single critical point for the three thermodynamic variables
$(P,V,T)$, which is displayed in the following equation:
\begin{equation}\label{puntocriticoVDW}
V_c \, = \,  3 b n  \quad ; \quad T_c\, = \,  \frac{8 a}{27 b \, R}
\quad \Rightarrow \quad P_c \, = \, \frac{a}{27 b^2}
\end{equation}
\par
As can be seen, the critical temperature and pressure depend only on the parameters $a,b$, while the critical volume also depends on the number of moles.  If we introduce dimensionless variables
$\mathcal{T},\mathfrak{v},\mathfrak{p}$ defined as the ratios of
$T,V,P$ with respect to their critical values:
\begin{equation}\label{adimvariW}
    T \, = \, \mathcal{T} \, T_c \quad ; \quad \quad V\, =
    \,\mathfrak{v} \, V_c \quad ; \quad P \, = \, \mathfrak{p} \,
    P_c
\end{equation}
in terms of such variables van der Waals equation of state becomes universal:
\begin{equation}\label{funzionestudio}
    \mathfrak{p} \, = \, \frac{8 \mathcal{T}}{3
    \mathfrak{v}-1}-\frac{3}{\mathfrak{v}^2} \,\equiv \,
    \mathfrak{p}_W\left(\mathcal{T}, \mathfrak{v}\right)
\end{equation}
\par
\begin{figure}
\begin{center}
\includegraphics[width=12cm]{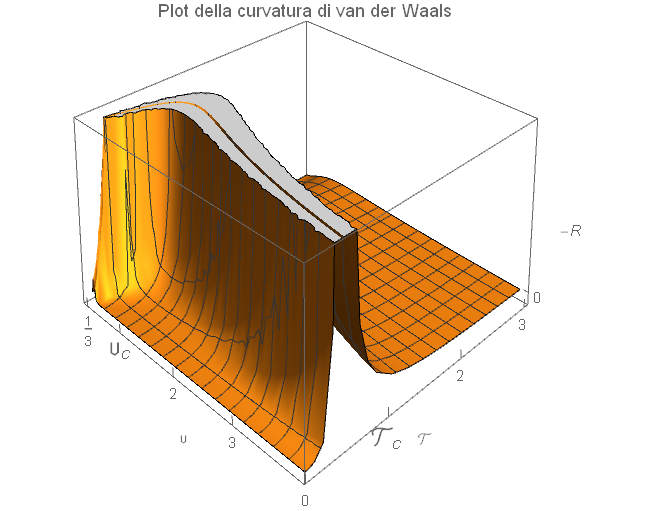}
\includegraphics[width=12cm]{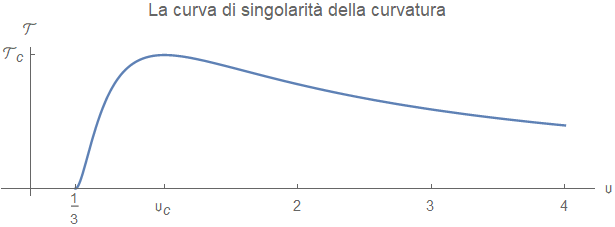}
 \caption{\label{curvoilvallo} The first image shows the two dimensional surface traced by the curvature scalar  (\ref{intrinsicRVDW})
 in the three dimensional space spanned by the coordinates $\{\mathfrak{v},\mathcal{T},-\mathcal{R}\}$.
 One easily sees the singularity line along which thermodynamic curvature
 diverges to  $-\infty$. Such a line projected onto the plane
 $\{\mathfrak{v},\mathcal{T}\}$ is displayed in the second image.}
\end{center}
\end{figure}

It is very much interesting to rewrite the function $R_{VDW}(T,V)$ in
terms of the dimensionless variables introduced in eq.
(\ref{adimvariW}). We get:
\begin{equation}\label{intrinsicRVDW}
    R_{VDW}(T,V)\, = \, \frac{1}{6 \, n\, R} \,
    \mathcal{R}(\mathcal{T},\mathfrak{v})\, \equiv \, -\frac{(1-3 \mathfrak{v})^2 \left(8 \mathcal{T} \mathfrak{v}^3-9
   \mathfrak{v}^2+6 \mathfrak{v}-1\right)}{\left(-4 \mathcal{T} \mathfrak{v}^3+9
   \mathfrak{v}^2-6 \mathfrak{v}+1\right)^2}
\end{equation}
The plot of function $\mathcal{R}(\mathcal{T},\mathfrak{v})$  is shown in fig.\ref{curvoilvallo}. The singularity line is simply  the vanishing locus in the
$\{\mathfrak{v},\mathcal{T}\}$ plane for the denominator  in
eq.(\ref{intrinsicRVDW}).

\section{The Example of the K\"ahlerian Manifold $\mathcal{M}^{[2,2]}$ with Non Trivial Paint Group}
\label{belohorizonte}
 Within the Tits Satake Universality Class (\ref{r2class}) we choose the explicit case $q=2$ in order to illustrate how the essential items in the formulation of generalized thermodynamics {\`a} la Souriau can be written in Paint Group covariant fashion and hence extended to the whole class.
 The  symmetric space $\mathcal{M}^{[2,2]}$ of our example has dimension 8, which, therefore, is also the dimension of the corresponding solvable group $\mathcal{S}_8$ and of the corresponding solvable Lie algebra $Solv_8$:
 \begin{eqnarray}\label{laborroto}
 \mathcal{M}^{[2,2]}&=&\frac{\mathrm{SO(2,4)}}{\mathrm{SO(2) \times SO(4)}} \nonumber\\
  \text{dim} \left[\mathcal{M}^{[2,2]}\right]& = & 8 \, = \, \text{dim} [\mathcal{S}_{[2,2]}] \, = \, \text{dim} [Solv_{[2,2]}]
 \end{eqnarray}
 The chosen basis of generators of the solvable Lie algebra are displayed in table \ref{baldovino} (for more details on their construction
 and normalization see \cite{pgtstheory}).
\begin{table}[htb]
  \centering
  \begin{alignat*}{7}
   T_1 & =  & \left(
\begin{array}{cccccc}
 1 & 0 & 0 & 0 & 0 & 0 \\
 0 & 0 & 0 & 0 & 0 & 0 \\
 0 & 0 & 0 & 0 & 0 & 0 \\
 0 & 0 & 0 & 0 & 0 & 0 \\
 0 & 0 & 0 & 0 & 0 & 0 \\
 0 & 0 & 0 & 0 & 0 & -1 \\
\end{array}
\right) &\quad ;  \quad & T_2& =  & \left(
\begin{array}{cccccc}
 0 & 0 & 0 & 0 & 0 & 0 \\
 0 & 1 & 0 & 0 & 0 & 0 \\
 0 & 0 & 0 & 0 & 0 & 0 \\
 0 & 0 & 0 & 0 & 0 & 0 \\
 0 & 0 & 0 & 0 & -1 & 0 \\
 0 & 0 & 0 & 0 & 0 & 0 \\
\end{array}
\right) \nonumber\\
   T_3 & =  & \left(
\begin{array}{cccccc}
 0 & \frac{1}{\sqrt{2}} & 0 & 0 & 0 & 0 \\
 0 & 0 & 0 & 0 & 0 & 0 \\
 0 & 0 & 0 & 0 & 0 & 0 \\
 0 & 0 & 0 & 0 & 0 & 0 \\
 0 & 0 & 0 & 0 & 0 & -\frac{1}{\sqrt{2}} \\
 0 & 0 & 0 & 0 & 0 & 0 \\
\end{array}
\right) &\quad ;  \quad   & T_4& =  & \left(
\begin{array}{cccccc}
 0 & 0 & 0 & 0 & \frac{1}{\sqrt{2}} & 0 \\
 0 & 0 & 0 & 0 & 0 & -\frac{1}{\sqrt{2}} \\
 0 & 0 & 0 & 0 & 0 & 0 \\
 0 & 0 & 0 & 0 & 0 & 0 \\
 0 & 0 & 0 & 0 & 0 & 0 \\
 0 & 0 & 0 & 0 & 0 & 0 \\
\end{array}
\right) \nonumber\\
  T_5 & =  & \left(
\begin{array}{cccccc}
 0 & 0 & \frac{1}{\sqrt{2}} & 0 & 0 & 0 \\
 0 & 0 & 0 & 0 & 0 & 0 \\
 0 & 0 & 0 & 0 & 0 & -\frac{1}{\sqrt{2}} \\
 0 & 0 & 0 & 0 & 0 & 0 \\
 0 & 0 & 0 & 0 & 0 & 0 \\
 0 & 0 & 0 & 0 & 0 & 0 \\
\end{array}
\right) &\quad ;  \quad  & T_6& =  & \left(
\begin{array}{cccccc}
 0 & 0 & 0 & \frac{1}{\sqrt{2}} & 0 & 0 \\
 0 & 0 & 0 & 0 & 0 & 0 \\
 0 & 0 & 0 & 0 & 0 & 0 \\
 0 & 0 & 0 & 0 & 0 & -\frac{1}{\sqrt{2}} \\
 0 & 0 & 0 & 0 & 0 & 0 \\
 0 & 0 & 0 & 0 & 0 & 0 \\
\end{array}
\right) \nonumber\\
   T_7 & =  & \left(
\begin{array}{cccccc}
 0 & 0 & 0 & 0 & 0 & 0 \\
 0 & 0 & \frac{1}{\sqrt{2}} & 0 & 0 & 0 \\
 0 & 0 & 0 & 0 & -\frac{1}{\sqrt{2}} & 0 \\
 0 & 0 & 0 & 0 & 0 & 0 \\
 0 & 0 & 0 & 0 & 0 & 0 \\
 0 & 0 & 0 & 0 & 0 & 0 \\
\end{array}
\right)&\quad ;  \quad  & T_8& =  & \left(
\begin{array}{cccccc}
 0 & 0 & 0 & 0 & 0 & 0 \\
 0 & 0 & 0 & \frac{1}{\sqrt{2}} & 0 & 0 \\
 0 & 0 & 0 & 0 & 0 & 0 \\
 0 & 0 & 0 & 0 & -\frac{1}{\sqrt{2}} & 0 \\
 0 & 0 & 0 & 0 & 0 & 0 \\
 0 & 0 & 0 & 0 & 0 & 0 \\
\end{array}
\right) \nonumber\\
 \end{alignat*}
  \caption{The generators of the solvable Lie algebra $Solv_8$}\label{baldovino}
\end{table}
Following the conventions and the theory exposed in \cite{pgtstheory,TSnaviga} a generic element of the solvable Lie algebra is parameterized as follows:
\begin{equation}\label{genelement}
  Solv_8 \, \ni \, \mathbf{X}(\boldsymbol{\Upsilon}) \, = \, \Upsilon^1 \,T_1\, + \, \Upsilon^2 \,T_2\, + \, \Upsilon^3 \,T_3\, + \,
  \Upsilon^4\,T_4\, + \, \Upsilon^{5,1}\,T_5\, + \, \Upsilon^{5,2}\,T_6\, + \,\Upsilon^{6,1}\,T_7\, + \, \Upsilon^{6,2}\,T_8
\end{equation}
 The reason for the special naming of the solvable coordinates $\boldsymbol{\Upsilon}$ is the distinction between the long roots (generators $T_{3,4}$ associated with roots $\alpha_{3,4}$) that have no multiplicity and the short ones that have multiplicity and transform in the fundamental representation of the Paint Group $\mathrm{G_{Paint}}$ (see \cite{pgtstheory} for details). For the chosen example the Paint Group is just $\mathrm{SO(2)}$ and the solvable generators $T_{5,6}$ form the doublet of painted roots $\alpha_5$, while the solvable generators $T_{7,8}$ form the  doublet  of painted roots $\alpha_6$ in the root system of $\so(2,3)\simeq \sym (4,\mathbb{R})$.
 \par
 Following the conventions and notations of  \cite{pgtstheory,TSnaviga}, the $\Sigma$ exponential map from the solvable Lie algebra to the solvable group yields the generic element of the solvable group manifold in  the  form presented in eq. (3.56) of \cite{pgtstheory}, that we repeat here for reader's convenience:
{\scriptsize
\begin{eqnarray}\label{exempli2}
  &\mathbb{L}(\boldsymbol\Upsilon)^{[2,1]} \, = \,& \nonumber\\
  & \left(
\begin{array}{cccccc}
 e^{\Upsilon _1} & \frac{e^{\Upsilon _1} \Upsilon _3}{\sqrt{2}} & \frac{1}{2}
   e^{\Upsilon _1} \left(\sqrt{2} U_1+\Upsilon _3 V_1\right) & \frac{1}{2}
   e^{\Upsilon _1} \left(\sqrt{2} U_2+\Upsilon _3 V_2\right) & -\frac{1}{8}
   e^{\Upsilon _1} \left(4 \mathbf{U}\cdot \mathbf{V}+\sqrt{2} \left(\Upsilon _3 \mathbf{V}^2-4 \Upsilon
   _4\right)\right) & -\frac{1}{4} e^{\Upsilon _1} \left(\mathbf{U}^2+2 \Upsilon _3
   \Upsilon _4\right) \\
 0 & e^{\Upsilon _2} & \frac{e^{\Upsilon _2} V_1}{\sqrt{2}} & \frac{e^{\Upsilon
   _2} V_2}{\sqrt{2}} & -\frac{1}{4} e^{\Upsilon _2} \mathbf{V}^2 & -\frac{e^{\Upsilon _2}
   \Upsilon _4}{\sqrt{2}} \\
 0 & 0 & 1 & 0 & -\frac{V_1}{\sqrt{2}} & -\frac{U_1}{\sqrt{2}} \\
 0 & 0 & 0 & 1 & -\frac{V_2}{\sqrt{2}} & -\frac{U_2}{\sqrt{2}} \\
 0 & 0 & 0 & 0 & e^{-\Upsilon _2} & -\frac{e^{-\Upsilon _2} \Upsilon _3}{\sqrt{2}}
   \\
 0 & 0 & 0 & 0 & 0 & e^{-\Upsilon _1} \\
\end{array}
\right)&\nonumber\\
\end{eqnarray}
}
\par
In eq.(\ref{exempli2}) we have used the notation $U_i = \Upsilon_{5,i}$ and  $V_i = \Upsilon_{6,i}$  (in our case $i=1,2$) which puts into evidence the existence of two Paint vectors $\mathbf{U},\mathbf{V}$ in the case $r=2$ and the Paint Group covariant structure of the solvable group element $\mathbb{L}(\boldsymbol\Upsilon)^{[2,s]}$. Indeed from eq.(\ref{exempli2}) it is immediate to deduce the general form of the matrix for any value of $q$.
\par
Starting from eq.(\ref{exempli2}) we easily calculate all the further required items necessary for our argumentation. To begin with we calculate the left-invariant $1$-form:
$$\Theta \equiv \mathbb{L}^{-1} d\mathbb{L}$$
 and then we project it onto the coset generators
in the orthogonal decomposition of the full $\mathbb{U}$ Lie algebra:
\begin{equation}\label{carnicchio}
  \so(2,4) \, = \, \underbrace{\so(2)\oplus\so(4)}_{\mathbb{H}} \oplus \mathbb{K}
\end{equation}
The list of $K$ generators is given in table \ref{clodoveo}.
\begin{table}[htb]
  \centering
  \begin{alignat*}{7}
   K_1 & =  &\left(
\begin{array}{cccccc}
 1 & 0 & 0 & 0 & 0 & 0 \\
 0 & 0 & 0 & 0 & 0 & 0 \\
 0 & 0 & 0 & 0 & 0 & 0 \\
 0 & 0 & 0 & 0 & 0 & 0 \\
 0 & 0 & 0 & 0 & 0 & 0 \\
 0 & 0 & 0 & 0 & 0 & -1 \\
\end{array}
\right) &\quad ;  \quad & K_2& =  &\left(
\begin{array}{cccccc}
 0 & 0 & 0 & 0 & 0 & 0 \\
 0 & 1 & 0 & 0 & 0 & 0 \\
 0 & 0 & 0 & 0 & 0 & 0 \\
 0 & 0 & 0 & 0 & 0 & 0 \\
 0 & 0 & 0 & 0 & -1 & 0 \\
 0 & 0 & 0 & 0 & 0 & 0 \\
\end{array}
\right) \nonumber\\
   K_3 & =  & \left(
\begin{array}{cccccc}
 0 & \frac{1}{\sqrt{2}} & 0 & 0 & 0 & 0 \\
 \frac{1}{\sqrt{2}} & 0 & 0 & 0 & 0 & 0 \\
 0 & 0 & 0 & 0 & 0 & 0 \\
 0 & 0 & 0 & 0 & 0 & 0 \\
 0 & 0 & 0 & 0 & 0 & -\frac{1}{\sqrt{2}} \\
 0 & 0 & 0 & 0 & -\frac{1}{\sqrt{2}} & 0 \\
\end{array}
\right) &\quad ;  \quad   & K_4& =  & \left(
\begin{array}{cccccc}
 0 & 0 & 0 & 0 & \frac{1}{\sqrt{2}} & 0 \\
 0 & 0 & 0 & 0 & 0 & -\frac{1}{\sqrt{2}} \\
 0 & 0 & 0 & 0 & 0 & 0 \\
 0 & 0 & 0 & 0 & 0 & 0 \\
 \frac{1}{\sqrt{2}} & 0 & 0 & 0 & 0 & 0 \\
 0 & -\frac{1}{\sqrt{2}} & 0 & 0 & 0 & 0 \\
\end{array}
\right) \nonumber\\
  K_5 & =  & \left(
\begin{array}{cccccc}
 0 & 0 & \frac{1}{\sqrt{2}} & 0 & 0 & 0 \\
 0 & 0 & 0 & 0 & 0 & 0 \\
 \frac{1}{\sqrt{2}} & 0 & 0 & 0 & 0 & -\frac{1}{\sqrt{2}} \\
 0 & 0 & 0 & 0 & 0 & 0 \\
 0 & 0 & 0 & 0 & 0 & 0 \\
 0 & 0 & -\frac{1}{\sqrt{2}} & 0 & 0 & 0 \\
\end{array}
\right)&\quad ;  \quad  & K_6& =  & \left(
\begin{array}{cccccc}
 0 & 0 & 0 & \frac{1}{\sqrt{2}} & 0 & 0 \\
 0 & 0 & 0 & 0 & 0 & 0 \\
 0 & 0 & 0 & 0 & 0 & 0 \\
 \frac{1}{\sqrt{2}} & 0 & 0 & 0 & 0 & -\frac{1}{\sqrt{2}} \\
 0 & 0 & 0 & 0 & 0 & 0 \\
 0 & 0 & 0 & -\frac{1}{\sqrt{2}} & 0 & 0 \\
\end{array}
\right)\nonumber\\
   K_7 & =  & \left(
\begin{array}{cccccc}
 0 & 0 & 0 & 0 & 0 & 0 \\
 0 & 0 & \frac{1}{\sqrt{2}} & 0 & 0 & 0 \\
 0 & \frac{1}{\sqrt{2}} & 0 & 0 & -\frac{1}{\sqrt{2}} & 0 \\
 0 & 0 & 0 & 0 & 0 & 0 \\
 0 & 0 & -\frac{1}{\sqrt{2}} & 0 & 0 & 0 \\
 0 & 0 & 0 & 0 & 0 & 0 \\
\end{array}
\right)&\quad ;  \quad  & K_8& =  & \left(
\begin{array}{cccccc}
 0 & 0 & 0 & 0 & 0 & 0 \\
 0 & 0 & 0 & \frac{1}{\sqrt{2}} & 0 & 0 \\
 0 & 0 & 0 & 0 & 0 & 0 \\
 0 & \frac{1}{\sqrt{2}} & 0 & 0 & -\frac{1}{\sqrt{2}} & 0 \\
 0 & 0 & 0 & -\frac{1}{\sqrt{2}} & 0 & 0 \\
 0 & 0 & 0 & 0 & 0 & 0 \\
\end{array}
\right) \nonumber\\
 \end{alignat*}
  \caption{The coset generators of the of $\mathrm{SO(2,4)}/\mathrm{SO(2) }\times \mathrm{SO(4)}$}\label{clodoveo}
\end{table}
Since the $K_i$ are normalized in such a way that $\mathrm{Tr}(K_i\cdot K_j)\, = \, \delta_{ij}$ we can immediately calculate the vielbein as:
\begin{equation}\label{cardigan}
  V^i \, = \, \text{Tr} \left(K_i \Theta \right) \quad ; \quad i\, = \, 1,\dots , 8
\end{equation}
The vielbein are, as they should, linear combinations, with constant coefficient of the left-invariant $1$-forms $e^i$, defined by the decomposition:
\begin{equation}\label{defiforme}
  \Theta \, = \, \sum_{i=1}^8 \, e^i \, T_i
\end{equation}
The latter have the following explicit appearance
\begin{eqnarray}
\label{mc1formeA}
  e^1 &=& \mathrm{d}\Upsilon_1  \nonumber\\
  e^2  &=& \mathrm{d}\Upsilon_2 \nonumber\\
  e^3 &=& \mathrm{d}\Upsilon_3+\Upsilon_3 (\mathrm{d}\Upsilon_1-\mathrm{d}\Upsilon_2) \nonumber\\
  e^4 &=& \frac{1}{4} \left(\Upsilon_{6,1}^2 (-\mathrm{d}\Upsilon_3)+\Upsilon_3 \Upsilon_{6,1}^2
   \mathrm{d}\Upsilon_2-2 \sqrt{2} \Upsilon_{6,1} \mathrm{d}\Upsilon_{5,1}-\Upsilon_{6,2}^2
   \mathrm{d}\Upsilon_3+\Upsilon_3 \Upsilon_{6,2}^2 \mathrm{d}\Upsilon_2-2 \sqrt{2} \Upsilon_
   {6,2} \mathrm{d}\Upsilon_{5,2}\right.\nonumber\\
   &&\left.+\left(-\Upsilon_3 \Upsilon_{6,1}^2-2 \sqrt{2} \Upsilon_{5,1} \Upsilon_{6,1}-\Upsilon_3 \Upsilon_{6,2}^2-2 \sqrt{2} \Upsilon_{5,2} \Upsilon_{6,2}+4 \Upsilon_4\right) \mathrm{d}\Upsilon_1+4 \mathrm{d}\Upsilon_4+4 \Upsilon_4
   \mathrm{d}\Upsilon_2\right)\nonumber\\
\end{eqnarray}
\begin{eqnarray}
\label{mc1formeB}
  e^{5,1} &=& \mathrm{d}\Upsilon_{5,1}+\frac{\Upsilon_{6,1} (\mathrm{d}\Upsilon_3-\Upsilon_3 \mathrm{d}\Upsilon_2
   )}{\sqrt{2}}+\left(\Upsilon_{5,1}+\frac{\Upsilon_3 \Upsilon_{6,1}}{\sqrt{2}}\right) \mathrm{d}\Upsilon_1 \nonumber\\
  e^{5,2} &=& \mathrm{d}\Upsilon_{5,2}+\frac{\Upsilon_{6,2} (\mathrm{d}\Upsilon_3-\Upsilon_3 \mathrm{d}\Upsilon_2)}{\sqrt{2}}+\left(\Upsilon_{5,2}
  +\frac{\Upsilon_3 \Upsilon_{6,2}}{\sqrt{2}}\right) \mathrm{d}\Upsilon_1 \nonumber\\
  e^{6,1} &=& \mathrm{d}\Upsilon_{6,1}+\Upsilon_{6,1} \mathrm{d}\Upsilon_2 \nonumber\\
  e^{6,2} &=& \mathrm{d}\Upsilon_{6,2}+\Upsilon_{6,2} \mathrm{d}\Upsilon_2
\end{eqnarray}
One finds the following relation between the vielbein and the left invariant $1$-forms:
\begin{equation}\label{carmelitano}
  V^i \, = \, \nu^i_A \, e^A\quad ; \quad \nu \, = \, \left(
\begin{array}{cccccccc}
 1 & 0 & 0 & 0 & 0 & 0 & 0 & 0 \\
 0 & 1 & 0 & 0 & 0 & 0 & 0 & 0 \\
 0 & 0 & \frac{1}{2} & 0 & 0 & 0 & 0 & 0 \\
 0 & 0 & 0 & \frac{1}{2} & 0 & 0 & 0 & 0 \\
 0 & 0 & 0 & 0 & \frac{1}{2} & 0 & 0 & 0 \\
 0 & 0 & 0 & 0 & 0 & \frac{1}{2} & 0 & 0 \\
 0 & 0 & 0 & 0 & 0 & 0 & \frac{1}{2} & 0 \\
 0 & 0 & 0 & 0 & 0 & 0 & 0 & \frac{1}{2} \\
\end{array}
\right)
\end{equation}
This result determines the form of the $\kappa$ matrix in this case and consequently
the expression of the $\mathrm{SO(2,4)}$  invariant metric on the symmetric space (\ref{laborroto}) in terms of the left-invariant $1$-forms. Indeed we have:
\begin{eqnarray}\label{barnabone}
  ds^2 & =& \kappa_{AB}\, \, e^A \times e^B \nonumber\\
  \kappa & \equiv & \nu^T\cdot\nu \, = \, \left(
\begin{array}{cccccccc}
 1 & 0 & 0 & 0 & 0 & 0 & 0 & 0 \\
 0 & 1 & 0 & 0 & 0 & 0 & 0 & 0 \\
 0 & 0 & \frac{1}{4} & 0 & 0 & 0 & 0 & 0 \\
 0 & 0 & 0 & \frac{1}{4} & 0 & 0 & 0 & 0 \\
 0 & 0 & 0 & 0 & \frac{1}{4} & 0 & 0 & 0 \\
 0 & 0 & 0 & 0 & 0 & \frac{1}{4} & 0 & 0 \\
 0 & 0 & 0 & 0 & 0 & 0 & \frac{1}{4} & 0 \\
 0 & 0 & 0 & 0 & 0 & 0 & 0 & \frac{1}{4} \\
\end{array}
\right)
\end{eqnarray}
to be compared with the result obtained in eq.(\ref{crisippo}) for the case of the maximally split master example $\mathrm{SL(3,\mathbb{R})/SO(3)}$.  As in all the other cases the left-invariant $1$-forms satisfy a set of  Maurer Cartan equations:
\begin{equation}\label{maurocartus}
\begin{array}{lcl}
 \mathrm{d}e^1 & = & 0 \\
 \mathrm{d}e^2 & = & 0 \\
 \mathrm{d}e^3+\frac{1}{2} (2 e^{1}\wedge e^{3}-2 e^{2}\wedge e^{3}) & = & 0 \\
 \mathrm{d}e^4+\frac{1}{2} \left(2 e^{1}\wedge e^{4}+2 e^{2}\wedge e^{4}-\sqrt{2} e^{5,1}\wedge
   e^{6,1}-\sqrt{2} e^{5,2}\wedge e^{6,2}\right) & = & 0 \\
 \mathrm{d}e^{5,1}+\frac{1}{2} \left(2 e^{1}\wedge e^{5,1}+\sqrt{2} e^{3}\wedge e^{6,1}\right) & = & 0 \\
 \mathrm{d}e^{5,2}+\frac{1}{2} \left(2 e^{1}\wedge e^{5,2}+\sqrt{2} e^{3}\wedge e^{6,2}\right) & = & 0 \\
 \mathrm{d}e^{6,1}+e^{2}\wedge e^{5,1} & = & 0 \\
 \mathrm{d}e^{6,2}+e^{2}\wedge e^{6,2} & = & 0 \\
\end{array}
\end{equation}
from which we read off the explicit value of the solvable Lie algebra structure constants $f^{\phantom{BC}A}_{BC}$ since their general form is that given in eq.s(\ref{Maurocarto}\ref{ciamblocca}).
\par
Observing eq.s (\ref{maurocartus}) we immediately see how they can be generalized to any manifold of the TS universality class introducing the Paint index that runs in the fundamental vector representation of the Paint Group
$\mathrm{G_{Paint}} \, = \, \mathrm{SO(q)}$. The Paint coariant transcription of the Maurer Cartan equation is the following:
\begin{equation}\label{maurocartus}
\begin{array}{lcl}
 \mathrm{d}e^1 & = & 0 \\
 \mathrm{d}e^2 & = & 0 \\
 \mathrm{d}e^3+\frac{1}{2} (2 e^{1}\wedge e^{3}-2 e^{2}\wedge e^{3}) & = & 0 \\
 \mathrm{d}e^4+\frac{1}{2} \left(2 e^{1}\wedge e^{4}+2 e^{2}\wedge e^{4}-\sqrt{2}
 \sum^{q}_{i=1} \, e^{5,i}\wedge e^{6,i}\right)& = & 0 \\
 \mathrm{d}e^{5,i}+\frac{1}{2} \left(2 e^{1}\wedge e^{5,i}+\sqrt{2} e^{3}\wedge e^{6,i}\right) & = & 0 \\
 \mathrm{d}e^{6,i}+e^{2}\wedge e^{5,i} & = & 0 \\
\end{array}
\end{equation}
and it has the same form as the Maurer Cartan equations on the Siegel plane, namely on the Tits Satake projection of the manifold. It suffices to restrict the Paint index $i$ to the first value $i=1$.
\subsection{The K\"ahler $2$-Form}
The reason why the symmetric spaces (\ref{r2class}) are all K\"ahler manifolds is the presence in the isotropy compact subgroup
$\mathrm{H_c }\, = \, \mathrm{SO(2) \times H^\prime}$ of the factor $\mathrm{SO(2)} \simeq \mathrm{U(1)}$ and the arrangement of the coset generator vector space $\mathbb{K}$ into a representation $(2\mid \mathbf{v})$ where $2$ is the doublet of $\mathrm{SO(2)}$ and  $\mathbf{v}$ is some irreducible representation of the other factor $ \mathrm{H^\prime}$. All coset manifolds where such situation is realized are K\"ahler manifolds, since the generator of $\mathrm{SO(2)}$ in the $\mathbb{K}$ representation of $\mathrm{H}_c$ can be identified with the complex structure and leads to the explicit expression of the closed K\"ahler $2$-form. In our specific case (\ref{laborroto}) the $\mathrm{SO(2)}$-generator in the fundamental representation of $\mathrm{SO(2,4)}$ is the following one:
\begin{equation}\label{Xci}
  X^c \, = \, \left(
\begin{array}{cccccc}
 0 & \frac{1}{2} & 0 & 0 & -\frac{1}{2} & 0 \\
 -\frac{1}{2} & 0 & 0 & 0 & 0 & \frac{1}{2} \\
 0 & 0 & 0 & 0 & 0 & 0 \\
 0 & 0 & 0 & 0 & 0 & 0 \\
 \frac{1}{2} & 0 & 0 & 0 & 0 & -\frac{1}{2} \\
 0 & -\frac{1}{2} & 0 & 0 & \frac{1}{2} & 0 \\
\end{array}
\right)
\end{equation}
and its representation on the space of $K_i$ generators and hence on the vielbein $V^i$ is the following one:
\begin{equation}\label{XcOnK}
 J^c \, =\, \left(
\begin{array}{cccccccc}
 0 & 0 & -\frac{1}{\sqrt{2}} & \frac{1}{\sqrt{2}} & 0 & 0 & 0 & 0 \\
 0 & 0 & \frac{1}{\sqrt{2}} & \frac{1}{\sqrt{2}} & 0 & 0 & 0 & 0 \\
 \frac{1}{\sqrt{2}} & -\frac{1}{\sqrt{2}} & 0 & 0 & 0 & 0 & 0 & 0 \\
 -\frac{1}{\sqrt{2}} & -\frac{1}{\sqrt{2}} & 0 & 0 & 0 & 0 & 0 & 0 \\
 0 & 0 & 0 & 0 & 0 & 0 & -1 & 0 \\
 0 & 0 & 0 & 0 & 0 & 0 & 0 & -1 \\
 0 & 0 & 0 & 0 & 1 & 0 & 0 & 0 \\
 0 & 0 & 0 & 0 & 0 & 1 & 0 & 0 \\
\end{array}
\right)
\end{equation}
The matrix $J^c$
is obtained from the adjoint action of $X^c$ on the $K_i$ coset generators, namely from:
\begin{equation}\label{aggiungoazio}
  (J^c)_{ij} \, = \, \ft 12 \text{Tr} \left( \left[X^c \, , \, K_i\right] \cdot K_j\right)
\end{equation}
and it squares to minus the identity:
\begin{equation}\label{carolonagalbani}
  J^c\cdot J^c \, = \, - \, \mathbf{1}_{8\times 8}
\end{equation}
hence it acts as a complex structure on the cotangent bundle (implying the same for the tangent bundle). Correspondingly the K\"ahler $2$-form can be written as:
\begin{equation}\label{formaggiomagro}
  \mathbf{K} \, = \, \sum_{a=1}^8 \sum_{b=1}^8 \, J^c_{ab} V^a\wedge V^b \, = \, -\frac{e^{1}\wedge e^{3}}{\sqrt{2}}+\frac{e^{1}\wedge e^{4}}{\sqrt{2}}+\frac{e^{2}\wedge
   e^{3}}{\sqrt{2}}+\frac{e^{2}\wedge e^{4}}{\sqrt{2}}-\frac{1}{2} \,\sum_{i=1}^{q} e^{5,i}\wedge
   e^{6,i}
\end{equation}
where, once again we have utilized the example under investigation to put the Paint invariance structure into evidence. In the present case the index $i$ takes only two values $i=1,2$ but the formula (\ref{formaggiomagro})
applies to all values of $q$, namely to the entire Tits Satake universality class. In particular for $q=1$, eq.(\ref{formaggiomagro}) coincides, up to an overall factor with eq.(\ref{siegelkaller}) corresponding to the Siegel plane. One  easily verifies that $\mathbf{K}$ is closed and of maximal rank
\begin{equation}\label{domenicano}
  \mathrm{d} \mathbf{K}  \, = \, 0 \quad ; \quad \mathbf{K}\wedge\mathbf{K}\wedge \mathbf{K}\wedge\mathbf{K} \, = \, \text{const} \,
  e^1 \wedge e^2 \wedge \dots \wedge e^{6,2}
\end{equation}
 just as a consequence of the Maurer Cartan equations (\ref{maurocartus}).
\par
The manifold (\ref{laborroto}) equipped with the K\"ahler $2$-form becomes a symplectic manifold $(\mathcal{M}^{[2,2]},\mathbf{K})$ and because of metric equivalence we can say that the solvable Lie group manifold $\mathcal{S}_{[2,2]}$ is a symplectic manifold
$(\mathcal{S}_{[2,2]},\mathbf{K})$.
\subsection{The Hamiltonian Vector Fields and Their Moment-Maps}
On the group manifold $\mathcal{S}_{[2,2]}$ we have both the left translations and the right ones and correspondingly we have the left-invariant vector fields $\mathbf{t}_A^{[L]}$ that generate right translations and the right invariant ones $\mathbf{t}_A^{[R]}$
that generate left translations. Both set of vector fields satisfy the solvable Lie algebra (\ref{ciamblocca}) with the structure constants defined by eq.(\ref{maurocartus}):
\begin{equation}\label{comancho}
  \left[\mathbf{t}_B^{[L]}\, , \, \mathbf{t}_C^{[L]}\right]\, = \, f_{BC}^{\phantom{BC}A} \, \mathbf{t}_A^{[L]} \quad ; \quad
  \left[\mathbf{t}_B^{[R]}\, , \, \mathbf{t}_C^{[R]}\right]\, = \, f_{BC}^{\phantom{BC}A} \, \mathbf{t}_A^{[R]}
\end{equation}
The metric (\ref{barnabone}) is invariant only with respect to the left-translations that are generated by the right-invariant vector fields, not with respect to both, since it is the metric of a coset manifold $\mathrm{U/H}$. Correspondingly only $\mathbf{t}_A^{[R]}$ are  Killing vectors for the K\"ahler metric and consequently the K\"ahler $2$-form $\mathbf{K}$ admits as symplectic Killing vector fields only the set  $\mathbf{t}_A^{[R]}$ :
\begin{eqnarray}\label{rinocefalo}
 0& = &  \ell_{\mathbf{t}_A^{[R]}}\, \mathbf{K} \, \equiv\,  i_{\mathbf{t}_A^{[R]}} \underbrace{\mathrm{d}\mathbf{K}}_{= \, 0} \, + \, \mathrm{d} \left(i_{\mathbf{t}_A^{[R]}} \mathbf{K} \right) \nonumber \\
 \null &\Downarrow& \nonumber\\
 i_{\mathbf{t}_A^{[R]}} \mathbf{K}  & = & \mathrm{d}\underbrace{\boldsymbol{\mathfrak{P}}_A(\boldsymbol{\Upsilon})}_{\text{moment map}}
\end{eqnarray}
The second of eq.s (\ref{rinocefalo}) corresponds to the definition of the moment map functions $\boldsymbol{\mathfrak{P}}_A(\boldsymbol{\Upsilon})$, such that
\begin{eqnarray}\label{momentimappo2}
  \boldsymbol{\mathfrak{P}} \quad : \quad \mathbf{k} & \longrightarrow & \boldsymbol{\mathfrak{P}}_\mathbf{k}\left(\boldsymbol{\Upsilon}\right) \in \mathbb{C}^{\infty} \left(\mathcal{S}_{[2,2]}\right)\nonumber\\
 \forall f(\boldsymbol{\Upsilon})\in \mathbb{C}^{\infty}\left(\mathcal{S}_{[2,2]}\right) \quad : \quad
\mathbf{k} f &= & \left\{\boldsymbol{\mathfrak{P}}_\mathbf{k}\, , \, f\right\} \, \equiv \, \mathbf{K} \left(\mathbf{k}\, , \, \mathbf{X}_f \right) \nonumber\\
\mathbf{X}_f & \equiv & \boldsymbol{\pi}^{\alpha\beta}(\boldsymbol{\Upsilon})\, \frac{\partial f}{\partial \Upsilon^\alpha}\, \frac{\partial}{\partial \Upsilon^\beta} \quad ; \quad \boldsymbol{\pi}^{\alpha\beta} \, \equiv \, \left(\mathbf{K}^{-1}\right)^{\alpha\beta}
\end{eqnarray}
 which are the exact analogues of eq.s (\ref{momentimappo},\ref{obbligohamil},\ref{hamilfreccia}). The difference making explicit the
 clearcut distinction advocated in section \ref{cromatillo} is that the symplectic manifold now is the very solvable Lie group manifold, rather than the total space of its tangent bundle, the moment maps are functions of the solvable coordinates $\Upsilon^\alpha$ rather than of the canonical momenta and the symplectic form is the K\"ahler $2$-form $\mathbf{K}$.
 \par
 The very important point is that the moment maps $\mu_A(\boldsymbol{\Upsilon})$ that solve the differential equation in the second line of eq.(\ref{rinocefalo}) have the closed form expression for all manifolds of the Tits Satake Universality class (\ref{r2class}), discussed in section \ref{deagostini} and presented in eq.(\ref{gelindoelapecora}).
 \par
We leave to a future publication the explicit calculation of the moment-maps and the study of the partition function that will closely follow the calculations already performed in the case of the Siegel plane. The main motivation for the present appendix was to show that all the structures analyzed in the main text for the Tits Satake projection of the entire universality class have obvious extensions to all members of the class by introducing Paint Group indices in the appropriate places.
\vspace{6pt}
\newpage

\end{document}